# Theory of the Integer and Fractional Quantum Hall Effects

Shosuke SASAKI

Center for Advanced High Magnetic Field Science, Graduate School of Science, Osaka University, 1-1 Machikaneyama, Toyonaka, Osaka 560-0043, Japan

## Preface

The Hall resistance in the classical Hall effect changes continuously with applied magnetic field. So, the discovery of a stepwise change of the Hall resistance in an ultra-thin layer of a MOSFET was a big surprise. The Hall resistance, $R_\mathrm{H}$ was found to take the following discrete values,

$$R_\mathrm{H} = R_\mathrm{K}/\nu \qquad (1.\mathrm{a})$$
$$R_\mathrm{K} = 25812.807\,4555(59)\,\Omega \qquad (1.\mathrm{b})$$

where, $R_\mathrm{K}$ is called the von Klitzing constant and $\nu$ is the filling factor that takes integer values. It was found later that $\nu$ is not only restricted to integers but also takes specific fractional numbers.

After these discoveries, a great deal of efforts has been made to elucidate the origin of the integer and fractional quantum Hall effects (IQHE and FQHE). The many electrons inside an ultra-thin layer construct a quasi-two-dimensional (quasi-2D) system because the quantum state along the direction perpendicular to the thin layer is confined to the single ground state at low temperatures. The IQHE and FQHE are observed in several materials namely Si, GaAs, graphene and so on. Thus the QHE appears independently of materials and is caused only by many electrons in a thin conducting layer.

If the Coulomb interaction between electrons is neglected, the eigen-value problem of the 2D Hamiltonian can be solved exactly and the eigen-states are described by the Landau states. Therein the energy gaps appear only in integer filling factors and then the FQHE doesn't appear. So the FQHE should be derived from the total Hamiltonian with the Coulomb interaction between electrons.

In the classical Hall device (3D system) both the applied magnetic field and the electric field due to the Hall voltage act on electrons. The magnetic force is balanced with the electric force in a steady state. Accordingly the total force becomes zero. This cancelation yields the relation between the magnetic field strength and the Hall voltage. Thereby the measurement of the Hall voltage determines the magnetic field strength as well known. That is to say the average orbit of electron is a straight line (not a circle).

In the quantum Hall system (2D system) many articles express the electron orbital by a circle. This view yields some confusion. The circular orbital is caused by missing of the electric potential gradient along the direction of Hall voltage. In this book we take the electric potential gradient into consideration.

The quasi-2D electron problem with the Coulomb interaction between electrons cannot be solved exactly. So we are obliged to use some approximations for clarifying the FQHE. Laughlin introduced a quasi-particle with fractional charge and its wave function. He obtained the wave function (Laughlin function) by using the variational-method. Jain introduced a quasi-particle called the composite fermion (CF) which is an electron capturing even number of magnetic flux quanta. The number of the flux quanta depends upon the filling factor. That is to say there are many types of the composite fermion (capturing 2 flux quanta, 4 flux quanta and so on). There has, however, been no direct evidence for the existence of these quasi particles. Some physicists might argue that the discovery of FQHE manifests itself the existence of the quasi particles. It is merely repeating the same thing. At least every CF wave function should be approximately expressed by the wave function of many electrons. The CF theory is lacking in this description.

On the other hand Tao and Thouless investigated the FQHE using normal electrons without any quasi-particle. However they were not able to identify the filling factors that give stable states. We extended the Tao-Thouless theory and have succeeded in finding the most uniform electron configuration in the Landau orbitals. The configuration is uniquely determined at any filling factor so as to minimize the expectation value of the total Hamiltonian.

The Coulomb interaction acts between two electrons, and produces the quantum transitions. That is to say, an electron pair in Landau orbitals transfers to the other empty orbitals. Therein the Coulomb interaction depends upon only the relative distance. Accordingly the two electrons move in a pair so as to satisfy the momentum conservation law along the current direction. (Note: the conservation law doesn't hold in the directions perpendicular to the current, because other potentials exist.) The number of the allowed transitions varies abruptly with changing the filling factor because of Pauli's exclusion principle. The number takes a maximum at specific filling factors $v_0$ such as, $v_0 = 2j/(2j+1)$, $1/(2j+1)$, $(j+1)/(2j+1)$, $j/(2j+1)$ etc. When the filling factor $v$ deviates from $v_0$, the transition number decreases discontinuously. So the transition number takes a maximum at $v_0$ with a gap. Because the pairing energy is negative, the pairing energy has a lowest value with accompanying an energy gap at each specific filling factor $v_0$. Accordingly the quasi-2D electron system is confined to these filling factors. The discontinuous form of the pairing energy versus filling factor produces the Hall plateaus which are observed on the Hall resistance curve. Then the Hall resistance becomes equal to $R_K/v_0$ at the specific fractional numbers $v_0$. The difference between the experimental value and $R_K/v_0$ is extremely small. For example the relative uncertainty is $\pm 3.3 \times 10^{-8}$ at the filling factor 1/3.

This accuracy at the specific filling factors will be clarified in this book. 1) Thousands of the Landau wave functions are overlapped with each other. 2) The number of the Coulomb transitions decreases discontinuously by an infinitesimal deviation from $v_0$ as clarified in chapter 8 with taking account of the momentum dependence of the pair energy. Thereby the Hall resistance is precisely confined to $R_K/v_0$ at the specific fractional numbers $v_0$.

In a high magnetic field, all the electron spins are directed antiparallel to the applied magnetic field. At lower magnetic fields, the difference between the numbers of up and down spins depends upon the applied magnetic field strength. Then the wide plateaus and small shoulders appear on the experimental polarization curves versus magnetic field strength.

The most uniform electron configuration in the Landau orbitals has spin degeneracy ie., many different spin arrangements are possible for a given configuration of electrons. Therefore the many-electron states with different spin arrangements have the same expectation value of the Coulomb interaction. The off-diagonal parts of the Coulomb interaction produce quantum transitions between these degenerate ground states. We have succeeded to solve exactly the eigen-value problem of the partial Hamiltonian which is composed of the strongest and second strongest interactions among the degenerate ground states. Then the eigen-solutions yield the wide plateaus on the polarization curve. Furthermore the Peierls instability exists in this system. The spin Peierls effect produces the small shoulders. The theoretical results are in good agreement with the experimental data.

Thus the IQHE and FQHE are explained by a standard treatment of interacting quasi-2D electron gas without assuming any quasi-particle in this book.

The author expresses his heartfelt appreciation for encouragement of Professor Koichi Katsumata, Professor Hidenobu Hori, Professor Yasuyuki Kitano, Professor Masayuki Hagiwara and Professor Takeji Kebukawa. Especially Professor Katsumata has given me important suggestions for improving my description in this book.

The friendly co-operation of the Editorial Office of Nova Science Publishers is gratefully acknowledged.

Also I wish to express appreciate for support of my wife Makiko Sasaki and my sons Takashi Sasaki and Takuma Sasaki. I cannot complete this book without their support.

# Contents

# 1. Formulation of the problem

  The integer quantum Hall effect (IQHE) was discovered by K. von Klitzing in the inversion layer of a metal-oxide-semiconductor field-effect transistor (MOSFET) [1-3]. The quantum Hall device is shown schematically in Fig.1.1. The length of the device is 400 μm, the width 50 μm and the thickness of the inversion layer is about $10^{-2}$ μm. A magnetic field is applied perpendicularly to the surface of the device and the electric current flows between the source and drain. The electron density was controlled by adjusting the gate voltage. The Hall voltage and the current were measured at low temperatures. Then the Hall resistance $R_H$ is obtained as [Hall voltage]/[current]. K. von Klitzing discovered a plateau in the Hall resistance curve for a wide range of gate voltage. The experimental value of $R_H$ is $6453.17 \pm 0.02 \Omega$ in the plateau. The value is close to $h/(4e^2)$, where $h$ is Planck's constant and ($-e$) is the charge of electron. Thus K. von Klitzing clarified that the quantum Hall effect gives an accurate method for determining the fine-structure constant.

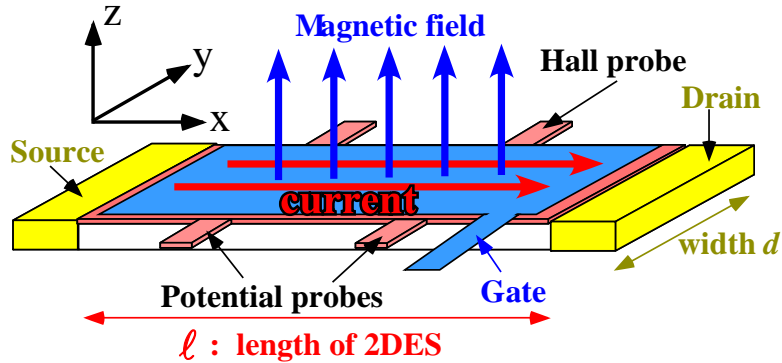

Fig.1.1 Quantum Hall Device

  Thereafter many plateaus in the Hall resistance curve have been observed at $2\pi\hbar/(\nu \cdot e^2)$, where $\nu$ is an integer and $\hbar = h/(2\pi)$. The Hall resistance for $\nu = 1$, $R_K = 2\pi\hbar/e^2 = 25812.8074555(59)\Omega$ is called the von Klitzing constant and is used for the international resistance standard. The phenomena are named the integer quantum Hall effects (IQHE).
  Quantum Hall devices have been improved so as to have higher electron mobility. D. C. Tsui, H. L. Stormer *et al*. [4-9] found many plateaus in the Hall resistance curve at $2\pi\hbar/(\nu \cdot e^2)$ with the values of $\nu = 4/5, 2/3, 3/5, 4/7, 4/9, 3/7, 2/5, 1/3, \cdots$ . This phenomenon is called the fractional quantum Hall effect (FQHE). The experimental values are drawn by the black curves in Fig.1.2 (reference [9]: H. L. Stormer, *Nobel lecture*). The upper panel of Fig.1.2 shows the Hall resistance $R_H$ versus magnetic field strength and the lower panel indicates the diagonal resistance $R$ which is the usual electric resistance between two potential probes in Fig.1.1.

Fig.1.2 Hall resistance $R_H$ and diagonal resistance $R$

The IQHE and FQHE have been observed in quantum Hall devices composed of several materials Si, GaAs and so on, at low temperatures 0.035 K, 0.4 K, 1.5 K etc. Furthermore it is surprising that the quantum Hall effect has been observed in graphene at a room temperature [10]. Thus the QHE is a phenomenon independent of materials.

Any quantum Hall device has an ultra-thin layer where electrons flow. Because the thickness is ultra small, only one ground state is effective along the direction perpendicular to the thin layer at low temperatures and all the other excited states have negligibly small probability. Accordingly the electron system may be approximated to a two-dimensional (2D) one. Graphene, a sheet of bonded carbon atoms, provides us with an ideal 2D electron system. Thus we can take a 2D electron system as a working model for the calculation of the quantum Hall effects.

As shown by the dashed lines in Fig.1.2, the diagonal resistance is almost zero on the wide plateaus of the Hall resistance namely at $\nu = 2, 1, 2/3, 2/5$ and $1/3$. The direction of Hall voltage is perpendicular to the current direction, and so the Hall resistance yields no ohmic heating. On the other hand the diagonal resistance yields ohmic heating. Because the diagonal resistance is almost zero at the Hall resistance confinements, the QHE is a pure quantum process without thermal excitations, electron scatterings and so on.

In the typical experiments of IQHE the Hall voltage is larger than about $10^{-4}$ Volts and the diagonal (potential) voltage is less than $10^{-11}$ Volts. Also in the FQHE the Hall voltage is extremely large compared with the diagonal (potential) voltage at the Hall resistance confinement. So we cannot ignore the gradient of the electric potential along the y-direction of Fig.1.1. Consequently the total Hamiltonian for many electrons should be composed of three kinds of interactions namely the strong magnetic field interaction, the Coulomb interactions between electrons, and the electric field produced by the Hall voltage. We take the three kinds of interactions into consideration in this book.

Many theories ignore the electric field along the Hall voltage. Then the orbital of electron is expressed as a circle. This picture leads some misunderstandings in the investigation of the FQHE.

(Note: <u>3D Hall system</u>: The magnetic and electric fields work on the electrons in a 3D Hall system. Therein the electric force due to the Hall voltage cancels the force by the magnetic field. So the electron moves on a straight line along the x-direction. The canceling condition gives the relation between the Hall voltage and the magnetic field strength. The relation makes it possible to determine the magnetic field strength by measuring the Hall voltage. Thus the electric force plays an important role in 3D Hall phenomenon.)

In this chapter we study the basic formulation of the quantum Hall system where we take account of the electric field along the Hall voltage and ignore the Coulomb interaction between electrons. Then the eigen energy problem of electron can be exactly solved and the wave function is expressed by the Landau solution. The wave function is the product of three wave functions of the x-, y- and z-directions. The wave function along the x-direction (current direction) is the plain wave form. That is to say the ballistic movement of electron appears in the current direction.

**1.1 Potential in a quantum Hall device**

Let us take the coordinate axes as in Fig. 1.1, in which the current flows along the x-axis, the Hall voltage emerges along the y-axis, and the magnetic field is applied along the z-axis.

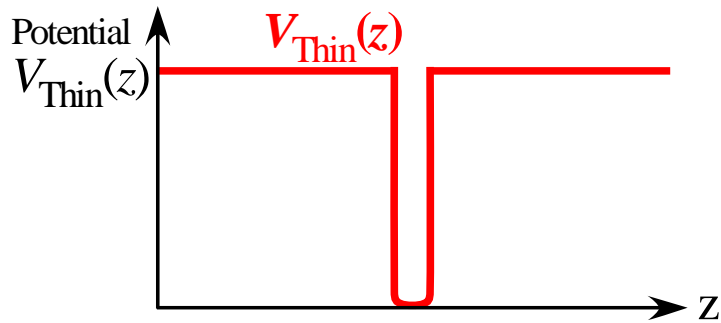

Fig.1.3 Potential of z-direction (Potential width is very narrow)

Any quantum Hall device has an ultra-thin conducting layer. Accordingly the potential, $V_{Thin}(z)$, along the z-axis has a very high barrier outside the thin layer as in Fig.1.3. This potential yields a very large excitation-energy from the ground state in the z-direction. The quantum Hall effects are observed in such a low temperature that the probability of the ground state is almost 100% for the z-direction.

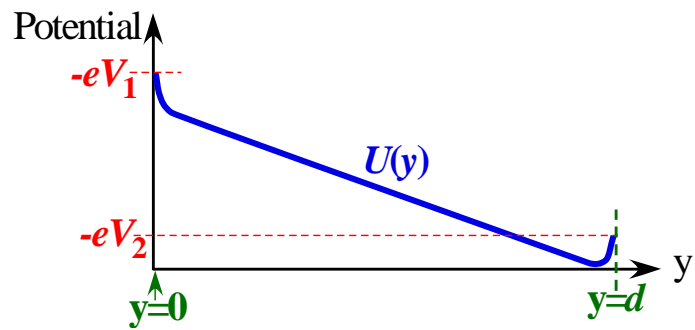

Fig.1.4 Potential $U(y)$

In Fig.1.4 we draw schematically the potential, $U(y)$, along the y-axis. Here the potential curve turns steeply upward near the both edges so that the electrons are confined to the conducting region. The voltage is denoted as $V_1$ and $V_2$ at $y=0$ and $y=d$, respectively. The difference, $V_2 - V_1$, is determined by measuring the voltage between the Hall and potential probes. At the Hall resistance confinement the value $V_2 - V_1$ is extremely larger than the potential (diagonal) voltage between potential probes. Therefore we examine the nonzero case of $V_2 - V_1$ in the investigation of QHE. The shape of the potential $U(y)$ is studied in more details in chapter 12.

We first study the wave function of single electron. In the Landau gauge the vector potential, $\mathbf{A}$, has the components as

$$\mathbf{A} = (-yB, 0, 0), \tag{1.1}$$

where $B$ is the strength of the magnetic field. This vector potential certainly satisfies the following fundamental equations:

$$\mathrm{rot}\,\mathbf{A} = \left(\frac{\partial A_z}{\partial y} - \frac{\partial A_y}{\partial z}, \frac{\partial A_x}{\partial z} - \frac{\partial A_z}{\partial x}, \frac{\partial A_y}{\partial x} - \frac{\partial A_x}{\partial y}\right) = (0, 0, B)$$

$$\mathrm{div}\,\mathbf{A} = \frac{\partial A_x}{\partial x} + \frac{\partial A_y}{\partial y} + \frac{\partial A_z}{\partial z} = 0$$

The Hamiltonian, $H_0$, of a single electron is obtained as follows:

$$H_0 = \frac{(\mathbf{p} + e\mathbf{A})^2}{2m^*} + U(y) + V_{\mathrm{Thin}}(z), \tag{1.2}$$

where $m^*$ is an effective mass of electron and $\mathbf{p} = (p_x, p_y, p_z)$ is the electron momentum. The effective mass $m^*$ differs from material to material [11] and the value in GaAs is about 0.067 times the free electron mass.

## 1.2 Single-electron states

If the Coulomb interaction between electrons is neglected, the eigen-energy problem for the electron system is exactly solved under the uniform magnetic field. The eigen-equation is given by,

$$H_0 \psi = E \psi. \tag{1.3a}$$

From Eqs.(1.1), (1.2) and (1.3a) we obtain the eigen-value problem as,

$$\left[\frac{1}{2m^*}\left\{\left(-i\hbar\frac{\partial}{\partial x}-eBy\right)^2-\hbar^2\frac{\partial^2}{\partial y^2}-\hbar^2\frac{\partial^2}{\partial z^2}\right\}+U(y)+V_{\text{Thin}}(z)\right]\psi=E\psi \qquad (1.3b)$$

This equation has no potential term depending on x and so the eigen-state along the x-axis is described by a plane wave. Also the term containing z is separated and so the wave function $\psi$ is expressed as,

$$\psi(x,y,z)=\sqrt{\frac{1}{\ell}}\exp(ikx)\varphi(y)\phi(z), \qquad (1.4)$$

where $\ell$ is the length of the device along the x-axis. Therein the wave number $k$ satisfies the periodic boundary condition;

$$k=(2\pi/\ell)\times J, \quad J:\text{integer} \qquad (1.5a)$$

Eq.(1.5a) is rewritten in terms of the momentum $p$ as

$$p=(2\pi\hbar/\ell)\times J. \qquad (1.5b)$$

Hereafter we use the symbol $p$ for the x-component of the electron momentum throughout in this book (namely we abbreviate $p_x$ as $p$). The wave function $\phi$ along the z-axis satisfies the following equation,

$$\left[-\frac{\hbar^2}{2m^*}\frac{\partial^2}{\partial z^2}+V_{\text{Thin}}(z)\right]\phi(z)=\lambda\phi(z) \qquad (1.6)$$

where $\lambda$ is the eigen-energy of the wave function $\phi$.

The excitation energy along the z-axis is very large because the potential $V_{\text{Thin}}(z)$ is very narrow. Consequently, the probability of finding the excited states in the z-direction is extremely small at a low temperature. In the case of graphene, the probability is expected to be still small even at room temperature. Therefore, we may take only the ground state for the z-direction in this book. That is to say, $\phi(z)$ indicates the ground state and $\lambda$ is the ground state energy. Substituting Eqs.(1.6) and (1.4) into Eq.(1.3b) we have,

$$\left[\frac{(k\hbar-eBy)^2}{2m^*}-\frac{\hbar^2}{2m^*}\frac{\partial^2}{\partial y^2}+U(y)\right]\sqrt{\frac{1}{\ell}}\exp(ikx)\varphi(y)=(E-\lambda)\sqrt{\frac{1}{\ell}}\exp(ikx)\varphi(y) \qquad (1.7)$$

We define an effective potential, $G(y)$, as,

$$G(y)=\frac{(k\hbar-eBy)^2}{2m^*}+U(y) \qquad (1.8)$$

Then, we get the eigen-equation for the y-direction as

$$\left[-\frac{\hbar^2}{2m^*}\frac{\partial^2}{\partial y^2}+G(y)\right]\varphi(y)=(E-\lambda)\varphi(y) \tag{1.9}$$

We will examine the shape of $G(y)$ below. When the magnetic field is strong, the first term of the right-hand-side in Eq.(1.8) dominates. Then the effective potential $G(y)$ takes a minimum at $y=\alpha_J$ where $\alpha_J$ is given by

$$\alpha_J = k\hbar/(eB) = p/(eB) = [2\pi\hbar/(\ell eB)]J. \tag{1.10}$$

Thus the quantity $\alpha_J$ of the y-direction is related to the momentum $p$ along the x-direction as in Eq.(1.10). As will be clarified below in Eq.(1.17), $\alpha_J$ is the centre position of the wave function along the y-direction. The spreading region of the wave function along the y-axis is about 10.5 nm at the magnetic field strength of 6 T as will be estimated below in Eq.(1.19). Because this distribution is very narrow in comparison with the device width, the potential $U(y)$ may be approximated by $U(\alpha_J)$.

(Note: A more precise treatment of this term)
Because the wave function along the y-axis has a very narrow width (about 10.5 nm), the y dependence of $U(y)$ can be approximated with the linear function in the neighborhood of $y \approx \alpha_J$ as follows;

$$U(y) \approx U(\alpha_J) + \beta(y-\alpha_J)$$

where $\beta$ represents the derivative at $y=\alpha_J$. Then we obtain

$$G(y) \approx (e^2 B^2/(2m^*))(y-\alpha_J)^2 + U(\alpha_J) + \beta(y-\alpha_J)$$

(Note: we can take account of the quadratic y-dependence in $U(y)$ near $y \approx \alpha_J$ by applying the same procedure mentioned above.)

The function $G(y)$ takes a minimum at $\alpha'_J$ which deviates by a small value $-b$ from the original value $\alpha_J$ as follows:

$$\alpha'_J = \alpha_J - b \qquad b = \beta m^*/(e^2 B^2)$$

$$G(y) \approx (e^2 B^2/(2m^*))(y-\alpha'_J)^2 + U(\alpha_J) - \tfrac{1}{2}\beta b$$

Thus the minimum position of the effective potential is $y=\alpha'_J$ which deviates by $-b$ from $\alpha_J$. The function shape is the same as for $b=0$ because the deviation $-b$ and the residual term $-\frac{1}{2}\beta b$ are constant. Therefore the eigenvalue-problem in the precise potential is solved by the same procedure as in the approximate form mentioned below.

(End of the Note)

For simplicity we use the value $\alpha_J$. Then $G(y)$ is approximated as

$$G(y) \approx \frac{1}{2} m^* \omega^2 (y - \alpha_J)^2 + U(\alpha_J), \tag{1.11}$$

where

$$\omega = \frac{eB}{m^*} \tag{1.12}$$

(Note: We can easily obtain the more precise results by replacing $\alpha_J$ with $\alpha'_J$ and also by adding the residual constant energy in all the chapters of this book.)

Substitution of Eq.(1.11) into Eq.(1.9) yields the eigen-value problem in the y-direction as,

$$\left[ -\frac{\hbar^2}{2m^*} \frac{\partial^2}{\partial y^2} + \frac{1}{2} m^* \omega^2 (y - \alpha_J)^2 \right] \varphi(y) = (E - \lambda - U(\alpha_J)) \varphi(y). \tag{1.13}$$

Equation (1.13) is rewritten as,

$$\left[ -\frac{\hbar^2}{2m^*} \frac{\partial^2}{\partial y^2} + \frac{1}{2} m^* \omega^2 (y - \alpha_J)^2 \right] \varphi(y) = \varepsilon \varphi(y), \tag{1.14a}$$

where

$$\varepsilon = (E - \lambda - U(\alpha_J)) \tag{1.14b}$$

This eigen-value problem has the same form as that of a harmonic oscillator. The eigen-state, $\varphi_{L,J}(y)$, and the eigen-value, $\varepsilon_L$ are given respectively, by,

$$\varphi_{L,J}(y) = u_L H_L\left( \sqrt{\frac{m^* \omega}{\hbar}} (y - \alpha_J) \right) \cdot \exp\left( -\frac{m^* \omega}{2\hbar} (y - \alpha_J)^2 \right) \tag{1.15}$$

$$\varepsilon_L = \hbar \omega \left( L + \frac{1}{2} \right) \quad (L = 0,1,2,3,\cdots) \tag{1.16}$$

where $H_L$ is the Hermite polynomial of $L$-th degree and $u_L$ is the normalization constant. We call $L$ the Landau level number. Substitution of Eq.(1.15) into Eq.(1.4) gives the single-electron wave function $\psi_{L,J}$ as follows,

$$\psi_{L,J}(x,y,z) = \sqrt{\frac{1}{\ell}} \exp(ikx) u_L H_L\left(\sqrt{\frac{m^*\omega}{\hbar}}(y-\alpha_J)\right) \exp\left(-\frac{m^*\omega}{2\hbar}(y-\alpha_J)^2\right)\phi(z) \quad (1.17)$$

The eigenenergy $E$ is derived from Eqs. (1.14b) and (1.16) as

$$E_{L,J} = \lambda + U(\alpha_J) + \hbar\omega\left(L+\frac{1}{2}\right) \qquad (L=0,1,2,3,\cdots) \quad (1.18)$$

The wave function, Eq.(1.17), has the same form as that of the Landau state except for the z-direction. The eigenenergy $E_{L,J}$ is a sum of the three terms, namely, the energy $\lambda$ in the z-direction, the potential energy $U(\alpha_J)$ in the y-direction and the Landau energy with the level number $L$. Eq.(1.17) expresses that $\alpha_J$ is the centre position of the wave function along the y-direction.

The distribution of the wave function is estimated here: Because the Hermite polynomial $H_L$ for $L=0$ is equal to 1, the probability density $\psi^*_{L,J}\psi_{L,J}$ is proportional to the following Gaussian form in the y-direction as

$$\exp\left(-\frac{m^*\omega}{2\hbar}(y-\alpha_J)^2\right)\exp\left(-\frac{m^*\omega}{2\hbar}(y-\alpha_J)^2\right) = \exp\left(-\frac{m^*\omega}{\hbar}(y-\alpha_J)^2\right)$$

When the magnetic field strength is 6 T, the width of the Gaussian, $\Delta y$ in the y-direction becomes

$$\Delta y \approx \sqrt{\frac{\hbar}{m^*\omega}} = \sqrt{\frac{\hbar}{eB}} \approx 10.5\,\text{nm} \quad (1.19)$$

Therefore this value $10.5\,\text{nm}$ indicates the spreading width $\Delta y$ of the wave function.

Next we calculate the number of states for a given $L$. We note that the value of $\alpha_J$ satisfies $0 \leq \alpha_J \leq d$ as in Fig.1.5. Then, using Eq.(1.10) we have,

$$0 \leq J \leq \frac{eB\ell}{2\pi\hbar}d. \quad (1.20a)$$

So, the number of states for a given $L$ is,

$$\frac{eB\ell}{2\pi\hbar}d. \quad (1.20b)$$

In Eq.(1.17) the Hermite polynomial $H_L$ with $L=0$ is equal to 1 and $u_L$ is a constant. So, the wave function is a product of the plane wave function along the x-axis and the Gaussian function along the y-axis. Figure 1.5 schematically expresses the wave function-shape by separate lines in order to be visible in the figure, although the robes of the wave functions with $J-1$, $J$ and $J+1$ actually overlap with each other. These lines are equally spaced with a distance given by $[2\pi\hbar/(\ell eB)]$ due to Eq.(1.10). The overlap of the wave functions will be studied in details in chapter 8.

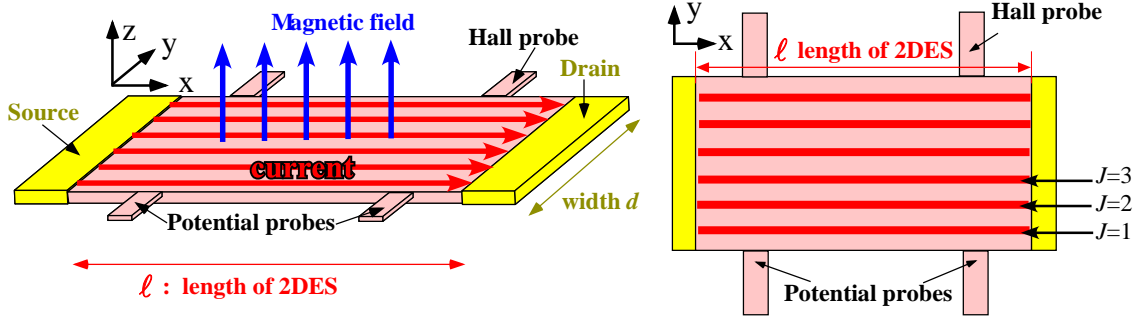

Fig.1.5 Schematic figure of Landau states

## 1.3 Energy differences between various states

The single electron states have each eigen-energies respectively. We estimate the rough values of the energy differences as follows:

### 1.3.1 Energy difference between Landau levels

The energy difference between the neighbouring two Landau levels is equal to $(\hbar eB/m^*)$ which is derived from Eq.(1.18) using Eq.(1.12).

$$\Gamma_{\text{Landau}} = \hbar\omega = \hbar eB/m^* \tag{1.21}$$

As already mentioned above, the effective mass $m^*$ for GaAs is 0.067 times the electron mass $m$ in vacuum. We roughly estimate the energy difference $\Gamma_{\text{Landau}}$ for $B=6[\text{T}]$ as

$$\Gamma_{\text{Landau}}/k_B = \hbar eB/(m^* k_B) \approx 120.5[\text{K}] \quad \text{for} \quad m^* = 0.067\,m, \ B=6[\text{T}] \tag{1.22}$$

where $k_B$ is the Boltzmann constant.

### 1.3.2 Energy difference between up and down spins

The Zeeman energy, $E_Z$, is given by

$$E_Z = \sum_{i=1,2,3\cdots} \mu_B g^* B \tfrac{1}{2}\sigma_i^z = \sum_{i=1,2,3\cdots} g^*\big(\hbar eB/(2m)\big)\tfrac{1}{2}\sigma_i^z \tag{1.23}$$

where $g^*$ is the effective g-factor, $(1/2)\sigma^z$ is the z-component of electron spin operator ($\sigma^z$: Pauli matrix) and $\mu_B$ is the Bohr magneton. The effective g-factor for GaAs is about 0.22 times the g-factor of electron in vacuum, namely, $g^* \approx 0.44$. The energy difference between the two Zeeman levels is equal to $g^*(\hbar eB/(2m))$ as

$$\Gamma_{\text{Spin}}/k_B = g^*(\hbar eB/(2mk_B)) \approx 1.8[\text{K}] \qquad \text{for} \quad g^* = 0.44, \quad B = 6[\text{T}] \quad (1.24)$$

Comparison of Eq.(1.22) with Eq.(1.24) yields that the energy difference between $L$ and $L+1$ Landau levels is about 67 times the Zeeman split energy for GaAs.

### 1.3.3 Energy by Coulomb interaction between electrons

The expectation value of the Coulomb interaction is named the classical Coulomb energy. The electron configuration in the Landau orbitals is uniquely determined at any filling factor so as to have the minimum classical Coulomb energy. This property is proven in chapter 3. Accordingly the ground state with the electron configuration has the lowest value in the Landau energy, the lowest value in Zeeman energy and the lowest value in the classical Coulomb energy. The Coulomb interaction acts on any electron pair and the strength is dependent on the relative distance only. Also there is no potential along the x-direction as in Eqs.(1.1) and (1.2). So the two electrons satisfy the momentum conservation law along the current direction (x-direction). The residual Coulomb interaction yields quantum transitions of electron pairs. We investigate the transition number versus filling factor in details in chapters 4-5. The number of allowed transitions takes local maximum at specific filling factors $\nu_0$ and the number decreases discontinuously by an infinitesimal deviation from $\nu_0$. This discontinuous change produces the energy gap via the Coulomb interaction which is studied in chapters 4-8. The gap value depends on quantum Hall devices. The experimental value $\Gamma_{\text{gap}}$ is discussed in chapter 5. The typical value of the energy gap is

$$\Gamma_{\text{gap}}/k_B \approx 0.03 \sim 3[\text{K}] \qquad (1.25)$$

As in Eqs.(1.24) and (1.25) the energy gap in the FQHE has a magnitude of the same order as in the Zeeman splitting energy. Therefore the spin polarizations of FQH states depend on the magnetic field strength in a quasi-2D many-electron system. The dependences are very complicated which are examined in chapter 9.

### 1.3.4 Energy levels of single electron

We show the energy levels of the single electron system in Fig.1.6 where the Landau energy levels are expressed by left lines and the Zeeman splitting is drawn by the arrow pairs. Then all the energy levels are expressed as in right lines.

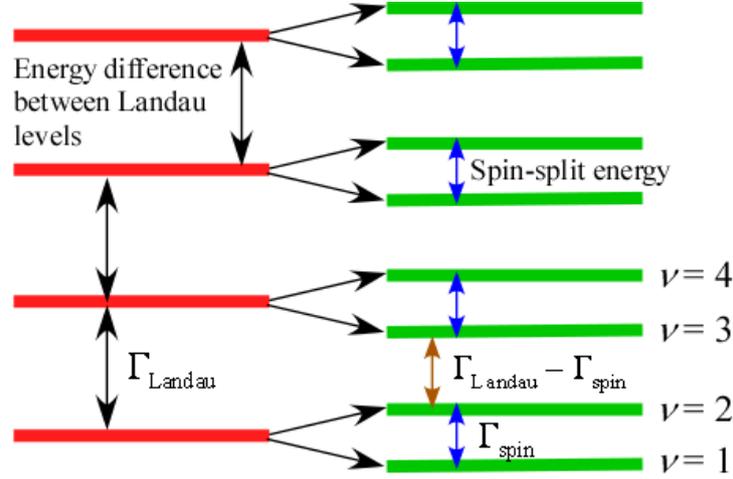

Fig.1.6 Single electron energy levels including spin separations

Next we define the filling factor $\nu$ as,

$\nu =$ [total number of electrons] / [total number of Landau states with $L=0$]   (1.23)

Some examples are given below.
(Case 1) When each of the $J$ Landau states with $L=0$ is occupied by an electron with down spin (the spin pointing opposite to $\boldsymbol{B}$), $\nu=1$.
(Case 2) When each of the $J$ Landau states with $L=0$ is occupied by two electrons with up and down spins, $\nu=2$.
(Case 3) Similarly, when each of the $J$ Landau states with $L=0$ is occupied by two electrons with up and down spins and each of the $J$ Landau states with $L=1$ is occupied by an electron with down spin, $\nu=3$.

The states with integer filling factor $\nu = 1, 3, 5 \cdots$ have the excitation energy $\Gamma_{spin}$ and ones with $\nu = 2, 4, 6 \cdots$ have the excitation energy $(\Gamma_{Landau} - \Gamma_{spin})$ as seen in Fig.1.6. These excitation energies are very large in comparison with the thermal excitation energy under a temperature applied in the QHE experiments.  Therefore the states with integer filling factors are actually realized and are responsible for the integer quantum Hall effect.

If the filling factor $v$ is a fractional number in the region of $2n < v < 2n+1$, each of the Landau states with $L \leq n-1$ are occupied by two electrons with up and down spins and Landau states of $L = n$ are partially filled with electrons having down spins. (It is noteworthy that $L$ is counted from 0.) In this case both empty and filled states have the same Landau energy. So the lowest energy states are degenerate in the absence of the Coulomb interactions between electrons. Similarly the ground states in the case of $2n+1 < v < 2n+2$ are degenerate. Therefore we should take the Coulomb interactions between electrons into consideration for investigating the fractional quantum Hall effect.

The Coulomb interaction yields repulsive forces between electrons, and so the electrons are distributed as uniformly as possible among many Landau states. Thereby only one electron-configuration has a minimum expectation-value of Coulomb interaction at a given $v$. As an example, a $v = 1/3$ state is obtained by repeating the unit-configuration (empty, filled, empty) as in Fig.1.7a. A $v = 2/3$ state is obtained by repeating the unit-configuration (filled, empty, filled) as in Fig.1.7b.

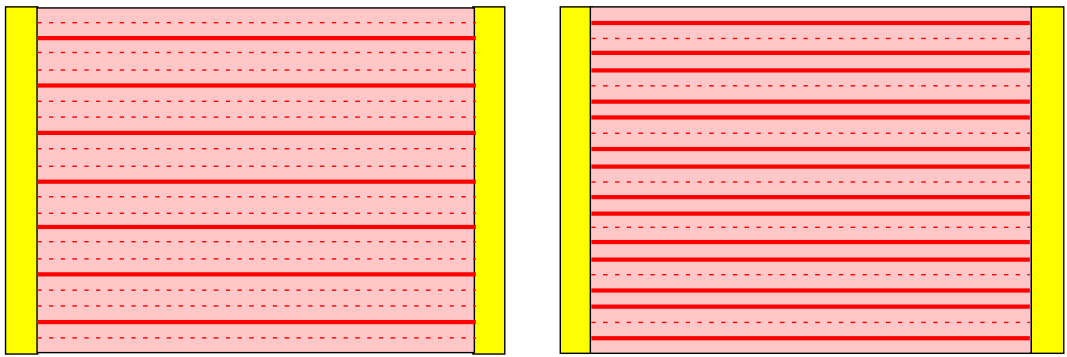

Fig.1.7a: Configuration pattern at $v = 1/3$    Fig.1.7b: Configuration pattern at $v = 2/3$

Bold lines indicate the Landau ground states filled with electron, and dashed lines empty states.

In a similar way, we can construct the most uniform filling pattern for any given fractional number $v$. The state with the most uniform electron-configuration constitutes a basis for the fractional quantum Hall effect. Quantum transitions from the state via the Coulomb interactions will be studied in Chapters 4-9.

## Chapter 2  Integer quantum Hall effect

Klaus von Klitzing discovered the integer quantum Hall effect (IQHE). The quantum Hall resistance $R_H$ is defined by [Hall voltage] / [current]. The experimental value takes the following value precisely:

$$R_H = R_K/\nu \tag{2.1a}$$

$$R_K = h/e^2 = 25\,812.807\,4555(59)\,\Omega \tag{2.1b}$$

where $R_K$ is the von Klitzing constant, $h$ the Planck constant, $e$ the elementary charge and $\nu$ is the filling factor with any integer value. The 2D electron system was studied theoretically by T. Ando, Y. Matsumoto and Y. Uemura in 1975. They calculated the electric conductivity by employing the self-consistent Born approximation and obtained the Hall conductivity as, $\sigma_{XY} = (nec/H)\left(\omega_c^2\tau^2/(1+\omega_c^2\tau^2)\right)$.

Then the Hall conductivity was predicted to be $\sigma_{XY} = -\left(e^2 N/(2\pi\hbar)\right)$ in the case of attractive scatterers and $\sigma_{XY} = -\left(e^2(N+1)/(2\pi\hbar)\right)$ in the case of repulsive scatterers at zero temperature [15], where, $N$ is the Landau level number. They have pointed out, for the first time, that the Hall conductivity is expressed by the natural constants only.

Before the discovery of IQHE, a standard of electrical resistance had some ambiguity. At that time measurements of resistance included some error which was caused by the lack of long-term stability and world-wide uniformity in the resistance standard. The value of the quantum Hall resistance is extremely accurate as in Eq.(2.1a, b) and then the phenomenon has been applied to a resistance standard from 1990. This astonishing accuracy means that the IQHE should be explained rigorously. It was pointed out for example, in the reference [16], that "The many theoretical models explain various aspects of the QHE, at least in a qualitative way. A complete theory which conclusively explains e.g. the remarkable accuracy of the QHR is still missing". In this chapter the IQHE is studied step by step.

At an integer filling factor the excitation is possible only to the higher levels. The minimum excitation energy is $\Gamma_{Landau} - \Gamma_{spin}$ or $\Gamma_{spin}$ for a spin-flipping excitation in

Fig.1.6. For the case of non spin-flipping, the minimum excitation energy is equal to $\Gamma_{\text{Landau}}$. Then the energy difference between an IQH state and its excited state is very large, compared with the thermal excitations, under a strong magnetic field. Accordingly the IQH state is extremely stable at a low temperature.

Even if we take the Coulomb interactions into consideration, the perturbation energy for any IQH state is very small because of the large denominator appearing in the perturbation calculation. (The denominator is the energy difference between the ground state and an excited state which is equal to or larger than $\Gamma_{\text{Landau}}$ because of non-flipping of spin in Coulomb transitions.). Consequently neglect of the Coulomb interactions is a good approximation for an integer filling factor.

## 2.1 Calculation of the Hall resistance

There are two kinds of velocity namely the phase velocity and the group velocity. It is well known in quantum physics that the velocity of a charge is the group velocity $u_G$, not phase velocity. Each electron has a wave function with the length $\ell$ in the current direction. So each electron carries an amount of charge per unit time given by,

$$\text{(charge per unit time)} = -eu_G/\ell \tag{2.2}$$

where an electron charge is $-e$. The number of Landau states inside the momentum region from $p$ to $p + dp$ is equal to

$$\text{(number of Landau states)} = \frac{\ell}{2\pi\hbar} dp \tag{2.3}$$

which is derived from Eq.(1.5b). We express the current inside the momentum region $p \sim p + dp$ by the symbol $dI$, which is given by

$$dI = \text{(charge per unit time)} \times \text{(number of electrons)}$$

In the case of $\nu = 1$, the number of electrons is equal to the number of Landau states and therefore $dI$ is the product of (2.2) and (2.3) as

$$dI = -\frac{eu_G}{\ell} \frac{\ell}{2\pi\hbar} dp = -\frac{eu_G}{2\pi\hbar} dp \tag{2.4a}$$

Therein we reexpress $p$ to the coordinate $y$ by making use of Eq. (1.10) in order to know the current distribution along the y-direction as follows;

$$dI = -\frac{eu_G}{2\pi\hbar}dp = -\frac{e^2 B u_G}{2\pi\hbar}d\alpha_J \tag{2.4b}$$

where $\alpha_J$ indicates the centre position of the y-direction for each electron wave function and is related to the momentum $p$ as

$$p = k\hbar = eB\alpha_J \tag{2.5}$$

We express the eigen-energy (1.18) by the symbol $f(p)$.

$$E_{L,J} = f(p) = \lambda + U(\alpha_J) + \hbar\omega\left(L + \frac{1}{2}\right) \qquad (L = 0,1,2,3,\cdots) \tag{2.6}$$

The group velocity $u_G$ along the x-direction is defined as

$$u_G = \frac{dE_{L,J}}{dp} = \frac{df(p)}{dp} \tag{2.7}$$

The group velocity is expressed with the potential from Eqs.(2.5), (2.6) and (2.7) as

$$u_G = \frac{df(p)}{dp} = \frac{dU(\alpha_J)}{dp} = \frac{dU(\alpha_J)}{d\alpha_J}\frac{d\alpha_J}{dp} = \frac{1}{eB}\frac{dU(\alpha_J)}{d\alpha_J} \tag{2.8}$$

This equation is substituted into Eq.(2.4b) and the result is

$$dI = -\frac{e^2 B}{2\pi\hbar}\frac{1}{eB}\frac{dU(\alpha_J)}{d\alpha_J}d\alpha_J = -\frac{e}{2\pi\hbar}\frac{dU(\alpha_J)}{d\alpha_J}d\alpha_J \tag{2.9a}$$

We obtain the total electric current by integration of Eq.(2.9) as

$$I = \int_0^d \left(-\frac{e}{2\pi\hbar}\frac{dU(\alpha_j)}{d\alpha_j}\right)d\alpha_j = -\frac{e}{2\pi\hbar}[U(y)]_{y=0}^d \tag{2.9b}$$

The potential difference at $y=0$ and $y=d$ is measured between the potential probe and the Hall probe in Fig.1.4 as

$$U(y=0) = -eV_1, \quad U(y=d) = -eV_2 \tag{2.10}$$

Substitution of Eq.(2.10) into (2.9b) yields the total current as

$$I = -\frac{e}{2\pi\hbar}[-eV_2 + eV_1] = \frac{e^2}{2\pi\hbar}(V_2 - V_1) \tag{2.11}$$

In the above discussion, we have neglected the effects of electron scattering by lattice vibrations, impurities, and so on. This neglect is allowed due to the experimental fact mentioned in Chapter 1, namely, the usual (diagonal) resistance is practically zero at the IQHE as in Fig. 1.2. Equation (2.11) gives the Hall resistance $R_H$ at $\nu = 1$ as follows:

$$R_H = \frac{(V_2 - V_1)}{I} = \frac{2\pi\hbar}{e^2} \qquad \text{(for } \nu = 1\text{)} \tag{2.12}$$

Thus the Hall resistance $R_H$ at $\nu = 1$ is described by the natural constants only. It is noteworthy that the total current does not depend upon the function-form of the potential $U(y)$, but depends on only the voltage between at both edges of the device (see Fig.1.4). Furthermore the total current does not depend upon the device sizes $\ell$ and $d$. Thus the relation (2.12) has been obtained. A rigorous proof will be given in section 2.3.

For the filling factor larger than 1, the calculation is done similarly. The number of electrons inside the region from $y$ to $y + dy$ is

$$\text{(number of electrons inside } y \sim y + dy\,) = \frac{\nu \cdot eB\ell}{2\pi\hbar} dy \quad \text{(for any integer } \nu\text{)} \tag{2.13}$$

which gives

$$dI = -\frac{\nu \cdot e}{2\pi\hbar} \frac{dU(y)}{dy} dy \tag{2.14}$$

$$I = \frac{\nu \cdot e^2}{2\pi\hbar}(V_2 - V_1) \tag{2.15}$$

Thus the Hall resistance $R_H$ for any integer value of $\nu$ is expressed as

$$R_H = \frac{(V_2 - V_1)}{I} = \frac{2\pi\hbar}{\nu \cdot e^2} \qquad \text{(for any integer } \nu\text{)} \tag{2.16}$$

**2.2 Distribution of current**

M. Büttiker investigated the distribution of electric current in a quantum Hall device [17]. We review his investigation in the case of $\nu = 1$. A current element, $dI$ through a width $dy$ is obtained from Eq.(2.9a) at the filling factor $\nu = 1$ as,

$$dI = -\frac{e}{2\pi\hbar} \frac{dU(y)}{dy} dy$$

Therefore electric current density namely current per unit width is given by

$$\frac{dI}{dy} = -\frac{e}{2\pi\hbar} \frac{dU(y)}{dy} \tag{2.17}$$

We have shown a typical potential curve along the y-direction in Fig.1.4, where the large potential barriers at both edges confine electrons inside a conduction layer. The derivative $dU(y)/dy$ at $y = 0$ is negative and $dU(y)/dy$ at $y = d$ is positive as easily seen in Fig.1.4. The absolute values of these two derivatives are very large to confine electrons inside the device. In the central region of the device, $dU(y)/dy$ is approximately equal to

$$dU(y)/dy \approx (eV_1 - eV_2)/d \tag{2.18}$$

We schematically draw $dU(y)/dy$ in Fig.2.1.

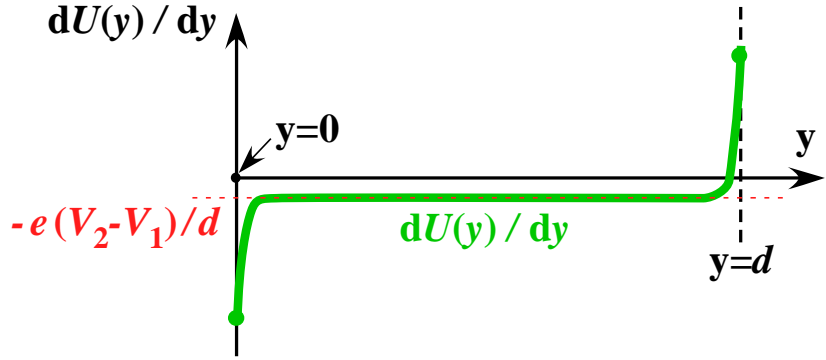

Fig.2.1 Derivative value $dU(y)/dy$

The current distribution is proportional to $-dU(y)/dy$ as in Eq.(2.17). By using Fig.2.1 we obtain the current density as shown in Fig.2.2. The absolute value of electric current density, $|dI/dy|$ becomes large at both edges of the device. The current distribution has been theoretically predicted by Büttiker [17]. The large current densities near both edges are named the edge currents. There are many experimental studies on the edge currents [18]. The edge current near $y = 0$ flows parallel to the x-axis and that near $y = d$ does anti-parallel to the x-axis. The current density near the central region of the device is directed parallel to the x-axis, the magnitude of which is given approximately as,

$$dI/dy \approx -(e/(2\pi\hbar))(eV_1 - eV_2)/d \tag{2.19}$$

The absolute value of $dI/dy$ at the central region is smaller than that at the both edges as in Fig.2.2.

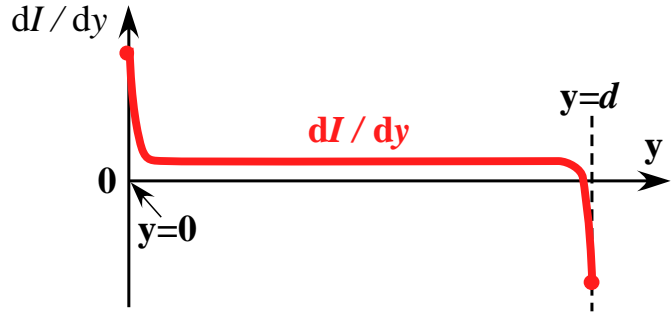

Fig.2.2 Current density d*I*/d*y*

This distribution of the current is also illustrated on a quantum Hall device in Fig.2.3. Therein the current distribution is drawn by bold arrows.

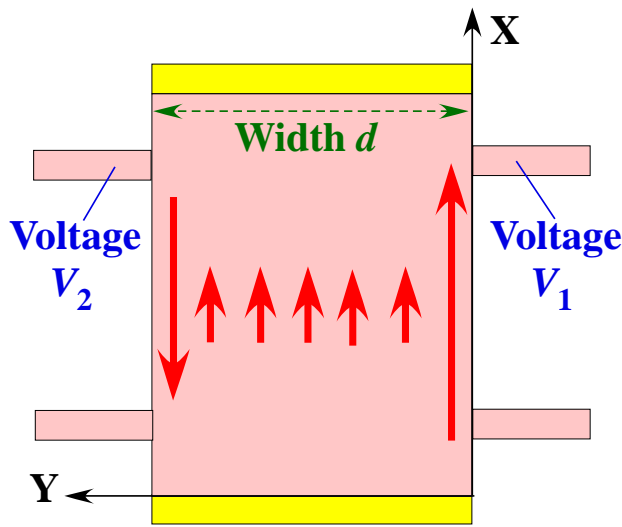

Fig.2.3 Distribution of current density
Current density is expressed by **bold arrows**

**2.3 A rigorous derivation of the Hall resistance**

The experimental value of Hall resistance at $\nu = 1$ is $25\,812.807\,4555(59)\,\Omega$ which is quite accurate. This fact means that the relation (2.16) should be derived rigorously. In an actual system, we do not know the function-form of the exact eigen-energy $E(p_x)$ because the function form of the potential $U(y)$ is unknown and the Coulomb interactions between electrons are not treated exactly, and so on. Nevertheless we can rigorously derive Eq.(2.16) as follows:

We re-study the electric current. The electric current per electron is $-eu_G/\ell$, where $-e$ and $u_G$ are the electron charge and the group velocity respectively. The electron number is $(\ell/2\pi\hbar)dp$ inside the momentum region $p \sim p + dp$. Therefore the current $dI$ inside the momentum region $p \sim p + dp$ is given by the product of $-eu_G/\ell$ and $(\ell/2\pi\hbar)dp$ as follows:

$$dI = -\frac{eu_G}{2\pi\hbar}dp \tag{2.20}$$

The relation between $dI$ and $dp$ is independent of the length $\ell$. So, even if the value of $\ell$ varies in different points of $y$, the relation (2.20) is correct. Accordingly we can apply the relation (2.20) to actual quantum Hall devices.

We introduce the exact eigen-energy $E(p)$, although the exact form of $E(p)$ is unknown. Thereby the group velocity $u_G$ is expressed as

$$u_G = \frac{dE(p)}{dp} \tag{2.21}$$

Substitution of Eq.(2.21) into Eq.(2.20) yields

$$dI = -\frac{e}{2\pi\hbar}\frac{dE(p)}{dp}dp \tag{2.22}$$

We obtain the total electric current by integrating Eq.(2.22) as

$$I = \int_{\text{edge1}}^{\text{edge2}} \left(-\frac{e}{2\pi\hbar}\frac{dE(p)}{dp}\right)dp = -\frac{e}{2\pi\hbar}[E(p)]_{\text{edge1}}^{\text{edge2}} \tag{2.23}$$

Therein the integration is carried out from the momentum $p_1$ at the edge 1 to $p_2$ at edge 2. Consequently the total current is described by the exact electron energies at edge 1 and edge 2 as follows:

$$I = -\frac{e}{2\pi\hbar}(E(\text{edge 2}) - E(\text{edge 1})) \tag{2.24}$$

Thus the total current is expressed by the difference $E(\text{edge 2}) - E(\text{edge 1})$ of the exact electron energies at the edges 2 and 1. The energy difference can be determined by measuring the voltage between the potential probe and the Hall probe.

$$E(\text{edge 2}) - E(\text{edge 1}) = (-eV_2) - (-eV_1) = -e(V_2 - V_1) \tag{2.25}$$

where $-e$ is exactly equal to the electron charge and $V_2 - V_1$ is the voltage between the Hall probe and the potential probe. Substitution of this equation into (2.24) gives

$$I = \frac{e^2}{2\pi\hbar}(V_2 - V_1) \tag{2.26}$$

Thus the total current is independent of the function-form of the potential and the sizes $\ell$ and $d$ of the device. Therefore the Hall resistance $R_H$ at $\nu = 1$ is exactly equal to

$$R_H = \frac{(V_2 - V_1)}{I} = \frac{2\pi\hbar}{e^2} \qquad \text{(for } \nu = 1\text{)} \tag{2.27}$$

As explained above, when the filling factor is $\nu$, the number of electrons is $\nu$ times that for $\nu = 1$. Therefore the total current is equal to

$$I = -\frac{\nu e}{2\pi\hbar}\left(E(\text{edge2}) - E(\text{edge1})\right) \tag{2.28}$$

The actual energy difference $\left(E(\text{edge2}) - E(\text{edge1})\right)$ is equal to the voltage between the Hall probe and the potential probe as in the case $\nu = 1$:

$$E(\text{edge2}) - E(\text{edge1}) = -e(V_2 - V_1) \qquad \text{(for any value of } \nu\text{)} \tag{2.29}$$

Substitution of Eq. (2.29) into Eq. (2.28) yields the Hall resistance as

$$R_H = \frac{(V_2 - V_1)}{I} = \frac{2\pi\hbar}{\nu e^2} \qquad \text{(for any value of } \nu\text{)} \tag{2.30}$$

Thus the relation (2.30) between the Hall resistance and the filling factor $\nu$ has been derived rigorously.

For any integer $\nu$ the energy gap is $\Gamma_{\text{spin}}$ or $(\Gamma_{\text{Landau}} - \Gamma_{\text{spin}})$ which are large in a strong magnetic field. When the spin doesn't flip, the excitation energy is $\Gamma_{\text{Landau}}$ which is very large as in Eq.(1.22). Therefore the ground state with an integer filling factor can be realized at low temperatures by adjusting the magnetic field strength or gate voltage. Therefore the Hall resistance given by Eq.(2.30) appears actually for any integer $\nu$.

When the filling factor $\nu$ is a fractional number, the Landau levels are partially filled with electrons. If we ignore the Coulomb interaction between electrons, the energy gap disappears and then the Hall resistance at a fractional filling factor $\nu$ is not confined. So we should take the Coulomb interaction between electrons into consideration. In the later chapters we examine the quasi-2D electron system with the Coulomb interaction.

## Chapter 3 Coulomb energy in the FQH states

In this chapter we study the many-electron system in a quantum Hall device and find a many-electron state with a minimum expectation value of the Coulomb interaction at any filling factor of $\nu$.

### 3.1 Total Hamiltonian and complete set of many-electron states

The total Hamiltonian $H_T$ of the many-electron system is given by

$$H_T = \sum_{i=1}^{N} H_0(x_i, y_i, z_i) + \sum_{i=1}^{N-1} \sum_{j>i}^{N} \frac{e^2}{4\pi\varepsilon\sqrt{(x_i - x_j)^2 + (y_i - y_j)^2 + (z_i - z_j)^2}} \qquad (3.1)$$

where $N$ is the total number of electrons, $(x_i, y_i, z_i)$ is the position of the $i$-th electron, $\varepsilon$ is the permittivity of the device. The Hamiltonian $H_0(x_i, y_i, z_i)$ is given by Eq.(1.2) for the single electron system. In chapter 1, we have already obtained the eigenstates and eigenenergies of $H_0(x, y, z)$. The second term in the right hand side of Eq.(3.1) indicates the Coulomb interactions between electrons.

Arbitrary many-electron state can be expressed by superposing of the many-electron wave-functions which are composed of direct products of single-electron wave functions. The single electron wave function is expressed by the symbol $\psi_{L_j, p_j}(x_j, y_j, z_j)$ which is given by Eq.(1.17). The $j$-th wave function is specified by a Landau level number $L_j$ and a momentum $p_j$. Accordingly the many-electron state is specified by a set of Landau level numbers $L_1, L_2, \cdots, L_N$ and momenta $p_1, p_2, \cdots, p_N$. The complete set of many-electron states is composed of the Slater determinant as

$$\Psi(L_1, \cdots, L_N; p_1, \cdots, p_N) = \frac{1}{\sqrt{N!}} \begin{vmatrix} \psi_{L_1, p_1}(x_1, y_1, z_1) & \cdots & \psi_{L_1, p_1}(x_N, y_N, z_N) \\ \vdots & & \vdots \\ \psi_{L_N, p_N}(x_1, y_1, z_1) & \cdots & \psi_{L_N, p_N}(x_N, y_N, z_N) \end{vmatrix} \qquad (3.2)$$

This state is the eigenstate of $\sum_{i=1}^{N} H_0(x_i, y_i, z_i)$. The expectation value of the total Hamiltonian is denoted as $W(L_1, \cdots, L_N; p_1, \cdots, p_N)$ which is given by:

$$W(L_1, \cdots, L_N; p_1, \cdots, p_N) = \langle \Psi(L_1, \cdots, L_N; p_1, \cdots, p_N) | H_T | \Psi(L_1, \cdots, L_N; p_1, \cdots, p_N) \rangle$$

$$W(L_1,\cdots,L_N;p_1,\cdots,p_N)=\sum_{i=1}^{N}E_{L_i}(p_i)+C(L_1,\cdots,L_N;p_1,\cdots,p_N) \tag{3.3}$$

where $E_{L_i}(p_i)$ is given by Eq.(1.18) which is the eigen-energy of $H_0(x_i,y_i,z_i)$ for the Landau level number $L_i$ and the momentum $p_i$. In Eq.(3.3), $C$ is the **expectation value of the Coulomb interaction** defined by

$$\begin{aligned}C(L_1,\cdots,L_N;p_1,\cdots,p_N)=\int\cdots\int\Psi(L_1,\cdots,L_N;p_1,\cdots,p_N)^{*}\times\\ \times\sum_{i=1}^{N-1}\sum_{j>i}^{N}\frac{e^2}{4\pi\varepsilon\sqrt{(x_i-x_j)^2+(y_i-y_j)^2+(z_i-z_j)^2}}\times\\ \times\Psi(L_1,\cdots,L_N;p_1,\cdots,p_N)\mathrm{d}x_1\mathrm{d}y_1\mathrm{d}z_1\cdots\mathrm{d}x_N\mathrm{d}y_N\mathrm{d}z_N\end{aligned} \tag{3.4}$$

Hereafter we call $C(L_1,\cdots,L_N;p_1,\cdots,p_N)$ "**classical Coulomb energy**".

In this chapter we investigate the case of the filling factor $\nu<1$. (The case of $\nu>1$ can be derived easily from the method in this chapter and also is studied in chapter 5.) We also restrict our investigation to the case of a low temperature and a strong magnetic field. Accordingly all the electron spins turn to the opposite direction of the magnetic field. (In Chapter 9, we will examine the case of a weak magnetic field where the many-electron state is constructed by a mixture of up and down spins.) The ground states of $\sum_{i=1}^{N}H_0(x_i,y_i,z_i)$ is specified by the quantum numbers $(L_1,\cdots,L_N)=(0,\cdots,0)$ for $\nu<1$. All the Landau states with $L=0$ are partially filled with electrons. Various electron-configurations in the Landau states are possible. If the electrons are distributed most uniformly in the Landau states, then the many-electron state has the minimum expectation value of the Coulomb interaction.

### 3.2 Many-electron state with minimum classical Coulomb energy

In this chapter we consider the case where all the electrons have down spins and $L=0$ because the magnetic field is strong and the filling factor is smaller than 1. So the Landau level numbers $L_1,\cdots,L_N$ are not shown, for simplicity as follows:

$$\Psi(0,\cdots 0;p_1,\cdots,p_N)\to\Psi(p_1,\cdots,p_N) \tag{3.5}$$
$$W(0,\cdots 0;p_1,\cdots,p_N)\to W(p_1,\cdots,p_N) \tag{3.6}$$
$$C(0,\cdots 0;p_1,\cdots,p_N)\to C(p_1,\cdots,p_N) \tag{3.7}$$

Thus the many-electron state at $\nu<1$ is specified by the set of momenta $p_1,\cdots,p_N$.

We can distinguish the many-electron state with $p_1, \cdots, p_N$ from the other one with $p'_1, \cdots, p'_N$. This momentum set is related to the centre positions $\alpha_1, \cdots, \alpha_N$ by the relation (1.10).

If the distribution ($\alpha_1, \cdots, \alpha_N$) of electron-positions is the most uniform one among possible distributions, then the state has the minimum classical Coulomb energy. We explain how to find the most uniform distribution ($\alpha_1, \cdots, \alpha_N$). First we examine two examples as follows: the most uniform distributions for $\nu = 1/3$ and $\nu = 2/3$ are schematically shown respectively in Fig.3.1a and b where the straight bold lines indicate the Landau states filled with an electron and the dashed lines empty states. (Note: The line schematically shows the Landau wave function namely the plane wave along the x-axis. The y-coordinate is the peak position of the Gaussian function along the y-axis. So the lines are equally spaced with the interval $[2\pi\hbar/(\ell eB)]$ due to Eq.(1.10).) Therein the current flows along the x-direction and the Hall voltage appears along the y-direction. (see Figs.1.1 and 1.5.) The definition of these directions, x and y, is used throughout this book. It is easily understood that the filling patterns of Figs.3.1a and 3.1b are the most uniform electron-configurations at $\nu = 1/3$ and $\nu = 2/3$ respectively. So these electron configurations have the minimum classical Coulomb energies respectively.

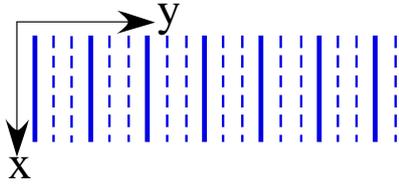 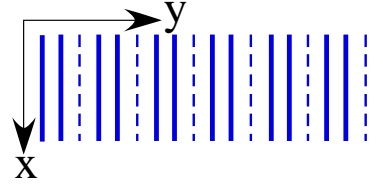

Fig.3.1a: Filling factor 1/3       Fig.3.1b: Filling factor 2/3

We next study how to find the electron-configuration with the minimum classical Coulomb energy for any filling factor. First, we examine an example of $\nu = 3/5$. We classify the electron configurations into the following two cases:

Case 1: In the whole region, three electrons exist in every sequential 5 Landau states. The filling factor becomes 3/5 because any region with sequential 5 Landau states is partially filled with 3 electrons.
Case 2: The average filling factor is equal to 3/5 in the whole system. Two electrons exist in some sequential 5 Landau states, and four electrons exist in some sequential 5 states and so on.

The Coulomb energy of Case 1 is smaller than that of Case 2 because the filling pattern of Case 1 is more uniform than that of Case 2. So we search the filling pattern of Case 1 to find the ground state of the many-electron system. The searching process is divided into two steps as follows:

Step 1: We examine filling patterns inside sequential 5 Landau orbitals at $\nu = 3/5$.
Step 2: We next consider the connection of the filling patterns.

(Step 1)
There are 10 filling patterns for sequential 5 Landau orbitals with $\nu = 3/5$. All the filling patterns are drawn in Fig.3.2. Each pattern is called "unit-configuration".

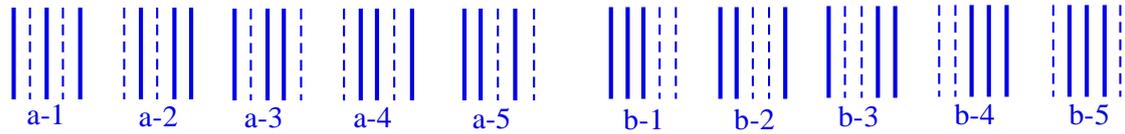

a-1　　a-2　　a-3　　a-4　　a-5　　　b-1　　b-2　　b-3　　b-4　　b-5

Fig.3.2　All unit-configurations for electron occupation in five orbitals

The five unit-configurations a-1, a-2, a-3, a-4, a-5 give the same electron-configuration by repeating of themselves as shown in Fig.3.3a except for both ends. The differences at the boundaries can be neglected in a quantum Hall system with a macroscopic number of electrons. Similarly the five unit-configurations b-1, b-2, b-3, b-4 and b-5 in Fig.3.2 give the configuration shown in Fig.3.3b. If we compare these two configurations, the electron-configuration of Fig.3.3a has a lower Coulomb energy than that of Fig.3.3b.

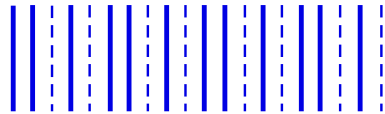　　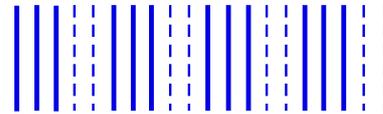

Fig.3.3a:　Filling factor 3/5　　　Fig.3.3b:　Filling factor 3/5

(Step 2)
We examine the connections between different unit-configurations in Fig. 3.2. Figure 3.4 shows five connections of (a-1 and a-2), (a-1 and a-3), (a-1 and a-4) and (a-1 and a-5). All the connections have green areas where 2 electrons or 4 electrons exist inside the sequential five states.

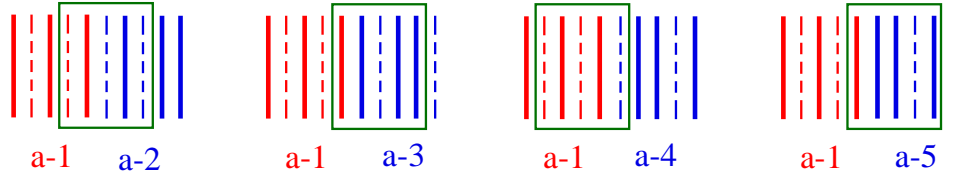

a-1　　a-2　　　　a-1　　a-3　　　　a-1　　a-4　　　　a-1　　a-5

Fig.3.4　Connections between different unit-configurations

Thus these connections belong to Case 2. Therefore the connections between different unit-configurations have a classical Coulomb energy larger than that of Fig.3.3a. Consequently we can obtain the most uniform configuration by repeating of the single unit-configuration in Fig.3.3a.

For any fractional filling factor $v = r/q$, $r$ electrons exist in sequential $q$ Landau states everywhere. The number of filling patters is $q!/(r!(q-r)!)$. We can write all the patterns. By comparing the patterns with each other we find the filling pattern with the most uniform distribution. Of course there are $q$ equivalent unit-configurations as in Fig.3.2a. We have done this procedure for the filling factors $v = r/q$ with $q = 2, 3, 4, 5, 6, 7$ and $8$. The results are shown in Fig.3.5 which expresses the unit-configurations with the minimum classical Coulomb energy. Therein only the representative unit-configurations are shown among equivalent unit-configurations. That is to say, the unit-configuration a-1 represents all the equivalent unit-configurations a-1, a-2, a-3, a-4 and a-5 for $v = 3/5$.

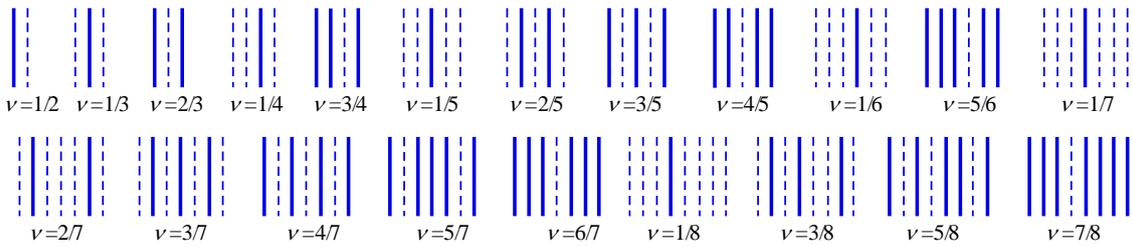

$v=1/2$　$v=1/3$　$v=2/3$　$v=1/4$　$v=3/4$　$v=1/5$　$v=2/5$　$v=3/5$　$v=4/5$　$v=1/6$　$v=5/6$　$v=1/7$

$v=2/7$　$v=3/7$　$v=4/7$　$v=5/7$　$v=6/7$　$v=1/8$　$v=3/8$　$v=5/8$　$v=7/8$

Fig.3.5　Most uniform unit-configurations of electron occupation

For confirmation, we show some examples with higher Coulomb energy in Fig. 3.6. Comparison of Fig.3.5 with Fig.3.6 reveals that the unit-configurations in Fig.3.5 have the classical Coulomb energy lower than that in Fig.3.6

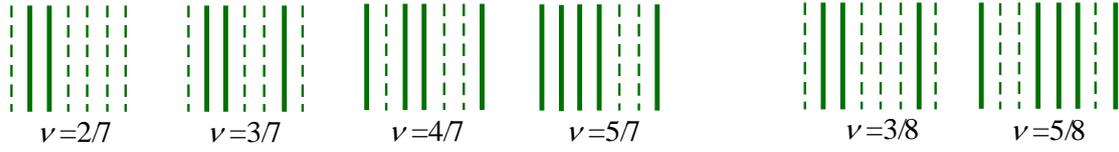

$v=2/7$    $v=3/7$    $v=4/7$    $v=5/7$      $v=3/8$    $v=5/8$

Fig.3.6    Unit-configurations with higher energy

Thus only one unit-configuration has the minimum classical Coulomb energy among all the possible unit-configurations of electrons. Then the whole configuration with the minimum classical Coulomb energy is obtained by repeating the representative unit-configuration for a given $v$. This conclusion is proven as follows;

(Proof): We consider the most uniform electron-configuration with $v = r/q$. Therein $r$ electrons should exist inside sequential $q$ Landau states everywhere. First we take sequential $q$ Landau states (orbitals) in the whole configuration. Next we consider the following new region: one Landau state is removed from the left end of the original region and one Landau state is added to the right end. Then we get a new region. Therein $r$ electrons should exist inside the new region in order to have the minimum classical Coulomb energy. This requirement yields the following two conditions:
   (1) If the removed state is empty, the added state should be empty.
   (2) If the removed state is filled with an electron, the added state should be also filled with an electron.

Thus we get a new region from left to right, one after another. Thereby we can reproduce the whole region of the many-electron configuration.

Consequently the whole electron-configuration with the minimum classical Coulomb energy is obtained by repeating only one unit-configuration with the most uniform filling pattern. (End of Proof)

## 3.3 Momentum set specifying the many-electron state with the minimum classical Coulomb energy

The central position $\alpha_J$ along the y-direction is related to the momentum of the x-direction as in Eqs.(1.10) and (1.5b). That is to say, both central position $\alpha_J$ and momentum $p$ depend on the same integer $J$ as follows;

$$\alpha_J = \frac{2\pi\hbar}{eB\ell}J \quad \text{and} \quad p = \frac{2\pi\hbar}{\ell}J \tag{3.8}$$

Accordingly any electron-configuration is given by a set of momenta (The correspondence is one to one mapping).

Let us consider three examples of Figs.3.1b, 3.3a and b. The many-electron state with the configuration of Fig.3.1b is specified by the momentum set as

$$p_{2j+1} = p_1 + \frac{2\pi\hbar}{\ell}3j, \quad p_{2j+2} = p_1 + \frac{2\pi\hbar}{\ell}(3j+1), \text{ for } j = 0, 1, 2, 3, \cdots, \tag{3.9}$$

where $p_1$ indicates the minimum momentum among all the momenta of electrons. We next consider two configurations at $\nu = 3/5$. Figure 3.3a show the first configuration specified by the momentum set $(p_1, \cdots, p_N)$ as

$$p_{3j+1} = p_1 + \frac{2\pi\hbar}{\ell}5j, \quad p_{3j+2} = p_1 + \frac{2\pi\hbar}{\ell}(5j+1), \quad p_{3j+3} = p_1 + \frac{2\pi\hbar}{\ell}(5j+3)$$
$$\text{for } j = 0, 1, 2, 3, \cdots. \tag{3.10}$$

The second configuration is shown in Fig.3.3b which is specified by the momentum set $(p'_1, \cdots, p'_N)$ as

$$p'_{3j+1} = p'_1 + \frac{2\pi\hbar}{\ell}5j, \quad p'_{3j+2} = p'_1 + \frac{2\pi\hbar}{\ell}(5j+1), \quad p'_{3j+3} = p'_1 + \frac{2\pi\hbar}{\ell}(5j+2)$$
$$\text{for } j = 0, 1, 2, 3, \cdots. \tag{3.11}$$

Thus arbitrary electron-configuration is described by the corresponding momentum set. The many-electron state has the expectation value $W(p_1, \cdots, p_N)$ of the total Hamiltonian $H_T$. As explained in section 3.2, all the Landau level numbers are zero for $\nu < 1$. Use of Eq.(3.3) yields the difference between $W(p_1, \cdots, p_N)$ and $W(p'_1, \cdots, p'_N)$ to be

$$W(p_1, \cdots, p_N) - W(p'_1, \cdots, p'_N) = C(p_1, \cdots, p_N) - C(p'_1, \cdots, p'_N) \tag{3.12}$$

Here $C(p_1, \cdots, p_N)$ is smaller than $C(p'_1, \cdots, p'_N)$ because the configuration of Fig.3.3a is more uniform than that of Fig.3.3b. Accordingly, we get

$$W(p_1, \cdots, p_N) < W(p'_1, \cdots, p'_N) \tag{3.13}$$

Consequently the momentum set of Eq.(3.10) has the minimum expectation value of $H_T$ at $\nu = 3/5$. In this way we can obtain the momentum set with the minimum classical Coulomb energy for arbitrary filling factor. The function-form of the classical Coulomb energy versus filling factor $\nu$ will be investigated in the next section.

### 3.4 $\nu$-dependence of the minimum classical Coulomb energy

Equation (1.20b) gives the number of states with a given Landau level number as;

$$\frac{eB\ell}{2\pi\hbar}d$$

Therefore, the macroscopic charge density of electrons at the filling factor $\nu$ is given by $\nu \times (-e) \times (\text{number of states})/(\ell d)$ which is

$$\text{macroscopic charge density } \sigma = -\frac{e^2 B}{2\pi\hbar}\nu \qquad (3.14)$$

The black curve in the upper panel of Fig.1.2 expresses the experimental data of Hall resistance $R_H$ divided by the Klitzing constant $R_K = 2\pi\hbar/e^2$. Therein the red line shows the average dependence of $R_H/R_K$. The ratio $R_H/R_K$ is equal to $1/\nu$. Accordingly the red line indicates that $1/\nu$ is almost proportional to the magnetic field strength $B$ except the plateau regions. That is to say, $B\nu$ is nearly equal to a constant value and then the charge density $\sigma$ is independent of $B$. So we can approximate the macroscopic Coulomb energy $C_{\text{Macroscopic}}(\sigma)$ to be a constant value in the experiment of Fig.1.2. When we examine the microscopic charge-distribution, we find a new difference of the classical Coulomb energy by the electron configuration.

Accordingly we need to study how the classical Coulomb energy depends on the microscopic electron-configuration. As an example, we first examine the many-electron state at $\nu = 2/3$. The electron-configuration with the minimum classical Coulomb energy is illustrated in Fig.3.7. Therein the bold lines express the occupied orbitals with electron and the dashed lines indicate the empty orbitals.

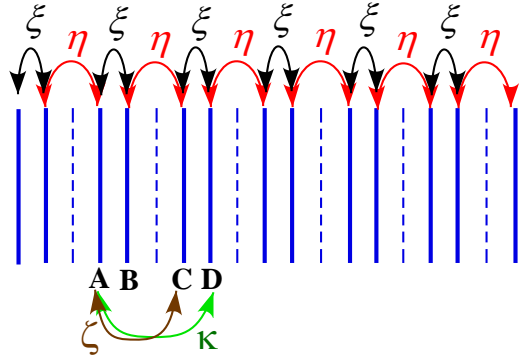

Fig.3.7 Classical Coulomb energies between electrons at $\nu = 2/3$

The electron pair located at the orbitals AB is one example of the nearest-electron-pairs. Two electrons placed at B and C show the second nearest pair. The electron pair at A and C is the third nearest pair. The pair at A and D is the fourth nearest pair and so on. The classical Coulomb energy between two electrons is expressed by the symbols $\xi, \eta, \varsigma$ and $\kappa$ respectively as in Fig.3.7. Therein $\xi$ is the largest one of the classical Coulomb pair energy, $\eta$ is the second largest, $\varsigma$ is the third largest, $\kappa$ is the fourth largest and so on.

The classical Coulomb energy between pair (A, C) is weakened by the screening (shielding) effect of electron B. Also the classical Coulomb energy between pair (A, D) is weakened by the screening effect of electrons B and C. Accordingly the $\nu$-dependence of the classical Coulomb energy mainly comes from the first nearest and the second nearest pairs. The number of the more distant pairs (third, fourth, fifth and so on) are enormous many. The total number of electron pairs is $N(N-1)/2$. On the other hand the total number of the first and second nearest pairs is $N$. Accordingly the residual energies (namely the sum of all the more distant pair energies) may be approximated by the macroscopic Coulomb energy $C_{\text{Macroscopic}}(\sigma)$ which is a constant value mentioned above. On the other hand the sum of the first and second nearest pair-energies is strongly dependent upon the filling factor. The $\nu$-dependence is examined for various filling factors as follows:

(Case of $\nu = 2/3$)

We can ignore the boundary effect in both ends for a macroscopic electron-number $N$. Then the number of the nearest pairs is $N/2$ and the number of the second nearest pairs is $N/2$ in the configuration of Fig.3.7. Then the sum of the classical Coulomb energies between the nearest pairs is equal to $\xi \cdot N/2$ and the sum of the energies between the second nearest pairs is equal to $\eta \cdot N/2$. Accordingly the total classical Coulomb energy $C(p_1, \cdots, p_N)$ is approximated by

$$C(p_1, \cdots, p_N) \approx [(\xi/2) + (\eta/2)]N + C_{\text{Macroscopic}}(\sigma) \qquad \text{at} \quad \nu = 2/3 \qquad (3.15)$$

Next we estimate the classical Coulomb energy for other cases.

(Case of $\nu = 3/5$)

The electron-configuration with the minimum classical Coulomb energy at $\nu = 3/5$ is shown in Fig.3.8.

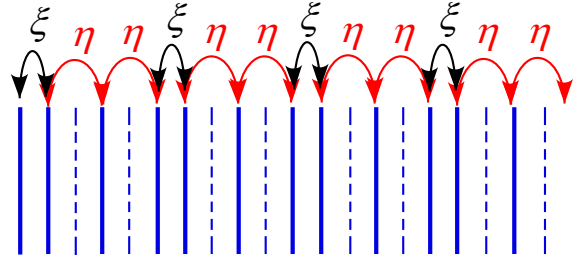

Fig.3.8 Classical Coulomb energies between electrons at $\nu = 3/5$

The number of the nearest pairs is equal to $(1/3)N$ and the number of the second nearest pairs is equal to $(2/3)N$ at $\nu = 3/5$. Then the total classical Coulomb energy is

$$C(p_1,\cdots,p_N) \approx [(\xi/3)+(2\eta/3)]N + C_{\text{Macroscopic}}(\sigma) \quad \text{at } \nu = 3/5 \quad (3.16)$$

(Case of $\nu = 4/7$)
Figure 3.9 show the electron-configuration with the minimum classical Coulomb energy at $\nu = 4/7$.

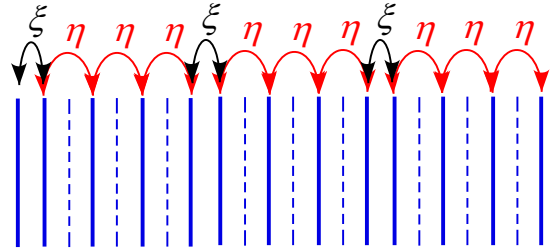

Fig.3.9 Classical Coulomb energies between electrons at $\nu = 4/7$

The number of the nearest pairs is equal to $(1/4)N$ and the number of the second nearest pairs is equal to $(3/4)N$ at $\nu = 4/7$. Then the total classical Coulomb energy is

$$C(p_1,\cdots,p_N) \approx [(\xi/4)+(3\eta/4)]N + C_{\text{Macroscopic}}(\sigma) \quad \text{for } \nu = 4/7 \quad (3.17)$$

(Case of $\nu = 5/7$)
The electron-configuration with the minimum classical Coulomb energy for $\nu = 5/7$ is shown in Fig.3.10.

Fig.3.10 Classical Coulomb energies between electrons at $v = 5/7$

The number of the nearest pairs is equal to $(3/5)N$ and the number of the second nearest pairs is equal to $(2/5)N$ at $v = 5/7$. Then the total classical Coulomb energy is

$$C(p_1,\cdots,p_N) \approx [(3\xi/5)+(2\eta/5)]N + C_{\text{Macroscopic}}(\sigma) \quad \text{at } v = 5/7 \quad (3.18)$$

(Any case of $1/2 < v < 1$)

We calculate the classical Coulomb energy for a general case of $v = r/q$ ($\tfrac{1}{2} < v < 1$).

As proven in Sec.3.2, the electron-configuration with the minimum classical Coulomb energy is constructed by repeating the representative unit-configuration where $r$ electrons exist in sequential $q$ Landau orbitals (states). The number of empty orbitals per unit-configuration is $q - r$. All the empty orbitals are separated by one or more filled-orbitals at $\tfrac{1}{2} < v < 1$. That is to say, all the empty orbitals are isolated as seen in Figs.3.7-3.10. Therefore the $q - r$ second-nearest pairs exist per unit-configuration. The total numbers of the first and the second nearest pairs is equal to the total number of electrons as easily seen in Figs.3.7-3.10. Therefore the number of nearest pairs becomes $r - (q - r) = 2r - q$ per unit-configuration. The total number of nearest electron pairs is equal to $(2r - q)(N/r)$ and the total number of second nearest electron pairs is equal to $(q - r)(N/r)$ for the filling factor $v = r/q$. Consequently the total classical Coulomb energy is obtained as

$$\begin{aligned} C(p_1,\cdots,p_N) &= [(\xi(2r-q)/r)+(\eta(q-r)/r)]N + C_{\text{Macroscopic}}(\sigma) \\ &= [(2\xi - \eta)-((\xi-\eta)q/r)]N + C_{\text{Macroscopic}}(\sigma) \end{aligned} \quad \text{at } v = r/q$$

This equation is rewritten by using $v$ and then the result becomes

$$C(p_1,\cdots,p_N) = [(2\xi - \eta)-(\xi-\eta)/v]N + C_{\text{Macroscopic}}(\sigma) \quad \text{at } v = r/q \quad (3.19)$$

Equations (3.3) and (3.19) yield the expectation value of the total Hamiltonian as

$$W(p_1,\cdots,p_N) = \sum_{i=1}^{N} E_0(p_i) + [(2\xi - \eta)-(\xi-\eta)/v]N + C_{\text{Macroscopic}}(\sigma)$$

$$\text{at } \nu = r/q \tag{3.20}$$

The single electron eigenenergy $E_0(p_i)$ is given by the following form as in Eq.(1.18).

$$E_0(p_i) = \lambda + U(\alpha_i) + \hbar eB/(2m^*) \tag{3.21}$$

Substitution of Eq.(3.21) into Eq.(3.20) yields the expectation value of the total Hamiltonian as:

$$\begin{aligned} W(p_1,\cdots,p_N) &= \sum_{i=1}^{N}\left(\lambda + U(\alpha_i) + \hbar eB/(2m^*)\right) \\ &\quad + N(2\xi - \eta) - N(\xi - \eta)/\nu + C_{\text{Macroscopic}}(\sigma) \end{aligned} \tag{3.22a}$$

We put together the constant part as follows;

$$W(p_1,\cdots,p_N) = \left[f + \hbar eB/(2m^*) - (\xi - \eta)/\nu\right]N + C_{\text{Macroscopic}}(\sigma) \tag{3.22b}$$

Therein $f$ is the constant value as

$$f = \lambda + \overline{U} + 2\xi - \eta \tag{3.23a}$$

$$\overline{U} = \sum_{i=1}^{N} U(\alpha_i)/N \tag{3.23b}$$

where $\overline{U}$ is the average of the potential along the y-direction.

(Note: We prove that the value $\overline{U}$ is invariant via the Coulomb transition as follows: The electron momenta before the transition are $p_i$ and $p_j$ which are related to the centre positions $\alpha_i$ and $\alpha_j$ as in Eq.(1.10). Also ($p'_i, p'_j$) and ($\alpha'_i, \alpha'_j$) express the momenta and the positions after the transition. The total momentum conserves via the Coulomb transition. So we obtain the equation $p_i + p_j = p'_i + p'_j$. This equation gives the relation $\alpha_i + \alpha_j = \alpha'_i + \alpha'_j$. The width $\Delta y$ of the wave function is very small namely $\Delta y \approx 10.5$ nm at 6T as in Eq.(1.19). So the effective quantum transition satisfies $|\alpha_i - \alpha'_i| \leq \Delta y$. Therefore the electric potential $U(y)$ is approximated to the linear form in the overlapped region of the electron wave functions. Accordingly we obtain the relation as;

$$U(\alpha_i) + U(\alpha_j) \approx U(\alpha'_i) + U(\alpha'_j) \tag{3.24}$$

Thus the value $\overline{U}$ can be treated as a constant in the quantum transitions.)

Eq.(3.22b) gives the function form of $W$ which depends linearly upon $1/\nu$. The proportional coefficient $-(\xi-\eta)N$ is negative, because the classical Coulomb energy between the first nearest electron pair is larger than that between the second nearest pair. Accordingly the expectation value of the total Hamiltonian, $W$, changes continuously with $1/\nu$ as shown in Fig.3.11.

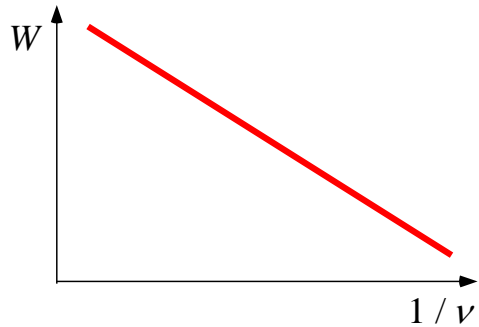

Fig.3.11 Expectation value $W$ of the total Hamiltonian

Thus the classical Coulomb energy has no energy-gap. Accordingly the classical Coulomb energy cannot produce the plateaus of Hall resistance. In the present theory, the confinement of the Hall resistance comes from another reason, which will be studied in Chapters 4-8.

Since the quasi-2D electron problem with the Coulomb interaction cannot be solved exactly, we are obliged to use some approximations in the investigation of the FQHE. Laughlin introduced a quasi-particle with fractional charge and obtained the wave function by using the variational method [19, 20]. Haldane and Halperin extended the scheme [21-23]. Jain proposed a theory to explain the experimental data on FQHE [24]. He introduced a quasi-particle called the composite fermion which is an electron bound to even number of magnetic flux quanta. These theories have used the many different types of quasi-particles. On the other hand, there is a theory used normal electrons without any quasi-particle. The theory is investigated by Tao and Thouless. They calculated the perturbation energy via the Coulomb interaction between electrons [25, 26]. We develop the Tao-Thouless theory in Chapters 3-5. The investigations has been published in Ref.[27-33]. The most uniform configuration has been examined in [31]. The present theory [27-33] is compared with the traditional theories [19-24] in Chapter 12.

# Chapter 4  Binding energy of electron pair

This chapter is devoted to calculate the perturbation energy via the Coulomb transition. The Coulomb interaction depends upon only the relative coordinate between electrons. The property derives the momentum conservation along the x-direction.

In this chapter we examine FQH states under the strong magnetic field and so all the spins of electrons are antiparallel to the magnetic field. (We will examine the other case namely the case of weak magnetic field in Chapter 9.)

## 4.1 Momentum conservation in Coulomb interaction

The complete set for the quasi-2D electron system is composed of the wave function $\Psi(L_1,\cdots,L_N;p_1,\cdots,p_N)$ defined by the Slater determinant (3.2). The total Hamiltonian (3.1) is re-expressed by using this complete set. We describe the diagonal part of $H_T$ by $H_D$ which is defined by

$$H_D = \sum_{L_1,\cdots,L_N} \sum_{p_1,\cdots,p_N} |\Psi(L_1\cdots L_N;p_1,\cdots,p_N)\rangle W(L_1\cdots L_N;p_1,\cdots,p_N)\langle\Psi(L_1\cdots L_N;p_1,\cdots,p_N)|$$

(4.1)

where $W(L_1\cdots L_N;p_1,\cdots,p_N)$ is the diagonal matrix element of $H_T$ as Eq. (3.3). The ground state in $H_D$ is identified by the momentum set which gives the most uniform electron-configuration. The configuration is determined uniquely at any fractional filling factor as verified in the previous chapter. This uniqueness means that the ground state in $H_D$ is not degenerate.

The total Hamiltonian is divided into two parts $H_D$ and $H_I$. The latter part $H_I$ is defined as

$$H_I = H_T - H_D.$$  (4.2)

where $H_I$ is constructed by the off-diagonal elements only. Next the properties of the matrix elements $\langle\Psi(L'_1\cdots L'_N;p'_1,\cdots,p'_N)|H_I|\Psi(L_1\cdots L_N;p_1,\cdots,p_N)\rangle$ will be examined below. The quantum numbers $(L'_j,p'_j)$ and $(L_j,p_j)$ identify the final and the initial states of the $j$-th electron, respectively. When the Coulomb interaction acts between two electrons $i$ and $j$, the other electrons have the same momenta before and after the transition. That is to say the matrix elements include the product of the Kronecker delta functions as

$$\prod_{s\neq i, s\neq j} \delta(L'_s - L_s)\delta(p'_s - p_s)$$

where the symbol $\delta(a-b)$ is used instead of the Kronecker delta function $\delta_{a,b}$. Accordingly the matrix element of $H_I$ is

$$\langle \Psi(L'_1 \cdots L'_N; p'_1, \cdots, p'_N)|H_I|\Psi(L_1 \cdots L_N; p_1, \cdots, p_N)\rangle$$

$$= \prod_{s\neq i, s\neq j} \delta(L'_s - L_s)\delta(p'_s - p_s) \times \quad (4.3a)$$

$$\left\langle \psi_{L'_i,p'_i} \times \psi_{L'_j,p'_j} \left| \frac{e^2}{4\pi\varepsilon\sqrt{(x_i-x_j)^2+(y_i-y_j)^2+(z_i-z_j)^2}} \right| \psi_{L_i,p_i} \times \psi_{L_j,p_j} \right\rangle$$

where $\psi_{L_i,p_i} \times \psi_{L_j,p_j}$ indicates the electron-pair wave function with the quantum numbers $(L_i, p_i)$ and $(L_j, p_j)$. When the final state is identical to the initial state, the matrix element is zero as

$$\langle \Psi(L_1 \cdots L_N; p_1, \cdots, p_N)|H_I|\Psi(L_1 \cdots L_N; p_1, \cdots, p_N)\rangle = 0$$

$$\text{for } (L'_n = L_n \quad p'_n = p_n \quad n = 1, 2, \cdots, N) \quad (4.3b)$$

The x component of the total momentum conserves in this system as

$$p'_i + p'_j = p_i + p_j \quad (4.4)$$

because the Coulomb interaction depends upon only the relative coordinate between electrons.

In chapters 3-8, we study the FQHE under a strong magnetic field where all the electrons have the same spin-direction antiparallel to the magnetic field for $\nu \leq 1$. When an electron pair transfers to other states via the Coulomb interaction at $\nu \leq 1$, the final two momenta $p'_i$ and $p'_j$ should be different from the momenta of all the other electrons because of the Pauli exclusion principle. That is to say the Coulomb transition is allowed to empty states only. The number of transitions is dependent upon the filling factor. We examine the property in details below.

## 4.2 Perturbation energy of a nearest electron pair

The electron-configurations with the minimum classical Coulomb energy are illustrated for the filling factors $\nu = 1/2, 2/3, 3/5, 4/7, 5/9, 6/11, 7/13$ in Fig.4.1 where bold lines indicate the Landau states filled with electron and dashed lines the empty states. The electron-configuration at $\nu = 1/2$ is produced by repeating the unit-configuration of (filled, empty). This configuration is the most uniform distribution of electrons at $\nu = 1/2$. Figure 4.1 also shows the electron-configurations with the filling factors of $\nu = j/(2j-1)$ for $j = 2,3,4,5,6,7$. These configurations give the most uniform distribution of electrons.

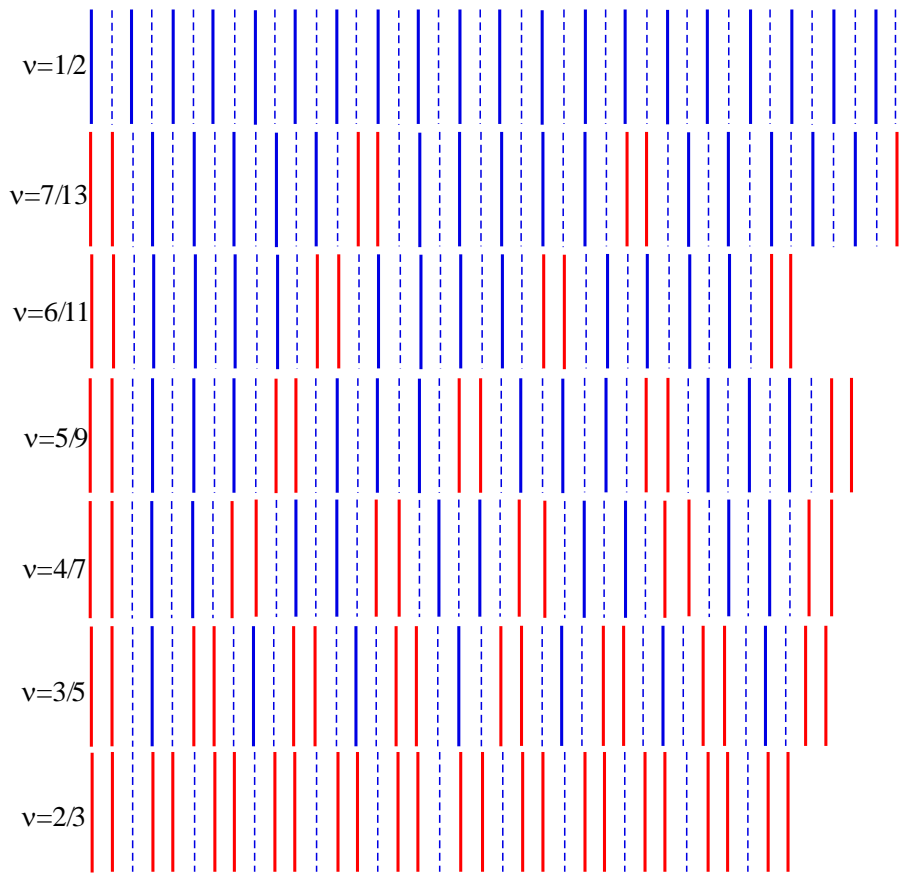

Fig.4.1 Most uniform electron configurations
Red lines indicate nearest neighbor electron pairs and dashed lines express empty states.

The upper most electron-configuration on Fig.4.1 has no electron pair placed in the nearest Landau orbitals. On the other hand the configurations other than $\nu=1/2$ have nearest neighbouring electron pairs which are shown by red lines. We call the electron pair placed in the nearest neighbouring orbitals "**nearest electron pair**".

All the configurations in Fig.4.1 are the ground state of $H_D$ at $\nu=1/2, 7/13, 6/11, 5/9, 4/7, 3/5$ and $2/3$, respectively. The electron configuration with the minimum energy of $H_D$ is only one pattern as it has been proven in chapter 3. Accordingly we can apply a perturbation method in non-degenerate ground state in order to calculate the binding energy.

We calculate the perturbation energy of the *nearest electron pair* at $\nu=7/13$ as an example. The Landau states denoted by A, B, C, D, E and F in Fig.4.2 are characterized by the momenta $p_A, p_B, p_C, p_D, p_E$ and $p_F$ respectively.

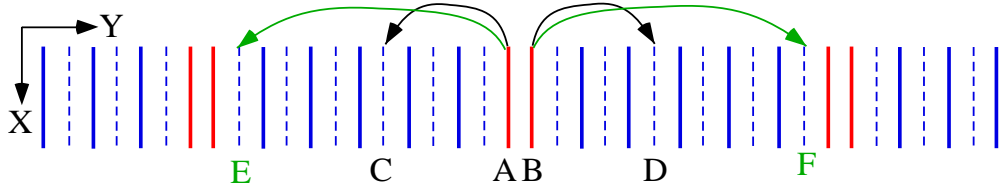

Fig.4.2  Coulomb transitions from nearest neighbour electron pair AB at $\nu=7/13$

We show a quantum transition from the electron pair AB via the Coulomb interaction by black arrow pair in Fig.4.2. When the electron at the site A transfers to the fifth orbital to the left, the momentum $p_A$ decreases by $5\times 2\pi\hbar/\ell$ due to Eq.(1.10). After the transition, the new momentum becomes $p_C$ as

$$p_C = p_A - 5\times 2\pi\hbar/\ell. \qquad (4.5)$$

The electron at the site B transfers to the site D with $p_D$. Because the total momentum conserves during the quantum transition, the following relation is satisfied:

$$p_C + p_D = p_A + p_B \qquad (4.6)$$

Equations (4.5) and (4.6) yield the momentum $p_D$ as

$$p_D = p_B + 5\times 2\pi\hbar/\ell. \qquad (4.7)$$

Consequently the electron at B transfers from its original orbital to the fifth orbital to the right as shown in Fig.4.2. This transition is allowed because the fifth orbital is empty. Also the *nearest electron pair* AB can transfer to the states E and F, and the momenta after the transition are given by
$$p_E = p_A - 11(2\pi\hbar/\ell) \tag{4.8a}$$
$$p_F = p_B + 11(2\pi\hbar/\ell) \tag{4.8b}$$
We have examined two examples mentioned above. The electron pair AB can transfer to all the empty orbitals as easily seen in Fig.4.2. The momenta after transition are denoted by $p'_A, p'_B$ and then the total momentum conservation is expressed by
$$p'_A + p'_B = p_A + p_B \tag{4.9}$$
Accordingly the momenta after the transition are given by
$$p'_A = p_A - \Delta p \tag{4.10a}$$
$$p'_B = p_B + \Delta p \tag{4.10b}$$

where the momentum transfer $\Delta p$ at $\nu = 7/13$ takes the values as

$$\Delta p = (13j+1)2\pi\hbar/\ell, (13j+3)2\pi\hbar/\ell, (13j+5)2\pi\hbar/\ell, (13j+7)2\pi\hbar/\ell,$$
$$(13j+9)2\pi\hbar/\ell \text{ and } (13j+11)2\pi\hbar/\ell \quad \text{for } j = 0, \pm 1, \pm 2, \pm 3, \cdots \tag{4.11}$$

because the electrons are allowed to transfer to all the empty orbitals. The second order perturbation energy of the *nearest electron pair* AB via the Coulomb interaction is given by

$$\varsigma_{\nu=7/13} = \sum_{\Delta p \text{ all allowed momenta at } \nu=7/13} \frac{\langle p_A, p_B | H_I | p'_A, p'_B \rangle \langle p'_A, p'_B | H_I | p_A, p_B \rangle}{W_G - W_{\text{excite}}(p_A \to p'_A, p_B \to p'_B)} \tag{4.12}$$

where $W_G$ is the ground state energy of $H_D$ at $\nu = 7/13$. It is noted that $\Delta p$ takes the values of Eq.(4.11). In Eq.(4.12) the denominator is negative because the ground state energy $W_G$ is the lowest of all. Then the second order perturbation energy is negative.

$$\varsigma_{\nu=7/13} < 0 \tag{4.13}$$

We have discussed the quantum transitions from pair AB. Also all the *nearest electron pairs* are able to transfer to all the empty orbitals. Therefore the perturbation energy of any nearest pair has the same perturbation energy as Eq.(4.12).

We examine another example $\nu = 2/3$. Figure 4.3 shows the electron-configuration with the minimum classical Coulomb energy.

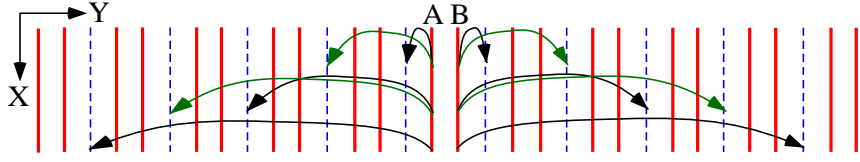

Fig.4.3   Coulomb transitions from nearest neighbour electron pair at $\nu = 2/3$

The momenta after the transition are denoted by $p'_A = p_A - \Delta p$   $p'_B = p_B + \Delta p$ where the momentum transfer $\Delta p$ takes the values as

$$\Delta p = (3j+1)2\pi\hbar/\ell \quad \text{for } j = 0, \pm1, \pm2, \pm3, \cdots \quad \text{at } \nu = 2/3 \quad (4.14)$$

because of the Pauli exclusion principle.

The second order perturbation energy of the *nearest electron pair* is given by

$$\varsigma_{\nu=2/3} = \sum_{\Delta p = (3j+1)2\pi\hbar/\ell \text{ for } j=0,\pm1,\pm2,\cdots} \frac{\langle p_A, p_B | H_I | p'_A, p'_B \rangle \langle p'_A, p'_B | H_I | p_A, p_B \rangle}{W_G - W_{\text{excite}}(p_A \to p'_A, p_B \to p'_B)} \quad (4.15)$$

In order to calculate the perturbation energy for various filling factors systematically, we introduce the following summation $Z$ as:

$$Z = -\sum_{\Delta p \neq 0, -2\pi\hbar/\ell} \frac{\langle p_A, p_B | H_I | p'_A, p'_B \rangle \langle p'_A, p'_B | H_I | p_A, p_B \rangle}{W_G - W_{\text{excite}}(p_A \to p'_A, p_B \to p'_B)} \quad (4.16)$$

where the momentum transfer $\Delta p$ takes all values $(2\pi\hbar/\ell)\times$integer except $\Delta p = 0$ and $\Delta p = -2\pi\hbar/\ell$. If the momentum transfer $\Delta p$ is equal to 0 or $-2\pi\hbar/\ell$ then the transferred state is the same state as the initial one. Accordingly the transition matrix element is zero because the diagonal element of $H_I$ is zero. Therefore the state is eliminated in the summation of (4.16).

The ground state energy $W_G$ is lower than $W_{\text{excite}}$ and so the denominator is negative.

In the definition of $Z$, (-1) is multiplied in the right-hand side of Eq.(4.16). Accordingly $Z$ is positive. [27-33]

Let us compare $Z$ with $\varsigma_{\nu=2/3}$. The summation over $\Delta p$ in $\varsigma_{\nu=2/3}$ is performed in $3\times 2\pi\hbar/\ell$ step. On the other hand the summation in $Z$ is performed with the interval $2\pi\hbar/\ell$. Then we get, in a macroscopic scale ($\ell \to \infty$)

$$\varsigma_{\nu=2/3} = -\frac{1}{3}Z \quad (4.17)$$

Thus we can express the perturbation energy of the *nearest electron pair* by $Z$. The perturbation energy depends upon $\ell$ and $B$ namely device size and magnetic field strength. These dependences are included in the summation $Z$. The total number of *nearest electron pairs* is $(1/2)N$ at $\nu = 2/3$. Accordingly the total perturbation energy $E_{\text{nearest pair}}$ of all the *nearest electron pairs* is equal to

$$E_{\text{nearest pair}} = \frac{1}{2} N \times \varsigma_{\nu=2/3} = -\frac{1}{6} ZN \qquad \text{for} \quad \nu = 2/3 \qquad (4.18)$$

Similar calculation leads the second order perturbation energy of the *nearest electron pair* $\varsigma_{\nu=7/13}$ at $\nu = 7/13$ as follows;

$$\varsigma_{\nu=7/13} = -\frac{6}{13} Z \qquad (4.19)$$

which is derived from the property that the number of empty states is 6 per sequential 13 Landau states. The total perturbation energy $E_{\text{nearest pair}}$ of all the *nearest electron pairs* is

$$E_{\text{nearest pair}} = \varsigma_{\nu=7/13} \times \frac{N}{7} = -\frac{1}{7} \times \frac{6}{13} ZN = -\frac{6}{91} ZN \qquad \text{for} \quad \nu = 7/13 \quad (4.20)$$

because the total number of the *nearest electron pairs* is $(1/7)N$ at $\nu = 7/13$.

Next we examine the case of $\nu = j/(2j-1)$ for arbitrary integer $j$. When $j$ becomes large, $\nu = j/(2j-1)$ approaches 1/2. So we choose the energy at $\nu = 1/2$ as a reference. The total energy at $\nu = 1/2$ is equal to

$$E(\nu = 1/2) \approx g(\tfrac{1}{2})N + \left[ f + \hbar eB/(2m^*) - (\xi - \eta)/\nu \right] N + C_{\text{Macroscopic}}(\sigma) \qquad (4.21)$$

where the last two terms on the right hand side of Eq.(4.21) represent the eigenenergy of $H_D$ obtained by Eq.(3.22b). The residual term $g(\tfrac{1}{2})N$ indicates the total perturbation energy via the interaction $H_I$ at $\nu = 1/2$. Because the $\nu = 1/2$ state has no nearest electron pair, the term $g(\tfrac{1}{2})N$ is the total perturbation energy from all the non-nearest pairs.

As is evident from Fig.4.1, nearest electron pairs exist in the case other than $\nu = 1/2$ and so the energy difference between $\nu = j/(2j-1)$ and $\nu = 1/2$ comes from the nearest electron pairs. Accordingly the total energy at the filling factor $\nu$ is

approximately equal to

$$E(\nu) \approx E_{\text{nearest pair}} + g(\tfrac{1}{2})N + \left[f + \hbar eB/(2m^*) - (\xi - \eta)/\nu\right]N + C_{\text{Macroscopic}}(\sigma) \quad (4.22)$$

where $E_{\text{nearest pair}}$ is the nearest pair energy. Hereafter we calculate the perturbation energy of the *nearest electron pairs* at $\nu = j/(2j-1)$ for arbitrary integer $j$. As easily seen from Fig.4.1, the most uniform configuration has one *nearest electron pair* (red pair in Fig.4.1) and $j-1$ empty states in the unit-configuration at $\nu = j/(2j-1)$. Because the *nearest electron pair* can transfer to all the empty states, the perturbation energy $\varsigma_{\nu = j/(2j-1)}$ per *nearest electron pair* is equal to

$$\varsigma_{\nu = j/(2j-1)} = -\frac{j-1}{(2j-1)}Z$$

The total number of *nearest electron pairs* is $1/j$ times the total number of electrons. Therefore the total perturbation energy of the *nearest electron pairs* is obtained as

$$E_{\text{nearest pair}} = \frac{1}{j}N \times \varsigma_{\nu = j/(2j-1)} = -\frac{1}{j} \times \frac{j-1}{(2j-1)}ZN \quad \text{for} \quad \nu = j/(2j-1) \quad (4.23a)$$

There is no *nearest electron pair* at $\nu = 1/2$ and so $E_{\text{enearest pair}}$ is

$$E_{\text{nearest pair}} = 0 \qquad \text{at} \quad \nu = 1/2 \qquad (4.23b)$$

The perturbation energy of *nearest electron pairs* per electron is equal to

$$\frac{E_{\text{nearest pair}}}{N} = -\frac{1}{j} \times \frac{j-1}{(2j-1)}Z \qquad \text{for} \quad \nu = j/(2j-1) \quad (4.23c)$$

where the boundary effects of both ends in a quantum Hall system have been ignored for a macroscopic number of $N$. The value of $E_{\text{enearest pair}}$ is listed on Table 4.1 for several filling factors.

Table 4.1  Energy of *nearest electron pairs* per electron at $\nu = j/(2j-1)$

| $\nu$ | $E_{\text{nearest pair}}/N$ |
|---|---|
| 1/2 | 0 |
| 2/3 | -(1/6) Z |
| 3/5 | -(2/15) Z |
| 4/7 | -(3/28) Z |
| 5/9 | -(4/45) Z |
| 6/11 | -(5/66) Z |
| 7/13 | -(6/91) Z |
| 8/15 | -(7/120) Z |

We comment briefly how the number of allowed quantum-transitions depends on the filling factor.

(Short comment)
As will be clarified in the next chapter, the perturbation energy is not a continuous function of $\nu$. At the specific filling factors of $\nu = $ 2/3, 3/5, 4/5, 4/7, 5/9, $\cdots$, the nearest electron pairs are **<u>allowed</u>** to transfer to **<u>all</u>** the empty orbitals. For almost all the other fractional filling factors, the quantum transitions from nearest electron pairs are **<u>forbidden</u>** to **<u>some</u>** empty orbitals. The forbidden or allowed transitions are caused by the combined effects of the most uniform electron-configuration, Pauli's exclusion principle and the total momentum conservation in the x-direction. At the specific filling factors, the nearest pairs can transfer to all the empty orbitals. When the filling factor changes from this specific value, the number of allowed transitions abruptly decreases. This decreasing of the transitions produces the increase of the perturbation energy because the second order perturbation energy is negative for the ground state. Thus the perturbation energy is not a smooth function of $\nu$. It will be shown in the latter chapter that this discontinuous structure yields the plateaus in the Hall resistance curve. The reader who is interested in knowing the discontinuous structure of the perturbation energy may skip the latter sections and proceed to the next chapter.

The discontinuous structure appears in all orders of the perturbation energy and so exists in the exact solution. The higher order calculation will be examined in Chapter 6.

### 4.3 Binding energy at the filling factors $(2j)/(2j+1)$ and $(2j-1)/(2j)$

In this section, we study the states with the filling factors $\nu = (2j)/(2j+1)$ and $\nu = (2j-1)/(2j)$. The former has an odd number of the denominator and the latter has an even number of the denominator. We first examine an example $\nu = 4/5$ which belongs to the former case for $j = 2$. Figure 4.4 shows the most uniform electron-configuration at $\nu = 4/5$.

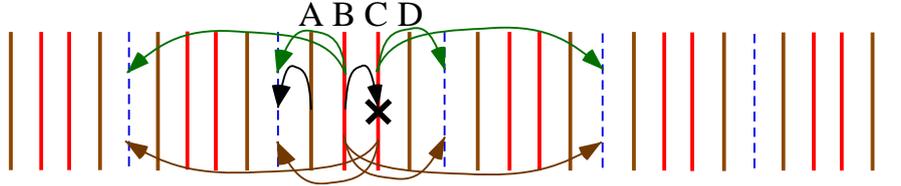

Fig.4.4　Allowed transitions of nearest electron pairs at $\nu = 4/5$

There are three nearest electron pairs AB, BC and CD in the unit-configuration. When the electron at site A is transferred to the first nearest orbital to the left in Fig.4.4, the electron at B should be transferred to the first orbital to the right because of the momentum conservation. However that orbital is already filled with electron and the transition is forbidden by the Fermi-Dirac statistics. Accordingly the nearest electron pairs AB and CD cannot transfer to any empty state. Only the electron pair BC can transfer to all the empty states. The total number of empty states is 1/5 times the number of all orbitals. Accordingly the pairs AB, BC and CD have the perturbation energy of the second order as

$$\varsigma_{AB} = 0, \quad \varsigma_{BC} = -(1/5)Z, \quad \varsigma_{CD} = 0$$

The number of the nearest electron pairs like BC is 1/4 times the total number of electrons $N$. Then we obtain the perturbation energy of all the nearest electron pairs $E_{\text{nearest pair}}$ as

$$E_{\text{nearest pair}}/N = -(1/5)Z\frac{N}{4}\frac{1}{N} = -(1/20)Z \quad \text{for} \quad \nu = 4/5 \tag{4.24}$$

Next, we discuss the case of $\nu = 6/7$. We draw the most uniform electron-configuration at $\nu = 6/7$ in Fig.4.5.

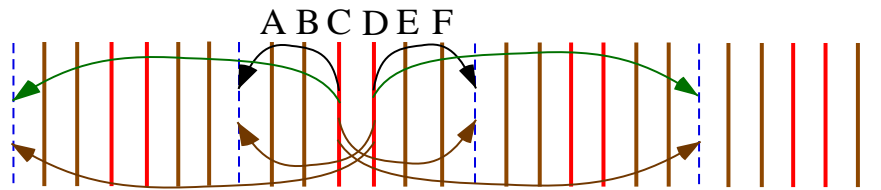

Fig.4.5　Allowed transitions of nearest electron pairs at $\nu = 6/7$

Only the electron pair CD in the unit-configuration can transfer to all the empty states. The total number of the empty states is 1/7 times the total number of orbitals. Accordingly the pair CD has the perturbation energy of the second order as

$$\varsigma_{CD} = -(1/7)Z$$

which gives

$$E_{\text{nearest pair}}/N = -(1/7)Z\frac{N}{6}\frac{1}{N} = -(1/42)Z \quad \text{for} \quad \nu = 6/7 \tag{4.25}$$

For the case of arbitrary integer $j$, the perturbation energy of the second order for $\nu = 2j/(2j+1)$ is given by multiplying the pair energy $-(1/(2j+1))Z$ and the total number of the pairs $N/(2j)$. Thereby

$$E_{\text{nearest pair}}/N = -(1/(2j+1))Z\frac{N}{2j}\frac{1}{N} = -\frac{1}{2j(2j+1)}Z \quad \text{for} \quad \nu = \frac{2j}{2j+1} \tag{4.26}$$

Next we calculate the perturbation energy with $\nu = (2j-1)/(2j)$, the denominator of which is an even integer. Fig.4.6 shows the most uniform configuration of electrons in the case of $j = 2$.

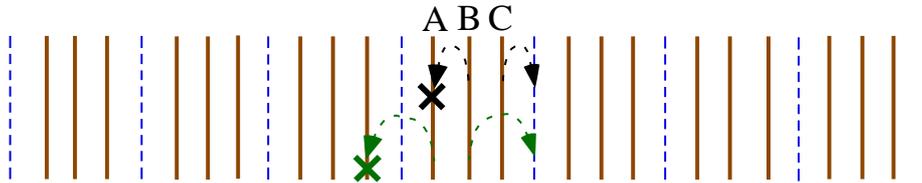

Fig.4.6 Nothing of transition from nearest electron pairs at $\nu = 3/4$

In this case, the *nearest electron pairs* AB and BC cannot transfer to all empty orbitals. Accordingly the perturbation energy of the *nearest electron pairs* is zero:

$$E_{\text{nearest pair}}/N = 0 \quad \text{for} \quad \nu = 3/4 \tag{4.27}$$

We illustrate the most uniform configurations for $j = 3$ and $4$ in Figs.4.7 and 4.8, respectively.

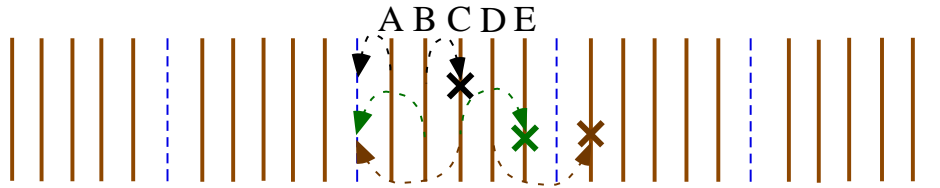

Fig.4.7 Nothing of transition from nearest electron pairs at $\nu = 5/6$

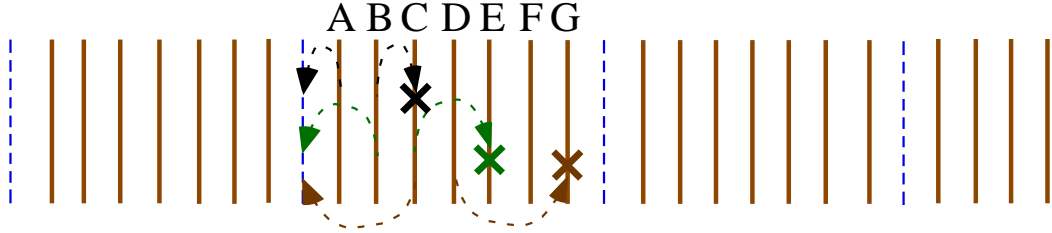

Fig.4.8 Nothing of transition from nearest electron pairs at $\nu = 7/8$

These figures show that all the quantum transitions from the *nearest electron pairs* are forbidden. Accordingly the energies of all the *nearest electron pairs* are zero at the filling factors $\nu = 5/6$, and $7/8$.

$$E_{\text{nearest pair}}/N = 0 \qquad \text{for} \qquad \nu = 5/6, \text{ and } 7/8 \qquad (4.28)$$

For the case of arbitrary integer $j$, the perturbation energies of all the nearest electron pairs at $\nu = (2j-1)/(2j)$ are equal to zero:

$$E_{\text{nearest pair}}/N = 0 \qquad \text{for} \qquad \nu = (2j-1)/(2j) \qquad (4.29)$$

Thus the *nearest electron pairs* at $\nu = (2j-1)/(2j)$ cannot transfer to all the empty states due to the combined effects of the most uniform electron-configuration, the Fermi-Dirac statistics of electrons and the momentum conservation [27-33].

Table 4.2 Second order perturbation energy of *nearest electron pairs* per electron for $\nu = (2j)/(2j+1)$ and $\nu = (2j-1)/(2j)$

| $\nu$ | $E_{\text{nearest pair}}/N$ |
|---|---|
| 1/2 | 0 |
| 2/3 | -(1/6) Z |
| 3/4 | 0 |
| 4/5 | -(1/20) Z |
| 5/6 | 0 |
| 6/7 | -(1/42) Z |
| 7/8 | 0 |

| 8/9 | -(1/72) Z |

Table 4.2 shows the second order perturbation energy of the *nearest electron pairs* at the filling factors with $\nu = (2j)/(2j+1)$ or $\nu = (2j-1)/(2j)$.

It is noteworthy that the *nearest pair* energies are zero in all order perturbation calculation for the filling factors $\nu = (2j-1)/(2j)$ because of no Coulomb transition. That is to say the exact binding energy of the *nearest electron pairs* is equal to zero at $\nu = (2j-1)/(2j)$.

### 4.4 Electron-Hole symmetry of nearest pairs

In this section we examine the case of the filling factor $\nu < \frac{1}{2}$ where the number of empty orbitals is greater than the number of orbitals occupied with electron. Hereafter we call empty orbital "hole". The electron-configuration at $\nu = \frac{1}{3}$ is shown in Fig.4.9.

This configuration is produced by repeating the unit-configuration (empty, filled, empty), which has the minimum energy of the classical Coulomb interaction.

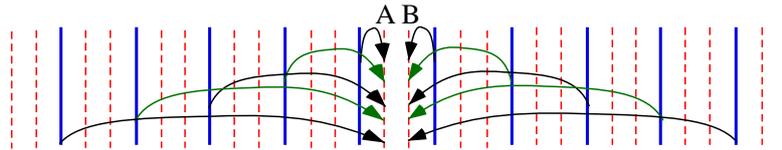

Fig.4.9   Coulomb transitions of nearest neighbor hole pair at $\nu = 1/3$

In this configuration, there is no *electron pair* placed in the nearest neighbour Landau orbitals. However there are *nearest hole pairs*. If we exchange electron for hole and hole for electron in Fig.4.9, we get the electron-configuration of $\nu = \frac{2}{3}$. It may be said that electron-hole symmetry appears between the $\nu = \frac{1}{3}$ and $\nu = \frac{2}{3}$ states.

We calculate the perturbation energy of the *nearest hole pair* AB which is denoted by $\varsigma_{AB}^{H}$. The hole pair AB is specified by the momenta $p_A$, $p_B$. The electron pair A'B' is also specified by the momenta $p'_A$, $p'_B$. Then, the electron pair A'B' transfers to the vacant orbitals at A and B as is illustrated by arrow pairs in Fig.4.9. Therein the momentum conservation is expressed as

$$p'_A = p_A - \Delta p \quad p'_B = p_B + \Delta p \tag{4.30a}$$

The momentum transfer $\Delta p$ takes the following values as

$$\Delta p = (3j+1)2\pi\hbar/\ell \quad \text{for} \quad j = 0, \pm 1, \pm 2, \pm 3, \cdots \tag{4.30b}$$

at $\nu = 1/3$. Then the second order perturbation energy of the hole pair AB is obtained by

$$\varsigma_{AB}^H = \sum_{\Delta p = (3j+1)2\pi\hbar/\ell \text{ for } j=0,\pm 1,\pm 2,\cdots} \frac{\langle p'_A, p'_B | H_I | p_A, p_B \rangle \langle p_A, p_B | H_I | p'_A, p'_B \rangle}{W_G - W_{excite}(p'_A \to p_A, p'_B \to p_B)} \tag{4.31}$$

Now we introduce the summation $Z_H$ as

$$Z_H = -\sum_{\Delta p \neq 0, -2\pi\hbar/\ell} \frac{\langle p'_A, p'_B | H_I | p_A, p_B \rangle \langle p_A, p_B | H_I | p'_A, p'_B \rangle}{W_G - W_{excite}(p'_A \to p_A, p'_B \to p_B)} \tag{4.32}$$

where the momentum transfer $\Delta p$ takes all the values $(2\pi\hbar/\ell) \times$ integer except $\Delta p = 0$ and $\Delta p = -2\pi\hbar/\ell$. The transferred states for $\Delta p = 0$ and $\Delta p = -2\pi\hbar/\ell$ are eliminated in the summation (4.32) because the diagonal element of $H_I$ is absent. The denominator in Eq.(4.32) is negative and so $Z_H$ is positive.

The perturbation energy of the *nearest hole pair*, $\varsigma_{AB}^H$, can be described by $Z_H$. Because the interval of momentum transfer is very small for the macroscopic size of the device, we obtain for $\nu = 1/3$

$$\varsigma_{AB}^H = -\frac{1}{3} Z_H \tag{4.33}$$

The total number of *nearest hole pairs* is equal to $(1/2)N_H$ at $\nu = 1/3$ where $N_H$ is the number of holes in the Landau states with $L = 0$. The perturbation energy $E_{\text{nearest pair}}$ is

$$E_{\text{nearest pair}} = \frac{1}{2} N_H \times \varsigma_{AB}^H = -\frac{1}{6} Z_H N_H \quad \text{for} \quad \nu = 1/3 \tag{4.34}$$

The perturbation energy of nearest hole pairs per hole is obtained as

$$E_{\text{nearest pair}} / N_H = -\frac{1}{6} Z_H \quad \text{for} \quad \nu = 1/3 \tag{4.35}$$

Comparison of Eqs.(4.16) and (4.32) reveals that $Z_H$ is close to $Z$:

$$Z_H \approx Z \quad \text{at the same strength of magnetic field} \tag{4.36}$$

This relation is derived from the fact that the right hand side of Eq.(4.32) has the same form as the right hand side of Eq. (4.16). So the value of $Z_H$ is almost equal to $Z$ at the same strength of magnetic field.

Also there is a symmetry between the filling factors $\nu$ and $1-\nu$ as will be explained below. One example is shown at $\nu = 3/5$ and $2/5$ as in Fig. 4.10.

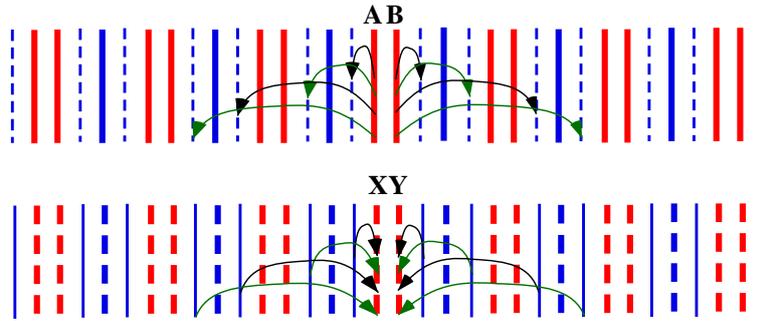

Fig.4.10   Electron hole symmetry at $\nu =$ 3/5 and 2/5

Electron orbitals are drawn by solid lines and hole orbitals are shown by dashed lines

As easily seen in this figure, the electron pair AB has two quantum transitions per unit-configuration and the hole pair XY has also two quantum transitions per unit-configuration. Therefore the electron pair AB has the perturbation energy $\varsigma_{AB}$ as

$$\varsigma_{AB} = (2/5)Z$$

The hole pair XY has the perturbation energy $\varsigma_{XY}^{H}$ as

$$\varsigma_{XY}^{H} = (2/5)Z_H$$

Therein the coefficient 2/5 comes from the ratio of the number of the allowed quantum transitions and the total number of Landau states. So the coefficient (2/5) at $\nu = 3/5$ is identical to the coefficient (2/5) at $\nu = 2/5$. Then the second order perturbation energy of all the *nearest hole pairs* at $\nu = 2/5$ is given by multiplying $\varsigma_{XY}^{H}$ and the total number of *nearest hole pairs*, $N_H/3$, as

$$E_{\text{nearest pair}} = \varsigma_{XY}^{H} \times \frac{1}{3} N_H = \frac{2}{15} Z_H N_H \qquad \text{for } \nu = 2/5 \qquad (4.37)$$

Next, we examine the quantum transitions at $\nu = 3/7$ which are drawn in Fig.4.11. All the nearest hole pairs (nearest dashed lines) can transfer to all electron states. That is to say all the electrons can transfer to the nearest hole pair AB. Counting of the transition number gives the second order perturbation energy of all the *nearest hole*

*pairs* per hole as

$$E_{\text{nearest pair}}/N_H = -\frac{3}{28}Z_H \qquad \text{for} \quad \nu = 3/7 \qquad (4.38)$$

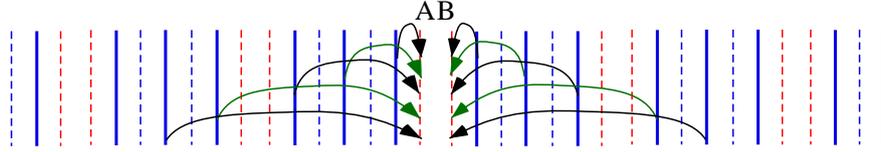

Fig.4.11  Coulomb transitions of nearest neighbour hole pair at $\nu = 3/7$

Consequently the number of the allowed transitions from the nearest electron pairs at $\nu$ $(1/2 < \nu < 1)$ is the same as that to the nearest hole pairs at the filling factor $1-\nu$. This symmetric property yields the following relation between the perturbation energies of the *nearest electron pairs* and the *nearest hole pairs*:

$$\left( E_{\substack{\text{nearest} \\ \text{hole pair}}}/N_H \right)_{\text{at filling factor } \nu} \approx \left( E_{\substack{\text{nearest} \\ \text{electron pair}}}/N \right)_{\text{at filling factor } (1-\nu)}$$

(4.39)

Then the perturbation energies of *nearest hole pairs* are listed in Table 4.3 where the centre column lists the energies of the nearest hole pairs per hole. We can confirm the electron-hole symmetry by comparing the centre column of Table 4.3 with Table 4.2. Hereafter "**nearest hole pair**" at $\nu < 1/2$ and "**nearest electron pair**" at $\nu > 1/2$ are simply named "**nearest pair**".

Table 4.3   Perturbation energy of *nearest hole pairs* per hole and per electron

| $\nu$ | $E_{\text{nearest pair}}/N_H$ | $E_{\text{nearest pair}}/N$ |
|---|---|---|
| 1/2 | 0 | 0 |
| 1/3 | -(1/6) $Z_H$ | -(1/3) $Z_H$ |
| 1/4 | 0 | 0 |
| 1/5 | -(1/20) $Z_H$ | -(1/5) $Z_H$ |
| 2/5 | -(2/15) $Z_H$ | -(1/5) $Z_H$ |
| 1/6 | 0 | 0 |
| 1/7 | -(1/42) $Z_H$ | -(1/7) $Z_H$ |
| 3/7 | $-(3/28) Z_H$ | $-(1/7) Z_H$ |
| 1/8 | 0 | 0 |
| 1/9 | $-(1/72) Z_H$ | $-(1/9) Z_H$ |
| 4/9 | $-(4/45) Z_H$ | $-(1/9) Z_H$ |
| 1/10 | 0 | 0 |
| 1/11 | $-(1/110) Z_H$ | $-(1/11) Z_H$ |
| 5/11 | $-(5/66) Z_H$ | $-(1/11) Z_H$ |
| 1/12 | 0 | 0 |
| 1/13 | $-(1/156) Z_H$ | $-(1/13) Z_H$ |
| 6/13 | $-(6/91) Z_H$ | $-(1/13) Z_H$ |
| 1/14 | 0 | 0 |

The rightmost column of Table 4.3 expresses the second order perturbation energies of nearest hole pairs per electron (not per hole). This energy indicates the energy of electron. Therefore the binding energy of electron is related to the value in the rightmost column of Table 4.3.

### 4.5 Character of the nearest electron pair and the nearest hole pair at the special filling factors $\nu = j/(2j\pm1)$, $\nu = j/(4j\pm1)$ and $\nu = (3j\pm1)/(4j\pm1)$

We calculate the nearest-pair energies $E_{\text{nearest pair}}$ for the filling factors $\nu = j/(2j\pm1)$, $\nu = j/(4j\pm1)$ and $\nu = (3j\pm1)/(4j\pm1)$. These filling factors approach $\nu = ½, ¼$ and $¾$, respectively in the limit of $j \to \infty$. We study the energies in the six regions of the following subsections.

### 4.5.1 (Region of $1/2 < \nu \leq 2/3$)

In the region of $1/2 < \nu \leq 2/3$, the most uniform electron-configuration is composed of two subunits $S_{11}$ and $S_{21}$ only, where $S_{11}$ is the configuration (filled, empty) and $S_{21}$ is (filled, filled, empty).

(Proof)

We consider a subunit $S_{12}$ (filled, empty, empty) which has the filling factor smaller than 1/2. Also a subunit $S_{31}$ is defined by (filled, filled, filled, empty) which has the filling factor larger than 2/3. If $S_{31}$ or $S_{12}$ is included in an electron-configuration, the configuration doesn't yield the most uniform distribution in $1/2 < \nu \leq 2/3$. Consequently the most uniform configuration includes the subunits $S_{11}$ and $S_{21}$ only in the region $1/2 < \nu \leq 2/3$.

Figure 4.12 shows the most uniform electron configuration at $\nu = 6/11$. The quantum transitions from the *nearest electron pair* AB are allowed to all the empty oribitals as in Fig.4.12. The electron at A can transfer to the left empty-orbitals as shown in black arrows ( $p'_A = p_A - \Delta p$, $p'_B = p_B + \Delta p$ for $\Delta p > 0$) and also to the right empty-orbitals as in green arrows ( $p'_A = p_A - \Delta p$, $p'_B = p_B + \Delta p$ for $\Delta p < 0$). These transitions conserve the total momentum in the x-direction certainly. Thus the electron pair AB is allowed to all the empty orbitals via the Coulomb interaction. This allowance to all the empty orbitals appears at any filling factor of $\nu = j/(2j-1)$ resulting in a large binding energy. We rewrite Eq.(4.23a) by replacing $j$ to $j+1$ as follows:

$$E_{\text{nearest pair}} = -\frac{1}{j+1} \times \frac{j}{(2j+1)} ZN \quad \text{for} \quad \nu = (j+1)/(2j+1) \qquad (4.40)$$

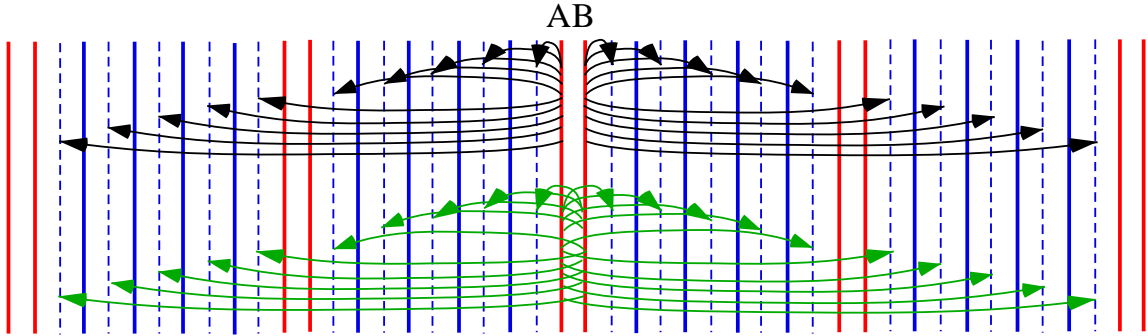

Fig.4.12  Allowed transitions from the nearest electron pair AB at $\nu = 6/11$

Substitution of $j = 5$ into Eq.(4.40) gives the pair energy $-(5/66)ZN$ at $\nu = 6/11$. When the filling factor changes from $\nu_0 = (j+1)/(2j+1)$ by an infinitesimally small value, then *nearest electron pairs* cannot transfer to some empty orbitals. Accordingly the number of the transitions decreases in the close vicinity of $\nu_0$. This property produces a discontinuous structure in the energy spectrum which will be investigated in the next chapter.

**4.5.2 (Region of $1/3 \leq \nu < 1/2$)**

In this region, the most uniform electron-configuration is composed of the subunits $S_{11}$(filled, empty) and $S_{12}$(filled, empty, empty) only. Figure 4.13 shows the quantum transitions of the hole pair AB at $\nu = 5/11$.

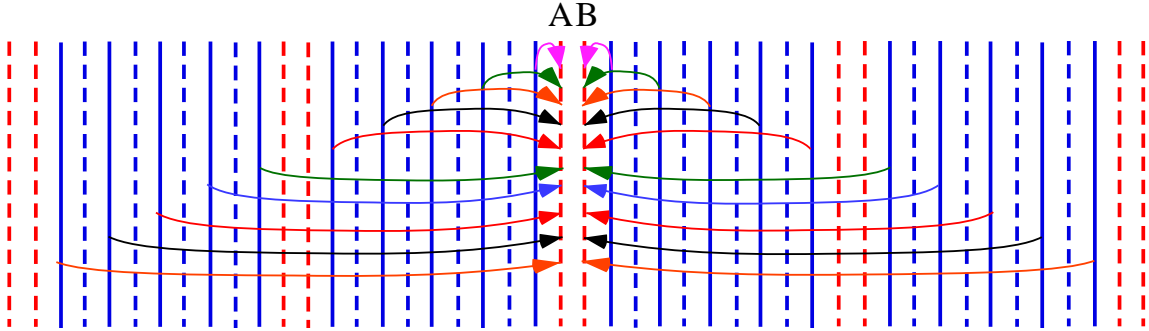

Fig.4.13   Allowed transitions of the nearest hole pair AB at $\nu = 5/11$

The *nearest hole pair* AB has the five allowed transitions for each unit-configuration composed of 11 Landau orbitals. That is to say the number of the allowed transitions from AB is 5/11 times all the orbitals. Also the number of *nearest hole pairs* is equal to 1/6 times the number of holes. Accordingly we obtain the perturbation energy as;

$$E_{\text{nearest pair}} = -\frac{1}{6} \times \frac{5}{11} Z_H N_H \qquad \text{for} \quad \nu = 5/11 \qquad (4.41)$$

Next we calculate the nearest-hole pair energies at $\nu = j/(2j+1)$. The electron-hole symmetry (4.39) expresses that the nearest-hole pair energy at $\nu = j/(2j+1)$ is equal to the nearest-electron pair energy at $1-\nu = (j+1)/(2j+1)$ given by Eq.(4.40). Then the perturbation energy of the *nearest hole pairs* is obtained as

$$E_{\text{nearest pair}} = -\frac{1}{j+1} \times \frac{j}{(2j+1)} Z_H N_H \quad \text{for} \quad \nu = j/(2j+1) \qquad (4.42)$$

The perturbation energy per hole is equal to

$$E_{\text{nearest pair}}/N_H = -\frac{1}{j+1} \times \frac{j}{(2j+1)} Z_H \quad \text{for} \quad \nu = j/(2j+1) \qquad (4.43)$$

The right hand side of this equation is a negative and the absolute value is a large value for small $j$. Therefore the state with $\nu = j/(2j+1)$ becomes stable.

### 4.5.3 (Region of $1/4 < \nu \leq 1/3$)

We count the number of quantum transitions from *nearest hole pairs* in the region of $1/4 < \nu \leq 1/3$. The most uniform electron-configuration is composed of subunits $S_{12}$(filled, empty, empty) and $S_{13}$(filled, empty, empty, empty) only. Figure 4.14 shows the most uniform electron-configuration at $\nu = 3/11$ which is an example of $\nu = j/(4j-1)$ for $j = 3$. All the electrons can transfer to the orbital pair AB as in Fig.4.14 where the arrows indicate the electron transitions. The movement of holes has the opposite direction against the arrows in Fig.4.14.

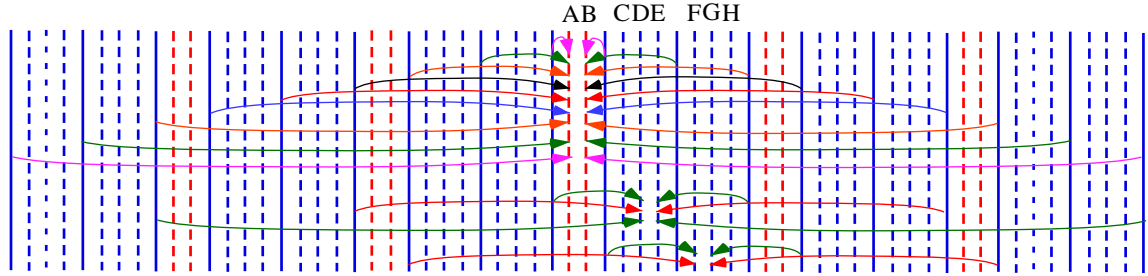

Fig.4.14   Allowed transitions of the nearest hole pair AB at $\nu = 3/11$

Three electron-pairs can transfer to the empty pair AB per unit configuration with eleven orbitals. Therefore the perturbation energy of the hole pair AB is given by

$$\varsigma_{AB}^{H} = -\frac{3}{11}Z_H \tag{4.44}$$

The *nearest hole pairs* DE and FG have the perturbation energies as follows;

$$\varsigma_{DE}^{H} = -\frac{1}{11}Z_H \quad \text{and} \quad \varsigma_{FG}^{H} = -\frac{1}{11}Z_H \tag{4.45}$$

The *nearest hole pairs* CD and GH have no allowed transition and therefore

$$\varsigma_{CD}^{H} = 0 \quad \text{and} \quad \varsigma_{GH}^{H} = 0 \tag{4.46}$$

The nearest hole pairs are expressed with nearest dashed-line pairs. The number of the pairs is 1/8 times the total number of holes namely $N_H$. Then the total perturbation energy of all the nearest hole pairs is equal to

$$E_{\text{nearest pair}} = \left(\varsigma_{AB}^{H} + \varsigma_{CD}^{H} + \varsigma_{DE}^{H} + \varsigma_{FG}^{H} + \varsigma_{GH}^{H}\right)\frac{N_H}{8} = -\frac{1}{8} \times \frac{5}{11} Z_H N_H \quad \text{for } \nu = 3/11 \tag{4.47}$$

The perturbation energy per hole is

$$E_{\text{nearest pair}}/N_H = -\frac{1}{8} \times \frac{5}{11} Z_H \qquad \text{for } \nu = 3/11 \tag{4.48}$$

We next examine the case of the filling factor $\nu = j/(4j-1)$. For any odd integer $j$, the *nearest hole pairs* have the following perturbation energy as

$$E_{\text{nearest pair}} = -\frac{1}{3j-1} \times \frac{1+3+\cdots+j+1+3+\cdots+(j-2)}{(4j-1)} Z_H N_H$$

$$E_{\text{nearest pair}} = -\frac{1}{3j-1} \times \frac{(1+j^2)/2}{(4j-1)} Z_H N_H \quad \text{for } \nu = j/(4j-1) \quad (4.49\text{a})$$

For any even integer $j$, the *nearest hole pairs* have the following perturbation energy as

$$E_{\text{nearest pair}} = -\frac{1}{3j-1} \times \frac{2+4+\cdots+j+2+4+\cdots+(j-2)}{(4j-1)} Z_H N_H$$

$$E_{\text{nearest pair}} = -\frac{1}{3j-1} \times \frac{j^2/2}{(4j-1)} Z_H N_H \quad \text{for } \nu = j/(4j-1) \quad (4.49\text{b})$$

Consequently the perturbation energy per hole is

$$E_{\text{nearest pair}}/N_H = -\frac{1}{3j-1} \times \frac{j^2+1}{2(4j-1)} Z_H \quad \text{for } \nu = j/(4j-1), \text{ odd integer } j \quad (4.50\text{a})$$

$$E_{\text{nearest pair}}/N_H = -\frac{1}{3j-1} \times \frac{j^2}{2(4j-1)} Z_H \quad \text{for } \nu = j/(4j-1), \text{ even integer } j \quad (4.50\text{b})$$

The *nearest hole pairs* shown by red dashed lines in Fig. 4.14 can transfer to all the electron orbitals. This means that the holes at the sites A and B are bound tightly at $\nu = j/(4j-1)$. So the $\nu = j/(4j-1)$ FQH states become stable.

### 4.5.4 (Region of $1/5 \leq \nu < 1/4$)

We examine the perturbation energy of the *nearest hole pair* with a filling factor in the region $1/5 \leq \nu < 1/4$. The most uniform electron-configuration is composed of subunits $S_{13}$(filled, empty, empty, empty) and $S_{14}$(filled, empty, empty, empty, empty) only. Figure 4.15 shows the most uniform electron-configuration at $\nu = 3/13$ which is an example of $\nu = j/(4j+1)$ for $j = 3$.

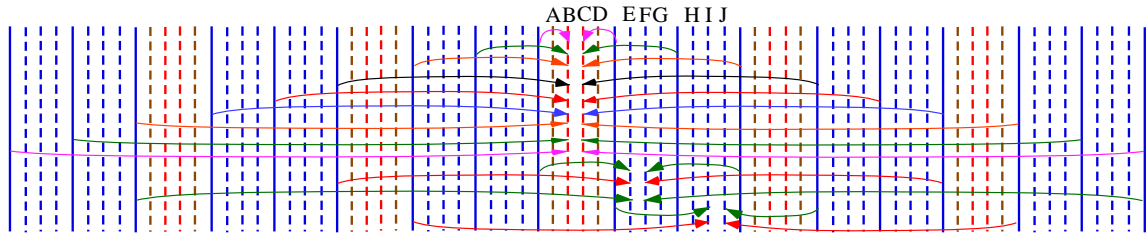

Fig.4.15  Allowed transitions of nearest hole pairs BC, EF and IJ at $\nu = 3/13$

All the electrons can transfer to the hole pair BC as illustrated by arrow pairs in Fig.4.15. Also a smaller number of electrons can transfer to the hole pairs EF and IJ. The total perturbation energy of all the nearest hole pairs is equal to

$$E_{\text{nearest pair}} = -\frac{1}{10} \times \left(\frac{3}{13} + \frac{1}{13} + \frac{1}{13}\right) Z_H N_H \quad \text{for} \quad \nu = 3/13 \quad (4.51)$$

The perturbation energy of nearest hole pairs per hole is

$$E_{\text{nearest pair}} / N_H = -\frac{1}{10} \times \frac{5}{13} Z_H \quad \text{for} \quad \nu = 3/13 \quad (4.52)$$

For any odd integer $j$, the *nearest hole pairs* have the following perturbation energy as

$$E_{\text{nearest pair}} = -\frac{1}{3j+1} \times \frac{j^2+1}{2(4j+1)} Z_H N_H \quad \text{for} \quad \nu = j/(4j+1) \text{ odd integer } j \quad (4.53a)$$

For any even integer $j$, the *nearest hole pairs* have the following perturbation energy as

$$E_{\text{nearest pair}} = -\frac{1}{3j+1} \times \frac{j^2}{2(4j+1)} Z_H N_H \quad \text{for} \quad \nu = j/(4j+1) \text{ even integer } j \quad (4.53b)$$

The perturbation energy per hole is

$$E_{\text{nearest pair}} / N_H = -\frac{1}{3j+1} \times \frac{j^2+1}{2(4j+1)} Z_H \quad \text{for} \quad \nu = j/(4j+1), \text{ odd integer } j \quad (4.54a)$$

$$E_{\text{nearest pair}} / N_H = -\frac{1}{3j+1} \times \frac{j^2}{2(4j+1)} Z_H \quad \text{for} \quad \nu = j/(4j+1), \text{ even integer } j \quad (4.54b)$$

In subsections 4.5.1-4.5.4 we have clarified that the *nearest electron* (or *hole*) *pair* has large binding energy at $\nu = j/(2j\pm1)$ and $\nu = j/(4j\pm1)$. In the next subsections 4.5.5-4.5.6 we will study the region $2/3 \leq \nu \leq 4/5$.

### 4.5.5 (Region of $2/3 \leq \nu < 3/4$)

We examine the case of $\nu = (3j-1)/(4j-1)$. Figure 4.16 shows the most uniform electron-configuration at $\nu = 8/11$ which is the case of $(3j-1)/(4j-1)$ for $j=3$. Therein the quantum transitions from the *nearest electron pairs* AB, DE and FG are illustrated by arrows. The electron pair AB are allowed to transfer to all the empty oribitals. The transition number is the three per unit-configuration.

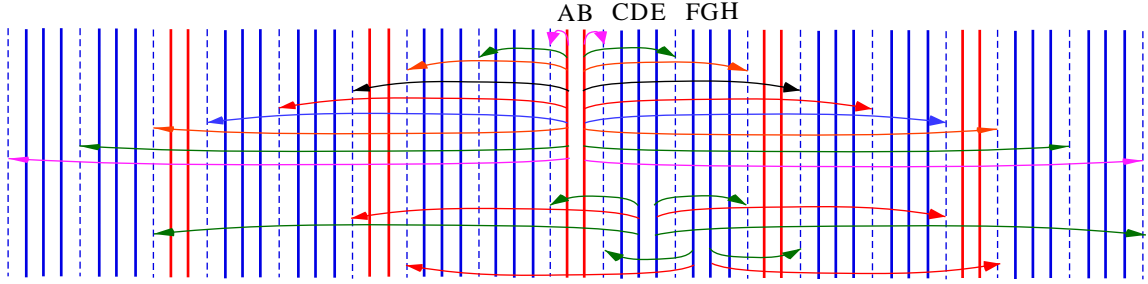

Fig.4.16  Allowed transitions from nearest electron pairs at $v = 8/11$

The electron pairs DE and FG can transfer to one empty orbital per unit-configuration. Then all the *nearest electron pairs* have the following perturbation energy.

$$E_{\text{nearest pair}} = -\frac{1}{8} \times \left(\frac{3}{11} + \frac{1}{11} + \frac{1}{11}\right) ZN = -\frac{5}{88} ZN \quad \text{for } v = 8/11 \quad (4.55)$$

Next we calculate the perturbation energy at $v = (3j-1)/(4j-1)$ for arbitrary integer $j$. For an odd integer $j$, all the *nearest electron pairs* have the following perturbation energy:

$$E_{\text{nearest pair}} = -\frac{1}{3j-1} \times \frac{j^2+1}{2(4j-1)} ZN \quad \text{for } v = (3j-1)/(4j-1) \quad (4.56a)$$

For an even integer $j$, the *nearest electron pairs* have the following perturbation energy:

$$E_{\text{nearest pair}} = -\frac{1}{3j-1} \times \frac{j^2}{2(4j-1)} ZN \quad \text{for } v = (3j-1)/(4j-1) \quad (4.56b)$$

The perturbation energy per electron is

$$E_{\text{nearest pair}}/N = -\frac{1}{3j-1} \times \frac{j^2+1}{2(4j-1)} Z$$
$$\text{for } v = (3j-1)/(4j-1), \text{ odd integer } j \quad (4.57a)$$

$$E_{\text{nearest pair}}/N = -\frac{1}{3j-1} \times \frac{j^2}{2(4j-1)} Z$$
$$\text{for } v = (3j-1)/(4j-1), \text{ even integer } j \quad (4.57b)$$

These nearest pair energies are negative and their absolute values are large. Accordingly the states with $\nu = (3j-1)/(4j-1)$ are stable.

**4.5.6 (Region of $3/4 < \nu \leq 4/5$)**

Figure 4.17 shows the quantum transitions from the *nearest electron pairs* BC, EF and IJ at $\nu = 10/13$ which is the case of $\nu = (3j+1)/(4j+1)$ for $j = 3$. The *nearest electron pairs* have the following perturbation energy.

$$E_{\text{nearest pair}} = -\frac{1}{10} \times \left(\frac{3}{13} + \frac{1}{13} + \frac{1}{13}\right) ZN = -\frac{5}{130} ZN \quad \text{for} \quad \nu = 10/13 \quad (4.58)$$

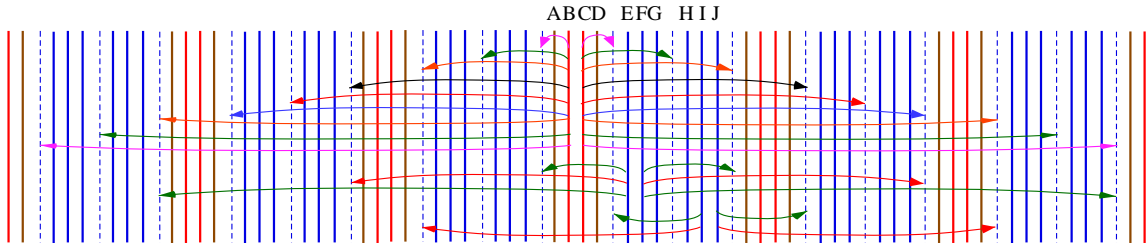

Fig.4.17　Allowed transitions from nearest electron pairs at $\nu = 10/13$

At $\nu = (3j+1)/(4j+1)$ with an odd integer $j$ the *nearest electron pairs* have the following perturbation energy:

$$E_{\text{nearest pair}} = -\frac{1}{3j+1} \times \frac{j^2+1}{2(4j+1)} ZN \quad \text{for} \quad \nu = (3j+1)/(4j+1) \quad (4.59a)$$

For an even integer $j$, the *nearest electron pairs* have the following perturbation energy:

$$E_{\text{nearest pair}} = -\frac{1}{3j+1} \times \frac{j^2}{2(4j+1)} ZN \quad \text{for} \quad \nu = (3j+1)/(4j+1) \quad (4.59b)$$

The perturbation energy per electron is

$$E_{\text{nearest pair}}/N = -\frac{1}{3j+1} \times \frac{j^2+1}{2(4j+1)} Z$$

$$\text{for} \quad \nu = (3j+1)/(4j+1), \text{ odd integer } j \quad (4.60a)$$

$$E_{\text{nearest pair}}/N = -\frac{1}{3j+1} \times \frac{j^2}{2(4j+1)} Z$$

$$\text{for } \nu = (3j+1)/(4j+1), \text{even integer } j \quad (4.60b)$$

These nearest pair energies are negative and their absolute values are large. Therefore the states with $\nu = (3j+1)/(4j+1)$ are stable.

In this chapter we have calculated the perturbation energy for the *nearest electron* (or *hole*) *pairs* at the filling factors $\nu = 1/(2j+1)$ (in Sec.4.4), $\nu = 2j/(2j+1)$ (in Sec.4.3), $\nu = j/(2j\pm 1)$, $\nu = j/(4j\pm 1)$, and $\nu = (3j\pm 1)/(4j\pm 1)$ (in Sec.4.5). Therein we can find the *nearest electron* (or *hole*) *pairs* transferring to all the empty (or occupied) orbitals. This property produces the large binding energy and then the states become stable. At the filling factors $\nu = 1/(2j)$ and $\nu = (2j-1)/(2j)$ (see Secs.4.4 and 4.3), the *nearest electron* (or *hole*) *pairs* cannot transfer to any empty (or occupied) orbitals. So the binding energy of the nearest electron (or hole) pairs is zero. This property yields no plateau of the Hall resistance at the filling factor of $\nu = 1/(2j)$ and $\nu = (2j-1)/(2j)$. In the next chapter we will examine the function-form of the energy versus filling factor. Then the function shows a discontinuous behaviour at the specific filling factors.

# Chapter 5 Valley, flat and peak structures in the energy spectrum

It is verified below that the energy of *nearest electron* (or *hole*) *pairs* takes a lowest value discontinuously at $\nu_0 = 1/(2j+1)$, $\nu_0 = 2j/(2j+1)$, $\nu_0 = j/(2j\pm1)$ and so on. That is to say the perturbation energy at the specific filling factors $\nu_0$ is lower than both limiting values from the left and right sides of $\nu_0$. We call this type of the discontinuity "**valley structure**". (Professor K. Katsumata advised me to use the name "valley".)

On the other hand the perturbation energy of *nearest electron* (or *hole*) *pairs* at $\nu = 1/2$ is continuously equal to the limiting value from its neighborhood. We call this case "**flat structure**".

The perturbation energy of *nearest electron* (or *hole*) *pairs* is zero at the filling factors of $\nu = 3/4,\ 1/4,\ 1/6,\cdots$ which is caused by the absence of the quantum transition as verified in Chapter 4. When the filling factor deviates from $\nu = 3/4,\ 1/4,\ 1/6,\cdots$, some of the *nearest electron* (or *hole*) *pairs* can transfer to the other orbitals. These allowed transitions yield the negative energy. Therefore the perturbation energy of the *nearest electron* (or *hole*) *pairs* at $\nu = 3/4,\ 1/4,\ 1/6,\cdots$ is higher than that in their neighbourhood. The case is named "**peak structure**". Thereby the probability of the state with $\nu = 3/4,\ 1/4,\ 1/6,\cdots$ is very small at a low temperature in comparison with that of the neighbourhood. This case yields the phenomenon similar to the case with the flat structure.

The valley structure is caused by the combined effects of the most uniform electron-configuration, the Fermi-Dirac statistics and the momentum conservation in the x-direction [30-33]. As will be shown in Chapter 7, the valley structure produces the plateaus in the Hall resistance when the magnetic field or gate voltage is varied.

We examine FQH states with $\nu < 1$ in sections 5.1-5.8 and $\nu > 2$ in section 5.9. The FQH states with $1 < \nu < 2$ are studied in section 5.11. It is noteworthy that Chapters 4-8 treat only the case of a strong magnetic field, where all the electron spins are directed opposite to the magnetic field at $\nu \leq 1$. The mixing of up and down spins will be investigated in Chapter 9.

## 5.1 Valley structure at the filling factor of $\nu = j/(2j\pm1)$

We examine the perturbation energy of the *nearest electron pairs* in the neighborhood of $\nu = j/(2j-1)$. As an example, we first study the case of $j = 2$ namely $\nu = 2/3$.

### 5.1.1 Valley structure at $\nu = 2/3$

The fractional number $\nu = (4s+1)/(6s+2)$ is smaller than 2/3 and approaches 2/3 in the limit of infinitely large $s$. We calculate the perturbation energy at $\nu = (4s+1)/(6s+2)$ and find the value in the limit $s \to \infty$.

As an example we consider the case of $s = 2$ that gives $\nu = 9/14$. We draw the electron configuration with the minimum classical Coulomb energy in Fig.5.1. This configuration is obtained by repeating the unit-configuration of (filled, empty, filled, filled, empty, filled, filled, empty, filled, filled, empty, filled, filled, empty). There are four *nearest electron pairs* AB, CD, EF and GH in the unit-configuration as seen in Fig.5.1.

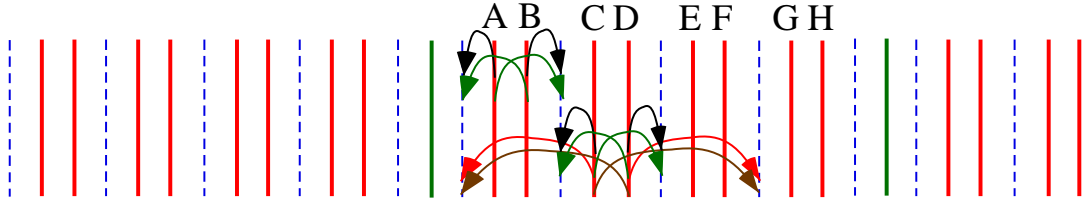

Fig.5.1 Coulomb transitions of nearest neighbour electron pairs at $\nu = 9/14$

We first examine the electron pairs AB and GH. Both electron pairs AB and GH can transfer to the two empty states per unit-configuration as in Fig.5.1. In the quantum transition from AB the momentum transfer $\Delta p$ is defined by $p'_B - p_B$ which takes the following values via the Coulomb transitions:

$$\Delta p = (14j-2)2\pi\hbar/\ell, (14j+1)2\pi\hbar/\ell \qquad \text{for } j = 0, \pm 1, \pm 2, \cdots \qquad (5.1)$$

That is to say the number of allowed transitions is two among the fourteen orbitals. Denoting the second order perturbation energy of the pair AB by the symbol $\varsigma_{AB}$, we obtain the perturbation energy as

$$\varsigma_{AB} = -(2/14)Z$$

where $Z$ have been already defined in Eq.(4.16).

The electron pairs CD and EF can transfer to the four empty states per unit-configuration. Then the perturbation energies of the pairs CD and EF, namely

$\varsigma_{CD}, \varsigma_{EF}$ are given by
$$\varsigma_{CD} = \varsigma_{EF} = -(4/14)Z$$

Similarly the perturbation energies of the pair GH, $\varsigma_{GH}$, is given by
$$\varsigma_{GH} = -(2/14)Z$$

Then the total perturbation energy of all the *nearest electron pairs* is
$$E_{\text{electron pair}} = [-(2/14)Z - (4/14)Z - (4/14)Z - (2/14)Z] \times (N/9)$$
$$= -(12/(14 \times 9))ZN \qquad \text{for } \nu = 9/14 \qquad (5.2)$$

where $N$ is the total number of electrons. Next we examine the case of $s = 3$ which gives the filling factor
$$\nu = (4s+1)/(6s+2) = 13/20 \ .$$

The most uniform configuration of electrons for the $\nu = 13/20$ state is shown in Fig.5.2

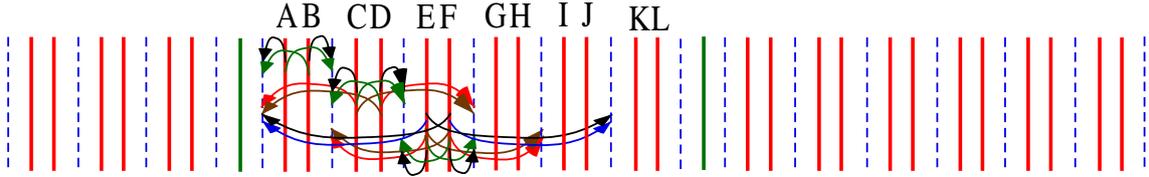

Fig.5.2 Coulomb transitions of nearest neighbour electron pairs at $\nu = 13/20$

In this electron configuration the six *nearest electron pairs* AB, CD, EF, GH, IJ and KL exist in the unit-configuration. These pairs have the perturbation energies as

$$\varsigma_{AB} = -(2/20)Z \ , \quad \varsigma_{CD} = -(4/20)Z \ , \quad \varsigma_{EF} = -(6/20)Z \ , \quad \varsigma_{GH} = -(6/20)Z \ ,$$

$$\varsigma_{IJ} = -(4/20)Z \ , \qquad \varsigma_{KL} = -(2/20)Z \qquad \text{at } \nu = 13/20$$

The sum of the perturbation energies of all the nearest electron pairs is equal to

$$E_{\text{nearest pair}} = [-(2/20)Z - (4/20)Z - (6/20)Z - (6/20)Z - (4/20)Z - (2/20)Z] \times (N/13)$$
$$= -(24/(20 \times 13))ZN \qquad \text{for } \nu = 13/20 \qquad (5.3)$$

Similarly we can calculate the sum of the perturbation energies at $\nu = (4s+1)/(6s+2)$ for arbitrary integer $s$. Therein the unit-configuration is composed of $6s+2$ Landau orbitals, $4s+1$ of which are occupied by electrons. Therefore the number of empty states is $2s+1$ per unit-configuration. Accordingly

$$E_{\text{nearest pair}} = \begin{bmatrix} -(2/(6s+2))Z - (4/(6s+2))Z \cdots - (2s/(6s+2))Z \cdots \\ -(4/(6s+2))Z - (2/(6s+2))Z \end{bmatrix} \times (N/(4s+1))$$

$$= -\frac{2s(s+1)}{(4s+1)(6s+2)}ZN$$

$$E_{\text{nearest pair}} = -\frac{2s^2+2s}{(4s+1)(6s+2)}ZN \qquad \text{for } \nu=(4s+1)/(6s+2) \qquad (5.4)$$

This result leads the limiting values as follows:

$$\nu = (4s+1)/(6s+2) \xrightarrow[s\to\infty]{} (2/3) \qquad (5.5a)$$

$$E_{\text{nearest pair}}/N \xrightarrow[\nu=(4s+1)/(6s+2),\ s\to\infty]{} -(1/12)Z \qquad \text{for } \nu=(2/3)-\varepsilon \qquad (5.5b)$$

where $\varepsilon$ indicates a positive infinitesimal and the symbol $(2/3)-\varepsilon$ means the limiting process from the left.
The previous result (4.18) gives

$$E_{\text{nearest pair}}/N = -(1/6)Z \qquad \text{at } \nu=2/3 \qquad (5.6)$$

When the filling factor $\nu$ approaches $2/3$ from the left, the energy of the *nearest electron pairs* per electron approaches $-(1/12)Z$ which is half of the energy at $\nu=2/3$. That is to say, the energy at $\nu=2/3$ is lower than the limiting value from the left with a gap.

We next calculate the limiting value from the right. The filling factor of $\nu=(4s-1)/(6s-2)$ is larger than $2/3$ and approaches $2/3$ in the limit of $s\to\infty$. The electron configuration for the case of $s=3$ is illustrated in Fig.5.3.

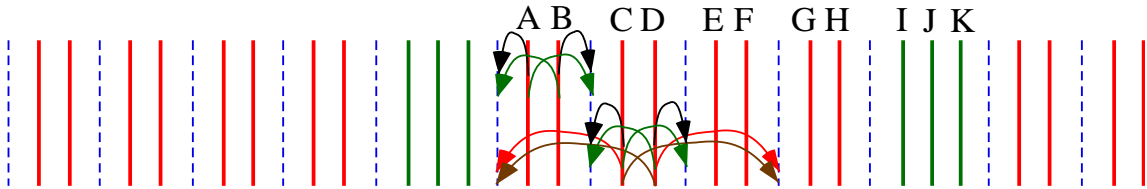

Fig.5.3 Coulomb transitions of nearest neighbour electron pairs at $\nu=11/16$

The second order perturbation energies of the nearest pairs AB, CD, EF, GH, IJ and JK are expressed by $\varsigma_{AB}$, $\varsigma_{CD}$, $\varsigma_{EF}$, $\varsigma_{GH}$, $\varsigma_{IJ}$ and $\varsigma_{JK}$ respectively which are

$$\varsigma_{AB} = \varsigma_{GH} = -(2/16)Z, \quad \varsigma_{CD} = \varsigma_{EF} = -(4/16)Z, \quad \varsigma_{IJ} = \varsigma_{JK} = 0$$

The sum of all the energies of the *nearest electron pairs* becomes

$$E_{\text{nearest pair}} = \left[-(2/16)Z - (4/16)Z - (4/16)Z - (2/16)Z\right] \times (1/11)N$$
$$= -(12/(16 \times 11))ZN \qquad \text{for } \nu = 11/16 \qquad (5.7)$$

Next we consider the case $s = 4$. Figure 5.4 shows the electron configuration with the minimum classical Coulomb energy at $\nu = 15/22$.

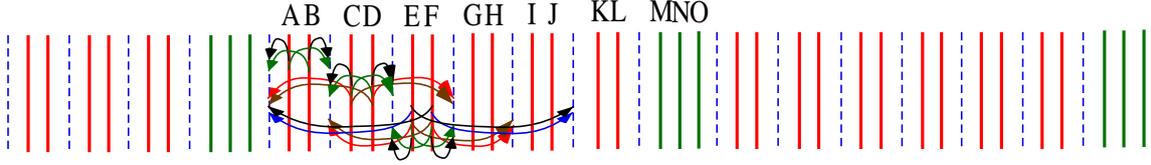

Fig.5.4 Coulomb transitions of nearest neighbour electron pairs at $\nu = 15/22$

The unit-configuration is composed of twenty-two sequential Landau states, fifteen of which are filled with electron as illustrated in Fig.5.4. There are eight *nearest electron pairs* AB, CD, EF, GH, IJ, KL, MN and NO in the unit-configuration. These pairs have the perturbation energy as

$$\varsigma_{AB} = \varsigma_{KL} = -(2/22)Z, \quad \varsigma_{CD} = \varsigma_{IJ} = -(4/22)Z,$$

$$\varsigma_{EF} = \varsigma_{GH} = -(6/22)Z, \quad \varsigma_{MN} = \varsigma_{NO} = 0, \quad \text{for } \nu = 15/22$$

Then the total perturbation energy of all the *nearest electron pairs* is equal to

$$E_{\text{nearest pair}} = \left[-2 \times (2/22)Z - 2 \times (4/22)Z - 2 \times (6/22)Z\right] \times (1/15)N$$
$$= -(24/(22 \times 15))ZN \qquad \text{for } \nu = 15/22 \qquad (5.8)$$

We calculate the perturbation energy at the filling factor $\nu = (4s-1)/(6s-2)$ for any integer $s$. The unit-configuration is composed of $6s-2$ Landau states, $4s-1$ of which are occupied by electron. Then the number of empty states is $2s-1$ per unit-configuration. Accordingly we get the energy for all the nearest pairs as

$$E_{\text{nearest pair}} = \left[-2 \times (2/(6s-2))Z - 2 \times (4/(6s-2))Z \cdots - 2 \times (2(s-1)/(6s-2))Z\right] \times (N/(4s-1))$$
$$= -\frac{2s(s-1)}{(4s-1)(6s-2)}ZN$$

$$E_{\text{nearest pair}} = -\frac{2s^2 - 2s}{(4s-1)(6s-2)} ZN \qquad \text{for } \nu = (4s-1)/(6s-2) \qquad (5.9)$$

The limiting values are

$$\nu = (4s-1)/(6s-2) \xrightarrow{s \to \infty} (2/3) \qquad (5.10a)$$

$$E_{\text{nearest pair}}/N \xrightarrow[\nu=(4s-1)/(6s-2),\ s\to\infty]{} -(1/12)Z \qquad \text{for} \quad \nu = (2/3) + \varepsilon \qquad (5.10b)$$

where the symbol $(2/3) + \varepsilon$ indicates the limiting process from the right. The energy of the *nearest electron pairs* per electron at $\nu = 2/3$ is equal to $-(1/6)Z$ which is twice of the limiting value $-(1/12)Z$ from the right. Therefore the $\nu = 2/3$ state has a lower energy than the limiting one from the right. Consequently Eqs.(5.5b), (5.10b) and (5.6) indicate that the perturbation energy is discontinuous at $\nu = 2/3$ and has a valley structure. This valley structure yields the stability of the state with $\nu = 2/3$.

Some readers might consider that the lowest property at $\nu = 2/3$ is caused by an even denominator of the fractional number $\nu = (4s \pm 1)/(6s \pm 2)$. We also examine the state at the filling factor $\nu = (4s \pm 1)/(6s \pm 1)$ with an odd denominator in Appendix. The results of the limiting values are the same as in this section.

### (Function form of total energy)

We express the function form of the total energy $E(\nu)$ in the neighborhood of $\nu = 2/3$. There are many non-nearest electron pairs, in addition to the nearest-neighboring ones. The energy of all the non-nearest electron pairs is denoted by the symbol $g(\nu)N$. We already used $g(1/2)$ in Eqs.(4.21) and (4.22). Then the total energy $E(\nu)$ is a sum of four terms, namely, the energy of all the nearest pairs, the energy of all the non-nearest pairs, the Landau energy and the classical Coulomb energy as follows: (Note: The classical Coulomb energy has been obtained in Eq.(3.22b) )

$$E(\nu) = E_{\text{nearest pair}} + g(\nu)N + [f + \hbar eB/(2m^*) - (\xi - \eta)/\nu]N + C_{\text{Macroscopic}}(\sigma) \qquad (5.11a)$$

Hereafter the total energy per electron is described by the symbol $\varepsilon(\nu)$ as

$$\varepsilon(\nu) = E(\nu)/N \qquad (5.11b)$$

There are contributions from the higher order perturbation which will be studied in the next chapter. The exact nearest-pair energy per electron, $\chi(\nu)$, is a sum of all order perturbation energies as

$$\chi(\nu) = \sum_{n=2,3,4,\cdots} \chi_n(\nu) \qquad (5.12a)$$

where the symbol $\chi_n(\nu)$ indicates the n-th order perturbation energy of all the nearest electron pairs per electron. The second order term $\chi_2(\nu)$ was already calculated in chapters 4 and 5. For confirmation we write the following relation:

$$E_{\text{nearest pair}}/N = \chi(\nu) \approx \chi_2(\nu) \tag{5.12b}$$

Eqs.(5.11a,b) and (5.12a,b) yield

$$\varepsilon(\nu) = \chi(\nu) + g(\nu) + [f + \hbar eB/(2m^*) - (\xi - \eta)/\nu] + C_{\text{Macroscopic}}/N \tag{5.13}$$

Eqs.(5.5b) (5.6) and (5.10b) are rewritten as

$$\chi_2\left(\tfrac{2}{3}\right) = -\tfrac{1}{6} Z \tag{5.14a}$$
$$\chi_2\left(\tfrac{2}{3} \pm \varepsilon\right) = -\tfrac{1}{12} Z \tag{5.14b}$$

Equations (5.14a) and (5.14b) express a discontinuous property of the energy at $\nu = 2/3$ [30-33].

### 5.1.2 Valley structure at $\nu = 3/5$

We examine the FQH states in the neighborhood of $\nu = 3/5$. The filling factor $\nu = (6s-1)/(10s-2)$ is larger than $3/5$ and has the limiting value $3/5$ for infinitely large integer $s$. An example is $\nu = 11/18$ for $s = 2$ whose most uniform electron configuration is illustrated in Fig.5.5.

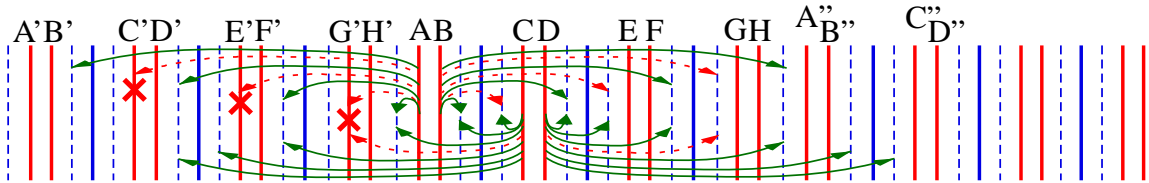

Fig.5.5 Coulomb transitions of nearest neighbour electron pairs at $\nu = 11/18$

The most uniform electron configuration is obtained by repeating the unit-configuration (filled, filled, empty, filled, empty, filled, filled, empty, filled, empty, filled, filled, empty, filled, empty, filled, filled, empty) in which 18 sequential Landau states are partially filled with 11 electrons. There are four *nearest electron pairs* AB, CD, EF and GH in the unit-configuration as in Fig.5.5. The electron pair AB can be transferred to the four sites in a unit-configuration shown by four arrow-pairs but cannot be transferred to the other sites shown by dashed arrow-pairs because of the momentum conservation and the Pauli exclusion principle. Also the pair CD is transferred to the six sites in a unit-configuration indicated by six solid arrow-pairs. Then the nearest pairs AB, CD, EF and GH have the second order perturbation energy as

$$\varsigma_{AB} = -(4/18)Z, \quad \varsigma_{CD} = -(6/18)Z, \quad \varsigma_{EF} = -(6/18)Z,$$

$$\varsigma_{GH} = -(4/18)Z, \qquad \text{for } \nu = 11/18 \qquad (5.15)$$

The perturbation energy of all the *nearest electron pairs* is equal to

$$\begin{aligned}E_{\text{nearest pair}} &= [-2\times(4/18)Z - 2\times(6/18)Z]\times(1/11)N \\ &= -(20/(18\times 11))ZN \qquad \text{for } \nu = 11/18\end{aligned} \qquad (5.16)$$

We examine the case of $s=3$, namely $\nu = (6s-1)/(10s-2) = 17/28$. Figure 5.6 shows the most uniform electron configuration. Therein the dashed arrow pairs indicate forbidden transitions and the other arrow pairs indicate allowed transitions. The nearest electron pairs AB and KL are allowed the six transitions per unit-configuration, the pairs CD and IJ eight and the pairs EF and GH ten transitions. Then the pair energies are given by

$$\varsigma_{AB} = \varsigma_{KL} = -(6/28)Z, \quad \varsigma_{CD} = \varsigma_{IJ} = -(8/28)Z,$$

$$\varsigma_{EF} = \varsigma_{GH} = -(10/28)Z, \qquad \text{for } \nu = 17/28$$

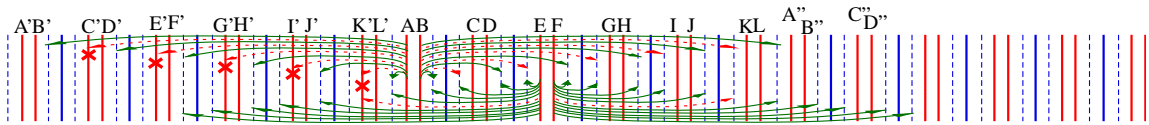

Fig.5.6 Coulomb transitions of nearest neighbour electron pairs at $\nu = 17/28$

Therefore the total transition energy of all the *nearest electron pairs* is equal to
$$\begin{aligned}E_{\text{nearest pair}} &= [-2\times(6/28)Z - 2\times(8/28)Z - 2\times(10/28)Z]\times(1/17)N \\ &= -(48/(28\times 17))ZN \qquad \text{for } \nu = 17/28\end{aligned}$$

We calculate the perturbation energy at the filling factor $\nu = (6s-1)/(10s-2)$ for any integer $s$. The unit-configuration is composed of $10s-2$ Landau states, $6s-1$ of which are filled with electron. Therefore the number of empty states is $4s-1$ per unit-configuration. Accordingly

$$E_{\text{nearest pair}} = \begin{bmatrix} -2\times(2s/(10s-2))Z - 2\times((2s+2)/(10s-2))Z \cdots\cdots \\ -2\times((4s-4)/(10s-2))Z - 2\times((4s-2)/(10s-2))Z \end{bmatrix} \times (1/(6s-1))N$$

$$= -\frac{s(6s-2)}{(6s-1)(10s-2)}ZN$$

$$E_{\text{nearest pair}} = -\frac{s(6s-2)}{(6s-1)(10s-2)}ZN \qquad \text{for } \nu = (6s-1)/(10s-2) \qquad (5.17)$$

The limiting values are

$$\nu = (6s-1)/(10s-2) \xrightarrow[s\to\infty]{} (3/5) \qquad (5.18a)$$

$$E_{\text{nearest pair}}/N \xrightarrow[\nu=(6s-1)/(10s-2),\ s\to\infty]{} -(1/10)Z \quad \text{for } \nu = (3/5)+\varepsilon \qquad (5.18b)$$

The previous result Eq.(4.23a) gives the perturbation energy at $\nu = 3/5$ as,

$$E_{\text{nearest pair}}/N = -\frac{1}{j}\times\frac{j-1}{(2j-1)}Z = -\frac{2}{15}Z \quad \text{at } \nu = 3/5 \qquad (5.18c)$$

The limiting value of the perturbation energy $-(1/10)Z$ is (3/4) times the nearest pair energy per electron at $\nu = 3/5$ and so discontinuous. Therefore the energy gap appears at $\nu = 3/5$.

Next we calculate the limiting value from the left. The filling factor $\nu = (6s+1)/(10s+2)$ is smaller than 3/5. First, we consider the case of $s=2$. The most uniform electron configuration at $\nu = 13/22$ is schematically drawn in Fig.5.7.

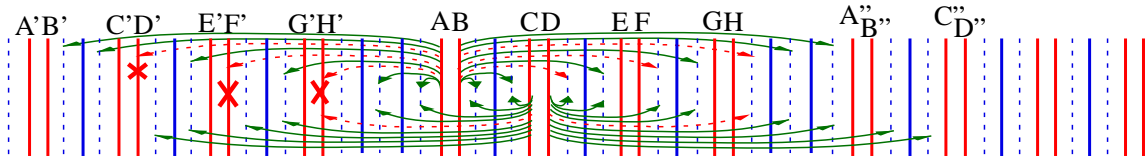

Fig.5.7 Coulomb transitions of nearest neighbour electron pairs at $\nu = 13/22$

There are four *nearest electron pairs* AB, CD, EF and GH in the unit-configuration. The electron pair AB and GH transfer to the six empty states and the pairs CD and EF also transfer to the eight empty states in a unit-configuration. These allowed transitions are illustrated by solid arrow pairs in Fig.5.7. The dashed arrow pairs indicate forbidden transitions. The perturbation energies are given by

$$\varsigma_{AB} = \varsigma_{GH} = -(6/22)Z, \quad \varsigma_{CD} = \varsigma_{EF} = -(8/22)Z \qquad \text{for } \nu = 13/22 \quad (5.19)$$

Then the sum of the perturbation energies for all the *nearest electron pairs* is equal to

$$\begin{aligned} E_{\text{nearest pair}} &= [-2\times(6/22)Z - 2\times(8/22)Z]\times(1/13)N \\ &= -(28/(13\times 22))ZN \qquad \text{for } \nu = 13/22 \end{aligned} \quad (5.20)$$

We calculate the perturbation energy for arbitrary integer $s$. At the filling factor $\nu = (6s+1)/(10s+2)$ the unit-configuration is composed of $10s+2$ sequential Landau states, $6s+1$ of which are occupied by electron. Therefore the number of empty states is $4s+1$ per unit-configuration. Accordingly

$$\begin{aligned} E_{\text{nearest pair}} &= \begin{bmatrix} -2\times((2s+2)/(10s+2))Z - 2\times((2s+4)/(10s+2))Z \cdots \\ \cdots -2\times(4s/(10s+2))Z \end{bmatrix} \times (1/(6s+1))N \\ &= -\frac{(6s+2)s}{(6s+1)(10s+2)}ZN \end{aligned}$$

$$E_{\text{nearest pair}} = -\frac{6s^2+2s}{(6s+1)(10s+2)}ZN \qquad \text{for } \nu = (6s+1)/(10s+2) \quad (5.21)$$

The limiting values are

$$\nu = (6s+1)/(10s+2) \xrightarrow[s\to\infty]{} 3/5 \quad (5.22\text{a})$$

$$E_{\text{nearest pair}}/N \xrightarrow[\nu=(6s+1)/(10s+2),\ s\to\infty]{} -(1/10)Z \quad \text{for } \nu = (3/5)-\varepsilon \quad (5.22\text{b})$$

The previous result (4.23a) gives

$$E_{\text{nearest pair}}/N = -\frac{1}{j}\times\frac{j-1}{(2j-1)}Z = -\frac{2}{3\times 5}Z = -\frac{2}{15}Z \quad \text{at } \nu = 3/5 \quad (5.22\text{c})$$

Thus Eqs.(5.18b), (5.22b) and (5.22c) indicate a discontinuous structure at $\nu = 3/5$. The energy of all the *nearest electron pairs* per electron is equal to $-2Z/15$ at $\nu = 3/5$ which is lower than the limiting value ($-Z/10$) from the both sides.

### 5.1.3 Valley structure at $\nu = j/(2j-1)$

In this subsection we examine the perturbation energies in the neighbourhood of $\nu = j/(2j-1)$ for any integer $j$. The filling factor $\nu = (j(2s)-1)/((2j-1)(2s)-2)$ approaches $\nu = j/(2j-1)$ in the limit of infinitely large $s$. As an example, we consider the case of $j = 4$, $s = 2$, where the filling factor is equal to

$$\nu = (j(2s)-1)/((2j-1)(2s)-2) = 15/26$$

The most uniform electron configuration is illustrated in Fig.5.8 where the allowed transitions are drawn with solid arrow pairs and the forbidden transitions with dashed arrow pairs.

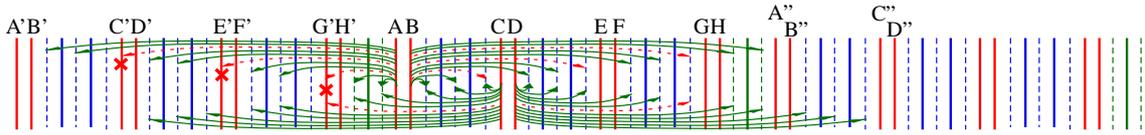

Fig.5.8 Coulomb transitions of nearest neighbour electron pairs at $\nu = 15/26$

The number of the allowed transitions is eight for the pairs AB and GH, and is ten for the pairs CD and EF per unit-configuration:

$$\varsigma_{AB} = \varsigma_{GH} = -(8/26)Z, \quad \varsigma_{CD} = \varsigma_{EF} = -(10/26)Z \quad \text{at} \quad \nu = 15/26$$

$$E_{\text{nearest pair}} = [-2\times(8/26)Z - 2\times(10/26)Z]\times(1/15)N = -\frac{36}{26\times15}ZN \quad \text{at} \quad \nu = 15/26$$

The procedure given above can be extended to arbitrary integers $j$ and $s$ as follows:
 Let us introduce three parameters, $\alpha$, $\beta$ and $\gamma$ to represent, respectively, the number of orbitals in the unit-configuration, the number of orbitals filled with electron, and the number of empty orbitals in the unit-configuration. These parameters are explicitly given at the filling factor $\nu = (j(2s)-1)/((2j-1)(2s)-2)$ by,

$$\alpha = ((2j-1)(2s)-2), \quad \beta = (j(2s)-1), \quad \gamma = ((j-1)(2s)-1) \quad (5.23)$$

By extending the calculation at $\nu = 15/26$ to any integer $s$, the total energy for all the *nearest electron pairs* for $j = 4$ is given by,

$$E_{\text{nearest pair}} = [-2\times((4s)/\alpha)Z - 2\times((4s+2)/\alpha)Z \cdots -2\times((6s-2)/\alpha)Z]\times\frac{N}{\beta} \quad (5.24)$$

Therein $(6s-2)$ is equal to $\gamma-1$ for $j=4$ and also $(4s)$ is equal to $\gamma+1-2s$.

Accordingly we obtain

$$E_{\text{nearest pair}} = [-2\times((\gamma+1-2s)/\alpha)Z - 2\times((\gamma+3-2s)/\alpha)Z\cdots - 2\times((\gamma-1)/\alpha)Z]\times\frac{N}{\beta}$$

$$E_{\text{nearest pair}} = [-2\times(\gamma+1-2s) - 2\times(\gamma+3-2s)\cdots - 2\times(\gamma-1)]\frac{Z}{\alpha}\times\frac{N}{\beta}$$

$$E_{\text{nearest pair}} = [-(2\gamma-2s)s]\frac{Z}{\alpha}\times\frac{N}{\beta} \qquad \text{at } \nu = \beta/\alpha \qquad (5.25)$$

Substitution of Eq.(5.23) into Eq.(5.25) yields

$$E_{\text{nearest pair}} = -\frac{((2j-3)2s-2)s}{((2j-1)(2s)-2)}\frac{ZN}{(j(2s)-1)} \qquad \text{at } \nu = \frac{(j(2s)-1)}{((2j-1)(2s)-2)} \qquad (5.26)$$

Thus we have obtained the perturbation energy of the nearest electron pairs for arbitrary integers of $j$ and $s$. Eq.(5.26) gives the limiting value from the right (for $s\to\infty$).

$$\nu = \beta/\alpha = (j(2s)-1)/((2j-1)(2s)-2) \xrightarrow{s\to\infty} j/(2j-1) \qquad (5.27a)$$

$$E_{\text{nearest pair}} = -\frac{((2j-3)2s-2)s}{((2j-1)(2s)-2)}\frac{ZN}{(j(2s)-1)} \xrightarrow{s\to\infty} -\frac{(2j-3)ZN}{2j(2j-1)} \quad \text{at } \nu = \frac{j}{2j-1}+\varepsilon \quad (5.27b)$$

From Eq.(4.23a)

$$E_{\text{nearest pair}} = -\frac{(j-1)ZN}{j(2j-1)} \qquad \text{at } \nu = \frac{j}{2j-1} \qquad (5.27c)$$

$$\chi_2\left(\tfrac{j}{2j-1}\right) = -\frac{(j-1)Z}{j(2j-1)} \qquad (5.27d)$$

where we have used Eq.(5.12b). The energy $E_{\text{nearest pair}}$ at $\nu = j/(2j-1)$ is compared with the limiting value, Eq.(5.27b), and we find the former is lower than the latter. The difference is equal to

$$\Delta E_{\text{nearest pair}} = E(\nu) - \lim_{\lambda\to\nu+\varepsilon} E(\lambda) = -\frac{ZN}{2j(2j-1)} \qquad \text{for } \nu = \frac{j}{2j-1} \qquad (5.28)$$

Next we calculate the limiting value from the left. We consider the filling factor $\nu = (j(2s)+1)/((2j-1)(2s)+2)$ which is smaller than $\nu = j/(2j-1)$. We introduce three parameters $\alpha', \beta', \gamma'$ defined by

$$\alpha' = ((2j-1)(2s)+2),\ \beta' = (j(2s)+1),\ \gamma' = ((j-1)(2s)+1) \qquad (5.29)$$

The parameter $\alpha'$ represents the number of orbitals in the unit-configuration, $\beta'$ the number of electrons in the unit-configuration and $\gamma'$ the number of empty orbitals in the unit-configuration, respectively. The filling factor is given by $\nu = \beta'/\alpha'$ :

$$\nu = \beta'/\alpha' = (j(2s)+1)/((2j-1)(2s)+2)$$

The sum of the perturbation energy for all the *nearest electron pairs* is equal to

$$E_{\text{nearest pair}} = \left[-(2\gamma' - 2s)s\right] \frac{Z}{\alpha'} \times \frac{N}{\beta'} \quad \text{at } \nu = \beta'/\alpha'$$

Substitution of Eq.(5.29) yields

$$E_{\text{nearest pair}} = -\frac{((2j-3)2s+2)s}{((2j-1)(2s)+2)} \frac{ZN}{(j(2s)+1)} \quad \text{at } \nu = \frac{(j(2s)+1)}{((2j-1)(2s)+2)} \quad (5.30)$$

The limiting value of Eq.(5.30) is

$$\nu = (j(2s)+1)/((2j-1)(2s)+2) \xrightarrow[s \to \infty]{} j/(2j-1) \quad (5.31a)$$

$$E_{\text{nearest pair}} = -\frac{((2j-3)2s+2)s}{((2j-1)(2s)+2)} \frac{ZN}{(j(2s)+1)} \xrightarrow[s \to \infty]{} -\frac{(2j-3)ZN}{2j(2j-1)} \quad \text{at } \nu = \frac{j}{2j-1} - \varepsilon \quad (5.31b)$$

This limiting value from the left is compared with the energy at $\nu = j/(2j-1)$ and we find the former is higher than the latter. The difference is given by

$$\Delta E_{\text{nearest pair}} = E(\nu) - \lim_{\lambda \to \nu - \varepsilon} E(\lambda) = -\frac{ZN}{2j(2j-1)} \quad \text{for } \nu = \frac{j}{2j-1} \quad (5.32)$$

Thus the discontinuity in the energy spectrum is clarified theoretically as in Eqs (5.28) and (5.32). If readers feel that the proofs mentioned above are a little complicated, the limiting value can be approximately evaluated by a computer.

We summarize the results:
1) All the *nearest electron pairs* are able to transfer to all the empty states at $\nu = j/(2j-1)$.
2) When the filling factor slightly deviates from $\nu = j/(2j-1)$, the *nearest electron pairs* are unable to transfer to some of the empty states.
3) Then the perturbation energy at $\nu = j/(2j-1)$ is lower than those at $\nu \pm \varepsilon$ as;

$$\Delta E_{\text{nearest pair}} = E(\nu) - \lim_{\lambda \to \nu \pm \varepsilon} E(\lambda) = -ZN/[2j(2j-1)] \quad \text{at } \nu = j/(2j-1).$$

Thus the valley structure is derived from the present theory. The discontinuity is caused by the combined effects of the most uniform electron configuration, the Fermi-Dirac statistics and the momentum conservation. As will be shown in Chapter 6, the discontinuity appears in all orders of the perturbation calculation.

### 5.1.4 Valley structure at $\nu = j/(2j+1)$

In this subsection, we examine the FQH state with $\nu = j/(2j+1)$ smaller than 1/2.

The $\nu = j/(2j+1)$ state is related to the $\nu' = (j+1)/(2j+1)$ state by the electron-hole symmetry as in Eq.(4.39). Equations (5.26) and (5.30) are rewritten by replacing $j \to j+1$:

$$E_{\text{nearest pair}} = -\frac{((2j-1)2s-2)s}{((2j+1)(2s)-2)} \frac{ZN}{((j+1)(2s)-1)} \quad \text{at } \nu' = \frac{((j+1)(2s)-1)}{((2j+1)(2s)-2)} \quad (5.33)$$

$$E_{\text{nearest pair}} = -\frac{((2j-1)2s+2)s}{((2j+1)(2s)+2)} \frac{ZN}{((j+1)(2s)+1)} \quad \text{at } \nu' = \frac{((j+1)(2s)+1)}{((2j+1)(2s)+2)} \quad (5.34)$$

The limiting values are obtained as

$$E_{\text{nearest pair}}/N \xrightarrow{s \to \infty} -\frac{2j-1}{2(j+1)(2j+1)} Z \quad \text{for } \nu' = \frac{j+1}{2j+1} \pm \varepsilon \quad (5.35)$$

where $\varepsilon$ (or $-\varepsilon$) indicates the limiting process from the right (or left).

From the electron-hole symmetry, the perturbation energy $E_{\text{nearest hole pair}}$ is obtained by replacing $Z \to Z_H$, $N \to N_H$ and $\nu' \to (1-\nu')$ in Eqs. (5.33) and (5.34) as follows:

$$E_{\text{nearest hole pair}} = -\frac{((2j-1)2s-2)s}{((2j+1)(2s)-2)} \frac{Z_H N_H}{((j+1)(2s)-1)} \quad \text{at } \nu = \frac{((j)(2s)-1)}{((2j+1)(2s)-2)} \quad (5.36)$$

$$E_{\text{nearest hole pair}} = -\frac{((2j-1)2s+2)s}{((2j+1)(2s)+2)} \frac{Z_H N_H}{((j+1)(2s)+1)} \quad \text{at } \nu = \frac{((j)(2s)+1)}{((2j+1)(2s)+2)} \quad (5.37)$$

In the limit of $s \to \infty$, both equations (5.36) and (5.37) have the same limiting value as

$$E_{\text{nearest hole pair}}/N_H \xrightarrow{\varepsilon \to 0} -\frac{2j-1}{2(j+1)(2j+1)} Z_H \quad \text{for } \nu = \frac{j}{2j+1} \pm \varepsilon \quad (5.38)$$

The perturbation energy of all the nearest hole pairs at $\nu = j/(2j+1)$ has been already obtained in Eq. (4.42) as

$$E_{\text{nearest hole pair}}/N_H = -\frac{j}{(j+1)(2j+1)} Z_H \quad \text{at } \nu = \frac{j}{2j+1} \quad (5.39)$$

Thus the energy spectrum versus filling factor is discontinuous at $\nu = j/(2j+1)$.

The energy per electron $\varepsilon(\nu) = E_{\text{total}}/N$ has been given by Eq.(5.13) as

$$\varepsilon(\nu) = \chi(\nu) + g(\nu) + [f + \hbar eB/(2m^*) - (\xi - \eta)/\nu] + C_{\text{Macroscopic}}/N$$

where $g(\nu)$ represents the energy of all the non-nearest pairs per electron via the Coulomb transitions. The first term $\chi(\nu)$ in the right hand side represents the energy of all the nearest pairs per electron, the second order term of which has been calculated as mentioned above. Using the second order term $\chi_2(\nu)$ instead of $\chi(\nu)$, the energy per electron at $\nu = (j+1)/(2j+1)$ is approximated by

$$\varepsilon(\nu) \approx -\frac{j \times Z}{(j+1)(2j+1)} + g(\nu) + \left[ f + \frac{\hbar eB}{2m^*} - \frac{(\xi - \eta)}{\nu} \right] + \frac{C_{\text{Macroscopic}}}{N}$$

$$\text{at } \nu = (j+1)/(2j+1) \qquad (5.40)$$

The energy gap (energy depth of the valley) is defined by

$$\Delta\varepsilon_-(\nu) = \varepsilon(\nu) - \lim_{\lambda \to \nu - \varepsilon} \varepsilon(\lambda) \qquad (5.41a)$$

$$\Delta\varepsilon_+(\nu) = \varepsilon(\nu) - \lim_{\lambda \to \nu + \varepsilon} \varepsilon(\lambda) \qquad (5.41b)$$

We will study the discontinuity of $g(\nu)$ in sections 5.9 and 5.10 which is small. So we ignore the discontinuity of $g(\nu)$ and then obtain the approximate form by substitute Eqs. (5.33), (5.34) and (5.40) into Eqs.(5.41a) and (5.41b), and obtain the energy gap (the energy depth of the valley) at $\nu = (j+1)/(2j+1)$ as

$$\Delta\varepsilon_+(\nu = (j+1)/(2j+1)) \approx -\frac{1}{2(j+1)(2j+1)} Z \qquad (5.42)$$

$$\Delta\varepsilon_-(\nu = (j+1)/(2j+1)) \approx -\frac{1}{2(j+1)(2j+1)} Z \qquad (5.43)$$

In the case of $\nu = j/(2j+1)$, the number of empty Landau orbitals is larger than that of filled Landau orbitals. The perturbation energy has been obtained as Eqs.(5.36) and (5.37). The energy per electron (not per hole) is obtained by multiplying $N_H/N$ to Eqs.(5.38) and (5.39) as follows;

$$E_{\text{nearest hole pair}}/N \xrightarrow{\varepsilon \to 0} -\frac{2j-1}{2j(2j+1)} Z_H \quad \text{for } \nu = \frac{j}{2j+1} \pm \varepsilon \qquad (5.44a)$$

$$E_{\text{nearest hole pair}}/N = -\frac{j}{j(2j+1)} Z_H = -\frac{Z_H}{(2j+1)} \quad \text{at } \nu = \frac{j}{2j+1} \qquad (5.44b)$$

Then the energy gap (energy depth of the valley) $\Delta\varepsilon_\pm(\nu)$ at $\nu = j/(2j+1)$ is equal to

$$\Delta\varepsilon_+(\nu = j/(2j+1)) = -\frac{Z_H}{2j(2j+1)} \qquad (5.45a)$$

$$\Delta\varepsilon_-(\nu = j/(2j+1)) = -\frac{Z_H}{2j(2j+1)} \qquad (5.45b)$$

It is noteworthy that Eqs.(5.45a) and (5.45b) indicate the energy depth of the valley per electron (not hole). When the energy depth is deep in comparison with the thermal excitation energy, the state with the filling factor $\nu = j/(2j \pm 1)$ becomes very stable.

## 5.2 Comparison of the theory with experimental data
### in the Neighbourhood of $\nu = j/(2j \pm 1)$

We list the energy gaps (energy depths of the valleys) at $\nu = j/(2j \pm 1)$ in Tables 5.1 and 5.2 which are derived from Eqs.(5.40) – (5.45).

Table 5.1   Energy gaps of *nearest electron pairs* per electron at $\nu = j/(2j-1)$

| $\nu$ | $E_{\text{nearest pair}}/N$ | $\nu$ | $\lim(E_{\text{nearest pair}}/N)$ | $\Delta\varepsilon_+(\nu) = \Delta\varepsilon_-(\nu)$ |
|---|---|---|---|---|
| 1/2 | 0 | $(1/2) \pm \varepsilon$ | 0 | 0 |
| 2/3 | $-(1/6)Z$ | $(2/3) \pm \varepsilon$ | $-(1/12)Z$ | $-(1/12)Z$ |
| 3/5 | $-(2/15)Z$ | $(3/5) \pm \varepsilon$ | $-(3/30)Z$ | $-(1/30)Z$ |
| 4/7 | $-(3/28)Z$ | $(4/7) \pm \varepsilon$ | $-(5/56)Z$ | $-(1/56)Z$ |
| 5/9 | $-(4/45)Z$ | $(5/9) \pm \varepsilon$ | $-(7/90)Z$ | $-(1/90)Z$ |
| 6/11 | $-(5/66)Z$ | $(6/11) \pm \varepsilon$ | $-(9/132)Z$ | $-(1/132)Z$ |
| 7/13 | $-(6/91)Z$ | $(7/13) \pm \varepsilon$ | $-(11/182)Z$ | $-(1/182)Z$ |
| 8/15 | $-(7/120)Z$ | $(8/15) \pm \varepsilon$ | $-(13/240)Z$ | $-(1/240)Z$ |

Table 5.2   Energy gaps of *nearest hole pairs* per electron at $\nu = j/(2j+1)$

| $\nu$ | $E_{\text{nearest pair}}/N_{\text{H}}$ | $E_{\text{nearest pair}}/N$ | $\nu$ | $\lim(E_{\text{nearest pair}}/N)$ | $\Delta\varepsilon_+(\nu) = \Delta\varepsilon_-(\nu)$ |
|---|---|---|---|---|---|
| 1/3 | $-(1/6)Z_{\text{H}}$ | $-(1/3)Z_{\text{H}}$ | $(1/3) \pm \varepsilon$ | $-(1/6)Z_{\text{H}}$ | $-(1/6)Z_{\text{H}}$ |
| 2/5 | $-(2/15)Z_{\text{H}}$ | $-(1/5)Z_{\text{H}}$ | $(2/5) \pm \varepsilon$ | $-(3/20)Z_{\text{H}}$ | $-(1/20)Z_{\text{H}}$ |
| 3/7 | $-(3/28)Z_{\text{H}}$ | $-(1/7)Z_{\text{H}}$ | $(3/7) \pm \varepsilon$ | $-(5/42)Z_{\text{H}}$ | $-(1/42)Z_{\text{H}}$ |
| 4/9 | $-(4/45)Z_{\text{H}}$ | $-(1/9)Z_{\text{H}}$ | $(4/9) \pm \varepsilon$ | $-(7/72)Z_{\text{H}}$ | $-(1/72)Z_{\text{H}}$ |
| 5/11 | $-(5/66)Z_{\text{H}}$ | $-(1/11)Z_{\text{H}}$ | $(5/11) \pm \varepsilon$ | $-(9/110)Z_{\text{H}}$ | $-(1/110)Z_{\text{H}}$ |
| 6/13 | $-(6/91)Z_{\text{H}}$ | $-(1/13)Z_{\text{H}}$ | $(6/13) \pm \varepsilon$ | $-(11/156)Z_{\text{H}}$ | $-(1/156)Z_{\text{H}}$ |
| 7/15 | $-(7/120)Z_{\text{H}}$ | $-(1/15)Z_{\text{H}}$ | $(7/15) \pm \varepsilon$ | $-(13/210)Z_{\text{H}}$ | $-(1/210)Z_{\text{H}}$ |

Table 5.1 indicates the discontinuous structure at the filling factors of $\nu = j/(2j-1)$. In the third column $\varepsilon$ indicates an infinitesimally small positive value, then $+\varepsilon$ indicates the limiting process from the right and $-\varepsilon$ indicates the limiting process from the left. Table 5.2 shows the discontinuity in the energy spectrum of the nearest hole pairs at $\nu = j/(2j+1)$. The rightmost columns in these two tables represent the energy gaps (energy depths of the valleys) per electron. The presence of the valley produces the Hall plateau as will be studied in Chapter 7.

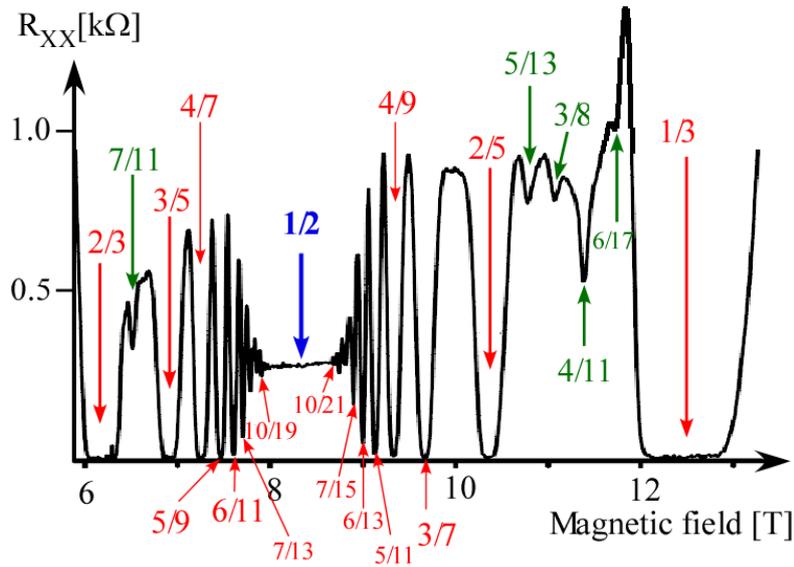

Fig.5.9 Many local minima of the diagonal resistance in the region of $2/3 \geq \nu \geq 1/3$

The experimental curve is given in reference [34, 35]

We compare the theoretical results with the experimental data in Fig.5.9 [34, 35]. The theoretical energy spectrum has the valleys with deep energy depths as in Tables 5.1 and 5.2. The deep valley yields the strong confinement to the ground state with the specific filling factor. Thereby the ground state cannot be excited by the electric current and so the diagonal resistance becomes nearly zero. The experimental curve of the diagonal resistance is almost zero at the filling factors $\nu = \frac{2}{3}, \frac{3}{5}, \frac{4}{7}, \frac{5}{9}, \frac{6}{11}$ and $\nu = \frac{5}{11}, \frac{4}{9}, \frac{3}{7}, \frac{2}{5}, \frac{1}{3}$. Thus the experimental data are in good agreement with the theoretical results.

Furthermore there is a small local minimum at $\nu = \frac{7}{11}, \frac{8}{13}, \frac{5}{13}, \frac{3}{8}, \frac{4}{11}, \frac{6}{17}$ shown by solid arrows in Fig.5.9. We will investigate the FQH states with $\nu = \frac{4}{11}, \frac{7}{11}, \frac{4}{13}, \frac{5}{13}, \frac{8}{13}, \frac{5}{17}, \frac{6}{17}$ in sections 5.5 and 5.6.

## 5.3 Flat structure at $v = 1/2$ and peak structure at $v = \frac{1}{2j}$ and $v = \frac{2j-1}{2j}$

We examine the fractional quantum Hall states in the neighborhood of $v = 1/2$. The filling factor $v = j/(2j-1)$ is larger than 1/2 and approaches 1/2 for an infinitely large $j$. The energy Eq.(4.23c) has the limiting value from the right of $v = 1/2$ as

$$v = j/(2j-1) \xrightarrow[j \to \infty]{} 1/2 \tag{5.46a}$$

$$E_{\text{nearest pair}}/N = -\frac{j-1}{j(2j-1)} Z \xrightarrow[v=j/(2j-1),\ j \to \infty]{} 0 \quad \text{for} \quad v = (1/2) + \varepsilon \tag{5.46b}$$

Equation (4.23b) indicates that the nearest pair energy is zero at $v = 1/2$ as

$$E_{\text{nearest pair}}/N = 0 \qquad \text{for} \quad v = 1/2 \tag{5.46c}$$

Thus the limiting energy from the right is equal to that at $v = 1/2$. Similarly we calculate the limiting value from the left of $v = 1/2$ which is already obtained in Eq.(5.44b). We write it again here:

$$E_{\text{nearest pair}}/N = -\frac{Z_H}{(2j+1)} \qquad \text{at} \quad v = \frac{j}{2j+1}$$

The limiting value from the left of $v = 1/2$ becomes zero as:

$$E_{\text{nearest pair}}/N = -\frac{Z_H}{(2j+1)} \xrightarrow[v=j/(2j+1),\ j \to \infty]{} 0 \quad \text{for} \quad v = (1/2) - \varepsilon \tag{5.46d}$$

Accordingly the energy of all the *nearest electron pairs* per electron is continuous at $v = 1/2$. The continuous property appears in the higher order perturbation as will be shown in chapter 6. That is to say the energy spectrum of the *nearest electron pairs* is continuous at $v = 1/2$ and we call it "flat structure".

We next examine the fractional quantum Hall states in the neighborhood of $v = 3/4$. The filling factor $v = (6s+1)/(8s+1)$ approaches 3/4 in the limit of $s \to \infty$. As an example we consider the case of $s = 2$. The most uniform configuration of electrons is illustrated in Fig.5.10. The unit-configuration has seventeen sequential Landau states, thirteen of which are filled with electron. There are four empty states inside each unit-configuration.

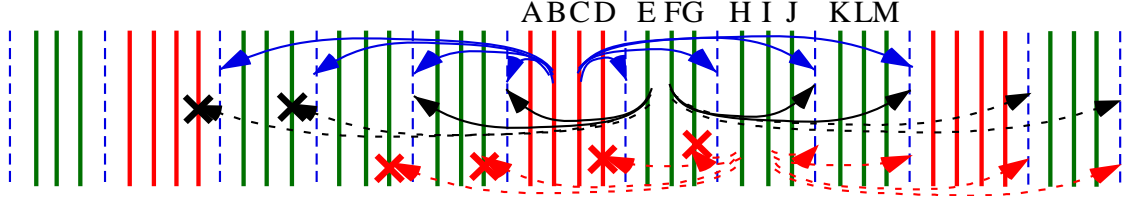

Fig.5.10 Allowed transitions of nearest electron pairs at $\nu = 13/17$

The *nearest electron pair* BC transfers to all the empty orbitals as drawn by blue arrows in Fig.5.10. The *nearest electron pair* EF (and LM) transfers to the two empty orbitals per unit-configuration as expressed by black arrows. The pair HI (and IJ) cannot transfer to any empty orbital as shown by dashed arrow pairs. The other pairs AB, CD, FG, KL cannot transfer to any empty orbital. Accordingly the perturbation energy of the *nearest electron pairs* at $\nu = 13/17$ is obtained as

$$\varsigma_{BC} = -(4/17)Z, \quad \varsigma_{EF} = -(2/17)Z, \quad \varsigma_{LM} = -(2/17)Z$$

$$\varsigma_{AB} = \varsigma_{CD} = \varsigma_{FG} = \varsigma_{HI} = \varsigma_{IJ} = \varsigma_{KL} = 0$$

Then the perturbation energy of all the *nearest electron pairs* is equal to

$$E_{\text{nearest pair}} = [-(2/17)Z - (4/17)Z - (2/17)Z] \times (1/13)N$$
$$= -(8/(17 \times 13))ZN \qquad \text{for } \nu = 13/17$$

We next calculate the perturbation energy at the filling factor $\nu = (6s+1)/(8s+1)$ for any integer $s$. The unit-configuration is composed of $8s+1$ Landau states, $6s+1$ of which are occupied by electron. Therefore the number of empty states is $2s$ per unit-configuration. Accordingly we obtain the total energy for the nearest pairs as

$$E_{\text{nearest pair}} = \begin{bmatrix} -(2/(8s+1))Z - (4/(8s+1))Z \cdots - ((2s)/(8s+1))Z \cdots \\ -(4/(8s+1))Z - (2/(8s+1))Z \end{bmatrix} \times (1/(6s+1))N$$
$$= -\frac{s(s+1) + (s-1)s}{(8s+1)(6s+1)} ZN$$

$$E_{\text{nearest pair}}/N = -\frac{2s^2}{(8s+1)(6s+1)} Z \qquad (5.47)$$

The filling factor and the energy of the nearest pairs approach in the limit of $s \to \infty$

$$\nu = (6s+1)/(8s+1) \xrightarrow[s \to \infty]{} 3/4$$

$$E_{\text{nearest pair}}/N \xrightarrow[s \to \infty]{} -\frac{1}{24} Z \qquad (5.48)$$

Thus the perturbation energy of the *nearest electron pairs* at $v = (6s+1)/(8s+1)$ is negative and the limiting value is $-(1/24)Z$ as in Eq.(5.48).

At $v = 3/4$, all the transitions from nearest electron pairs are forbidden [27-33]. Thereby the energy of all the nearest pairs is zero as in Eq.(4.27);

$$E_{nearest\ pair}/N = 0 \quad \text{at } v = 3/4 \quad (5.49)$$

Consequently the energy of all the *nearest electron pairs* at $v = 3/4$ is higher than that of the neighbourhood. So the state with $v = 3/4$ is unstable. Thus this case has a peak (discontinuous peak) in the energy spectrum. We call the energy spectrum "peak structure". At the peak no plateau appears in the Hall resistance curve as it will be studied in chapter 7.

For arbitrary integer $j$ we calculate the energy of the *nearest electron pairs* in the neighbourhood of $v = (2j-1)/(2j)$ by using a computer. The several results are written here.

(neighbourhood of $v = 5/6$) $\quad E_{nearest\ pair}/N = -\frac{5101}{304920} Z \quad$ at $v = \frac{504}{605}$

$\quad E_{nearest\ pair}/N = -\frac{501001}{30049020} Z \quad$ at $v = \frac{5004}{6005}$

$\quad E_{nearest\ pair}/N = -\frac{5101}{307142} Z \quad$ at $v = \frac{506}{607}$

$\quad E_{nearest\ pair}/N = -\frac{501001}{30071042} Z \quad$ at $v = \frac{5006}{6007}$

(neighbourhood of $v = 7/8$) $\quad E_{nearest\ pair}/N = -\frac{5101}{569742} Z \quad$ at $v = \frac{706}{807}$

$\quad E_{nearest\ pair}/N = -\frac{501001}{56097042} Z \quad$ at $v = \frac{7006}{8007}$

$\quad E_{nearest\ pair}/N = -\frac{5101}{572772} Z \quad$ at $v = \frac{708}{809}$

$\quad E_{nearest\ pair}/N = -\frac{501001}{56127072} Z \quad$ at $v = \frac{7008}{8009}$

The numerical results may give the limiting value as ;

$$E_{nearest\ pair}/N \longrightarrow -\tfrac{1}{60} Z \quad \text{for } v \to \tfrac{5}{6} \pm \varepsilon$$

$$E_{nearest\ pair}/N \longrightarrow -\tfrac{1}{112} Z \quad \text{for } v \to \tfrac{7}{8} \pm \varepsilon$$

Thus the energy of the *nearest electron pairs* at $v = (2j-1)/(2j)$ is larger than that of the neighbourhood. So the peak structure appears at $\nu = (2j-1)/(2j)$. Some examples are listed in Table 5.3.

Table 5.3  Comparison between *nearest electron pair* energy
at $\nu = (2j-1)/(2j)$ and in the neighbourhood

| $\nu$ | $E_{\text{nearest pair}}/N$ | $\nu$ | $\lim(E_{\text{nearest pair}}/N)$ |
|---|---|---|---|
| 1/2 | 0 | $(1/2) \pm \varepsilon$ | 0 |
| 3/4 | 0 | $(3/4) \pm \varepsilon$ | $-\frac{1}{24}Z$ |
| 5/6 | 0 | $(5/6) \pm \varepsilon$ | $-\frac{1}{60}Z$ |
| 7/8 | 0 | $(7/8) \pm \varepsilon$ | $-\frac{1}{112}Z$ |

Table 5.4  Comparison between *nearest hole pair* energy
at $\nu = 1/(2j)$ and in the neighbourhood

| $\nu$ | $E_{\text{nearest pair}}/N_{\text{H}}$ | $\nu$ | $\lim(E_{\text{nearest pair}}/N_{\text{H}})$ | $\nu$ | $\lim(E_{\text{nearest pair}}/N)$ |
|---|---|---|---|---|---|
| 1/4 | 0 | $(1/4) \pm \varepsilon$ | $-\frac{1}{24}Z_{\text{H}}$ | $(1/4) \pm \varepsilon$ | $-\frac{1}{8}Z_{\text{H}}$ |
| 1/6 | 0 | $(1/6) \pm \varepsilon$ | $-\frac{1}{60}Z_{\text{H}}$ | $(1/6) \pm \varepsilon$ | $-\frac{1}{12}Z_{\text{H}}$ |
| 1/8 | 0 | $(1/8) \pm \varepsilon$ | $-\frac{1}{112}Z_{\text{H}}$ | $(1/8) \pm \varepsilon$ | $-\frac{1}{16}Z_{\text{H}}$ |

From the electron-hole symmetry, the perturbation energies of the nearest hole pairs are easily calculated. The results are shown in Table 5.4 [27-33]. We find a "peak structure" in the energy spectrum. The *nearest electron pairs* (or *hole pairs*) at $\nu = (2j-1)/(2j)$ (or $\nu = 1/(2j)$) have the perturbation energy higher than in their neighborhood. Accordingly the states with $\nu = (2j-1)/(2j)$ (or $\nu = 1/(2j)$) are unstable. Thus the theoretical results yield no plateau in the Hall resistance curve at the filling factor with the peak structure. The results of the present theory are in good agreement with the experimental data.

**5.4 Valley structure at $\nu = (2j)/(2j+1)$ and $\nu = 1/(2j+1)$**

We compare the perturbation energy at $\nu_0 = (2j)/(2j+1)$ with that of the neighbourhood. The filling factor $v=((4j)s+1)/((4j+2)s+1)$ is an appropriate value for the neighbourhood of $v_0=(2j)/(2j+1)$ because $v$ approaches $v_0$ in the limit of infinitely large $s$. We examine an example for ($j=2$ $s=2$) which gives $v=17/21$. The most uniform configuration at $v=17/21$ is illustrated in Fig.5.11. The total number of empty states is 4/21 times the total number of orbitals.

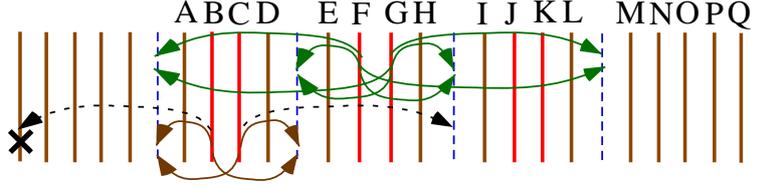

Fig.5.11 Allowed transitions of nearest electron pairs at $v=17/21$

The electron pair FG can transfer to all the empty states as shown by solid arrow pairs. The quantum transitions from the pair FG yield the perturbation energy as

$$\varsigma_{FG}=-(4/21)Z$$

The *nearest electron pairs* BC and JK transfer to two empty states per unit-configuration. Accordingly the pairs BC and JK have the perturbation energies as

$$\varsigma_{BC}=-(2/21)Z, \quad \varsigma_{JK}=-(2/21)Z$$

The perturbation energies of the remaining *nearest electron pairs* AB, CD, EF, GH, IJ, KL, MN, NO, OP, and PQ are zero because the transitions of these pairs are forbidden.

$$\varsigma_{AB}=\varsigma_{CD}=\varsigma_{EF}=\varsigma_{GH}=\varsigma_{IJ}=\varsigma_{KL}=\varsigma_{MN}=\varsigma_{NO}=\varsigma_{OP}=\varsigma_{PQ}=0$$

The sum of the perturbation energies of all the *nearest electron pairs* is

$$E_{nearest\ pair}=(\varsigma_{BC}+\varsigma_{FG}+\varsigma_{JK})\frac{N}{17}=-(8/21)Z\frac{N}{17}$$

$$E_{nearest\ pair}/N=-\frac{8}{21\times 17}Z \qquad \text{for} \quad v=17/21 \qquad (5.50)$$

Next we calculate the perturbation energy at the filling factor $v=(8s+1)/(10s+1)$ for any integer $s$. The unit-configuration is composed of $10s+1$ Landau orbitals, $8s+1$ of which are occupied by electron. Therefore the number of empty orbitals is $2s$ per unit-configuration. Accordingly we obtain

$$E_{nearest\ pair}=\begin{bmatrix}-(2/(10s+1))Z-(4/(10s+1))Z\cdots-((2s)/(10s+1))Z\cdots\\-(4/(10s+1))Z-(2/(10s+1))Z\end{bmatrix}\times(1/(8s+1))N$$

$$=-\frac{s(s+1)+(s-1)s}{(10s+1)(8s+1)}ZN$$

$$E_{\text{nearest pair}}/N = -\frac{2s^2}{(10s+1)(8s+1)}Z \tag{5.51}$$

The limiting values of the filling factor and the energy of the nearest pairs are equal to

$$\nu = (8s+1)/(10s+1) \xrightarrow[s\to\infty]{} 4/5$$

$$E_{\text{nearest pair}}/N \xrightarrow[s\to\infty]{} -\frac{1}{40}Z \qquad \text{in the limit from the right} \tag{5.52a}$$

The perturbation energy at $\nu = 4/5$ is listed in Table 4.2 as

$$E_{\text{nearest pair}}/N = -\frac{1}{20}Z \qquad \text{at } \nu = 4/5 \tag{5.52b}$$

Eqs. (5.52a) and (5.52b) reveals that a valley structure exists in the energy spectrum at $\nu = 4/5$.

The method mentioned above is generalized for arbitrary integer $j$, the limiting value of the energy for all the *nearest electron pairs* per electron is obtained as

$$E_{\text{nearest pair}}/N \xrightarrow[\nu\to 2j/(2j+1)\pm\varepsilon]{} -\frac{1}{2(2j)(2j+1)}Z \qquad \text{for } \nu = \frac{2j}{2j+1} \pm \varepsilon \tag{5.53a}$$

Eq.(4.26) gives the energy of all the *nearest electron pairs* per electron at $\nu = 2j/(2j+1)$ as

$$E_{\text{nearest pair}}/N = -\frac{1}{2j(2j+1)}Z \qquad \text{at } \nu = \frac{2j}{2j+1} \tag{5.53b}$$

Thus the limiting value from the both sides is half of the value at $\nu = (2j)/(2j+1)$. Then the energy gap (depth of the valley) is obtained as [30-33]

$$\Delta\varepsilon_+(\nu) = \Delta\varepsilon_-(\nu) = -\frac{1}{2(2j)(2j+1)}Z \qquad \text{for } \nu = \frac{2j}{2j+1} \tag{5.53c}$$

We express these values in Table 5.5.

Table 5.5  Comparison of *nearest electron pair* energies per electron at $\nu = 2j/(2j+1)$ and in its neighbourhood

| $\nu$ | $E_{\text{electron pair}}/N$ | $\nu$ | $\lim(E_{\text{nearest pair}}/N)$ | $\Delta\varepsilon_+(\nu) = \Delta\varepsilon_-(\nu)$ |
|---|---|---|---|---|
| 2/3 | -(1/6) Z | $(2/3) \pm \varepsilon$ | -(1/12) Z | -(1/12) Z |
| 4/5 | -(1/20) Z | $(4/5) \pm \varepsilon$ | -(1/40) Z | -(1/40) Z |
| 6/7 | -(1/42) Z | $(6/7) \pm \varepsilon$ | -(1/84) Z | -(1/84) Z |
| 8/9 | -(1/72) Z | $(8/9) \pm \varepsilon$ | -(1/144) Z | -(1/144) Z |

The energy of all the *nearest hole pairs* at $v = 1/(2j+1)$ is calculated on the basis of the electron-hole symmetry relation given in Eq.(4.39). The perturbation energy $E_{\text{nearest hole pair}}$ is obtained by replacing $Z \to Z_H$, $N \to N_H$ and $v \to (1-v)$ in Eqs. (5.53a) and (5.53b) as follows:

$$E_{\text{nearest pair}}/N_H = -\frac{1}{2j(2j+1)} Z_H \qquad \text{for} \quad v = \frac{1}{2j+1} \qquad (5.54a)$$

$$E_{\text{nearest pair}}/N_H \xrightarrow[v \to 2j/(2j+1)\pm\varepsilon]{} -\frac{1}{2(2j)(2j+1)} Z_H \qquad \text{for} \quad v = \frac{1}{2j+1} \pm \varepsilon \quad (5.54b)$$

The energies per electron (not hole) are given by multiplying $N_H/N = 2j$ to Eqs.(5.54a) and (5.54b);

$$E_{\text{nearest pair}}/N = -\frac{1}{(2j+1)} Z_H \qquad \text{for} \quad v = \frac{1}{2j+1} \qquad (5.55a)$$

$$E_{\text{nearest pair}}/N \xrightarrow[v \to 2j/(2j+1)\pm\varepsilon]{} -\frac{1}{2(2j+1)} Z_H \qquad \text{for} \quad v = \frac{1}{2j+1} \pm \varepsilon \qquad (5.55b)$$

Then the energy gap (depth of the valley) per electron becomes

$$\Delta\varepsilon_+(v) = \Delta\varepsilon_-(v) = -\frac{1}{2(2j+1)} Z_H \qquad \text{for} \quad v = \frac{1}{2j+1} \qquad (5.55c)$$

These results are listed in Table 5.6. [30-33]

Table 5.6 Comparison of *nearest hole pair* energies per hole and per electron at $v = 1/(2j+1)$ and in its neighbourhood

| $v$ | $E_{\text{nearest pair}}/N_H$ | $E_{\text{nearest pair}}/N$ | $v$ | $\lim(E_{\text{nearest pair}}/N)$ | $\Delta\varepsilon_+(v) = \Delta\varepsilon_-(v)$ |
|---|---|---|---|---|---|
| 1/3 | -(1/6) $Z_H$ | -(1/3) $Z_H$ | $(1/3) \pm \varepsilon$ | -(1/6) $Z_H$ | -(1/6) $Z_H$ |
| 1/5 | -(1/20) $Z_H$ | -(1/5) $Z_H$ | $(1/5) \pm \varepsilon$ | -(1/10) $Z_H$ | -(1/10) $Z_H$ |
| 1/7 | -(1/42) $Z_H$ | -(1/7) $Z_H$ | $(1/7) \pm \varepsilon$ | -(1/14) $Z_H$ | -(1/14) $Z_H$ |
| 1/9 | -(1/72) $Z_H$ | -(1/9) $Z_H$ | $(1/9) \pm \varepsilon$ | -(1/18) $Z_H$ | -(1/18) $Z_H$ |

It is noteworthy that the perturbation energy of all the *nearest hole pairs* per electron is more important quantity than that per hole. The values per electron are written in the third, fifth and sixth columns in Table 5.6.

Tables 5.5 and 5.6 indicate a discontinuity (depth of the valley) in the electron-energy spectrum at the filling factors of $v = 2j/(2j+1)$ and $v = 1/(2j+1)$. Therein $+\varepsilon$ indicates the limiting process from the right and $-\varepsilon$ does from the left. The electron-energy gaps (depths of the valley) in the energy spectrum are listed in the last columns of Tables 5.5 and 5.6. Accordingly the energy per electron at $\nu = 2j/(2j+1)$ (or $\nu = 1/(2j+1)$) is lower than that in their neighbourhood. Consequently these states become stable.

## 5.5 Specific filling factors with even number denominator

We examine the energies of the nearest electron pairs at the filling factors of $\nu = 5/8$ and $7/10$ which have each denominators of even number. The most uniform configuration of the many electron state at $\nu = 5/8$ is shown in Fig.5.12.

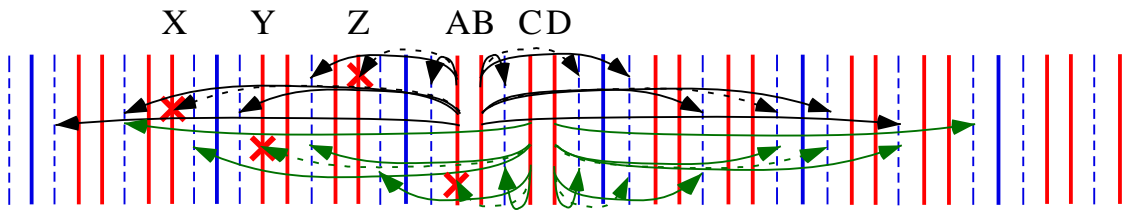

Fig.5.12 Allowed transitions of nearest electron pairs at $\nu = 5/8$

This configuration is composed of repeating a unit-configuration (filled, filled, empty, filled, filled, empty, filled, empty). Therein two *nearest electron pairs* AB and CD exists in the unit-configuration as shown in Fig.5.12. We examine allowed transitions from electron pair AB. When the electron at B transfers to the first orbital to the right, the electron at A transfers to the first orbital to the left because of the momentum conservation. When the electron at B transfers to the fourth orbital to the right, the electron at A is forbidden to transfer to the fourth orbital to the left because of the Pauli exclusion principle. Similarly the electron at A cannot transfer to the X orbital in Fig.5.12. That is to say the electron pair AB can transfer to the two empty orbital pairs per unit-configuration.

The transitions of the *nearest electron pair* CD are also allowed to the $1^{th}$, $6^{th}$, $9^{th}$, $14^{th}$, $17^{th}$ ··· orbitals. Accordingly the number of allowed transitions per unit-configuration is two. The number of allowed quantum transitions from the *nearest electron pairs* is smaller than the number of empty orbitals. This property is in contrast to that discussed in sections 5.1, 5.2 and 5.4, where the number of allowed transitions for the nearest electron pairs is equal to the total number of empty orbitals.

Thus the pairs AB and CD have the perturbation energy $\varsigma_{AB}$ and $\varsigma_{CD}$ as

$$\varsigma_{AB} = -(2/8)Z \quad \varsigma_{CD} = -(2/8)Z \tag{5.56}$$

The total perturbation energy of all the nearest electron pairs is equal to

$$E_{\text{nearest pair}} = (\varsigma_{AB} + \varsigma_{CD})(N/5) = -(1/10)ZN$$

$$E_{\text{nearest pair}}/N = -(1/10)Z \qquad \text{at } \nu = 5/8 \tag{5.57}$$

A similar situation occurs at $\nu = 7/10$ as illustrated in Fig.5.13.

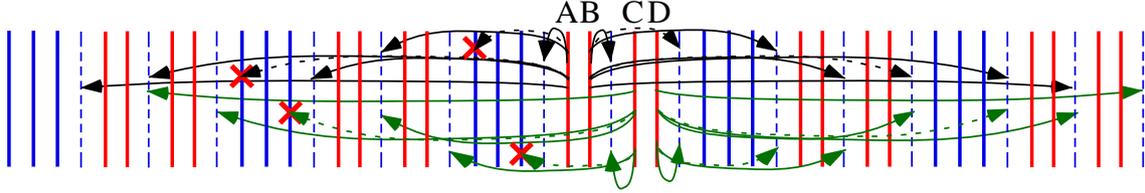

Fig.5.13 Allowed transitions of nearest electron pairs at $\nu = 7/10$

This configuration is composed of repeating a unit-configuration (filled, filled, empty, filled, filled, empty, filled, filled, filled, empty). Therein electron B can transfer to the $1^{th}$, $8^{th}$, $11^{th}$, $18^{th}$, $21^{th}$ ⋯ orbitals to the right in the quantum transitions from the pair AB. Also the transitions of the pair CD are allowed to the $1^{th}$, $8^{th}$, $11^{th}$, $18^{th}$, $21^{th}$ ⋯ orbitals. That is to say, the two transitions are allowed per unit-configuration. The pairs AB and CD have the perturbation energy as

$$\varsigma_{AB} = -(2/10)Z \quad \varsigma_{CD} = -(2/10)Z \tag{5.58}$$

Accordingly the total perturbation energy of all the nearest electron pairs is equal to

$$E_{\text{nearest pair}} = (\varsigma_{AB} + \varsigma_{CD})(N/7) = -(2/35)ZN$$

$$E_{\text{nearest pair}}/N = -(2/35)Z \qquad \text{for } \nu = 7/10 \tag{5.59}$$

Similarly we can calculate the energies for the nearest hole pairs at $\nu = 3/8$ and $\nu = 3/10$. The results of the calculation are listed in Tables 5.7 and 5.8.

Table 5.7  Energy of *nearest electron pairs* per electron
for the filling factors with an even number of denominator

| $\nu$ | $E_{\text{nearest pair}}/N$ | $\Delta\varepsilon_+(\nu) = \Delta\varepsilon_-(\nu)$ |
|---|---|---|
| 5/8 | -(1/10) Z | 0 |
| 7/10 | -(2/35) Z | 0 |

Table 5.8  Energy of *nearest hole pairs* per hole and per electron
for the filling factors with an even number of denominator

| $\nu$ | $E_{\text{nearest pair}}/N_H$ | $E_{\text{nearest pair}}/N$ | $\Delta\varepsilon_+(\nu) = \Delta\varepsilon_-(\nu)$ |
|---|---|---|---|
| 3/8 | -(1/10) $Z_H$ | -(1/6) $Z_H$ | 0 |
| 3/10 | -(2/35) $Z_H$ | -(2/15) $Z_H$ | 0 |

Next, we study the energy spectrum in the neighbourhood of $\nu = 5/8$. We have introduced the sub-units $S_{21}$ (filled, filled, empty) and $S_{11}$ (filled, empty) in Chapter 4, which are drawn again in Fig.5.14. Then the unit-configuration is $T_{5/8}=S_{21} + S_{21} + S_{11}$ as shown in the upper panel of Fig.5.14. Then the most uniform configuration of $\nu = 5/8$ is produced by repeating the unit-configuration $T_{5/8}$. In the neighbourhood of $\nu = 5/8$, the most uniform electron configuration should be composed of only the two types of sub-units $S_{21}$ and $S_{11}$, because the fractional number $5/8$ is larger than 1/2 and smaller than 2/3.

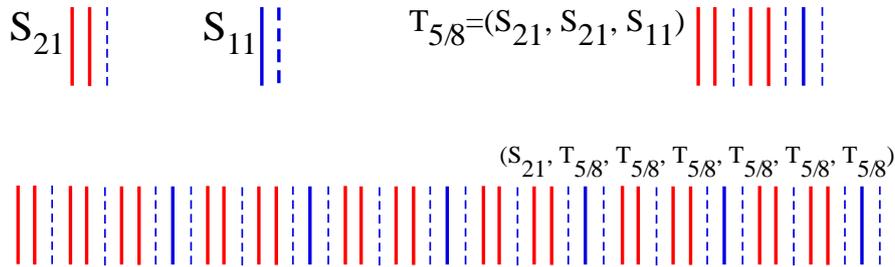

Fig.5.14 Sub-units and their combinations

The filling factor $\nu = (10s+2)/(16s+3)$ approaches $5/8$ in the limit of infinitely large value of $s$. The case of $s = 3$ gives the filling factor $\nu = 32/51$. The $\nu = 32/51$ state has the most uniform electron-configuration composed of one sub-unit $S_{21}$ and six sub-units $T_{5/8}$ namely $S_{21} + T_{5/8} + T_{5/8} + T_{5/8} + T_{5/8} + T_{5/8} + T_{5/8}$ which is shown in the lower panel of Fig.5.14. The unit-configuration has fifty-one sequential orbitals, thirty-two of which are filled with electron. This configuration is symmetric with respect to reflection with mirror plane at the centre of the pair CD as easily seen in Fig.5.15. Therefore the nearest pair CD can transfer to all the empty states and the number of the allowed transitions is nineteen per unit-configuration. Similarly the pair GH can transfer to seventeen empty states per unit configuration as shown by blue arrow pairs as in Fig.5.15.

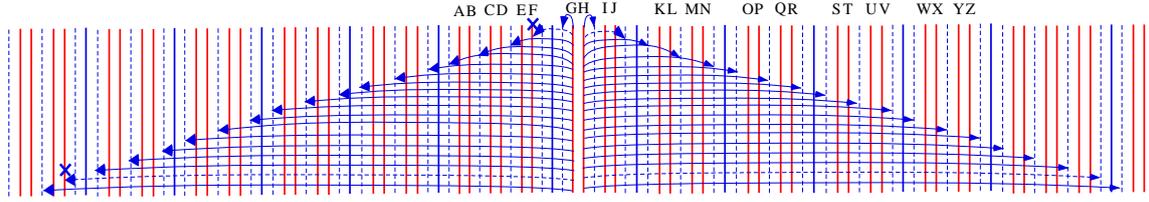

Fig.5.15 Allowed transitions of nearest electron pairs at $\nu = 32/51$

The unit-configuration includes thirteen nearest electron pairs AB, CD, EF, GH, IJ, KL, MN, OP, QR, ST, UV, WX and YZ. The thirteen pairs have the following perturbation energies:

$$\varsigma_{AB} = \varsigma_{EF} = -(7/51)Z, \quad \varsigma_{CD} = -(19/51)Z, \quad \varsigma_{GH} = \varsigma_{YZ} = -(17/51)Z \quad (5.60a)$$

$$\varsigma_{IJ} = \varsigma_{WX} = -(9/51)Z, \quad \varsigma_{KL} = \varsigma_{UV} = -(15/51)Z, \quad \varsigma_{MN} = \varsigma_{ST} = -(11/51)Z \quad (5.60b)$$

$$\varsigma_{OP} = \varsigma_{QR} = -(13/51)Z \quad (5.60c)$$

The sum of the perturbation energies is

$$E_{\text{nearest pair}} = (2 \times \varsigma_{AB} + \varsigma_{CD} + 2 \times \varsigma_{GH} + 2 \times \varsigma_{IJ} + 2 \times \varsigma_{KL} + 2 \times \varsigma_{MN} + 2 \times \varsigma_{OP}) \frac{N}{32}$$

$$= -((14 + 19 + 34 + 18 + 30 + 22 + 26)/51) Z \frac{N}{32}$$

$$E_{\text{nearest pair}}/N = -\frac{163}{51 \times 32} Z \quad \text{for} \quad \nu = 32/51 \quad (5.61)$$

Next we consider the case of arbitrary integer $s$. The unit-configuration at $\nu = (10s+2)/(16s+3)$ is composed of $(16s+3)$ Landau states which are partially occupied by $(10s+2)$ electrons. Therefore the number of empty states is $(6s+1)$ per unit-configuration. Accordingly we obtain

$$E_{\text{nearest pair}} = \begin{bmatrix} -((2s+1)/(16s+3))Z - ((2s+3)/(16s+3))Z \cdots - ((6s+1)/(16s+3))Z \cdots \\ -((2s+3)/(16s+3))Z - ((2s+1)/(16s+3))Z \end{bmatrix} \times (1/(10s+2))N$$

$$= -\frac{(2s+1)(4s+1) + (2s)(4s)}{(16s+3)(10s+2)} ZN$$

$$E_{\text{nearest pair}} = -\frac{16s^2 + 6s + 1}{(16s+3)(10s+2)} ZN = -\frac{16s^2 + 6s + 1}{160s^2 + 62s + 6} ZN$$

$$E_{\text{nearest pair}} = -\frac{16s^2 + 6.2s + 0.6 - 0.2s + 0.4}{160s^2 + 62s + 6} ZN = -\frac{1}{10} ZN + \frac{0.2s - 0.4}{160s^2 + 62s + 6} ZN$$

$$E_{\text{nearst pair}}/N = -\frac{1}{10} Z + \frac{0.2s - 0.4}{160s^2 + 62s + 6} Z \quad \text{for} \quad \nu = \frac{10s+2}{16s+3} \qquad (5.62)$$

We compare the perturbation energy of the *nearest pairs* at $\nu = (10s+2)/(16s+3)$ with that at $\nu = 5/8$ given in Table 5.7. Then we obtain the following inequality:

$$\left(E_{\text{nearest pair}}/N\right)_{\nu=(10s+2)/(16s+3)} > -\frac{1}{10}Z = \left(E_{\text{nearest pair}}/N\right)_{\nu=5/8} \quad \text{for} \quad s \geq 3 \qquad (5.63)$$

In the limit of infinitely large integer $s$ we get

$$\left(E_{\text{nearest pair}}/N\right)_{\nu=\frac{(10s+2)}{(16s+3)}} = -\frac{1}{10}Z + \frac{0.2s - 0.4}{160s^2 + 62s + 6}Z \xrightarrow{s \to \infty} -\frac{1}{10}Z \qquad (5.64)$$

That is to say, the energy spectrum of the nearest electron pairs is continuous at $\nu = 5/8$.

$$\Delta\varepsilon_+\left(\tfrac{5}{8}\right) = \varepsilon\left(\tfrac{5}{8}\right) - \lim_{\lambda \to (5/8)+\varepsilon} \varepsilon(\lambda) = 0 \qquad (5.65)$$

The property appears at $\nu = 5/8, 3/8, 7/10, 3/10$ as shown in Table 5.7 and 5.8 [30-33].

By examining Fig.5.12 in details we find that the second nearest pair BC can transfer to all the empty states as shown by arrows. Therefore the transition number of the second nearest pairs decreases discontinuously when $\nu$ deviates slightly from 5/8. A similar effect is seen at $\nu = 7/10$ as expressed with arrows in Fig.5.13. In section 5.9 we will study an effect induced from the electron pairs placed in the second neighbouring Landau orbitals. Therein small valleys appear at few filling factors. It is noteworthy that the small valley is not effective at the filling factors with the peak structure because the peak value is larger than the absolute value of the small valley energy.

## 5.6 The states with non-standard filling factors

In this section we examine the states with the filling factors $\nu = 7/11, 4/11, 4/13, 5/13,$ 5/17, 6/17 in Figs.5.16-21. These fractional numbers are named non-standard filling factors. It is difficult to explain the stability of these states by using the traditional theories. Jain has originally considered the multiflavor composite fermion model in which the composite fermions carrying different numbers of flux quanta coexist. Also many theorists have proposed their extended models, for example, Wojs et al. [36, 37], Smet [38], Peterson and Jain [39] and Pashitskii [40]. Pashitskii predicted new exotic fractions at $\nu = 5/14, 5/16,$ and 3/20 based on the conjecture proposed by Halperin. Therein free electrons and bound electron pairs coexist in the model. They used the different kinds of quasi-particles for non-standard filling factors. Thus the theoretical results are model dependent.

The present theory can calculate the perturbation energy of the nearest pairs for any FQH state. Let us start to calculate the energy for the non-standard filling factor $\nu = 7/11$ as an example. The most uniform electron configuration with $\nu = 7/11$ is schematically shown in Fig.5.16. The orbitals filled with electron are illustrated by solid lines and the empty orbitals by dashed lines.

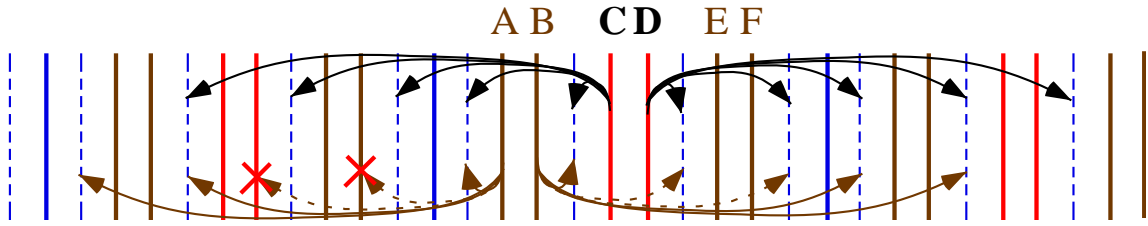

Fig.5.16 Coulomb transitions of nearest neighbor electron pairs at $\nu = 7/11$

There are three pairs of nearest electrons in unit-configuration (filled, empty, <u>filled, filled</u>, empty, <u>filled, filled</u>, empty, <u>filled, filled</u>, empty). Therein CD pair can transfer to all the empty orbitals. The pairs like CD are shown by red solid lines in Fig.5.16. The perturbation energy of the nearest pair CD is equal to

$$\varsigma_{CD} = -(4/11)Z \tag{5.66}$$

The *nearest electron pairs* AB and EF transfer to the two empty states per unit-configuration and have the perturbation energies $\varsigma_{AB}$ and $\varsigma_{EF}$ as follows:

$$\varsigma_{AB} = \varsigma_{EF} = -(2/11)Z \tag{5.67}$$

The sum of the energies for all the nearest pairs is given by

$$\left(E_{\text{nearest pair}}\right)_{\nu=7/11} = \left(-\frac{4}{11}Z - \frac{2}{11}Z - \frac{2}{11}Z\right)\frac{1}{7}N = -\frac{8}{77}ZN \quad \text{for} \quad \nu = 7/11 \quad (5.68a)$$

$$\left(E_{\text{nearest pair}}/N\right)_{\nu=7/11} = -\frac{8}{77}Z \approx -0.103896 \times Z \quad \text{for} \quad \nu = 7/11 \quad (5.68b)$$

We next consider the neighbourhood of $\nu = 7/11$. The filling factors are $\nu = (7s-2)/(11s-3)$ and $\nu = (7s+2)/(11s+3)$ which approach 7/11 in an infinitely large integer $s$. The most uniform electron configuration at $\nu = (7s \pm 2)/(11s \pm 3)$ is very complicated and so the perturbation energy of the nearest electron pairs is calculated by using a computer program. The results of the calculation for $s = 100$ are

$$\left(E_{\text{nearest pair}}/N\right)_{\nu=698/1097} = -\frac{74601}{765706}Z = -0.0974277 \times Z \quad \text{for} \quad \nu_{-} = 698/1097 \quad (5.69a)$$

$$\left(E_{\text{nearest pair}}/N\right)_{\nu=702/1103} = -\frac{75401}{774306}Z = -0.0973788 \times Z \quad \text{for} \quad \nu_{+} = 702/1103 \quad (5.69b)$$

We also calculate the case of $s = 1000$ the results of which are

$$\left(E_{\text{nearest pair}}/N\right) = -\frac{7496001}{76957006}Z = -0.097405 \times Z \quad \text{for} \quad \nu_{-} = 6998/10997 \quad (5.70a)$$

$$\left(E_{\text{nearest pair}}/N\right) = -\frac{7504001}{77043006}Z = -0.0974002 \times Z \quad \text{for} \quad \nu_{+} = 7002/11003 \quad (5.70b)$$

We compare the four values in Eqs.(5.69a,b) and (5.70a,b). Then we obtain the approximate value for the limiting values from both sides as

$$\lim_{\nu \to (7/11)\pm\varepsilon}\left(E_{\text{nearest pair}}/N\right) \approx -0.097403 \times Z \quad \text{for} \quad \nu_{\pm} = 7/11 \pm \varepsilon \quad (5.71a)$$

(The limiting value is probably equal to $-\frac{75}{770} \times Z$. The value may be clarified in the evaluation for $s = 10000$ although we cannot calculate it due to very long CPU-time.) The energy of the nearest electron pairs per electron is $(-\frac{8}{77}Z)$ at $\nu = 7/11$ as in Eq.(5.68). Therefore a small valley appears, the depth of which $\Delta\varepsilon_{\pm}$ is given by

$$\Delta\varepsilon_{\pm}(7/11) = \left(E_{\text{nearest pair}}/N\right)_{\nu=7/11} - \lim_{\nu\to(7/11)\pm\varepsilon}\left(E_{\text{nearest pair}}/N\right) \approx -0.006493 \times Z \quad (5.71b)$$

Next we examine the case of $\nu = 4/11$. We illustrate the most uniform electron configuration at $\nu = 4/11$ in Fig.5.17. For a quantum Hall device with a macroscopic size, we can ignore the effect of the boundaries namely both ends. Therein the electron configuration in Fig.5.17 has left-right symmetry at the centre of the pair CD. The nearest hole pairs like CD are expressed by red dashed lines. All the hole pairs with the red dashed lines can transfer to all the electron states.

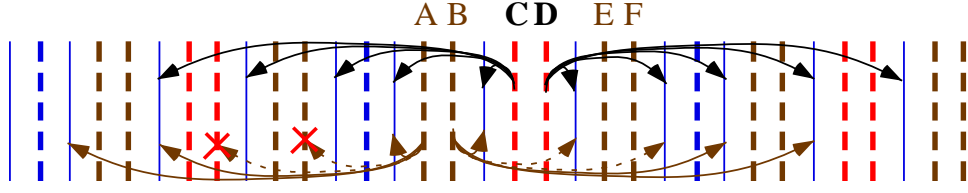

Fig.5.17 Coulomb transitions of nearest neighbor hole pairs at $\nu = 4/11$

The pair AB has no left-right symmetry and so there are some forbidden transitions to the electron states. Also the dashed pair EF cannot transfer to some of the electron states because of the momentum conservation and the Fermi-Dirac statistics. The perturbation energy of the nearest hole pairs is obtained by counting the number of allowed transitions as follows:

$$\varsigma_{CD} = -(4/11)Z_H, \quad \varsigma_{AB} = -(2/11)Z_H, \quad \varsigma_{EF} = -(2/11)Z_H$$

The sum of the energies for all the nearest pairs is given by

$$\left(E_{\text{nearest pair}}\right)_{\nu=4/11} = -\frac{8}{77}Z_H N_H \qquad \text{for } \nu = 4/11 \qquad (5.72a)$$

$$\left(E_{\text{nearest pair}}/N_H\right)_{\nu=4/11} = -\frac{8}{77}Z_H \qquad \text{for } \nu = 4/11 \qquad (5.72b)$$

$$\lim_{\nu \to (4/11)\pm\varepsilon}\left(E_{\text{nearest pair}}/N_H\right) \approx -0.097403 \times Z_H \qquad \text{for } \nu_\pm = 4/11 \pm \varepsilon \qquad (5.72c)$$

Equations (5.72b) and (5.72c) are re-expressed to the energy per electron (not hole) as

$$\left(E_{\text{nearest pair}}/N\right)_{\nu=4/11} = -\frac{8}{77}Z_H \frac{7}{4} = -\frac{2}{11}Z_H \approx -0.181818 \times Z_H \quad \text{for } \nu = 4/11 \quad (5.73a)$$

$$\lim_{\nu \to (4/11)\pm\varepsilon}\left(E_{\text{nearest pair}}/N\right) \approx -0.170455 \times Z_H \qquad \text{for } \nu_\pm = 4/11 \pm \varepsilon \qquad (5.73b)$$

The difference between Eqs.(5.73a) and (5.73b) gives the small energy gap (depth of the valley):

$$\Delta\varepsilon_\pm(4/11) = \left(E_{\text{nearest pair}}/N\right)_{\nu=4/11} - \lim_{\nu \to (4/11)\pm\varepsilon}\left(E_{\text{nearest pair}}/N\right) \approx -0.011363 \times Z_H \qquad (5.74)$$

These energy gaps (depths of the valley) at $\nu = 7/11$ and $4/11$ are smaller than that at $\nu = 2/3$. Similarly we calculate the energies of the nearest pairs at $\nu = 8/13$ and its neighbours as follows:

$$\left(E_{\text{nearest pair}}/N\right)_{\nu=8/13} = -\frac{11}{104}Z$$

$$\left(E_{\text{nearest pair}}/N\right)_{\nu=8003/13005} = -\frac{10508002}{104079015}Z$$

$$\left(E_{\text{nearest pair}}/N\right)_{\nu=8005/13008} = -\frac{2628251}{26032260}Z$$

Also the small valley appears at $\nu = 5/13$. Table 5.9 shows the depths of non standard filling factors.

Table 5.9 Energy gap of *nearest pairs* per electron at the filling factors 7/11, 4/11, 8/13 and 5/13

| $\nu$ | $E_{\text{nearest pair}}/N$ | $\Delta\varepsilon_+(\nu) = \Delta\varepsilon_-(\nu)$ |
|---|---|---|
| 7/11 | $-(8/77)Z$ | $-0.006493 \times Z$ |
| 4/11 | $-(2/11)Z_H$ | $-0.011363 \times Z_H$ |
| 8/13 | $-(11/104)Z$ | $-0.004808 \times Z$ |
| 5/13 | $-(11/65)Z$ | $-0.007691 \times Z_H$ |

There is a common property in these ground states. We explain it. The most uniform configurations at $\nu = 4/13, 5/13, 5/17$ and $6/17$ are schematically drawn in Figs.5.18-21, as follows;

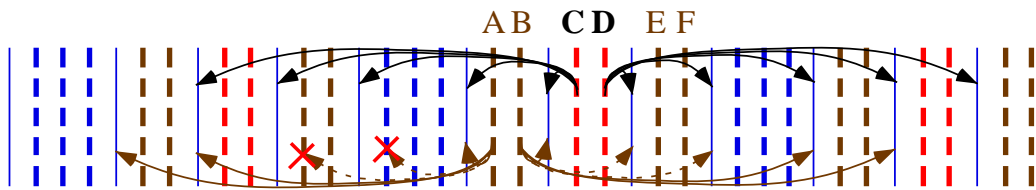

Fig.5.18 Coulomb transitions of nearest neighbor hole pairs at $\nu = 4/13$

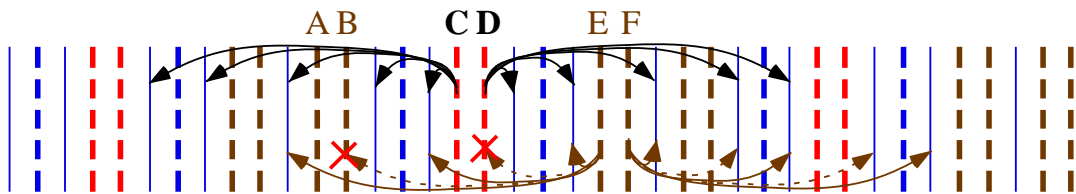

Fig.5.19 Coulomb transitions of nearest neighbor hole pairs at $\nu = 5/13$

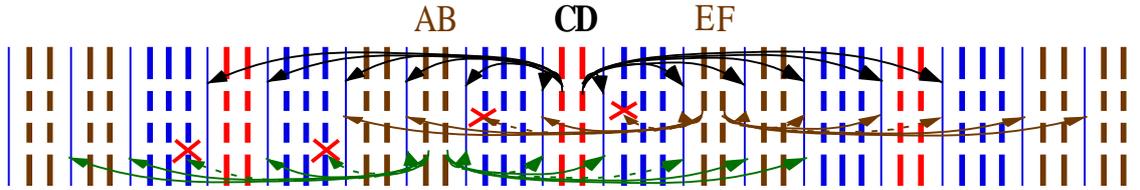
Fig.5.20 Coulomb transitions of nearest neighbor hole pairs at $\nu=5/17$

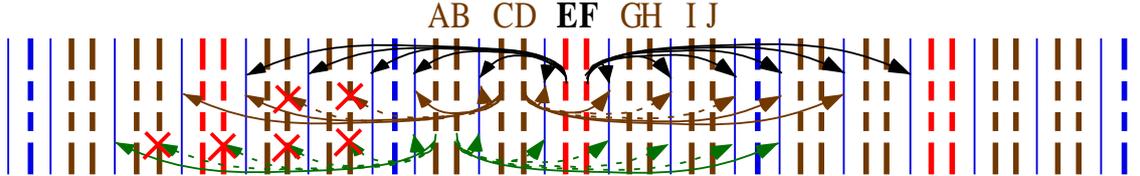
Fig.5.21 Coulomb transitions of nearest neighbor hole pairs at $\nu=6/17$

The nearest hole pairs CD with red colour can transfer to all the electron states at $\nu=$ 4/13, 5/13 and 5/17 as shown in Figs.5.18-20. Also the nearest red hole pair EF in Fig.5.21 can transfer to all the electron states. When the filling factor deviates slightly from the original value, the left-right symmetry with respect to the centre of CD (or EF for $\nu=6/17$) is lost. The left-right symmetry ensures that the hole pair can transfer to all the electron states. So the broken symmetry means that the number of allowed transitions becomes discontinuously small. Thus the small valleys appear in the energy spectrum at $\nu=7/11, 4/11, 4/13, 5/13, 5/17$ and $6/17$. The theoretical results are in good agreement with the experimental data [41].

### 5.7 Excitation-energy-gap in FQH states

We have used the term "energy gap (depth of the valley)" for the gap in the energy spectrum versus filling factor as in the previous sections. This gap (depth) produces a plateau in the Hall resistance curve which will be shown in Chapter 7. There is another gap which is named "excitation-energy-gap". The excitation-energy-gap is defined as the difference between the ground state energy and the first excited state energy. The excited state has the same filling factor as that of the ground state. We investigate the relation between "energy gap in the spectrum" and "excitation-energy-gap".

We first examine an excited state at $\nu=2/3$ as an example. The ground state with $\nu=2/3$ has the electron configuration as in Fig.4.3 where all the electrons are paired. When an electron in the ground state is excited, the electron moves to another empty orbital. So, one electron pair is lost and one empty orbital is occupied by the electron.

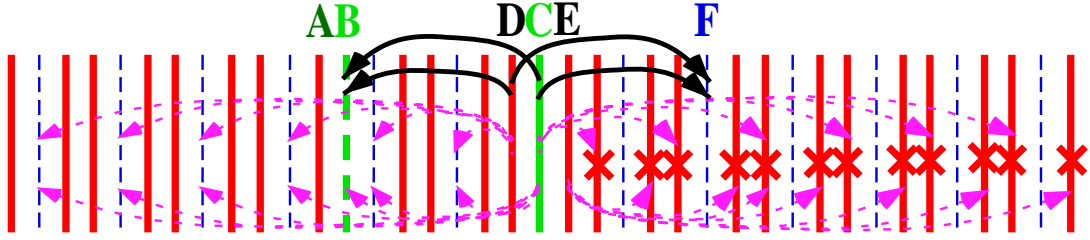

Fig.5.22 Electron configuration in excited state #1 at $\nu = 2/3$

We illustrate an example of the excited state. The electron in the orbital B is transferred to the orbital C by the excitation as in Fig.5.22. Then the landau orbital B becomes empty (bold dashed green line) and the Landau orbital C is filled with electron (bold solid green line) as expressed in Fig.5.22. We call this configuration "excited state #1". The excited state #1 has the filling factor $\nu = 2/3$ exactly same as that of the ground state. Therein the nearest pair AB disappears via this excitation and two nearest-electron-pairs DC and CE are newly created. The nearest pair DC can transfer to the two empty states as illustrated by black arrow pairs in Fig.5.22. The other transitions from the pair DC are forbidden as shown by pink color. All the transitions from the pair CE are forbidden because of the momentum conservation and the Pauli exclusion principle. In the case of the previous sections the transition number is enormously large and so the number two is negligibly small. Accordingly the excitation energy is caused by disappearance of the electron pair AB. We examine the details below.

The second order perturbation energy of the nearest pair DC is described by the symbol $\varsigma'_{DC}$, where the symbol prime ($\varsigma'$) indicates the perturbation energy in the excited state #1. The energy $\varsigma'_{DC}$ is given by

$$\varsigma'_{DC} = \sum_{\Delta p = 6(2\pi\hbar/\ell),\, -7(2\pi\hbar/\ell)} \frac{\langle p_D, p_C | H_I | p'_D, p'_C \rangle \langle p'_D, p'_C | H_I | p_D, p_C \rangle}{W_{state\#1} - W_{excite}(p_D \to p'_D, p_C \to p'_C)} \quad (5.75)$$

We introduce the following summation $Z'$ as:

$$Z' = -\sum_{\Delta p \neq 0,\, -2\pi\hbar/\ell}^{all} \frac{\langle p_D, p_C | H_I | p'_D, p'_C \rangle \langle p'_D, p'_C | H_I | p_D, p_C \rangle}{W_{state\#1} - W_{excite}(p_D \to p'_D, p_C \to p'_C)} \quad (5.76)$$

It is noteworthy that Eq. (5.76) is obtained by replacing $W_G$ by $W_{state\#1}$ into Eq.(4.16). The summation in (5.75) is carried out only for the two momenta. On the other hand, in Eq. (5.76), the summation is carried out for a large number of momenta. The Landau wave function spreads in the width $\Delta y$ along the y direction, the value of which was already calculated in (1.19) of chapter 1 as

$$\Delta y \approx \sqrt{\frac{\hbar}{m^*\omega}} = \sqrt{\frac{\hbar}{eB}} \approx 10.5 \text{ nm} \qquad \text{for the case of } B \approx 6\,[\text{T}] \qquad (5.77)$$

From Eq.(1.10), the distance between adjacent orbitals is,

$$\Delta \alpha = \frac{2\pi\hbar}{eB\ell} \approx 6.5 \times 10^{-4}\,[\text{nm}] \qquad \text{for } \ell = 1\,[\text{mm}] = 10^6\,[\text{nm}] \qquad (5.78)$$

Then a large number of single electron states exists inside the width $\Delta y$. The number of the states is

$$\frac{\Delta y}{\Delta \alpha} \approx \frac{\ell}{2\pi}\sqrt{\frac{eB}{\hbar}} \approx 2 \times 10^4 \qquad (5.79)$$

Accordingly about $2 \times 10^4$ wave functions overlap and so the Coulomb transitions are effective in the region as;

$$|p'_C - p_C| \leq \frac{2\pi\hbar}{\ell} \times 2 \times 10^4 \qquad (5.80)$$

The sum in Eq.(5.75) is done only for the two transitions and so the ratio $-\varsigma'_{DC}/Z'$ is about $10^{-4}$:

$$-\varsigma'_{DC}/Z' \approx 2/(2 \times 10^4) = 10^{-4} \qquad (5.81)$$

Because this value is negligibly small, we obtain the following result.

$$\varsigma'_{DC} \approx -10^{-4} \times Z' \approx 0 \qquad (5.82a)$$

All the transitions from the pair CE are forbidden and then the perturbation energy of the nearest pair CE is zero:

$$\varsigma'_{CE} = 0 \qquad (5.82b)$$

Furthermore the nearest pair AB in the ground state disappears in the excited state #1 as seen in Fig.5.22. From these considerations the excitation-energy $\Delta E_{\text{excitation \#1}}$ is given by

$$\Delta E_{\text{excitation \#1}} = \varsigma'_{DC} + \varsigma'_{CE} - \varsigma_{AB} \qquad (5.83)$$

where $\varsigma_{AB}$ indicates the summation (4.15), the result of which is equal to (4.17) as

$$\varsigma_{AB} = -(1/3)Z \qquad (5.84)$$

Substitution of Eqs.(5.82a, b) and (5.84) into Eq.(5.83) yields the excitation-energy-gap at $\nu = 2/3$ as

$$\Delta E_{\text{excitation \#1}} = \varsigma'_{DC} + \varsigma'_{CE} - \varsigma_{AB} \approx (1/3)Z \qquad \text{at } \nu = 2/3 \qquad (5.85)$$

This excitation-energy-gap $(1/3)Z$ is positive because $Z$ is a positive number.

The second example is the filling factor $\nu = 1/2$ shown in Fig.5.23. Therein the electron A moves to the orbital B from the pristine orbital. This excited state is named excited state #2. The excited state #2 has the two nearest electron pairs CB and BD.

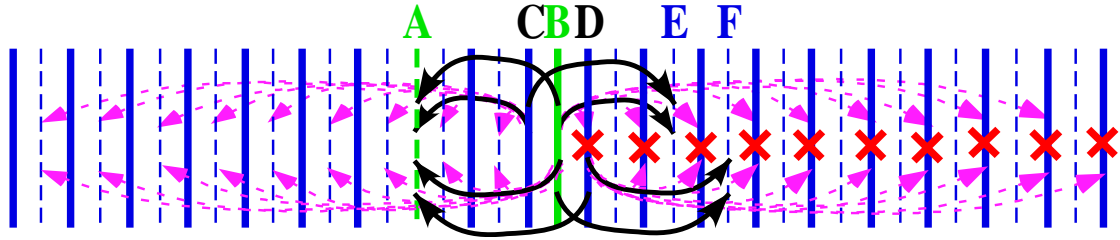
Fig.5.23 Electron configuration in excited state #2 at $\nu = 1/2$

The excitation energy $\Delta E_{\text{excitation \#2}}$ from the ground state to the state #2 is given by
$$\Delta E_{\text{excitation \#2}} = \varsigma''_{\text{CB}} + \varsigma''_{\text{BD}}$$
where $\varsigma''$ indicates the perturbation energy in the excited state #2. From the same argument as in Eq.(5.81), $\varsigma''_{\text{CB}}$ and $\varsigma''_{\text{BD}}$ are extremely small. Accordingly we obtain
$$\Delta E_{\text{excitation \#2}} = \varsigma''_{\text{CB}} + \varsigma''_{\text{BD}} \approx 0 \qquad \text{at } \nu = 1/2 \qquad (5.86)$$

The third example is the case of $\nu = 1/4$. When the hole at B transfers to the orbital C, the orbital B is filled with electron and the orbital C becomes empty. This transition yields the excited state #3 as in Fig.5.24.

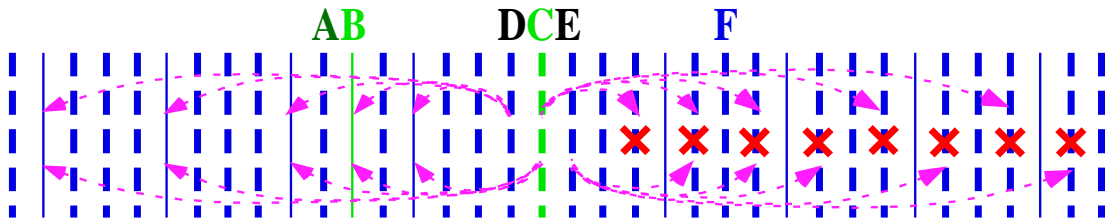
Fig.5.24 Electron configuration in excited state #3 at $\nu = 1/4$

The nearest hole pairs DC and CE are created and the two nearest-hole-pairs disappear from the ground state via this excitation process. Accordingly the excitation-energy $\Delta E_{\text{excitation \#3}}$ is given by

$$\Delta E_{\text{excitation \#3}} = \varsigma_{\text{DC}}^{\prime\prime\prime\ \text{hole}} + \varsigma_{\text{CE}}^{\prime\prime\prime\ \text{hole}} - 2\times \varsigma_{\nu=1/4}^{\text{hole}}$$

where $\varsigma_{\nu=1/4}^{\text{hole}}$ means the energy for the nearest hole pair in the ground state. The value of $\varsigma_{\nu=1/4}^{\text{hole}}$ is zero as shown in Table 5.4. All the quantum transitions from the hole-pairs DC and CE are forbidden as is illustrated by pink arrow pairs in Fig.5.24. Then the excitation-energy $\Delta E_{\text{excitation \#3}}$ is equal to zero as

$$\Delta E_{\text{excitation \#3}} = \varsigma_{\text{DC}}^{\prime\prime\prime\ \text{hole}} + \varsigma_{\text{CE}}^{\prime\prime\prime\ \text{hole}} - 2\times \varsigma_{\nu=1/4}^{\text{hole}} = 0 \qquad \text{at}\ \nu=1/4 \qquad (5.87)$$

Thus the excitation-energies at $\nu=1/2$ and $1/4$ are nearly equal to zero. From the similar investigation mentioned above, the excitation-energy is almost equal to zero in the cases of the flat structure and the peak structure namely at $\nu=1/(2j)$ and $\nu=(2j-1)/(2j)$.

On the other hand, a large excitation-energy-gap appears for the case of the valley structure at $\nu=1/(2j+1)$, $\nu=(2j)/(2j+1)$, $\nu=j/(2j\pm1)$ and so on. The large excitation-energy-gap confines electrons to the ground state and suppresses the scatterings of electron by impurities, lattice defects and lattice vibrations. Accordingly the diagonal resistance $R_{xx}$ is expected to be very small at these filling factors. The result of the present theory is compared with the experimental curve of $R_{xx}$ in the next section.

## 5.8 Comparison between the theory and the experimental data in a wider region of magnetic field

We have calculated the perturbation energies of the *nearest electron* (or *hole*) *pairs* at the filling factors of $\nu=1/(2j+1)$, $\nu=(2j)/(2j+1)$, $\nu=j/(2j\pm1)$, $\nu=j/(4j\pm1)$, $\nu=(3j\pm1)/(4j\pm1)$ $\nu=\frac{5}{8},\frac{3}{8},\frac{7}{10},\frac{3}{10},\frac{7}{11},\frac{4}{11},\frac{8}{13},\frac{5}{13}\cdots$, $\nu=1/(2j)$, $\nu=(2j-1)/(2j)$ and so on. Then we have obtained the three types of the energy spectrum, namely valley, flat and peak structures. The valley structure has been studied in sections 5.1, 5.2, 5.4. The small valley structure has been examined in section 5.6. The other structures (peak and flat structures) have been also investigated in sections 5.3, and 5.5. We compare the theoretical results with the experimental data in Fig.5.25.

1. **(Valley structure)**: At $\nu_1 = 1/(2j+1)$, $(2j)/(2j+1)$, $j/(2j\pm1)$, and $j/(4j\pm1)$, the nearest electron (or hole) pairs can transfer to all of the empty (or filled) orbitals. When the filling factor $\nu$ deviates from $\nu_1$, the repeating of the unit-sequence at $\nu_1$ is violated by the deviation $\nu - \nu_1$. So the nearest electron (or hole) pairs cannot transfer to some empty (or filled) orbitals. Thereby a valley appears at these filling factors. The energy gaps (depths of the valleys) are obtained in Tables 5.1, 5.2, 5.5 and 5.6. The following depths $\Delta\varepsilon_\pm\left(\frac{1}{3}\right) = -(1/6)Z_H$, $\Delta\varepsilon_\pm\left(\frac{1}{5}\right) = -(1/10)Z_H$ and $\Delta\varepsilon_\pm\left(\frac{2}{3}\right) = -(1/12)Z$ are the most deep three in our calculation. Then the electron scatterings are suppressed by the large excitation energies. Accordingly we can expect that the value of $R_{xx}$ is nearly equal to zero at $\nu = 1/3, 2/3, 1/5$. Experimentally $R_{xx}$ is almost zero in a wide range at the three filling factors as shown by blue arrows (The $\nu = 1/5$ state is out of range in Fig.5.25). The other filling factors of this type are shown by green and pink arrows for $\nu_1 = j/(2j\pm1)$ and $j/(4j\pm1)$, respectively. The experimental curve of $R_{xx}$ exhibits local minima at these filling factors. Thus the theoretical results are in good agreement with the experimental data.

2. **(Small valley structure)**: The red pairs of electrons (or holes) in Figs.5.16-21 can transfer to all the empty (or filled) orbitals at $\nu_2 = \frac{7}{11}, \frac{4}{11}, \frac{4}{13}, \frac{5}{13}, \frac{5}{17}, \frac{6}{17}$. However the brown pairs cannot transfer to some empty (or filled) orbitals. This property yields a small valley as calculated in Eqs. (5.71b), (5.74) and so on. The energy depth of the valley is very small; $-0.006493 \times Z$ at $\nu_2 = 7/11$ and $-0.011363 \times Z_H$ at $\nu_2 = 4/11$ and so on. The experimental value of $R_{xx}$ has small local-minima at these filling factors which are expressed by brown arrows in Fig.5.25. Thus our theoretical results are in agreement with the experimental data.

3. **(Peak structure)**: At $\nu_3 = 1/(2j)$ and $(2j-1)/(2j)$ for $j \geq 2$ all the *nearest electron* (or *hole*) *pairs* cannot transfer to any empty (or filled) orbital. Then all the nearest pairs have zero binding energy which is the highest value among the second order perturbation energies. In the neighbourhood of $\nu_3$ the nearest pair energies have finite negative values as in Tables 5.3 and 5.4. Accordingly the state at $\nu_3$ has a peak structure. The effect of more distant pairs cannot cancel the peak because the peak value is large. So the excitation-energy-gap is almost zero at $\nu_3$ as in the previous section 5.7, and then the electron scatterings are not suppressed. Consequently it is concluded that the diagonal resistance $R_{xx}$ is finite at these filling factors. The property is in agreement with the experimental data shown by red arrows in Fig.5.25.

4. **(Flat structure)**: At $\nu_4 = 1/2$ the allowed transition number from the *nearest*

*electron* (or *hole*) pairs is equal to that in the neighbourhood of $v_4$. Then the energy spectrum of the nearest pairs is continuous namely flat structure. Also we will examine the FQH states with $v = 5/2, 7/2 \cdots$ in section 5.9 and 5.10. Therein the specific aspects appear as in Figs.5.26-31, Fig.5.36 and Fig.5.39.

Also the allowed transition number from the *nearest electron* (or *hole*) pairs is equal to that in the neighbourhood of $v_5 = 5/8, 3/8, 7/10, 3/10$. So the more distant pairs are important to know the detail structure at $v_5$. When we add the perturbation energy of the second nearest pairs at $v_5 = 5/8, 3/8, 7/10, 3/10$, the small valley structure appears in the theoretical spectrum. These experimental data are shown by black arrows in Fig.5.25.

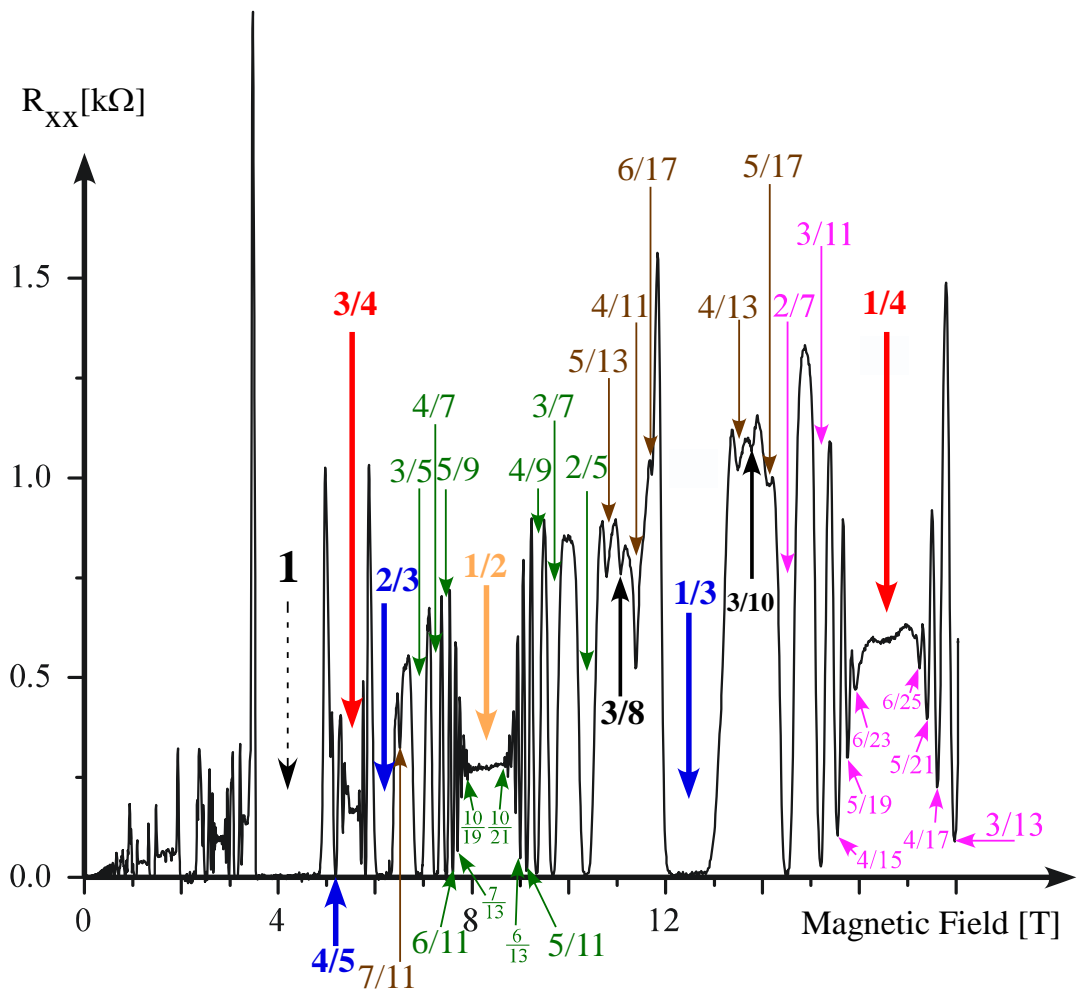

Fig.5.25 Diagonal resistance versus magnetic field in the region of $0 < B < 18$ [T]
Many types of the filling factors are drawn by the arrows with different colours. The experimental data are measured by W. Pan, H.L. Stormer, D.C. Tsui, L.N. Pfeiffer, K.W. Baldwin, and K.W. West, in Ref. [41].

Thus the present theory can explain the many filling factors with a local minimum in the diagonal resistance on the basis of the fundamental Hamiltonian of the normal electrons without any quasi particles. On the other hand the traditional theories have employed many quasi-particles with many different types. We will compare the present theory with the traditional theories in chapter 12.

Further investigation will be done for $\nu > 2$ in the next section 5.9 and for $1 < \nu < 2$ in the section 5.11. Section 5.10 is devoted to examine the effect of the more distant pairs. The valley structure is caused by the drastic number-decreasing of the allowed-transitions from the nearest pairs when the filling factor varies slightly from the original value. This drastic decrease is independent of the perturbation order. Therefore the discontinuous structure appears for all higher orders of the perturbation calculation which will be examined in Chapter 6. We will quantitatively compare the theoretical values of the valley depths with the experimental data in Chapter 7.

## 5.9 Pair energy of electrons placed in the second neighbouring Landau orbitals (Explanation of the Hall plateaus in the region $2 < \nu < 4$)

In the previous sections we have examined the pair energies of electrons (or holes) placed in the nearest neighbouring Landau orbitals. In this section, we investigate the effect of quantum transitions from electron pairs placed in the second neighbouring Landau orbitals. The pair energy of electrons (or holes) placed in the second nearest orbitals yields a small valley at the specific filling factors, which is less than about 0.1 times that of the nearest pairs.

### 5.9.1 Experimental data for $\nu > 2$

In this subsection we shortly study the experimental results for $\nu > 2$. The plateau at $\nu = 5/2$ attracts a great deal of attention because of a new FQH character. First, we review the experimental findings at $2 < \nu < 4$.

In the experiment by Pan, Du, Stormer, Tsui, Pfeiffer, Baldwin, and West [42, 43], the diagonal resistance exhibits a deep minimum at $\nu = 5/2$ and 7/2 as in Fig. 5.26. On the other hand, at $\nu = 9/2$ and 11/2 the diagonal resistance exhibits a strongly anisotropic behaviour. Therein $R_{XX}$ has a sharp peak while $R_{YY}$ is much smaller than $R_{XX}$ and has a minimum at $\nu = 9/2, 11/2$ as is seen in the dashed curve of Fig.5.26 where $R_{XX}$ and $R_{YY}$ mean the diagonal resistances along the x and y directions, respectively.

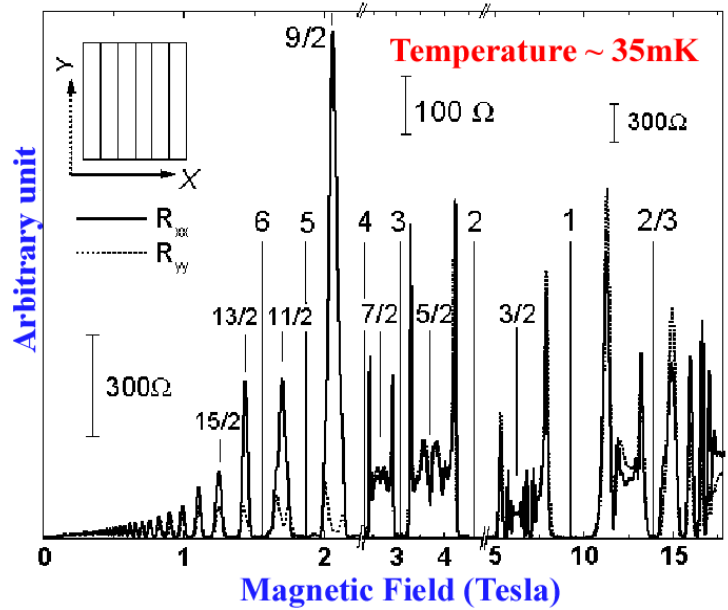

Fig.5.26 Diagonal resistance in wide region of magnetic field [42, 43]
The dashed curve indicates $R_{YY}$.

In the region $2 < \nu < 4$ the plateaus of Hall resistance have been found at $\nu = 5/2, 7/2$ (which have even number denominator) and at $\nu = 7/3$, 8/3, 11/5, 14/5, 16/5, 19/5 (which have odd number denominator) as seen in Fig.5.27. The experimental data are obtained by Eisenstein, Cooper, Pfeiffer, and West [44].

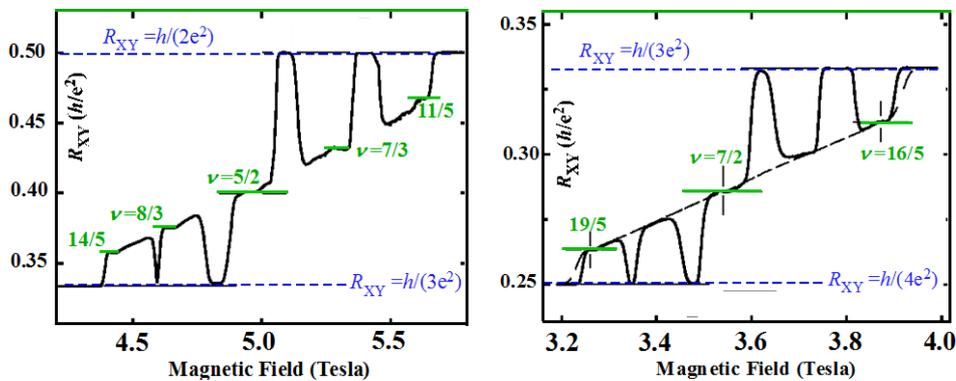

Fig.5.27 Behavior of Hall resistance in the region of $2 < \nu < 4$ [44]

The plateaus have the precise Hall resistance value. For example, the plateau at $\nu = 7/2$ has the Hall resistance value $2h/(7e^2)$ within the deviation of 0.015% as mentioned in [44].

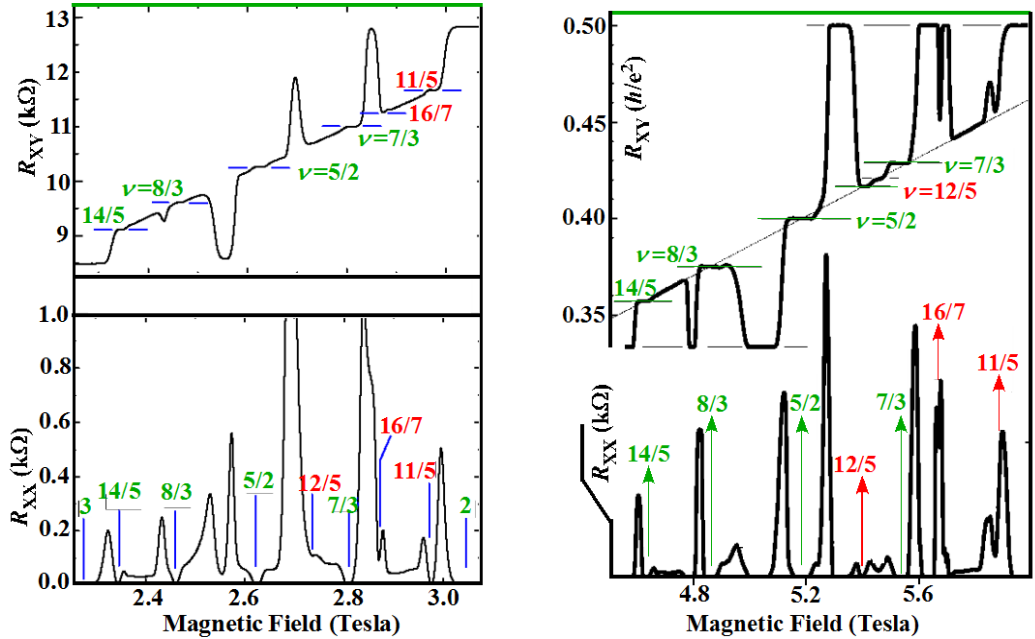

Fig.5.28 Hall resistance curve in the region $2 < \nu < 3$
The left panel is quoted from reference [45] and the right panel from [46]

We show other experimental data in Fig.5.28. The Hall resistance-curve in the left panel [45] of Fig.5.28 is different from that in the right panel [46]. This difference means that the shape of the Hall resistance versus magnetic field curve depends on the samples and the experimental conditions (magnetic field strength, temperature etc.). Especially the difference is large at $\nu = 16/7, 11/5$ and $12/5$.

When the magnetic field is tilted from the direction perpendicular to the 2D electron system, the Hall resistance plateau at $\nu = 5/2$ disappears as seen in Fig.5.29. On the other hand, the $\nu = 7/3, 8/3$ plateaus persist also under tilting of the magnetic field in Fig.5.29.

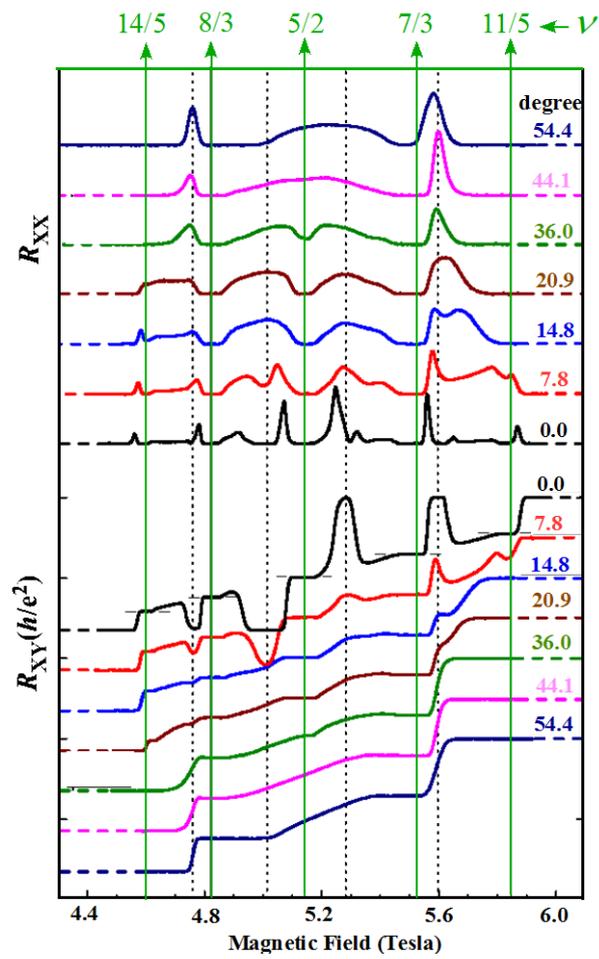

Fig.5.29 Tilt dependence of the Hall resistance and diagonal resistance [47]

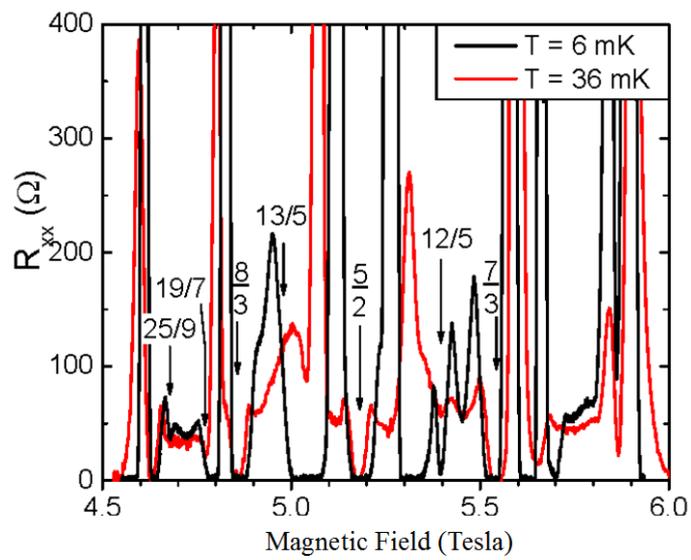

Fig.5.30 Temperature-dependence of the diagonal resistance [48]

The diagonal resistance curve depends on temperature as in Fig.5.30 where the red curve indicates the data at 36 mK and the black ones at 6 mK. Some local minima in the diagonal resistance curve disappear at 36 mK.

Many reserchers have measured the temperature dependence of $R_{xx}$ which gives the energy gaps from Arrhenius plots. The energy gaps are shown in Fig.5.31.

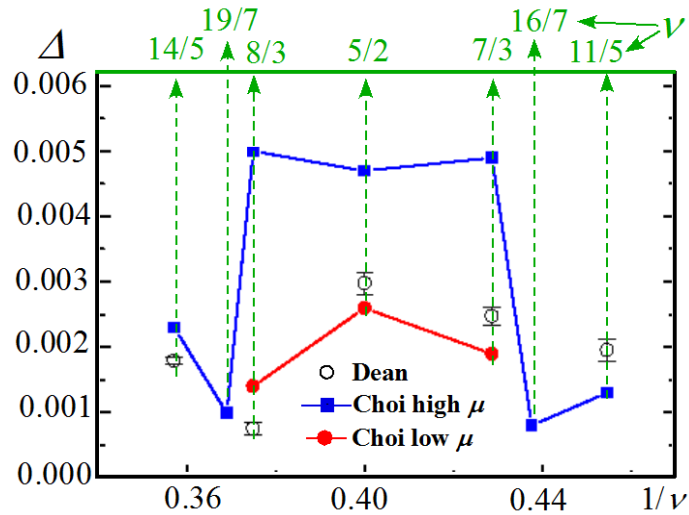

Fig.5.31 Energy gaps for the FQH states. Open circles are quoted from reference [45]. Solid squares and circles are quoted from reference [49]

In Fig.5.31 the energy gaps expressed by open crcles are obtained by Dean, Piot, Hayden, Sarma, Gervais, Pfeiffer, and West [45]. Also the data with the symbols of blue solid squares and red circles are mesured by Choi, Kang, Sarma, Pfeiffer, and West [49]. The energy gaps in the high mobility sample are drawn by the blue squares in Fig.5.31. The energy gaps at $\nu = 5/2, 8/3, 7/3$ etc. are listed in Table 5.10 which is measured by the experiment [49]. These values of the energy gap change from sample to sample at the same filling factor as easily seen in Fig. 5.31 and Table 5.10.

Table 5.10: Energy gap for the filling factors of 14/5, 19/7, 8/3, 5/2, 7/3, 16/7 and 11/5 in [49]

| ν | ν =14/5 | ν =19/7 | ν =8/3 | ν =5/2 | ν =7/3 | ν =16/7 | ν =11/5 |
|---|---|---|---|---|---|---|---|
| Sample A | 252 mK | 108 mK | 562 mK | 544 mK | 584 mK | 94 mK | 160 mK |
| Sample B | <60 mK | | 150 mK | 272 mK | 206 mK | | <40 mK |

The energy gaps for $\nu<1$ are plotted in Fig.5.32 which is quoted from the article [50]. Therein the $\nu=2/3$ energy gap is about 4.3 K. On the other hand the energy gap of $\nu=5/2$ state is about $0.272\sim 0.544\,\mathrm{K}$ as measured in the references [49]. Thus the energy gap in the region $2<\nu<3$ is about 1/10 times that in $\nu<1$. So we need to investigate the more detail structure in order to discuss the FQH states with $\nu>2$.

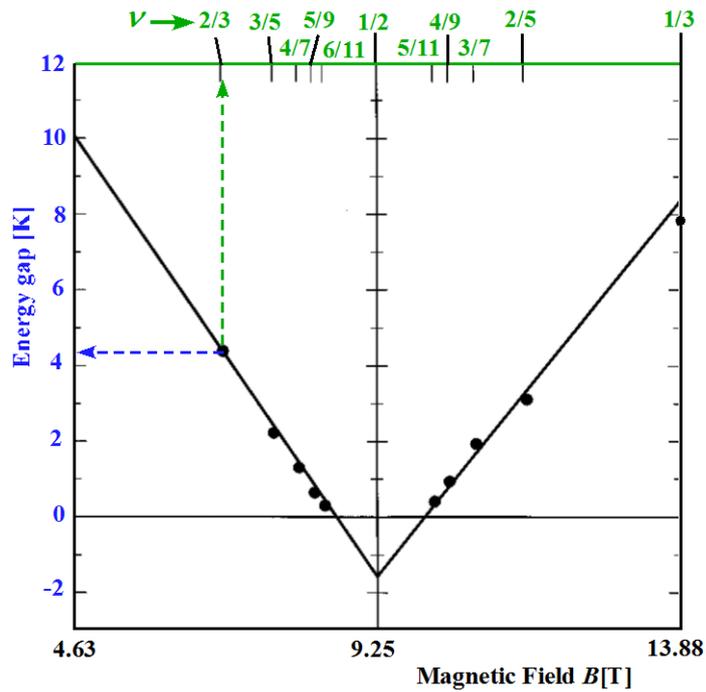

Fig.5.32 Energy gaps at $\nu=2/3,3/5,4/7,5/9,6/11,5/11,4/9,3/7,2/5$ in [50]

### 5.9.2 Theoretical approaches other than our theory for $\nu>2$

The experimental findings at $\nu>2$ have stimulated many physicists and then theoretical studies have been carried out. The theories assumed their own models to explain the plateaus of the Hall resistance at $\nu>2$ especially at $\nu=5/2$. Some of them are briefly reviewed below.

(The stripe states):
Fogler, Koulakov and Shklovskii have studied the ground state of a partially filled upper Landau level in a weak magnetic field. They have used the effective interaction [51] which was derived by Aleiner and Glazman in the 2D-electron system with high Landau levels, taking into account the screening effect by the lower fully occupied levels. Then they have found that the ground state is a charge density wave (CDW) state with a large period [52]. Moessner and Chalker studied a 2D-electron system with a fermion hardcore interaction and without disorder. They found a transition to both unidirectional and triangular charge-density wave states at finite temperatures [53]. Rezayi, Haldane and Yang numerically studied a 2D-electron system in magnetic field with a high Landau level half filled by electrons. In finite size systems with up to 12 electrons and torus geometry, they found a charge density wave ordering in the ground state. Their results show that the highest weight single Slater determinant has the occupation pattern 11111000001111100000 where 1 and 0 stand, respectively, for an occupied and an empty orbitals [54].

(The HR and MR states):
Haldane and Resayi investigated the pair state with spin-singlet [55]. They used a hollow core Hamiltonian. In the Landau level number $L=1$, the hollow core Hamiltonian has the first pseudopotential $V_1 > 0$ although the zeroth Haldane pseudopotential $V_0$ is zero. They found a ground state called HR state. Moore and Read are inspired by the structure of the HR state, and constructed the pair state, a p-wave $(p_x - ip_y)$ polarized state. They have described the FQH state in terms of conformal-field-theory [56]. The state is called the Moore-Read state (MR state). In the reference [57], Read wrote "the wavefunction $\psi_{MR}$ represents BCS [58] pairing of composite fermions. ⋯ One type are the charged vortices discussed above, with charge $1/(2q)$ which according to MR are supposed to obey nonabelian statistics." Greiter, Wen, and Wilczek investigated the MR state from the viewpoint of the composite fermion pair [59]. The statistics is an ordinary abelian fractional statistics.

(Numerical works)
 There are many numerical works. We will shortly review several works below:
 Morf argued the quantum Hall states at $\nu = 5/2$ by a numerical diagonalization [60]. He studied spin polarized and unpolarized states with $N \leq 18$ electrons. His result indicates that the 5/2 state is expected to be the spin-polarized MR state. Razayi and Haldane [61] confirmed Morf's results. Their results are based on numerical studies for up to 16 electrons in two geometries: sphere and torus. They found a first order phase transition from a striped phase to a strongly-paired state. They examined 12 electrons in a rectangular unit cell with the aspect ratio 0.5. They found the stripe state, the probability weight of which is 58% for the single Slater determinant state with the occupation pattern 000011110000111100001111. Also they found an evidence that the $\nu = 5/2$ state is derived from a paired state which is closely related to the MR polarized state or, more precisely, to the state obtained particle-hole (PH) symmetrisation of the MR state [61].

Thus there are many investigations for the $\nu = 5/2$ state. However the present author has some questions: (1) fractional charge, (2) small number of electrons in numerical calculations and so on. These problems are discussed in Chapter 12.

### 5.9.3 Explanation of 5/2 Plateau

In the previous sections and chapters we have ignored the discontinuous structure of the second-nearest and more distant pairs, because the pair-energies are expected to be smaller than those of the pairs placed in nearest neighboring Landau orbitals. The FQH state at $\nu > 2$ has an energy-gap smaller than that at $\nu \leq 2$ as shown in the subsection 5.9.1. So we need to take account of the contribution from the electron pairs placed in the second nearest and more distant Landau orbitals [62] in order to investigate FQH states with $\nu > 2$.

The energy difference between the Landau levels has been calculated in Eq.(1.21). The value $\Gamma_{\text{Landau}}/k_B$ is about 80.3[K] at $B = 4$[T] for GaAs which is very large in comparison with the gaps shown in Fig.5.31 and Table 5.10. Accordingly the mixing of higher levels ($L \geq 2$) in the ground state is negligibly small in the region $2 < \nu < 3$. Then all the Landau states with $L = 0$ are filled with electrons of up and down spins, and the Landau states with $L = 1$ are partially occupied by electrons.

The property of FQH states with $2 < \nu < 3$ is very different from that with $\nu < 2$ as:
1) The probability density of the Landau wave function is zero at the centre position for $L = 1$ but that is maximum for $L = 0$. Thus the shape of the $L = 1$ Landau wave function is quite different from that of $L = 0$. So the interactions between electrons in $L = 1$ are quite different from that in $L = 0$.
2) The $\nu$–dependence of the classical Coulomb energy at $\nu > 2$ is very small as will be studied at the subsection 5.9.7. The strength of $\nu$–dependence at $\nu = 5/2$ is 0.04 times that at $\nu = 1/2$ as will be estimated in Eqs.(5.118), (5.119a) and (5.119b). So the contribution from the second nearest pairs at $\nu > 2$ is more effective than that at $0 < \nu < 1$.
3) Furthermore the interactions between electrons in $L = 1$ are shielded by the many electrons in the lowest Landau level.

Thus the wave function and the interaction in a higher Landau level $L \geq 1$ are different from those in the lowest Landau level $L = 0$. On the other hand, the momentum conservation law and the Fermi-Dirac statistics of electrons are satisfied at $\nu > 2$ same as at $\nu \leq 2$. So we can use the same logic for the Coulomb transitions in the region $\nu > 2$.

Now we examine the case of $\nu = 5/2$. The most uniform electron configuration at $\nu = 5/2 = 2 + (1/2)$ is drawn in Fig.5.33 where only the orbitals with $L = 1$ are shown. All the orbitals with $L = 0$ are filled with electrons and these orbitals are not drawn in Fig.5.33 for simplicity. The electron pair CD in the Landau level $L = 1$ can transfer to all the empty states in $L = 1$ which are illustrated by the solid arrows in Fig.5.33. These transitions satisfy the momentum conservation law. Any electron pair placed in the second neighboring orbitals can also transfer to all the empty orbitals at $\nu = 5/2$.

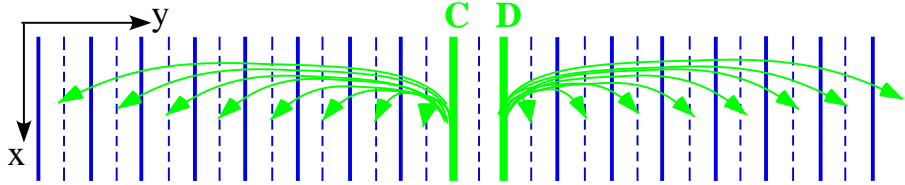

Fig.5.33 Most uniform configuration at $\nu = 5/2$

Allowed transitions are shown by arrows. Dashed lines indicate empty orbitals in the second Landau level $L = 1$

We calculate the perturbation energy for the electron pair placed in the second neighboring Landau orbitals. Therein we introduce the following integral $S$ for the Landau level $L = 1$.

$$S = - \sum_{\Delta p \neq 0, -4\pi\hbar/\ell} \frac{\langle L=1, p_C, p_D | H_I | L=1, p'_C, p'_D \rangle \langle L=1, p'_C, p'_D | H_I | L=1, p_C, p_D \rangle}{W_G - W_{\text{excite}}(p_C \to p'_C, p_D \to p'_D)}$$

$$p_D = p_C + 4\pi\hbar/\ell$$
$$p'_C = p_C - \Delta p, \quad p'_D = p_D + \Delta p \quad (5.88)$$

where the momentum-changes, $0$ and $-4\pi\hbar/\ell$, have been eliminated because the diagonal matrix element of $H_I$ is zero. It should be noted here that the summation $S$ is positive because the denominator in Eq.(5.88) is negative.

Then the perturbation energy $\varsigma_{CD}$ of the pair CD is expressed by using $S$ as

$$\varsigma_{CD} = -(1/2)S \quad (5.89)$$

Therein the factor 1/2 comes from the fact that the number of allowed transitions is equal to the number of the empty orbitals which is half of the total Landau orbitals with $L=1$. Let us count the number of electron pairs like CD. We define the total number of electrons placed in $L=1$ by the symbol $N_{L=1}$. Then the total number of the pairs like CD is equal to $N_{L=1}$. Accordingly the total energy of these pairs is given by

$$E^{pair}_{\nu=5/2} = -(1/2)SN_{L=1} \qquad (5.90)$$

The summation $S$ depends on the thickness, size, shape and material in the quasi-2D electron system. The z-component of the electron wave function depends on the thickness and the x-component depends on the device size and shape. Also the effective mass of electron and the permittivity depend on the material of the device. Accordingly the classical Coulomb energy $W$ and the transition matrix element in Eq.(5.88) vary with changing the quantum Hall device. That is to say the value of $S$ varies from sample to sample. Furthermore the $L=0$ electrons yield the screening effects for the classical Coulomb energy. Thus there are many unknown effects in a detail calculation of $S$. Accordingly the value of $S$ is treated as a parameter.

In order to clarify the stability of the $\nu=5/2$ state, we examine the $\nu=78/31$ state which is close to 5/2 as

$$\nu = 78/31 = 2.5161\cdots \qquad (5.91)$$

This filling factor is different from $\nu=5/2$ by about 0.6%. The most uniform electron configuration at $\nu=78/31=2+16/31$ is illustrated in Fig.5.34 where the Landau orbitals with $L=0$ are not shown, for simplicity. The electron pair CD can transfer to any empty orbital as shown by green arrows. On the other hand the pair IJ can transfer to three sites per unit configuration as shown by blue arrows in Fig.5.34. Therefore the second order energy is given by

$$\varsigma'_{CD} = -(15/31)S \qquad \text{for } \nu=78/31 \qquad (5.92)$$
$$\varsigma'_{IJ} = -(3/31)S \qquad \text{for } \nu=78/31 \qquad (5.93)$$

Similarly

$$\varsigma'_{DE} = -(13/31)S,\ \varsigma'_{EF} = -(11/31)S,\ \varsigma'_{FG} = -(9/31)S,\ \varsigma'_{GH} = -(7/31)S,$$
$$\varsigma'_{HI} = -(5/31)S,\ \varsigma'_{JA} = -(1/31)S, \quad \text{for } \nu=78/31 \qquad (5.94)$$

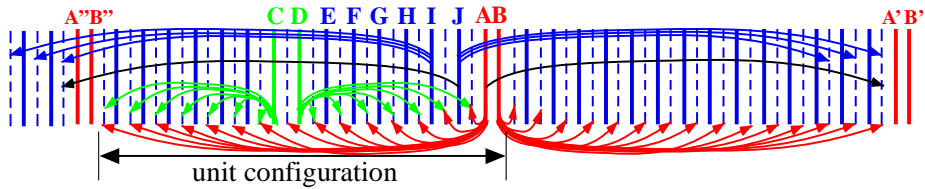

Fig.5.34 Quantum transitions at $\nu=78/31$ state

The electron pairs, AB, A'B', A"B", etc. are placed in the nearest Landau orbitals as seen in Fig.5.34. In order to calculate the energy of these pairs, we introduce the following summation:

$$T = -\sum_{\Delta p \neq 0, -2\pi\hbar/\ell} \frac{\langle L=1, p_A, p_B | H_I | L=1, p'_A, p'_B \rangle \langle L=1, p'_A, p'_B | H_I | L=1, p_A, p_B \rangle}{W_G - W_{excite}(p_A \to p'_A, p_B \to p'_B)} \quad (5.95)$$

where

$$p_B = p_A + 2\pi\hbar/\ell \quad (5.96)$$

This summation $T$ is also treated as a parameter like $S$. The nearest pair AB can transfer to all the empty states as shown by the red arrows in Fig.5.34 and the number of the empty states is 15 per unit configuration. The pair energy of AB is given by

$$\varsigma'_{AB} = -(15/31)T, \qquad \text{for} \quad \nu = 78/31 \quad (5.97)$$

We calculate the total energy from all the electron pairs placed in the first and second neighboring Landau orbitals with $L=1$:

$$E^{pair}_{\nu=78/31} = (\varsigma'_{AB} + \varsigma'_{CD} + 2\varsigma'_{DE} + 2\varsigma'_{EF} + 2\varsigma'_{FG} + 2\varsigma'_{GH} + 2\varsigma'_{HI} + 2\varsigma'_{IJ} + 2\varsigma'_{JA})(N_{L=1}/16)$$

$$E^{pair}_{\nu=78/31} = -((15/31)T + (113/31)S) \cdot (N_{L=1}/16) \quad (5.98)$$

If $E^{pair}_{\nu=78/31}$ is equal to $E^{pair}_{\nu=5/2}$, then the $\nu = 78/31$ state is mixing into the $\nu = 5/2$ state with equal probability caused by the thermal transitions. In this case the Hall resistance value deviates from $2h/(5e^2)$ by about 0.3% which is half of the difference between $\nu = 78/31$ and 5/2. So we examine the magnitude of $S$ and $T$. The difference of the pair energies is derived from Eqs.(5.90) and (5.98) as

$$E^{pair}_{\nu=5/2}/N^{L=1}_{\nu=5/2} - E^{pair}_{\nu=78/31}/N^{L=1}_{\nu=78/31} = (T - 9S)(15/496) \quad (5.99)$$

We discuss the following two cases 1 and 2.

(Case 1)  $9S \gg T$

In this case the perturbation energy of the first and second pairs per electron is

$$E^{pair}_{\nu=5/2}/N^{L=1}_{\nu=5/2} \ll E^{pair}_{\nu=78/31}/N^{L=1}_{\nu=78/31} \qquad \text{for} \quad S \gg T/9 \quad (5.100)$$

Accordingly the pair energy at $\nu = 5/2$ is sufficiently lower than that in the neighborhood of $v=5/2$. When the energy difference

$$-\left(E^{pair}_{v=5/2}/N^{L=1}_{v=5/2} - E^{pair}_{v=78/31}/N^{L=1}_{v=78/31}\right) \gg \text{Boltzmann constant} \times \text{Temperature}$$

then the $v=5/2$ state is confined and the Hall plateau appears at $v=5/2$.

(Case 2) $9S \ll T$

In this case, the $v=5/2$ Hall plateau does not appear because the pair energy at $v=5/2$ is higher than that in its neighborhood.

Thus the FQH state is sensitive to the relative value of $S$ and $T$ which are dependent on the materials, thickness, device shape, and so on. In the experiments [42-49] and [63-65], we also see that the $v=5/2$ Hall plateau is the sample dependent phenomena. So these phenomena can be understood by the present theory on the basis of the normal Hamiltonian without any quasi particle.

### 5.9. 4. Appearance or disappearance of the plateau at $v=5/2$ and $1/2$

For example, the $v=5/2$ and $7/2$ Hall plateaus do not exist on the red curve of Hall conductance in the article [63] as seen in Fig.5.35. This result has been obtained by Dean, Young, Cadden-Zimansky, Wang, Ren, Watanabe, Taniguchi, Kim, Hone & Shepard. On the other hand, the experimental results in Figs.5.26-31 indicate the appearance of 5/2 and 7/2 Hall plateaus. Thus appearance or disappearance of the $v=5/2$ and 7/2 plateaus seems to depend upon the samples used in the experiments.

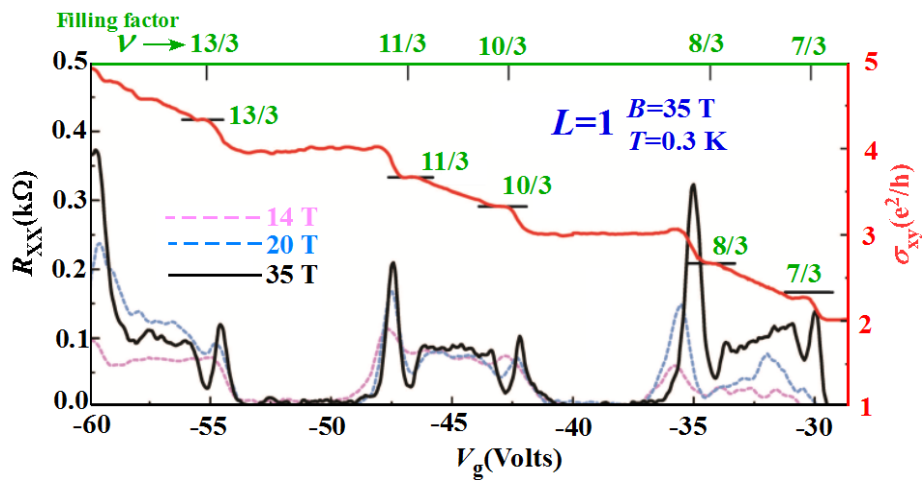

Fig.5.35 Plateaus of Hall conductance and local minima of diagonal resistance in the experimental results of the reference [63]

Similar phenomena have been found at $v=1/2$ in graphene samples as follows:

Figure 5.36 shows the appearance of the Hall plateau at $v=1/2$. Fig.5.36 was obtained by Dong-Keun Ki, Vladimir I. Fal'ko & Alberto F. Morpurgo. Also, Fig.5.37 was obtained by Kirill I. Bolotin, Fereshte Ghahari, Michael D. Shulman, Horst L. Stormer & Philip Kim where the $v=1/2$ Hall plateau appears at the point B of Fig.5.37.

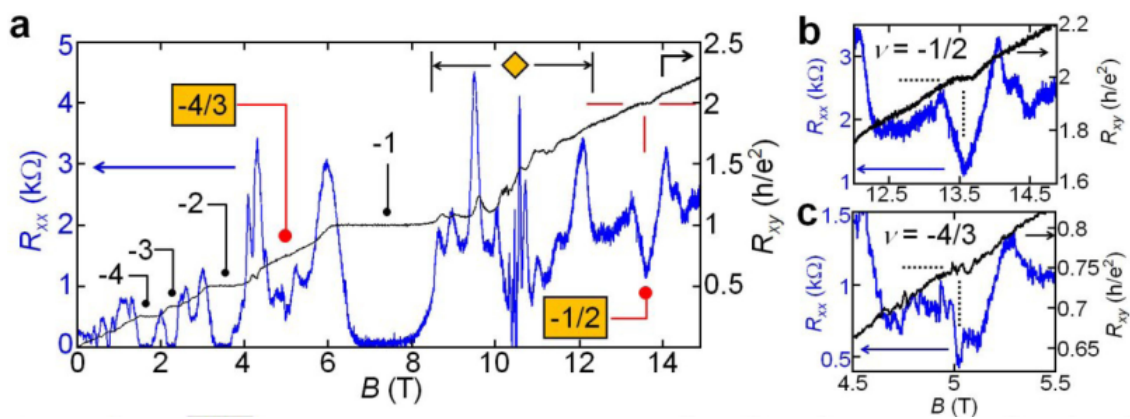

Fig.5.36 Plateaus of Hall resistance and local minima of diagonal resistance in the experimental results of the reference [64]

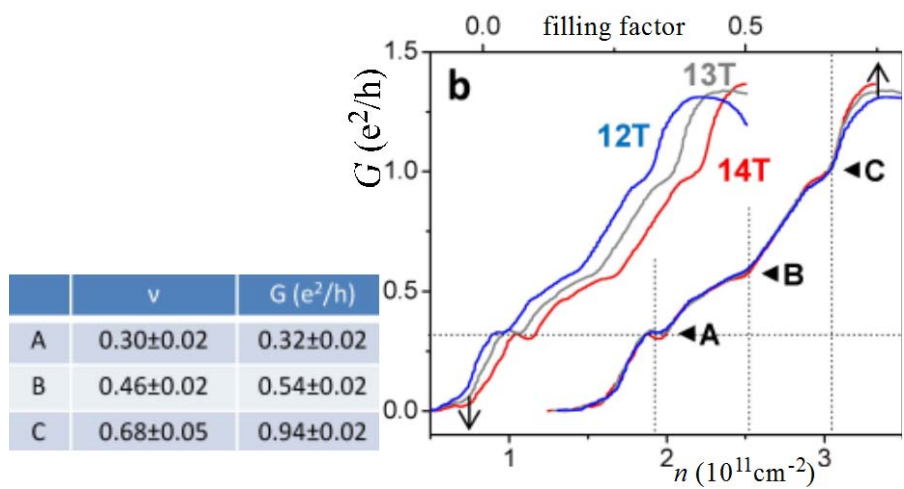

Fig.5.37 Plateaus of Hall conductance in the experimental results of the reference [65]

On the other hand, the plateau of the Hall resistance disappears at $v=1/2$ as already examined in Figs.5.9 and 5.25. Also the disappearance at $v=3/2$ can be seen in Fig.5.38 [63]. Thus different samples show different behaviours for the Hall plateau at the filling factor with an even number denominator.

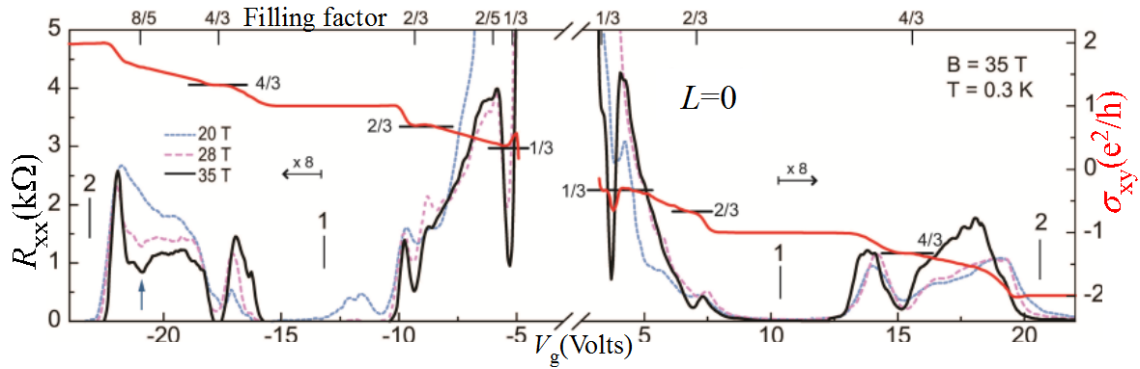

Fig.5.38 Plateaus of Hall conductance and local minima of diagonal resistance in the experimental results of the reference [63]

The difference has been also found in the GaAs samples. When the quantum Hall device has a wide quantum well, a $v = 1/2$ Hall plateau appears as shown in the experiments [66-68]. For example Shabani, Yang Liu, Shayegan, Pfeiffer, West, and Baldwin [67, 68] have observed the plateau of the Hall resistance at $v = 1/2$ as seen on the upper curve in Fig.5.39.

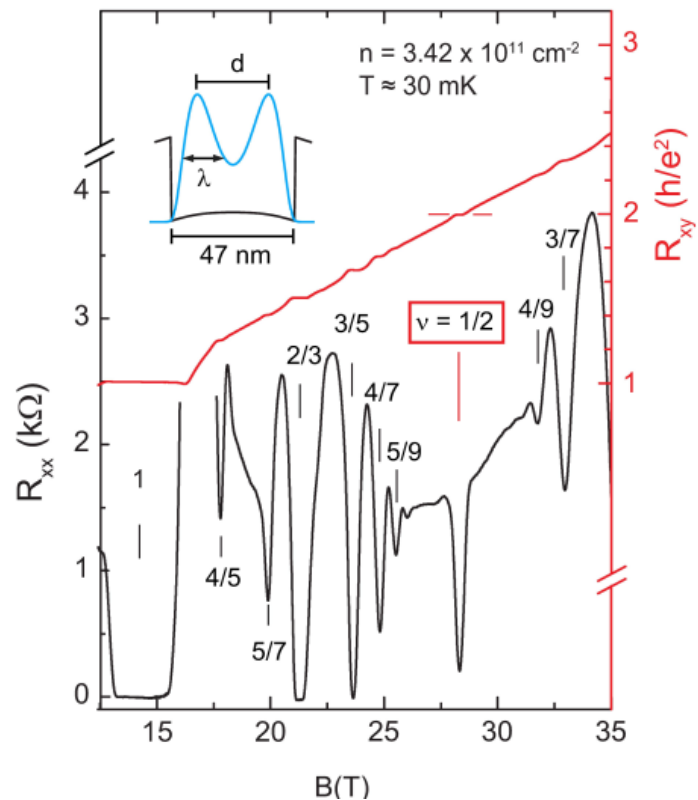

Fig.5.39 Plateaus of Hall resistance and local minima of diagonal resistance in the experimental results of the reference [67, 68]

Thus the Hall plateau at $v = 5/2$ and $v = 1/2$ depends on the samples used in the experiments. In the present theory this property is caused by the values of $S$ and $T$ which are also dependent upon the samples. Next we will calculate the pair energies for the filling factors $2.5 < v < 3$ and $2 < v < 2.5$ in the subsections 5.9.5 and 5.9.6, respectively [62]. Similar calculation for $v > 3$ can be performed by using the method of this section. The pair energy spectrum takes locally lowest at the filling factor $v = 7/2, 10/3, 11/3, 16/5$ and so on. These theoretical results are in agreement with the experimental data.

### 5.9.5 FQH states at filling factors $2.5 < v < 3$

In the region of $2.5 < v < 3$, the plateaus of Hall resistance are observed at the fractional filling factors $v = 8/3, 14/5, 19/7$ and so on. We examine the states in this subsection.

Figure 5.40 shows the most uniform configuration at $v = 8/3 = 2 + (2/3)$ where two types of electron pairs, AB and CD exist (where Landau orbitals with $L = 0$ are not drawn for simplicity). The pair AB represents the first nearest electron pair and the pair CD the second nearest one. Both the pairs AB and CD can transfer to all the empty orbitals with $L = 1$. The allowed transitions are shown by arrows in Fig.5.40.

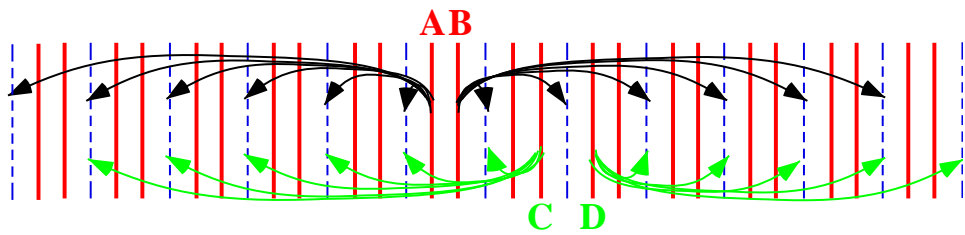

Fig.5.40 Quantum transitions at $v = 8/3$ state

The number of empty orbitals with $L = 1$ is 1/3 of the Landau orbitals with $L = 1$. Therefore the pair energies are given by

$$\varsigma_{AB} = -(1/3)T \qquad \text{for } v = 8/3 \qquad (5.101a)$$
$$\varsigma_{CD} = -(1/3)S \qquad \text{for } v = 8/3 \qquad (5.101b)$$

Accordingly the total energy of the electron pairs placed in the first and second neighboring Landau orbitals with $L=1$ is

$$E_{\nu=8/3}^{pair} = \varsigma_{AB} \times (N_{L=1}/2) + \varsigma_{CD} \times (N_{L=1}/2)$$

Substitution of Eqs.(5.101a and b) yields

$$E_{\nu=8/3}^{pair} = -(T/6 + S/6)N_{L=1} \qquad (5.102)$$

We examine the pair energy in the limit from the right and left to $\nu = 8/3$. Such a limiting value has already been calculated in section 5.4. Using the same method we obtain the limiting values from the right and left side to $\nu = 8/3$ as

$$\lim_{\nu \to (8/3)+\varepsilon} E_\nu^{pair} = -(T/12 + S/12)N_{L=1} \qquad (5.103a)$$

$$\lim_{\nu \to (8/3)-\varepsilon} E_\nu^{pair} = -(T/12 + S/12)N_{L=1} \qquad (5.103b)$$

Therefore an energy valley appears as follows:

$$\Delta E_{\nu=8/3}^{pair} = E_{\nu=8/3}^{pair} - \lim_{\nu \to (8/3)\pm\varepsilon} E_\nu^{pair} = -\frac{1}{12}(T+S)N_{L=1} \qquad (5.104)$$

Thus the FQH state has a deep valley at $\nu = 8/3$.

The $\nu = 14/5, 18/7, 19/7$ states have the most uniform configuration as shown in Figs.5.41, 5.42 and 5.43, respectively. The allowed transitions are schematically drawn by arrows for the first and second nearest electron pairs, respectively.

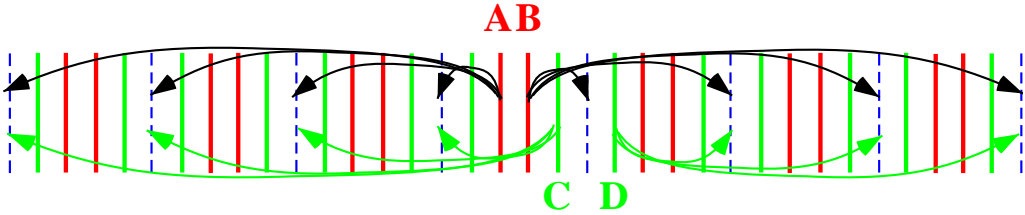

Fig.5.41 Quantum transitions at $\nu = 14/5$ state

The pair energy of AB and CD is given, respectively by

$$\varsigma_{AB} = -(1/5)T \qquad \text{for } \nu = 14/5 \qquad (5.105a)$$
$$\varsigma_{CD} = -(1/5)S \qquad \text{for } \nu = 14/5 \qquad (5.105b)$$

Then we obtain the total pair energy from the electron pairs placed in the first and second neighboring orbitals with $L=1$ as

$$E_{\nu=14/5}^{pair} = -(T/20 + S/20)N_{L=1} \qquad (5.106)$$

Figure 5.42 shows the allowed transitions of the pairs AB and CD at the filling factor $\nu = 18/7$.

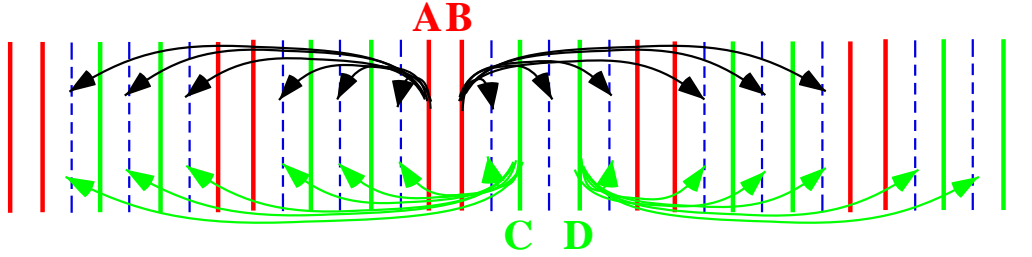

Fig.5.42 Quantum transitions at $\nu = 18/7$ state

The number of the empty orbitals is 3/7 times that of the Landau orbitals with $L=1$. Accordingly the pair energy of AB and CD is given, respectively

$$\varsigma_{AB} = -(3/7)T \qquad \text{for} \quad \nu = 18/7 \qquad (5.107a)$$
$$\varsigma_{CD} = -(3/7)S \qquad \text{for} \quad \nu = 18/7 \qquad (5.107b)$$

Then we obtain

$$E^{pair}_{\nu=18/7} = -(3T/28 + 3S/28)N_{L=1} \qquad (5.108)$$

Next we count the number of allowed transitions of the pairs AB and CD at $\nu = 19/7$. The electron pairs AB and CD in Fig.5.43 can transfer to all the empty Landau orbitals with $L=1$.

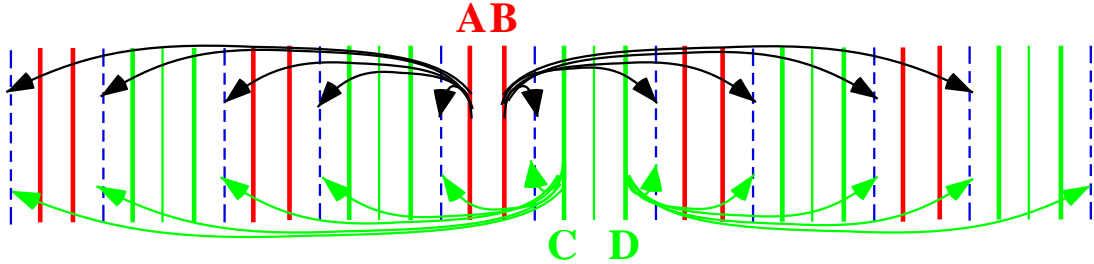

Fig.5.43 Quantum transitions at $\nu = 19/7$ state

Since the number of the allowed transitions is two per unit configuration, the pair energy of AB and CD is given, respectively,

$$\varsigma_{AB} = -(2/7)T \qquad \text{for} \quad \nu = 19/7 \qquad (5.109a)$$
$$\varsigma_{CD} = -(2/7)S \qquad \text{for} \quad \nu = 19/7 \qquad (5.109b)$$

which gives

$$E^{pair}_{\nu=19/7} = -(2T/35 + 2S/35)N_{L=1} \qquad (5.110)$$

Thus the electron pairs AB and CD can transfer to all the empty orbitals at $\nu = 5/2$, 8/3, 14/5, 18/7, 19/7, and therefore the pair energy becomes very low, resulting in a large binding-energy.

Next we examine the values of $S$ and $T$ which vary from sample to sample. Our calculations mentioned above have treated the case for no impurity-effect and absolutely zero temperature. So we should compare the theoretical results with the experimental data in a high mobility sample under an ultra low temperature as in Fig.5.31 (see reference [49]).

Eqs.(5.90) and (5.102) give the theoretical pair energy per electron:

$$E^{pair}_{\nu=5/2}/N_{L=1} = -(S/2) \quad (5.111a)$$

$$E^{pair}_{\nu=8/3}/N_{L=1} = -(T/6 + S/6) \quad (5.111b)$$

The experimental value of the energy gap at $\nu = 5/2$ is nearly equal to that at $\nu = 8/3$ for the high mobility sample in the experiment [58] as found in Fig.5.31 and Table 5.10. This experimental property is derived from the following condition between $S$ and $T$:

$$\text{(condition):} \quad T \approx 2S \quad (5.112)$$

At $\nu = 5/2$, 8/3, 14/5 and 19/7 the theoretical ratio of the pair energies is obtained by employing Eqs. (5.90), (5.102), (5.106) and (5.110) as follows:

$$\begin{aligned}&\left|E^{pair}_{\nu=5/2}/N_{L=1}\right|:\left|E^{pair}_{\nu=8/3}/N_{L=1}\right|:\left|E^{pair}_{\nu=14/5}/N_{L=1}\right|:\left|E^{pair}_{\nu=19/7}/N_{L=1}\right| \\ &= (S/2):(T/6+S/6):(T/20+S/20):(2T/35+2S/35)\end{aligned} \quad (5.113)$$

When the condition (5.112) is satisfied, the theoretical ratio of the pair energies becomes

$$(S/2):(T/6+S/6):(T/20+S/20):(2T/35+2S/35) = 1:1:(3/10):(12/35) \quad (5.114)$$

The experimental ratios are obtained from the experimental data of the energy gap for the high mobility sample in Fig.5.31 and Table 5.10.

$$0.0047 : 0.005 : 0.0023 : 0.001 = 0.94 : 1 : 0.46 : 0.2 \quad (5.115)$$

By comparing these ratios the theoretical ratio (5.114) is almost equal to the experimental ratio (5.115). Thus the present theory explains reasonably well the experimental data for $2.5 \leq \nu < 3$.

**5.9.6 FQH states at filling factors** $2 < \nu < 2.5$

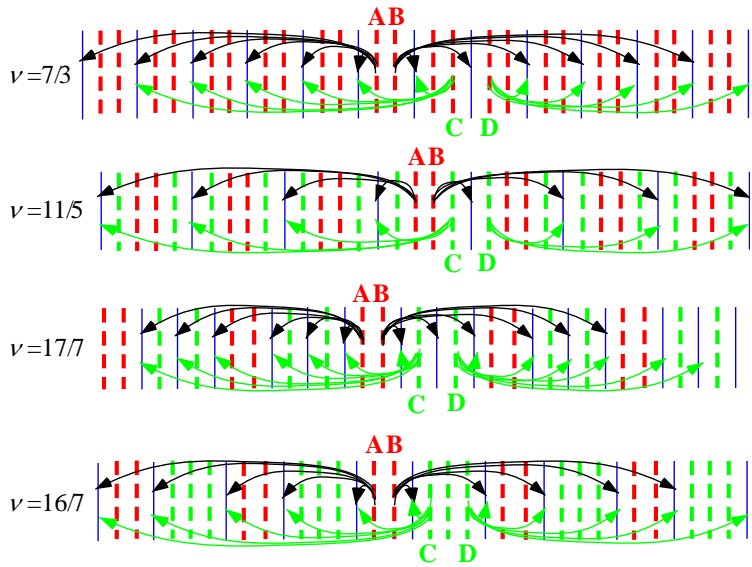

Fig.5.44 Quantum transitions at $\nu = 7/3$, 11/5, 17/7, 16/7 states

Dashed lines indicate empty Landau orbitals with $L=1$ and solid lines are orbitals filled with electron.

The most uniform configurations at $\nu = 7/3$, 11/5, 17/7 and 16/7 are schematically drawn in Fig.5.44. The hole-pairs AB and CD can transfer to all the electron states in $L=1$ as easily seen in Fig.5.44 (where the Landau orbitals with $L=0$ are not drawn for simplicity).

Figure 5.45 shows one to one correspondence of the allowed transitions between $\nu = 8/3$ and 7/3. The one to one correspondence yields the symmetry of the allowed transition numbers. That is to say the number of allowed transitions of the hole-pairs at $\nu = 7/3$ is equal to that of the electron-pairs at $\nu = 8/3$. In subsection 5.9.5 we have calculated the pair energies at $\nu = 8/3$, 14/5, 18/7 and 19/7 in Eqs. (5.102-110). So we can obtain the pair energies at $\nu = 7/3$, 11/5, 17/7 and 16/7 by using the symmetry between hole and electron. The absolute values of the pair energies is large and therefore these states become stable.

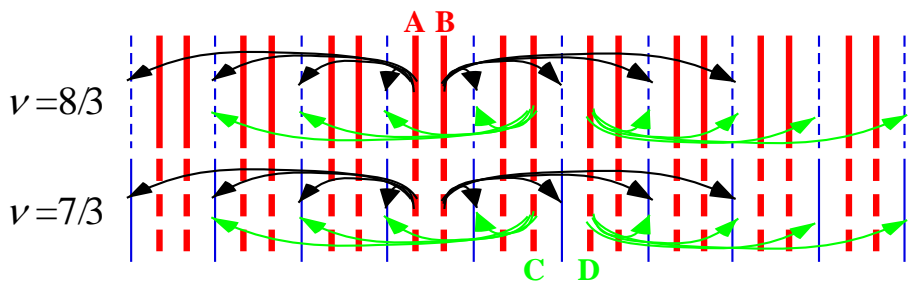

Fig.5.45 Comparison of allowed transitions between at $\nu = 8/3$ and at 7/3

Thus the present theory gives the fact that the state with $\nu = 5/2$, 7/3, 8/3, 11/5, 14/5, 16/7, 17/7, 18/7 and 19/7 is stable in the region $2 < \nu < 3$. The corresponding experimental data [45] shows the plateaus of the Hall resistance at $\nu = 5/2$, 7/3, 8/3, 11/5, 14/5, and 16/7 as in the left panel of Fig.5.28. Thereby the theoretical results are in agreement with the experimental data.

### 5.9.7 Comparison of classical Coulomb energies between $0 < \nu < 1$ and $2 < \nu < 3$

Equation (5.13) gives the energy per electron for FQH states where $\chi(\nu)$ is the pair energy between electrons placed in the nearest neighbouring Landau orbitals and $g(\nu)$ is the pair energy between electrons in the second-nearest and more distant Landau orbitals. We write again the energy per electron in the quasi 2D electron system:

$$\varepsilon(\nu) = \chi(\nu) + g(\nu) + (\hbar e/(2m^*))B - a/\nu + b \qquad (5.116)$$

where

$$a = (\xi - \eta) \qquad (5.117a)$$

$$b = f + (C_{\text{Macroscopic}}/N) \qquad (5.117b)$$

The energy increase $\Delta\varepsilon(\nu)$ by a small increase of $\nu$ is given by

$$\Delta\varepsilon(\nu) = \Delta\chi(\nu) + \Delta g(\nu) + (a/\nu^2)\Delta\nu \qquad (5.118)$$

where the last term has the following coefficients at $\nu = 2.5$ and 0.5:

$$(a/\nu^2) = 0.16 \times a \quad \text{at} \quad \nu = 2.5 \qquad (5.119a)$$

$$(a/\nu^2) = 4 \times a \quad \text{at} \quad \nu = 0.5 \qquad (5.119b)$$

Therefore the contribution of the last term at $\nu = 2.5$ is 0.04 times that at $\nu = 0.5$ for the same variation $\Delta\nu$. Accordingly the remaining terms $\Delta\chi(\nu)$ and $\Delta g(\nu)$ are more effective at $2 < \nu < 3$ than at $0 < \nu < 1$. So we have examined the energy valley of $\Delta g(\nu)$ in this section 5.9 and succeed to explain the plateaus of the Hall resistance in the region $2 < \nu < 3$.

### 5.10 Further investigation for the total energy of 2D electron system
       (*Energy of electron pair placed in more distant Landau orbitals*)

We examine the structure of the exact total energy in more details [62].

The total energy $E_\mathrm{T}$ of the quasi-2D electron system is the sum of the eigen-energy $W$ and the pair-energy $E^\mathrm{pair}$ via the interaction $H_\mathrm{I}$ as follows:

$$E_\mathrm{T} = W + E^\mathrm{pair} \tag{5.120}$$

where $W$ is the eigen-energy of $H_\mathrm{D}$ given by Eq.(3.3) as

$$W = \sum_{i=1}^{N} E_{L_i}(p_i) + C(L_1,\cdots,L_N; p_1,\cdots,p_N) \tag{5.121}$$

Substitution of Eq.(1.18) yields the following equation:

$$W = C(L_1,\cdots,L_N; p_1,\cdots,p_N) + N\lambda + \sum_{i=1}^{N} U(p_i/(eB)) + \sum_{i=1}^{N}(\hbar eB/m^*)(L_i + \tfrac{1}{2}) \tag{5.122}$$

The Landau energies between different levels are estimated for GaAs in Eq.(1.21):

$$(E_1 - E_0)/k_\mathrm{B} = (E_2 - E_1)/k_\mathrm{B} \approx 80.3[\mathrm{K}] \qquad \text{for } B = 4\,\mathrm{T} \tag{5.123}$$

which is extremely large in comparison with the experimental values of the energy gaps. (The values are smaller than 1 [K] at $2 < \nu < 3$ as in Table 5.10.)

In order to make our discussions clear, we consider the FQH states with $2 < \nu < 3$ as an example. In this case the probability of higher Landau levels $L \geq 2$ can be ignored in the exact ground state. Therein all the Landau states with $L = 0$ are occupied by electrons with up and down spins and the Landau states with $L = 1$ are partially occupied by electrons. The number of electrons in the Landau level $L$ is expressed by the symbol $N_\nu^L$ and the number of Landau orbitals by $N^\mathrm{orbital}$. The ratio $N_\nu^L/N^\mathrm{orbital}$ is given for $2 < \nu < 3$ as follows:

$$N_\nu^{L=0}/N^\mathrm{orbital} = 2 \qquad \text{in the ground state at } 2 < \nu < 3$$

(5.124a)

$$N_\nu^{L=1}/N^\mathrm{orbital} = \nu - 2 \qquad \text{in the ground state at } 2 < \nu < 3 \tag{5.124b}$$

$$N_\nu^{L=2}/N^\mathrm{orbital} = 0 \qquad \text{in the ground state at } 2 < \nu < 3 \tag{5.124c}$$

$$N_\nu^{L>2}/N^\mathrm{orbital} = 0 \qquad \text{in the ground state at } 2 < \nu < 3$$

(5.124d)

where $N^\mathrm{orbital}$ depends on the magnetic field strength and the device size but is independent of $L$. The total number $N$ of electrons is

$$N = N_\nu^{L=0} + N_\nu^{L=1} \qquad \text{in the ground state at } 2 < \nu < 3 \tag{5.125}$$

Substitution of Eqs.(5.124a,b,c,d) and (5.125) into Eq.(5.122) gives the eigen-energy $W$ of $H_D$ as

$$W = C(L_1,\cdots,L_N;p_1,\cdots,p_N) + N\lambda + \sum_{i=1}^{N} U(p_i/(eB)) + \tfrac{1}{2}(\hbar eB/m^*)N_\nu^{L=0} + \tfrac{3}{2}(\hbar eB/m^*)N_\nu^{L=1}$$

in the ground state at $2 < \nu < 3$ (5.126)

Next we investigate the pair energy which is caused by the quantum transitions via $H_1$. The electron pairs in the ground state with $2 < \nu < 3$ have been classified into the following three types:

(First type)   Both electrons in the pair are placed in the orbitals with $L = 0$ only.
(Second type)  One electron is placed in $L = 0$ and the other in $L = 1$.
(Third type)   Both electrons in the pair are placed in $L = 1$ only.

These pair energies are described by the symbols $E_{L=0}^{\text{pair}}$, $E_{L=0 \text{ and } 1}^{\text{pair}}$ and $E_{L=1}^{\text{pair}}$, respectively. The total energy of all the electron pairs is

$$E^{\text{pair}} = E_{L=0}^{\text{pair}} + E_{L=0 \text{ and } 1}^{\text{pair}} + E_{L=1}^{\text{pair}} \quad \text{in the ground state with } 2 < \nu < 3 \quad (5.127)$$

Therein the pair energies $E_{L=0}^{\text{pair}}$ and $E_{L=0 \text{ and } 1}^{\text{pair}}$ are negligibly small because of the following reason: As in Eqs.(5.88) an d (5.95) the perturbation energy is composed of the terms with the denominator $W_G - W_{\text{excite}}$. The denominator is very large in the first and second types because the quantum transition includes the excitation from $L = 0$ to $L = 1$. (Note: $W_{\text{excite}} - W_G$ includes the Landau excitation energy $\hbar eB/m^*$.) On the other hand the denominator is very small in the third type because the transition arises between the electron pairs with $L = 1$ only. (Note: $W_{\text{excite}} - W_G$ doesn't include the Landau excitation energy $\hbar eB/m^*$.) Therefore we may ignore the pair energy belonging to the first and second types.

$$E_{L=0}^{\text{pair}} \approx 0 \text{ and } E_{L=0 \text{ and } 1}^{\text{pair}} \approx 0 \quad \text{in the ground state with } \nu > 2 \quad (5.128)$$

In the third type the electron pair in $L = 1$ transfers to empty orbitals in $L = 1$ at $2 < \nu < 3$. Accordingly the energy difference of $W$ between the ground and the intermediate state is derived from the difference of the classical Coulomb energies. So the difference (denominator) is very small and the perturbation energy (which is negative) becomes very low. We consider any electron (or hole) pair placed in Landau orbitals with $L=1$. As an example, we examine the case of $v=8/3$. Figure 5.46 schematically shows the electron pairs at $v=8/3$. The electron pairs IL, HM and GN posses the total momentum same as that of the pair JK. So these pairs can transfer to all the empty orbitals as easily seen in Fig.5.46.

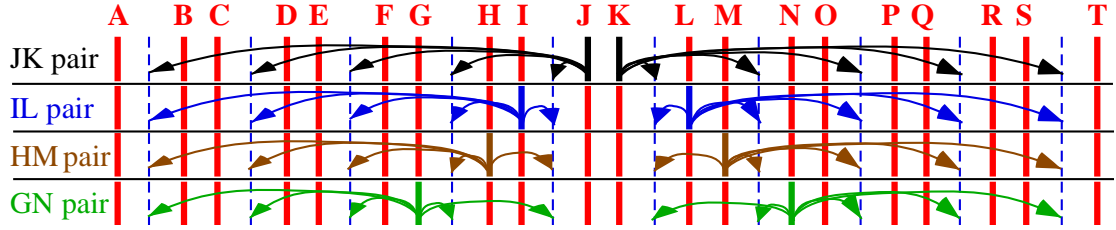

Fig.5.46 Various electron pairs with the same total momentum at $v=8/3$

Dashed lines indicate empty orbitals and solid lines indicate filled orbitals in the Landau level $L=1$. Allowed transitions from the electrons J and K are shown by black arrow pairs, from IL by blue, from HM by brown and from GN by green.

The momenta of the electrons at G, H, I, J, K, L, M and N are described by $p_G, p_H, p_I, p_J, p_K, p_L, p_M$ and $p_N$, respectively. Then the total momenta of the electron pairs take the same value due to Eq.(1.10) as

$$p_{total} = p_G + p_N = p_H + p_M = p_I + p_L = p_J + p_K \quad (5.129)$$

The energies of the pairs GN, HM, IL and JK are expressed systematically by using a symbol $\varsigma_v^{L=1}(p_{total}, j)$ where $p_{total}$ and $j$ indicate the total momentum and the distance between the pair.

$$\varsigma_{JK} = \varsigma_{v=8/3}^{L=1}(p_{total}, j=1) \quad (5.130a)$$

$$\varsigma_{IL} = \varsigma_{v=8/3}^{L=1}(p_{total}, j=5) \quad (5.130b)$$

$$\varsigma_{HM} = \varsigma_{v=8/3}^{L=1}(p_{total}, j=7) \quad (5.130c)$$

$$\varsigma_{GN} = \varsigma_{v=8/3}^{L=1}(p_{total}, j=11) \quad (5.130d)$$

Therein the momentum of each electron is given as

$$p_J = \tfrac{1}{2}(p_{total} - 1 \times 2\pi\hbar/\ell), \quad p_K = \tfrac{1}{2}(p_{total} + 1 \times 2\pi\hbar/\ell) \quad (5.131a)$$

$$p_{\text{I}} = \tfrac{1}{2}(p_{\text{total}} - 5 \times 2\pi\hbar/\ell), \quad p_{\text{L}} = \tfrac{1}{2}(p_{\text{total}} + 5 \times 2\pi\hbar/\ell) \qquad (5.131\text{b})$$

$$p_{\text{H}} = \tfrac{1}{2}(p_{\text{total}} - 7 \times 2\pi\hbar/\ell), \quad p_{\text{M}} = \tfrac{1}{2}(p_{\text{total}} + 7 \times 2\pi\hbar/\ell) \qquad (5.131\text{c})$$

$$p_{\text{G}} = \tfrac{1}{2}(p_{\text{total}} - 11 \times 2\pi\hbar/\ell), \quad p_{\text{N}} = \tfrac{1}{2}(p_{\text{total}} + 11 \times 2\pi\hbar/\ell) \qquad (5.131\text{d})$$

Thus any momentum-pair $(p_{\text{V}}, p_{W})$ is related to $p_{\text{total}}$ and $j$ as

$$p_{\text{V}} = \tfrac{1}{2}(p_{\text{total}} - j \times 2\pi\hbar/\ell), \quad p_{\text{W}} = \tfrac{1}{2}(p_{\text{total}} + j \times 2\pi\hbar/\ell) \qquad (5.132\text{a})$$

$$p_{\text{total}} = p_{\text{V}} + p_{\text{W}}, \qquad j = (p_{\text{W}} - p_{\text{V}})/(2\pi\hbar/\ell) \qquad (5.132\text{b})$$

Because both momenta $p_{\text{V}}$ and $p_{\text{W}}$ should be equal to $(2\pi\hbar/\ell) \times$ integer, the values of $p_{\text{total}}$ and $j$ are classified to the following two cases:

Case 1: $\quad p_{\text{total}} = (2\pi\hbar/\ell) \times (\text{odd integer}) \quad \text{for} \quad j = (\text{odd integer}) \qquad (5.133\text{a})$

Case 2: $\quad p_{\text{total}} = (2\pi\hbar/\ell) \times (\text{even integer}) \quad \text{for} \quad j = (\text{even integer}) \qquad (5.133\text{b})$

We have already shown the case of odd integer $j$ in Fig.5.46.
Next we examine the case of even integer $j$. Figure 5.47 shows quantum transitions with even integers $j$ given by Eq.(5.133b).

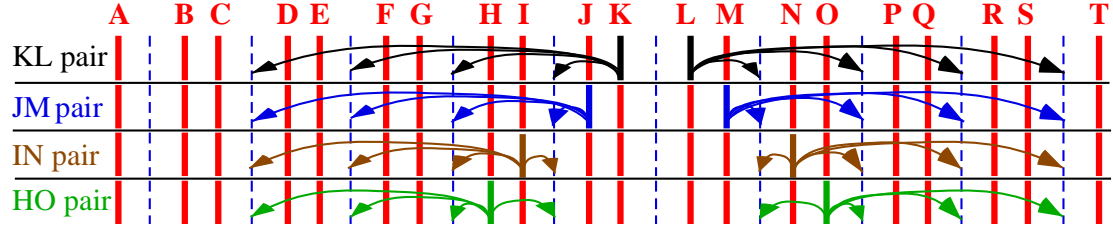

Fig.5.47 Various electron pairs with the same total momentum at $\nu = 8/3$

Dashed lines indicate empty orbitals and solid lines indicate filled orbitals in the Landau level $L = 1$. Allowed transitions from the electrons K and L are shown by black arrow pairs, from JM by blue, from IN by brown and from HO by dark green.

The electron pairs KL, JM, IN and HO indicate the cases of $j = 2, 4, 8$ and 10 respectively. All the electron pairs can transfer to all the empty orbitals as in Fig.5.47. The pair energies are described as

$$\varsigma_{\text{KL}} = \varsigma_{\nu=8/3}^{L=1}(p_{\text{total}}, j = 2) \qquad (5.134\text{a})$$

$$\varsigma_{\text{JM}} = \varsigma_{\nu=8/3}^{L=1}(p_{\text{total}}, j = 4) \qquad (5.134\text{b})$$

$$\varsigma_{\text{IN}} = \varsigma_{\nu=8/3}^{L=1}(p_{\text{total}}, j = 8) \qquad (5.134\text{c})$$

$$\varsigma_{\text{HO}} = \varsigma_{\nu=8/3}^{L=1}(p_{\text{total}}, j=10) \tag{5.134d}$$

The total energy of all the electron pairs is described by the symbol $E^{\text{pair}}$ defined by Eq.(5.120). Use of Eqs.(5.127) and (5.128) gives

$$E^{\text{pair}} \approx E_{L=1}^{\text{pair}}(\nu) \qquad \text{in the ground state with } 2 < \nu < 3 \tag{5.135}$$

The energy $E_{L=1}^{\text{pair}}(\nu)$ is the sum of all the pair energies as:

$$E_{L=1}^{\text{pair}}(\nu) = \sum_{p_{\text{total}}, j} \varsigma_\nu^{L=1}(p_{\text{total}}, j) \qquad \text{in the ground state with } 2 < \nu < 3 \tag{5.136}$$

where the sum is curried out for all the values $p_{\text{total}}$ and $j$. Equations (5.120), (5.135) and (5.136) yield the total energy of the quasi-2D electron system:

$$E_{\text{T}} \approx W + \sum_{p_{\text{total}}, j} \varsigma_\nu^{L=1}(p_{\text{total}}, j) \qquad \text{in the ground state with } 2 < \nu < 3 \tag{5.137}$$

Substitution of Eq.(5.126) into Eq.(5.137) gives

$$E_{\text{T}} \approx C + N\lambda + \sum_{i=1}^{N} U(p_i/(eB)) + \tfrac{1}{2}(\hbar eB/m^*)N_\nu^{L=0} + \tfrac{3}{2}(\hbar eB/m^*)N_\nu^{L=1} + \sum_{p_{\text{total}}, j} \varsigma_\nu^{L=1}(p_{\text{total}}, j)$$

$$\text{in the ground state with } 2 < \nu < 3 \tag{5.138}$$

We express the pair energy per electron by the symbol $\xi_\nu^{L=1}(j)$ which is defined by

$$\xi_\nu^{L=1}(j) = \sum_{p_{\text{total}}} \varsigma_\nu^{L=1}(p_{\text{total}}, j) \Big/ N_\nu^{L=1} \qquad \text{in the ground state with } 2 < \nu < 3 \tag{5.139}$$

The exact pair energy is the sum of all order terms in the perturbation calculation as

$$\xi_\nu^{L=1}(j) = \sum_{n=2,3,4,\cdots} \xi_\nu^{L=1}(j;n) \tag{5.140}$$

where $\xi_\nu^{L=1}(j;n)$ indicates the n-th order of the perturbation energy. Substitution of Eqs.(5.139) and (5.140) into Eq.(5.138) yields

$$E_{\text{T}} = W + E^{\text{pair}} \approx C(L_1, \cdots, L_N; p_1, \cdots, p_N) + N\lambda + \sum_{i=1}^{N} U(p_i/(eB))$$

$$+ \tfrac{1}{2}(\hbar eB/m^*)N_\nu^{L=0} + \tfrac{3}{2}(\hbar eB/m^*)N_\nu^{L=1} + N_\nu^{L=1} \times \sum_{j=1,2,3,\cdots} \left( \sum_{n=2,3,4,\cdots} \xi_\nu^{L=1}(j;n) \right)$$

$$\text{in the ground state with } 2 < \nu < 3 \tag{5.141}$$

Therein the function form of $W$ is continuous with the change in $\nu$. On the other hand the pair energy $E^{\text{pair}}$ has a discontinuous form for $\nu$ at the specific filling factors $\nu = \nu_0$, because the number of the allowed transitions depends discontinuously upon $\nu$ at $\nu = \nu_0$. This discontinuous property produces the plateaus of the Hall resistance at the specific filling factors. We have already calculated the second order perturbation energies for $j = 1$ and 2. We list the results in Tables 5.11 and 5.12.

Table 5.11 Second order of the perturbation energy per electron for the electron pairs placed in the second neighboring Landau orbitals

| $\nu$ | 5/2 | 48/19 | 78/31 | 8/3 | 14/5 | 18/7 | 19/7 |
|---|---|---|---|---|---|---|---|
| $\xi_\nu^{L=1}(2;2)$ | $-(1/2)S$ | $-(41/190)S$ | $-(113/496)S$ | $-S/6$ | $-S/20$ | $-3S/28$ | $-2S/35$ |

Table 5.12 Second order of the perturbation energy per electron for the electron pairs placed in the nearest Landau orbitals

| $\nu$ | 5/2 | 48/19 | 78/31 | 8/3 | 14/5 | 18/7 | 19/7 |
|---|---|---|---|---|---|---|---|
| $\xi_\nu^{L=1}(1;2)$ | 0 | $-(9/190)T$ | $-(15/496)T$ | $-T/6$ | $-T/20$ | $-3T/28$ | $-2T/35$ |

Next we study the energies $\xi_{\nu=14/5}^{L=1}(j)$ of the electron-pairs placed in more distant neighbouring Landau orbitals with $j \geq 3$. Figure 5.48 shows the most uniform configuration at $\nu = 14/5$. The centre position between the nearest pair $A_1B_1$ is equal to that of the electron pair $A_nB_n$ for $n = 2,3,4,\cdots$. Then the total momentum of the pair $A_nB_n$ is equal to that of the pair $A_1B_1$. Therefore the electron pair $A_nB_n$ ($n = 1,2,3,4,\cdots$) can transfer to all the empty states as shown by the arrow pairs in Fig.5.48.

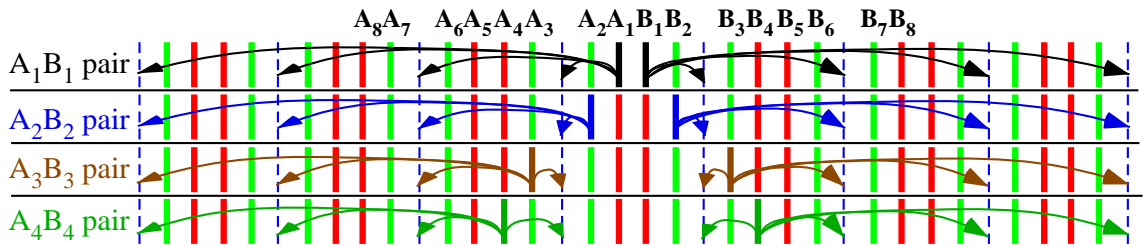

Fig.5.48 Various electron pairs with the same total momentum at $\nu = 14/5$

Dashed lines indicate empty orbitals and solid lines indicate filled orbitals in the Landau level $L = 1$. Allowed transitions from the electrons $A_1$ and $B_1$ are shown by black arrow pairs, from $A_2B_2$ by blue, from $A_3B_3$ by brown and from $A_4B_4$ by dark green.

Also the total momentum of the electron pair $C_1D_1$ in Fig.5.49 is equal to that of the pairs $C_nD_n$ (n=1, 2, 3, ...) and therefore the pair $C_nD_n$ can transfer to all the empty states except the orbital Y. This only one forbidden transition may be ignored in comparison with the enormously many allowed transitions.

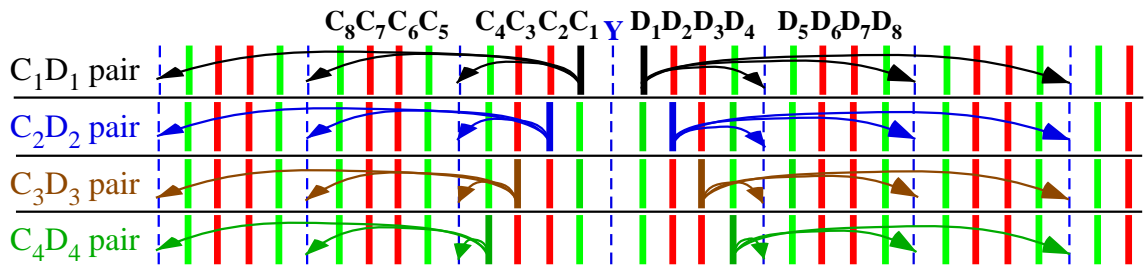

Fig.5.49 Various electron pairs with the same total momentum at $\nu = 14/5$

Dashed lines indicate empty orbitals and solid lines indicate filled orbitals in the Landau level $L = 1$. Allowed transitions from the electrons $C_1$ and $D_1$ are shown by black arrow pairs, from $C_2D_2$ by blue, from $C_3D_3$ by brown and from $C_4D_4$ by dark green.

Thus the further neighbouring electron (or hole) pairs with $j \geq 3$ can transfer to all the empty (or filled) orbitals at $\nu = 8/3$, 14/5, 7/3 and 11/5 as shown in Figs.5.46-5.49. The energies of these pairs with $j \geq 3$ are negative in the second order perturbation. Therefore the energies are accumulated to give a stronger binding energy and so the states become more stable.

## 5.11 Hall plateaus in the region $1 < \nu < 2$

When the direction of the magnetic field is upward, the Zeeman energy of down-spin is lower than that of up-spin for electrons. So, in the region $1 < \nu < 2$ all the Landau orbitals in the lowest level are occupied by the electrons with down-spin under a strong magnetic field. On the other hand the Landau orbitals are partially occupied by up-spins. (Note: We will examine the case of a weak magnetic field in Chapter 9. The FQH state with any filling factor $\nu$ is realized in a weak magnetic field by adjusting the gate voltage. In the case both down- and up-spin-electrons partially occupy the lowest Landau orbitals.)

The up-spin electron pair placed in the nearest orbitals can transfer to all the empty orbitals of up-spin at $\nu_0 = (4j+1)/(2j+1)$, $\nu_0 = (3j-1)/(2j-1)$ and so on. When the filling factor $\nu$ deviates from $\nu_0$, the number of allowed transitions decrease abruptly in comparison with that at $\nu_0$. This mechanism yields the energy gaps at $\nu_0 = (3j+1)/(2j+1)$, $(3j-1)/(2j-1)$, $(4j+1)/(2j+1)$, $(2j+2)/(2j+1)$ and so on. The energy gaps produce the fractional quantum Hall effect in the region $1 < \nu < 2$ as examined in Ref.[69]. (Note: We have studied the energy gaps for the specific filling factors in the regions $\nu < 1$ and $\nu > 2$ in the previous sections.)

We introduce the total number $M$ of the $L = 0$ orbitals and also express the number of electrons with down-spin and up-spin by $N_\downarrow$ and $N_\uparrow$ respectively. Then we obtain the following relations for $1 < \nu < 2$ as

$$N_\downarrow = M \tag{5.142a}$$

$$N_\uparrow < M \tag{5.142b}$$

$$N = N_\downarrow + N_\uparrow \tag{5.142c}$$

$$\nu = (N_\downarrow + N_\uparrow)/M \tag{5.142d}$$

## 5.11.1 Electron configurations and energy gaps for $3/2 < \nu < 2$

We examine the FQH states with $3/2 < \nu < 2$ in this sub-section [69]. As an example we consider the filling factor $\nu = 5/3$. Equation (5.142d) becomes

$$\nu = 1 + N_\uparrow/M = 5/3$$

$$N_\uparrow/M = 2/3 \quad \text{for} \quad \nu = 5/3 \tag{5.143}$$

The most uniform configuration of up-spin electrons is the repeat of (filled, empty, filled). Figure 5.50 shows the electron configuration in a 3D view. Therein the tilted lines with the x-direction express the Landau orbitals of the lowest level schematically. All the orbitals are filled with down-spin electrons under a strong magnetic field because of the Zeeman energy. The empty orbitals of up-spin are drawn by dashed blue lines in Fig.5.50. The up-spin electrons occupy the red-coloured orbitals. This electron configuration of up-spin has the minimum value of the classical Coulomb energy.

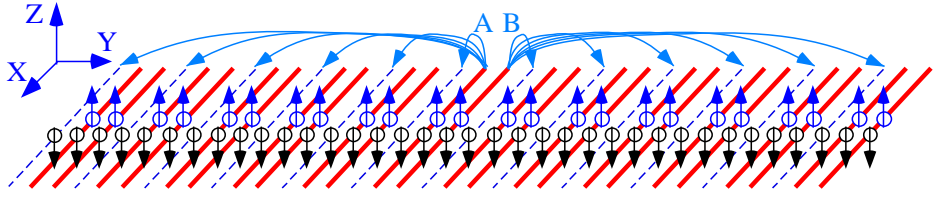

Fig.5.50 Electron configuration at $\nu = 5/3$

We examine the quantum transitions via the Coulomb interaction $H_I$. All the Coulomb transitions satisfy the momentum conservation along the x-axis. Figure 5.50 shows the quantum transitions from the electron pair AB with up-spin. The momenta at A and B are expressed by $p_A$ and $p_B$ respectively. The electron pair AB transfers to the orbitals with $p'_A$ and $p'_B$ after the transition. Then the momentum conservation gives the following relations:

$$p'_A = p_A - \Delta p \tag{5.144a}$$
$$p'_B = p_B + \Delta p \tag{5.144b}$$

where $\Delta p$ is the momentum transfer. All the allowed transitions from AB are illustrated by the blue allow-pairs in Fig.5.50. Accordingly the transfer momentum takes the following value:

$$\Delta p = (2\pi\hbar/\ell) \times (3n+1) \qquad n = 0, \pm 1, \pm 2, \pm 3, \pm 4, \cdots \tag{5.145}$$

We introduce the following summation $Z$ in order to obtain the perturbation energy of the electron pair AB.

$$Z = -\sum_{\Delta p \neq 0, -2\pi\hbar/\ell} \frac{\langle L=0, p_A, p_B | H_I | L=0, p'_A, p'_B \rangle \langle L=0, p'_A, p'_B | H_I | L=0, p_A, p_B \rangle}{W_G - W_{\text{excite}}(p_A \to p'_A, p_B \to p'_B)} \tag{5.146a}$$

$$p_A = p_B + 2\pi\hbar/\ell \tag{5.146b}$$
$$p'_A = p_A - \Delta p, \quad p'_B = p_B + \Delta p \tag{5.146c}$$

Therein the summation is carried out for all the values $\Delta p = (2\pi\hbar/\ell) \times$ integer except $\Delta p = 0$ and $-2\pi\hbar/\ell$. The elimination comes from disappearance of the diagonal matrix element of $H_I$. The summation $Z$ is positive, because the denominator of Eq.( 5.146a) is negative.

As shown in Fig.5.50, the transfer-momenta from AB (up-spin electron-pair) satisfies Eq.(5.145). The number of the transfer-momenta is 1/3 of the total orbitals. Accordingly the perturbation energy $\varsigma_{AB}$ of the pair AB is expressed by $Z$ as

$$\varsigma_{AB} = -(1/3)Z \qquad \text{at } \nu = 5/3 \tag{5.147}$$

because the momentum-interval, $2\pi\hbar/\ell$, is extremely small in a macroscopic size of a quantum Hall device. The total number of the nearest electron pairs with up-spin is 1/2 of $N_\uparrow$. Thereby we obtain the nearest pair energy per up-spin-electron $\varepsilon_\uparrow$ as

$$\varepsilon_\uparrow = \left(\varsigma_{AB} \times \tfrac{1}{2} N_\uparrow\right)/N_\uparrow = -(1/6)Z \quad \text{at} \quad \nu = 5/3 \quad (5.148)$$

When the filling factor deviates from $\nu = 5/3$, the electron configuration changes from the regular repeating of (filled, empty, filled). Then the number of the Coulomb transitions decreases abruptly because the changing disturbs the Coulomb transitions. As an example, the $\nu = 42/25$ state is illustrated in Fig.5.51 where the nearest orbitals with up-spin are illustrated by red and brown colours.

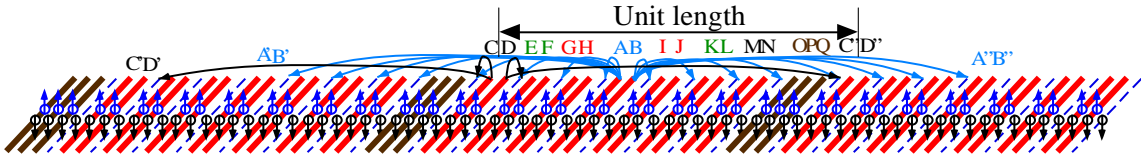

Fig.5.51 Electron configuration at $\nu = 42/25$

There are 9 nearest electron pairs, namely, AB, CD, EF, GH, IJ, KL, MN, OP and PQ in every unit sequence. The pair CD can transfer to two orbital pairs per unit length as shown by black arrow pairs. Accordingly the perturbation energy $\varsigma_{CD}$ is equal to

$$\varsigma_{CD} = -(2/25)Z \quad \text{at} \quad \nu = 42/25 \quad (5.149)$$

The pair AB can transfer to all the empty orbitals of up-spin and then the number of allowed transitions is eight per unit length. Therefore the perturbation energy $\varsigma_{AB}$ is

$$\varsigma_{AB} = -(8/25)Z \quad \text{at} \quad \nu = 42/25 \quad (5.150)$$

The other pairs have the perturbation energies as

$$\varsigma_{EF} = \varsigma_{KL} = -(4/25)Z, \quad \varsigma_{GH} = \varsigma_{IJ} = -(6/25)Z, \quad \varsigma_{MN} = -(2/25)Z$$

$$\varsigma_{OP} = \varsigma_{PQ} = 0 \quad \text{at} \quad \nu = 42/25 \quad (5.151)$$

The sum of the nearest electron pairs with up-spin is

$$F = \varsigma_{AB} + \varsigma_{CD} + \varsigma_{EF} + \varsigma_{GH} + \varsigma_{IJ} + \varsigma_{KL} + \varsigma_{MN} + \varsigma_{OP} + \varsigma_{PQ} = -(32/25)Z \quad (5.152)$$

The number of electrons with up-spin is seventeen in a unit sequence. Therefore the nearest pair energy per up-spin electron is

$$\varepsilon_\uparrow = -(32/(25\times 17))Z \quad \text{at} \quad \nu = 42/25 \tag{5.153}$$

When the filling factor $\nu$ is $(10s+2)/(6s+1)$ (s is a positive integer), the sum of the nearest-pair-energies inside the unit sequence is

$$F = -Z(2+\cdots+(2s-2)+2s+(2s-2)+\cdots+2)/(6s+1) = -(2s^2/(6s+1))Z \tag{5.154}$$

The filling factor $(10s+2)/(6s+1)$ is larger than 5/3. The number of up-spin-electrons inside a unit length is equal to $(4s+1)$ and therefore the pair energy per up-spin-electron is given by

$$\varepsilon_\uparrow = -(2s^2/((6s+1)(4s+1)))Z \quad \text{at} \quad \nu = (10s+2)/(6s+1) \tag{5.155}$$

When $s$ becomes infinitely large, $\varepsilon_\uparrow$ approaches

$$\lim_{\nu\to(5/3)+0}\varepsilon_\uparrow = -(1/12)Z \tag{5.156}$$

Next we consider the filing factor 38/23 which is smaller than 5/3. The most uniform configuration is illustrated in Fig.5.52.

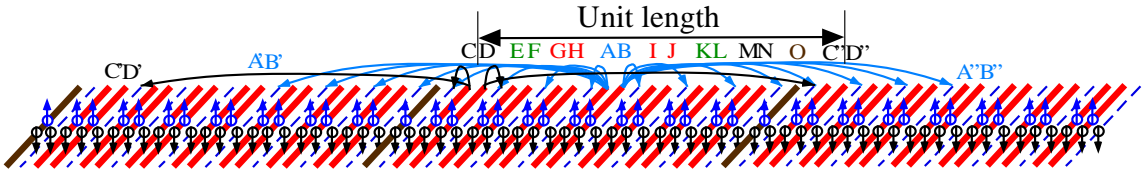

Fig.5.52 Electron configuration at $\nu = 38/23$

In this case, the sum of the nearest-pair-energies inside the unit sequence is

$$F = \varsigma_{AB} + \varsigma_{CD} + \varsigma_{EF} + \varsigma_{GH} + \varsigma_{IJ} + \varsigma_{KL} + \varsigma_{MN} = -(32/23)Z \tag{5.157}$$

At $\nu = (10s-2)/(6s-1)$, the sum of the nearest-pair-energies inside the unit sequence is

$$F = -[(2+\cdots+(2s-2)+2s+(2s-2)+\cdots+2)/(6s-1)]Z = -(2s^2/(6s-1))Z \tag{5.158}$$

Accordingly

$$\varepsilon_\uparrow = -(2s^2/((6s-1)(4s-1)))Z \quad \text{at} \quad \nu = (10s-2)/(6s-1) \tag{5.159}$$

$$\lim_{\nu\to(5/3)-0}\varepsilon_\uparrow = -(1/12)Z \tag{5.160}$$

Thus the energy gap appears between the energy at $\nu = 5/3$ and the limiting energy from the left and right sides:

$$\varepsilon_\uparrow(\nu = 5/3) - \lim_{\nu \to (5/3)\pm 0} \varepsilon_\uparrow = -(1/6)Z + (1/12)Z = -(1/12)Z \quad (5.161)$$

Table 5.13 shows the energy gaps in the fourth column at $\nu = (4j+1)/(2j+1)$.

Table 5.13 Energy gaps for $\nu = (4j+1)/(2j+1)$

| $\nu_0$ | $\varepsilon_\uparrow(\nu_0)$ | $\lim_{\nu \to \nu_0 \pm 0} \varepsilon_\uparrow(\nu)$ | $\varepsilon_\uparrow(\nu_0) - \lim_{\nu \to \nu_0 \pm 0} \varepsilon_\uparrow(\nu)$ |
|---|---|---|---|
| 5/3 | -(1/6) Z | -(1/12) Z | -(1/12) Z |
| 9/5 | -(1/20) Z | -(1/40) Z | -(1/40) Z |
| 13/7 | -(1/42) Z | -(1/84) Z | -(1/84) Z |
| 17/9 | -(1/72) Z | -(1/144) Z | -(1/144) Z |

We consider the other cases. Figure 5.53 shows the most uniform configuration of electrons at $\nu = 8/5$. Therein the x-, y-, z-directions are expressed at the upper-left of Fig.5.53.

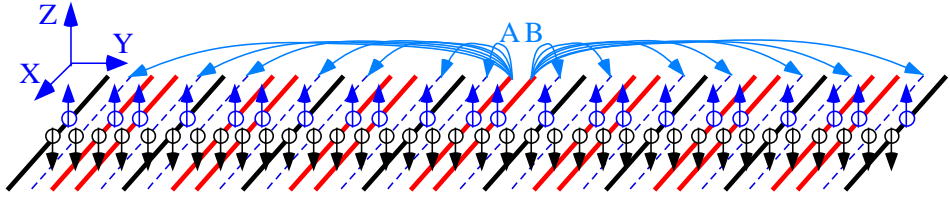

Fig.5.53 Electron configuration at $\nu = 8/5$

The unit configuration is composed of five Landau orbitals and three electrons with up-spin. The number of the allowed transitions is two per unit length. Accordingly the perturbation energy $\varsigma_{AB}$ of the pair AB with up-spin is obtained as

$$\varsigma_{AB} = -(2/5)Z \quad \text{at } \nu = 8/5 \quad (5.162)$$

The total number of the nearest electron pairs with up-spin is 1/3 times $N_\uparrow$. Therefore the nearest pair energy per up-spin-electron is

$$\varepsilon_\uparrow = (\varsigma_{AB} \times \tfrac{1}{3} N_\uparrow)/N_\uparrow = -(2/15)Z \quad \text{at } \nu = 8/5 \quad (5.163)$$

When the filling factor deviates from $\nu = 8/5$, the number of the Coulomb transitions decreases abruptly because the electron configuration at $\nu \neq 8/5$ disturbs the Coulomb transitions as seen in Fig.5.54 for $\nu = 43/27$.

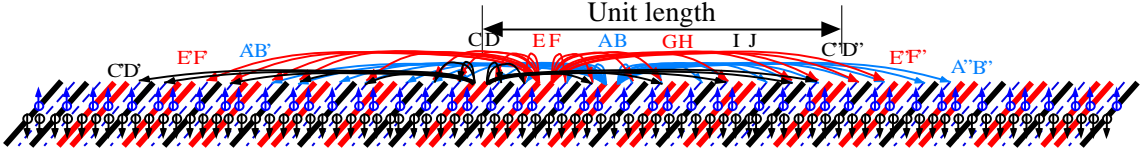

Fig.5.54 Electron configuration at $\nu = 43/27$

There are five up-spin-electron pairs placed in the nearest orbitals inside a unit length as in Fig.5.54. The number of allowed transitions is eleven for the pair AB, nine for EF and seven for CD in a unit length. Therefore the perturbation energies are obtained as follows:

$$\varsigma_{AB} = -(11/27)Z, \quad \varsigma_{EF} = \varsigma_{GH} = -(9/27)Z, \quad \varsigma_{CD} = \varsigma_{IJ} = -(7/27)Z \quad (5.164)$$

The sum of these pair energies is

$$F = \varsigma_{AB} + \varsigma_{CD} + \varsigma_{EF} + \varsigma_{GH} + \varsigma_{IJ} = -(43/27)Z \quad (5.165)$$

The number of electrons with up-spin is sixteen in a unit length and then the nearest pair energy per up-spin-electron is

$$\varepsilon_\uparrow = F/16 = -(43/(27 \times 16))Z \quad \text{at} \quad \nu = 43/27 \quad (5.166)$$

We examine more general cases of $\nu = (16s-5)/(10s-3)$. At the filling factor, the sum of the nearest-pair-energies inside a unit length is

$$F = -Z((2s+1) + (2s+3) \cdots + (4s-1) + (4s-3) + \cdots + (2s+1))/(10s-3) \\ = -((6s^2 - 4s + 1)/(10s-3))Z \quad (5.167)$$

Accordingly

$$\varepsilon_\uparrow = -[(6s^2 - 4s + 1)/((10s-3)(6s-2))]Z \quad \text{at} \quad \nu = (16s-5)/(10s-3) \quad (5.168)$$

$$\lim_{\nu \to (8/5)-0} \varepsilon_\uparrow = -(1/10)Z \quad (5.169)$$

Next we study the FQH state with $\nu = 11/7$. The most uniform configuration is illustrated in Fig.5.55.

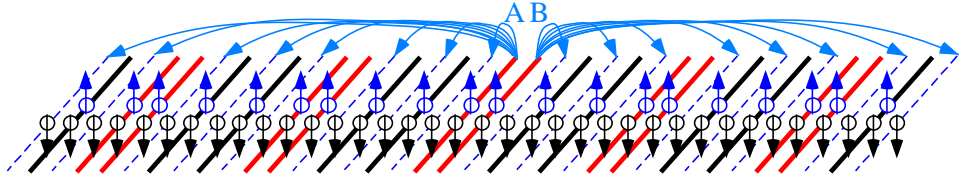

Fig.5.55 Electron configuration at $\nu = 11/7$

The perturbation energy of the pair AB is

$$\varsigma_{AB} = -(3/7)Z \qquad \text{at } \nu = 11/7 \qquad (5.170)$$

The number of the nearest electron pairs with up-spin is 1/4 of $N_\uparrow$. Therefore the nearest pair energy per up-spin-electron is

$$\varepsilon_\uparrow = \left(\varsigma_{AB} \times \tfrac{1}{4} N_\uparrow\right)/N_\uparrow = -(3/28)Z \qquad \text{at } \nu = 11/7 \qquad (5.171)$$

At $\nu = (3j-1)/(2j-1)$ the perturbation energy $\varsigma_{AB}$ for the nearest pair AB is obtained as

$$\varsigma_{AB} = -[(j-1)/(2j-1)]Z \text{ at } \nu = (3j-1)/(2j-1) \qquad (5.172)$$

The total number of the nearest electron pairs with up-spin is $1/j$ times $N_\uparrow$. Therefore the nearest pair energy per up-spin-electron is

$$\varepsilon_\uparrow = \left(\varsigma_{AB} \times \tfrac{1}{j} N_\uparrow\right)/N_\uparrow = -[(j-1)/(j(2j-1))]Z \qquad \text{at } \nu = (3j-1)/(2j-1) \quad (5.173)$$

The energies are listed in the second column of Table 5.14. Next we calculate the number of the allowed transitions in the neighbourhood of $\nu = (3j-1)/(2j-1)$. Then the energy gaps are shown in the fourth column of Table 5.14.

Table 5.14 Energy gaps for $\nu = (3j-1)/(2j-1)$

| $\nu_0$ | $\varepsilon_\uparrow(\nu_0)$ | $\lim_{\nu \to \nu_0 \pm 0} \varepsilon_\uparrow(\nu)$ | $\varepsilon_\uparrow(\nu_0) - \lim_{\nu \to \nu_0 \pm 0} \varepsilon_\uparrow(\nu)$ |
|---|---|---|---|
| 5/3 | -(1/6) Z | -(1/12) Z | -(1/12) Z |
| 8/5 | -(2/15) Z | -(1/10) Z | -(1/30) Z |
| 11/7 | -(3/28) Z | -(5/56) Z | -(1/56) Z |
| 14/9 | -(4/45) Z | -(7/90) Z | -(1/90) Z |

Thus the present theory yields the energy gaps at the specific filling factors as in Tables

5.13 and 5.14.

### 5.11.2 Electron configurations and energy gaps for $1 < \nu < 3/2$

We examine the FQH states with $1 < \nu < 3/2$ in this sub-section [69]. Four examples are shown in Figs.5.56-59 where the filling factors are $\nu = 4/5$, 6/5, 7/5 and 10/7, respectively. The electron configurations are illustrated in a 3D view where the down-spin electrons occupy all the Landau orbitals with the lowest level.

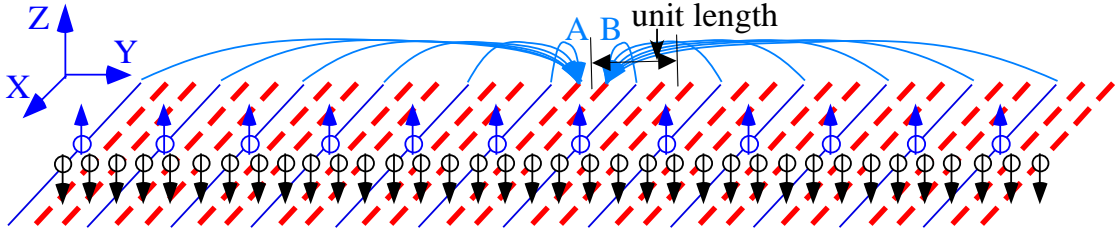

Fig.5.56 Electron configuration at $\nu = 4/3$

The most uniform electron configuration with up-spin is the repeat of the sequence (empty, filled, empty) at $\nu = 4/3$ as in Fig. 5.56. The empty orbitals for up-spin are shown by red dashed lines and the filled orbitals with up-spin by blue lines. The blue arrows express the quantum transitions to the empty orbitals AB (nearest hole pair). The symbol $\varsigma_{AB}^{H}$ means the perturbation energy via all the quantum transitions shown by blue arrow pairs. The nearest vacant-orbital-pair AB is specified by the momenta $p_A, p_B$. The electron pair A'B' is also specified by the momenta $p'_A, p'_B$. The electron pair A'B' transfers to the vacant orbitals at A and B. Therein the total momentum of the pair conserves in the Coulomb transition as

$$p'_A = p_A - \Delta p \quad p'_B = p_B + \Delta p \tag{5.174}$$

where the momentum transfer $\Delta p$ takes the following values as

$$\Delta p = (3j+1)2\pi\hbar/\ell \quad \text{for } j = 0, \pm 1, \pm 2, \pm 3, \cdots \tag{5.175}$$

at $\nu = 4/3$. Then the second order perturbation energy of the hole pair AB is given by

$$\varsigma_{AB}^{H} = \sum_{\Delta p = (3j+1)2\pi\hbar/\ell \text{ for } j=0,\pm 1,\pm 2,\cdots} \frac{\langle p'_A, p'_B | H_I | p_A, p_B \rangle \langle p_A, p_B | H_I | p'_A, p'_B \rangle}{W_G - W_{\text{excite}}(p'_A \to p_A, p'_B \to p_B)} \tag{5.176}$$

In order to evaluate the energy $\varsigma_{AB}^{H}$ we introduce the summation $Z_H$ as

$$Z_{\text{H}} = - \sum_{\Delta p \neq 0, -2\pi\hbar/\ell} \frac{\langle p'_{\text{A}}, p'_{\text{B}} | H_{\text{I}} | p_{\text{A}}, p_{\text{B}} \rangle \langle p_{\text{A}}, p_{\text{B}} | H_{\text{I}} | p'_{\text{A}}, p'_{\text{B}} \rangle}{W_{\text{G}} - W_{\text{excite}}(p'_{\text{A}} \to p_{\text{A}}, p'_{\text{B}} \to p_{\text{B}})} \quad (5.177)$$

where the momentum transfer $\Delta p$ takes all the values $(2\pi\hbar/\ell) \times$ integer except $\Delta p = 0$ and $\Delta p = -2\pi\hbar/\ell$. The transferred states for $\Delta p = 0$ and $\Delta p = -2\pi\hbar/\ell$ are eliminated in the summation (5.177) because the diagonal element of $H_{\text{I}}$ is absent. The denominator in Eq.( 5.177) is negative and so $Z_{\text{H}}$ is positive. (The value of $Z_{\text{H}}$ is nearly equal to $Z$ for the same magnetic field strength.) The interval of momentum transfer is very small for a macroscopic size of a device and therefore the perturbation energy, $\varsigma_{\text{AB}}^{\text{H}}$, can be expressed by $Z_{\text{H}}$ as

$$\varsigma_{\text{AB}}^{\text{H}} = -\frac{1}{3} Z_{\text{H}} \quad \text{for} \quad \nu = 4/3 \quad (5.178)$$

The electron configurations at $\nu = 6/5$, $7/5$ and $10/7$ are illustrated in Fig.5.57, 5.58 and 5.59, respectively.

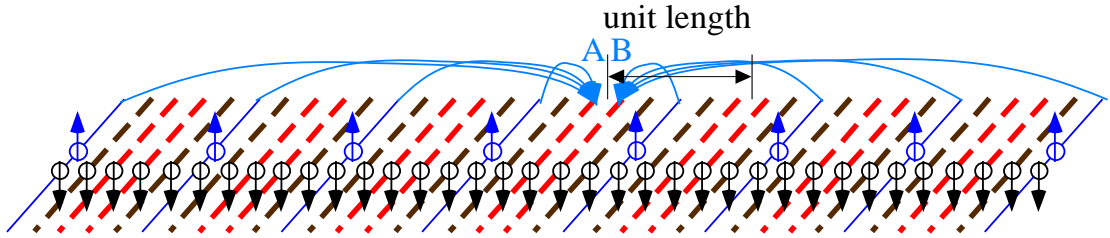

Fig.5.57 Electron configuration at $\nu = 6/5$

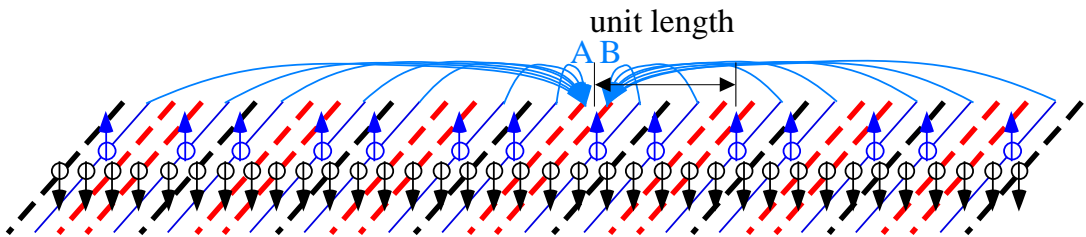

Fig.5.58 Electron configuration at $\nu = 7/5$

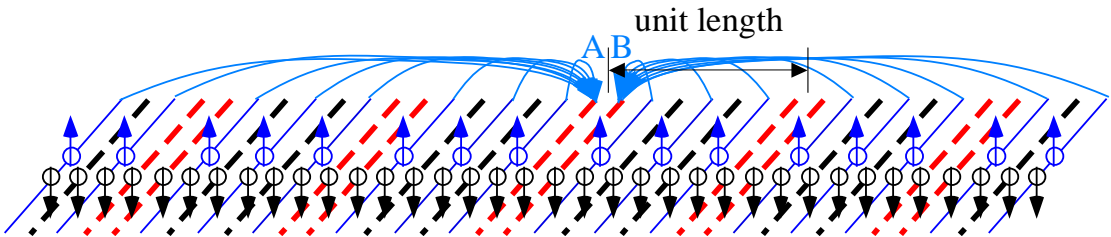

Fig.5.59 Electron configuration at $\nu = 10/7$

The perturbation energies, $\varsigma_{AB}^{H}$, are also obtained by making use of $Z_H$ as follows:

$$\varsigma_{AB}^{H} = -\frac{1}{5} Z_H \quad \text{for } \nu = 6/5 \tag{5.179a}$$

$$\varsigma_{AB}^{H} = -\frac{2}{5} Z_H \quad \text{for } \nu = 7/5 \tag{5.179b}$$

$$\varsigma_{AB}^{H} = -\frac{3}{7} Z_H \quad \text{for } \nu = 10/7 \tag{5.179c}$$

There are 1, 2, or 3 electrons in the unit length for the filling factor $\nu = 6/5$, 7/5, or 10/7, respectively. Therefore the energy per electron becomes $-(1/5)Z_H$, $-(1/5)Z_H$ and $-(1/7)Z_H$. We express the perturbation energy per electron by the symbol $\varepsilon_\uparrow(\nu_0)$ at $\nu = \nu_0$ which is listed in the second column of Table 5.15.

Table 5.15 Energy gaps for $\nu = (2j+2)/(2j+1)$ and $(3j+1)/(2j+1)$

| $\nu_0$ | $\varepsilon_\uparrow(\nu_0)$ | $\lim_{\nu \to \nu_0 \pm 0} \varepsilon_\uparrow(\nu)$ | $\varepsilon_\uparrow(\nu_0) - \lim_{\nu \to \nu_0 \pm 0} \varepsilon_\uparrow(\nu)$ |
|---|---|---|---|
| 4/3  | -(1/3) $Z_H$ | -(1/6) $Z_H$  | -(1/6) $Z_H$  |
| 6/5  | -(1/5) $Z_H$ | -(1/10) $Z_H$ | -(1/10) $Z_H$ |
| 7/5  | -(1/5) $Z_H$ | -(3/20) $Z_H$ | -(1/20) $Z_H$ |
| 8/7  | -(1/7) $Z_H$ | -(1/14) $Z_H$ | -(1/14) $Z_H$ |
| 10/7 | -(1/7) $Z_H$ | -(5/42) $Z_H$ | -(1/42) $Z_H$ |
| 10/9 | -(1/9) $Z_H$ | -(1/18) $Z_H$ | -(1/18) $Z_H$ |
| 13/9 | -(1/9) $Z_H$ | -(7/72) $Z_H$ | -(1/72) $Z_H$ |

The limiting values from both sides are calculated and written in the third column of Table 5.15. Subtractions of the limiting value from $\varepsilon_\uparrow(\nu_0)$ give the energy gaps which are listed in the fourth column of Table 5.15. Tables 5.13, 5.14 and 5.15 show the energy gaps at the filling factors. Thus the present theory can explain the confinement of the Hall resistance in the region of $1 < \nu < 2$.

### 5.11.3 Peak structure at the filling factors $\nu = (4j-1)/(2j)$ and $(2j+1)/(2j)$

We examine the $\nu = 7/4$ state. The filling factor has the denominator of even number. Figure 5.60 shows the most uniform configuration at $\nu = 7/4$.

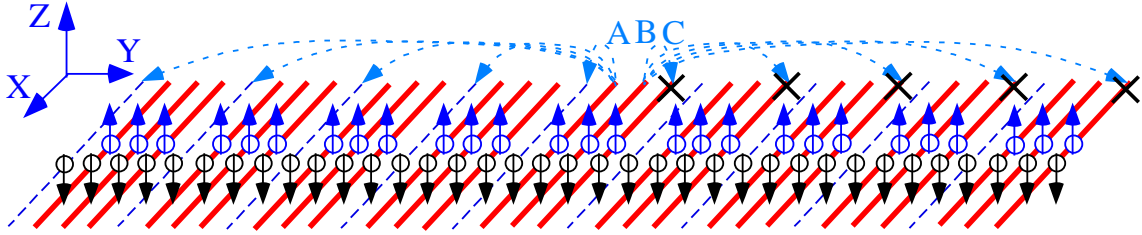

Fig.5.60 Electron configuration at $\nu = 7/4$

There are many electron pairs in Fig.5.60. The pair AB is an example of the nearest electron pair. The quantum transition via the Coulomb interaction conserves the total momentum. Accordingly the electron B should transfer to one orbital to the right when the electron A transfers to one orbital to the left. However the transformation to the right-direction is forbidden because the Landau orbital is already occupied by up-spin electron as in Fig.5.60. When the electron A transfers to the fifth orbital to the left, the electron B cannot transfer to the fifth orbital to the right because of the Pauli exclusion principle. Thus all the transitions from the nearest electron pairs are forbidden via the Coulomb interaction. Therefore the electron pair AB has no binding energy. Also all the quantum transition from the electron pair BC are forbidden. Accordingly all the nearest electron pairs have no binding energy.

Similarly all the nearest electron pairs have no binding energy at the filling factor $\nu_0 = (4j-1)/(2j)$.

$$\varepsilon_\uparrow(\nu_0) = 0 \qquad \text{at} \quad \nu_0 = (4j-1)/(2j) \qquad (5.180)$$

The energies are listed in the second column of Table 5.16.

Table 5.16  Comparison of *nearest electron pair* energies at $\nu_0 = (4j-1)/(2j)$ and in its neighbourhood

| $\nu_0$ | $\varepsilon_\uparrow(\nu_0)$ | $\lim_{\nu \to \nu_0 \pm 0} \varepsilon_\uparrow(\nu)$ | $\varepsilon_\uparrow(\nu_0) - \lim_{\nu \to \nu_0 \pm 0} \varepsilon_\uparrow(\nu)$ |
|---|---|---|---|
| 3/2 | 0 | 0 | 0 |

| 7/4 | 0 | $-\frac{1}{24}Z$ | $\frac{1}{24}Z$ |
| --- | --- | --- | --- |
| 11/6 | 0 | $-\frac{1}{60}Z$ | $\frac{1}{60}Z$ |
| 15/8 | 0 | $-\frac{1}{112}Z$ | $\frac{1}{112}Z$ |

We examine the energies of the nearest electron pairs in the neighbourhood of $\nu = (4j-1)/(2j)$. As an example for the neighbourhood of $j = 2$, we have calculated the number of allowed transitions by using a computer, then we obtain

$$\varepsilon_\uparrow(\nu) = -\tfrac{5101}{121706} Z \qquad \text{at } \nu = 1 + \tfrac{302}{403} \tag{5.181a}$$

$$\varepsilon_\uparrow(\nu) = -\tfrac{501001}{12017006} Z \qquad \text{at } \nu = 1 + \tfrac{3002}{4003} \tag{5.181b}$$

$$\varepsilon_\uparrow(\nu) = -\tfrac{5101}{123120} Z \qquad \text{at } \nu = 1 + \tfrac{304}{405} \tag{5.181c}$$

$$\varepsilon_\uparrow(\nu) = -\tfrac{501001}{12031020} Z \qquad \text{at } \nu = 1 + \tfrac{3004}{4005} \tag{5.181d}$$

The limiting value of the pair energy for $\nu \to (3/4)\pm 0$ is obtained as follows;

$$\lim_{\nu \to 3/4 \pm 0} \varepsilon_\uparrow(\nu) = -\tfrac{1}{24} Z \tag{5.182}$$

Next we examine the case of $\nu_0 = 5/6$.

$$\varepsilon_\uparrow(\nu) = -\tfrac{5101}{304920} Z \qquad \text{at } \nu = 1 + \tfrac{504}{605} \tag{5.183a}$$

$$\varepsilon_\uparrow(\nu) = -\tfrac{501001}{30049020} Z \qquad \text{at } \nu = 1 + \tfrac{5004}{6005} \tag{5.183b}$$

$$\varepsilon_\uparrow(\nu) = -\tfrac{5101}{307142} Z \qquad \text{at } \nu = 1 + \tfrac{506}{607} \tag{5.183c}$$

$$\varepsilon_\uparrow(\nu) = -\tfrac{501001}{30071042} Z \qquad \text{at } \nu = 1 + \tfrac{5006}{6007} \tag{5.183d}$$

For the limiting of $\nu \to (5/6)\pm 0$ the nearest pair energy approaches

$$\lim_{\nu \to 5/6 \pm 0} \varepsilon_\uparrow(\nu) = -\tfrac{1}{60} Z \tag{5.184}$$

As already calculated in Eq.(5.180), the nearest pair energy at $\nu_0 = (4j-1)/(2j)$ is zero.

Therefore the pair energy at $\nu_0 = (4j-1)/(2j)$ is higher than the energy in the neighbourhood of $v_0$. The peak values are listed in the fourth column of Table 5.16.

Similarly we calculate the pair energy of quantum transitions to the nearest empty orbitals at the filling factor $v = (2j+1)/(2j)$. The values are listed in Table 5.17. It is noteworthy that the values indicate the pair-energy per electron (not hole). That is to say the peak structure appears at the filling factor $v = (2j+1)/(2j)$ with an even number denominator.

Table 5.17 Comparison of *nearest hole pair* energies per electron at $v = (2j+1)/(2j)$ and in its neighbourhood

| $v_0$ | $\varepsilon_\uparrow(v_0)$ | $\lim_{v \to v_0 \pm 0} \varepsilon_\uparrow(v)$ | $\varepsilon_\uparrow(v_0) - \lim_{v \to v_0 \pm 0} \varepsilon_\uparrow(v)$ |
|---|---|---|---|
| 5/4 | 0 | $-\frac{1}{8}Z_H$ | $\frac{1}{8}Z_H$ |
| 7/6 | 0 | $-\frac{1}{12}Z_H$ | $\frac{1}{12}Z_H$ |
| 9/8 | 0 | $-\frac{1}{16}Z_H$ | $\frac{1}{16}Z_H$ |

Accordingly the FQH state is not stable at $v_0 = (4j-1)/(2j)$ and $v_0 = (2j+1)/(2j)$ for $j \geq 2$, because the energy is higher than that of the neighbourhood. So the Hall resistance confinement doesn't appear at $v_0 = (4j-1)/(2j)$ and $v_0 = (2j+1)/(2j)$ for $j \geq 2$. This logic can be extended to the FQHE at the filling factor larger than 2. Then the Hall resistance plateau doesn't appear at the filling factor $(\text{integer} + (4j-1)/(2j))$ and $(\text{integer} + (2j+1)/(2j))$ for $j \geq 2$. This property is in agreement with the experimental results.

## 5.12 Short comment

1) The present theory takes the electric potential along the Hall voltage into consideration (Traditional theories ignore it). So both Hall voltage and electric current are not zero. The ratio between them yields the Hall resistance properly.
2) The total momentum of electrons along the current direction conserves via the Coulomb interaction in FQHE. The conservation is satisfied in the present theory.
3) The FQHE at $v > 1$ have been explained by the discontinuous energy spectrum of the electron pairs as in sections 5.9 and 5.11. On the other hand the traditional FQH theories explained the FQHE at $v>1$ by superposing the IQH state of usual electrons with that of quasi-particles as will be discussed in Chapter 12. That is to say, all the electrons in a 2D system are classified into the following two groups: 1) the normal electrons and 2) the quasi-particles. This classification violates the anti-symmetric relation between electrons in the wave function. The lack of anti-symmetry may disturb the Fermi-Dirac statistics for all the electrons existing in the same thin layer.

4) The $v=5/2, 7/2$ FQHE and the nonstandard FQHE at $v = 7/11, 4/11, 4/13, 5/13, 5/17, 6/17$ have been explained on the same logic as the FQHE at $v = 1/3, 2/3, 1/5, 2/5, 3/5, 4/5$ and so on. The traditional theories have explained the three types of the FQHE by employing the different models respectively.

5) The Coulomb interaction acts between electrons in the 2D-electron system. So the binding energy belongs to electron pairs (not single electron). The present theory deals the pair energy properly. Thereby the phenomena of FQHE have been caused by the discontinuous dependence of the pair energy upon the filling factor $v$.

# Chapter 6 Effect of higher order perturbation and contribution from upper Landau levels

We will study the higher order perturbation energy due to $H_\mathrm{I}$ in this chapter. The third order terms are examined in section 6.1, the higher order term in section 6.2 and the contributions from the upper Landau levels in section 6.3.

## 6.1 Third order perturbation energy

As an example we study the third order perturbation at the filling factor of $\nu = 2/3$. One of the third order transitions is illustrated in Fig.6.1. The momenta of electrons A and B are denoted by $p_\mathrm{A}$ and $p_\mathrm{B}$. The momenta after the first transition are described by $p'_\mathrm{A}$ and $p'_\mathrm{B}$. The total momentum conserves as

$$p'_\mathrm{A} + p'_\mathrm{B} = p_\mathrm{A} + p_\mathrm{B} \tag{6.1}$$

The second transition yields the momenta $p''_\mathrm{A}$ and $p''_\mathrm{B}$ which satisfy the momentum conservation as

$$p''_\mathrm{A} + p''_\mathrm{B} = p_\mathrm{A} + p_\mathrm{B} \tag{6.2}$$

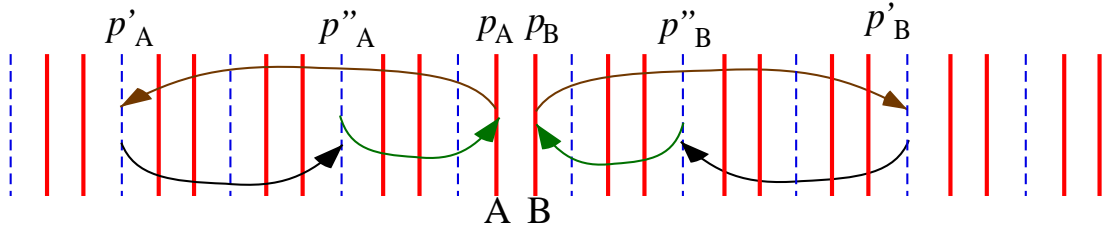

**Fig. 6.1** Third order perturbation of electron pair AB at $\nu = 2/3$

We introduce the following new summation $^3Z$ as

$$^3Z = -\sum_{\Delta p' \neq 0, -2\pi\hbar/\ell} \sum_{\Delta p'' \neq 0, -2\pi\hbar/\ell} \frac{\langle p_\mathrm{A}, p_\mathrm{B}|H_\mathrm{I}|p''_\mathrm{A}, p''_\mathrm{B}\rangle\langle p''_\mathrm{A}, p''_\mathrm{B}|H_\mathrm{I}|p'_\mathrm{A}, p'_\mathrm{B}\rangle\langle p'_\mathrm{A}, p'_\mathrm{B}|H_\mathrm{I}|p_\mathrm{A}, p_\mathrm{B}\rangle}{(W_\mathrm{G} - W_\mathrm{excite}(p_\mathrm{A} \to p'_\mathrm{A},\ p_\mathrm{B} \to p'_\mathrm{B}))(W_\mathrm{G} - W_\mathrm{excite}(p_\mathrm{A} \to p''_\mathrm{A},\ p_\mathrm{B} \to p''_\mathrm{B}))} \tag{6.3}$$

where $W_\mathrm{G}$ is the ground state energy of the Hamiltonian $H_\mathrm{D}$. Therein $p'_\mathrm{A}, p'_\mathrm{B}, p''_\mathrm{A}, p''_\mathrm{B}$ indicate the momenta in the two intermediate states. These four momenta are expressed by using the momentum transfers $\Delta p', \Delta p''$ as follows:

$$p'_\mathrm{A} = p_\mathrm{A} - \Delta p', \quad p'_\mathrm{B} = p_\mathrm{B} + \Delta p' \tag{6.4}$$
$$p''_\mathrm{A} = p_\mathrm{A} - \Delta p'', \quad p''_\mathrm{B} = p_\mathrm{B} + \Delta p'' \tag{6.5}$$

which are derived from the momentum conservation. In the summation $^3Z$, the momentum transfers $\Delta p', \Delta p''$ are able to take all of the values $(2\pi\hbar/\ell) \times$ integer except $\Delta p' = 0$, $\Delta p'' = 0$, $\Delta p' = -2\pi\hbar/\ell$ and $\Delta p'' = -2\pi\hbar/\ell$. The values $\Delta p' = 0$, $\Delta p'' = 0$, $\Delta p' = -2\pi\hbar/\ell$ and $\Delta p'' = -2\pi\hbar/\ell$ are eliminated in the summation of (6.3), because the diagonal matrix elements of $H_I$ are zero. Therefore the denominator in Eq.(6.3) is not zero.

The third order perturbation energy can be expressed by $^3Z$ systematically. As an example, we calculate the third order perturbation energy $^3\varsigma_{AB}$ of the nearest electron pair AB in Fig.6.1. Therein the momentum transfers $\Delta p', \Delta p''$ take the values as

$$\Delta p' = \cdots -5 \times 2\pi\hbar/\ell,\ -2 \times 2\pi\hbar/\ell,\ 2\pi\hbar/\ell,\ 4 \times 2\pi\hbar/\ell,\ 7 \times 2\pi\hbar/\ell,\ \cdots \quad (6.6)$$

$$\Delta p'' = \cdots -5 \times 2\pi\hbar/\ell,\ -2 \times 2\pi\hbar/\ell,\ 2\pi\hbar/\ell,\ 4 \times 2\pi\hbar/\ell,\ 7 \times 2\pi\hbar/\ell,\ \cdots \quad (6.7)$$

The interval of the momentum transfers is $3 \times 2\pi\hbar/\ell$ for the calculation of the third order perturbation energy $^3\varsigma_{AB}$ as in Eqs.(6.6) and (6.7). On the other hand the interval is $2\pi\hbar/\ell$ in $^3Z$. The interval value $2\pi\hbar/\ell$ is extremely small for a macroscopic size of $\ell$ and then we get

$$^3\varsigma_{AB} = -\left(\frac{1}{3}\right)^2 \,^3Z \qquad \text{at } \nu = 2/3 \qquad (6.8)$$

It is noteworthy that the factor $(1/3)^2$ comes from double summations of $\Delta p'$ and $\Delta p''$. Because the number of nearest electron pairs is $(1/2)$ times the total number $N$ of electrons, the third order perturbation energy $^3E_{\text{nearest pair}}$ of all the nearest electron pairs is expressed as

$$^3E_{\text{nearest pair}} = \frac{1}{2} N \left(^3\varsigma_{AB}\right) = -\frac{1}{2}\left(\frac{1}{3}\right)^2 \left(^3Z\right) N \qquad \text{at } \nu = 2/3 \qquad (6.9)$$

The energy per electron is

$$^3E_{\text{nearest pair}} / N = -\frac{1}{2}\left(\frac{1}{3}\right)^2 \,^3Z \qquad \text{at } \nu = 2/3 \qquad (6.10)$$

Similarly the nearest pair energy per electron in the third order at $\nu = 2j/(2j+1)$ is given by

$$^3E_{\text{nearest pair}}/N = -\frac{1}{2j}\left(\frac{1}{2j+1}\right)^2 {}^3Z \qquad \text{at } \nu = 2j/(2j+1) \qquad (6.11)$$

Also the energy of the nearest pairs per electron in the third order at $\nu = j/(2j-1)$ is given by

$$^3E_{\text{nearest pair}}/N = -\frac{1}{j}\left(\frac{j-1}{2j-1}\right)^2 {}^3Z \qquad \text{at } \nu = j/(2j-1) \qquad (6.12)$$

The electron-hole symmetry gives the following pair energies per hole

$$^3E_{\text{nearest pair}}/N_H = -\frac{1}{2j}\left(\frac{1}{2j+1}\right)^2 {}^3Z_H \qquad \text{at } \nu = 1/(2j+1) \qquad (6.13)$$

$$^3E_{\text{nearest pair}}/N_H = -\frac{1}{j}\left(\frac{j-1}{2j-1}\right)^2 {}^3Z_H \qquad \text{at } \nu = (j-1)/(2j-1) \qquad (6.14)$$

which are derived from Eqs.(6.11) and (6.12) respectively. Then the energies per electron (not hole) becomes

$$^3E_{\text{nearest pair}}/N = -\left(\frac{1}{2j+1}\right)^2 {}^3Z_H \qquad \text{at } \nu = 1/(2j+1) \qquad (6.15)$$

$$^3E_{\text{nearest pair}}/N = -\frac{1}{j-1}\left(\frac{j-1}{2j-1}\right)^2 {}^3Z_H \qquad \text{at } \nu = (j-1)/(2j-1) \qquad (6.16)$$

## 6.2 Higher order perturbation energy

We define the following summation $^nZ$ for the higher order calculations as

$$^nZ = -\sum_{\Delta p' \neq 0, -2\pi\hbar/\ell} \cdots \sum_{\Delta p^{n-1} \neq 0, -2\pi\hbar/\ell} \frac{\langle p_A, p_B | H_I | p^{n-1}{}_A, p^{n-1}{}_B \rangle \cdots \langle p''_A, p''_B | H_I | p'_A, p'_B \rangle \langle p'_A, p'_B | H_I | p_A, p_B \rangle}{(W_G - W_{\text{excite}}(p_A \to p'_A, p_B \to p'_B)) \cdots (W_G - W_{\text{excite}}(p_A \to p^{n-1}{}_A, p_B \to p^{n-1}{}_B))}$$
(6.17)

We can obtain the *n*-th order perturbation energy of the nearest electron pairs by making use of $^nZ$:

$$^nE_{\text{nearest pair}}/N = -\frac{1}{2}\left(\frac{1}{3}\right)^{n-1} {}^nZ \qquad \text{at} \quad \nu = 2/3 \tag{6.18}$$

It is noteworthy that we have used the symbol $Z$ instead of ${}^2Z$ and also $E_{\text{nearest pair}}$ instead of ${}^2E_{\text{nearest pair}}$ in chapters 4 and 5.

At $\nu = 2j/(2j+1)$ with arbitrary integer $j$ the $n$-th order perturbation energy per electron is given by

$$^nE_{\text{nearest pair}}/N = -\frac{1}{2j}\left(\frac{1}{2j+1}\right)^{n-1} {}^nZ \qquad \text{at} \quad \nu = 2j/(2j+1) \tag{6.19}$$

At the filling factor of $\nu = j/(2j-1)$, the $n$-th order perturbation energy per electron is obtained as

$$^nE_{\text{nearest pair}}/N = -\frac{1}{j}\left(\frac{j-1}{2j-1}\right)^{n-1} {}^nZ \qquad \text{at} \quad \nu = j/(2j-1) \tag{6.20}$$

The electron-hole symmetry gives the following pair energy per hole:

$$^nE_{\text{nearest pair}}/N_{\text{H}} = -\frac{1}{2j}\left(\frac{1}{2j+1}\right)^{n-1} {}^nZ_{\text{H}} \qquad \text{at} \quad \nu = 1/(2j+1) \tag{6.21}$$

$$^nE_{\text{nearest pair}}/N_{\text{H}} = -\frac{1}{j}\left(\frac{j-1}{2j-1}\right)^{n-1} {}^nZ_{\text{H}} \qquad \text{at} \quad \nu = (j-1)/(2j-1) \tag{6.22}$$

The sum of the perturbation energies for all orders is given by

$$E_{\text{nearest pair}}^{\text{all order}}/N = \sum_{n=2,3,\cdots}\left(-\frac{1}{2j}\left(\frac{1}{2j+1}\right)^{n-1} {}^nZ\right) \qquad \text{at} \quad \nu = 2j/(2j+1) \tag{6.23}$$

$$E_{\text{nearest pair}}^{\text{all order}}/N = \sum_{n=2,3,\cdots}\left(-\frac{1}{j}\left(\frac{j-1}{2j-1}\right)^{n-1} {}^nZ\right) \qquad \text{at} \quad \nu = j/(2j-1) \tag{6.24}$$

Also we get the sum of the perturbation energies in all orders for the hole pair

$$E_{\text{nearest pair}}^{\text{all order}} / N_{\text{H}} = \sum_{n=2,3,\cdots} \left( -\frac{1}{2j}\left(\frac{1}{2j+1}\right)^{n-1} {}^n Z_{\text{H}} \right) \quad \text{at} \quad \nu = 1/(2j+1) \qquad (6.25)$$

$$E_{\text{nearest pair}}^{\text{all order}} / N_{\text{H}} = \sum_{n=2,3,\cdots} \left( -\frac{1}{j}\left(\frac{j-1}{2j-1}\right)^{n-1} {}^n Z_{\text{H}} \right) \quad \text{at} \quad \nu = (j-1)/(2j-1) \qquad (6.26)$$

These energies of the *nearest hole pairs* per hole are re-expressed to the energies per electron as

$$E_{\text{nearest pair}}^{\text{all order}} / N = -\sum_{n=2,3,\cdots} \left( \left(\frac{1}{2j+1}\right)^{n-1} {}^n Z_{\text{H}} \right) \quad \text{at} \quad \nu = 1/(2j+1) \qquad (6.27)$$

$$E_{\text{nearest pair}}^{\text{all order}} / N = -\sum_{n=2,3,\cdots} \left( \frac{1}{j-1}\left(\frac{j-1}{2j-1}\right)^{n-1} {}^n Z_{\text{H}} \right) \text{at} \quad \nu = (j-1)/(2j-1) \qquad (6.28)$$

Next we examine the case of the filling factor $\nu = (2j-1)/(2j)$, the denominator of which is an even integer. There is no quantum transition from any *nearest electron* (or *hole*) *pair* at the filling factors of $\nu = (2j-1)/(2j)$ ( or $\nu = 1/(2j)$ ). Therefore the sum of the all order terms is equal to zero:

$$E_{\text{nearest pair}}^{\text{all order}} / N = 0 \quad \text{at} \quad \nu = (2j-1)/(2j) \quad \text{for nearest electron pairs} \qquad (6.29)$$

$$E_{\text{nearest pair}}^{\text{all order}} / N = 0 \quad \text{at} \quad \nu = 1/(2j) \quad \text{for nearest hole pairs} \qquad (6.30)$$

We have already examined the number of the allowed transitions for various filling factors in Chapters 4 and 5. The number of quantum transitions varies continuously at $\nu = 1/2$, 5/8. 7/10, discontinuously at $\nu = 1/(2j+1)$, $\nu = 2j/(2j+1)$, $\nu = j/(2j\pm1)$ and so on. The continuous (or discontinuous) structure is independent of the order of perturbation and is dependent on the filling factor only. Therefore the continuous or discontinuous structure is also maintained in the exact energy spectrum.

It is additionally noted here that (6.23-28) indicate that the higher order terms are multiplied by $(1/(2j+1))^{n-1}$ and $((j-1)/(2j-1))^{n-1}$. These multipliers are evaluated in the case of $j = 2$, $n = 3$ and 4 as follows:

$$(1/(2j+1))^{n-1} = (1/5)^2 = 0.04 \qquad \text{for} \quad j=2, \ n=3 \qquad (6.31a)$$

$$((j-1)/(2j-1))^{n-1} = (1/3)^2 = 0.111\cdots \quad \text{for} \quad j=2, \ n=3 \qquad (6.31b)$$

$$(1/(2j+1))^{n-1} = (1/5)^3 = 0.008 \qquad \text{for} \quad j=2,\ n=4 \qquad (6.31c)$$

$$((j-1)/(2j-1))^{n-1} = (1/3)^3 = 0.0370\cdots \qquad \text{for} \quad j=2,\ n=4 \qquad (6.31d)$$

These multipliers are small values and so the higher order contribution becomes very small. Consequently the second order contribution may be a main part of $E_{\text{nearest pair}}^{\text{all order}}$.

## 6.3 Contribution of upper Landau levels

We examine the contributions from upper Landau levels. The ground state of $H_D$ at $\nu < 2$ is composed of the Landau states with Landau level number $L = 0$ only. Accordingly all the orbitals with $L \geq 1$ are empty for the ground state at $\nu < 2$. Here we examine the contribution of the quantum transitions to the upper Landau levels. The transitions are classified into the following two cases:

(Case 1) One electron transfers to an orbital with $L = 0$ and the other electron transfers to an orbital with $L \geq 1$. Therein the one electron transfers from the momentum $p_1$ to $p_1'$ with $L = 0$ and then the other electron should transfer from $p_2$ to $p_2' = p_1 + p_2 - p_1'$ because the momentum conservation. All the orbitals in the upper Landau level $L \geq 1$ are empty. So the transition number is equal to the number of empty orbitals with $L = 0$. The number of empty orbitals with $L = 0$ varies continuously with changing of the filling factor $\nu$. Therefore the perturbation energies depend continuously upon the filling factor.

(Case 2) Both electrons in the electron pair transfer to the upper Landau levels. In this case the transition number is equal to the number of Landau orbitals with the upper level. So the transition number is constant.

Consequently the perturbation energy via the upper Landau levels ($L \geq 1$) does not produce a discontinuous structure in the energy spectrum. Therefore the energy gap is not derived from the contribution of the upper Landau levels. (Although the discussion was carried out for $\nu < 2$, the continuous property via the upper Landau levels in $\nu > 2$ is also the same as in $\nu < 2$.)

Furthermore the energy in the perturbation calculation is expressed by the Eq.(6.17). Therein the denominator of Eq.(6.17) is composed of the difference $W_{\text{excite}} - W_G$. The difference value for the upper Landau levels is extremely large in comparison with the case of the same Landau level. So the contribution from the upper Landau levels is negligibly small compared with that in the same level.

# Chapter 7 Plateaus of quantum Hall resistance

The expectation value of the total Hamiltonian is examined in chapter 3 and the pair energy via the Coulomb interaction is investigated in chapters 4-6. We obtain the total energy by adding these results. Then the total energy per electron $\varepsilon(\nu)$ on the quasi-2D electron system depends upon the filling factor $\nu$ discontinuously at the specific filling factors $\nu = \nu_0$ but continuously in other filling factors. This $\nu$-dependence yields the Hall resistance curve as will be clarified below [70].

## 7.1 Function-form of the total energy

The total energy $E(\nu)$ of the FQH state for $\nu < 1$ has been expressed as

$$E(\nu) = \chi(\nu)N + g(\nu)N + \left[f + \hbar eB/(2m^*) - (\xi - \eta)/\nu\right]N + C_{\text{Macroscopic}}(\sigma)$$

which was obtained by Eq.(5.11a) and Eq.(5.12b). Therein $\chi(\nu)N$ indicates the energy via the Coulomb transitions from all of the nearest pairs and $g(\nu)N$ indicates that from the rest of the pairs. The constant term $f$ was defined by Eq.(3.23a, b) as

$$f = \lambda + \overline{U} + 2\xi - \eta$$

where $\overline{U}$ is the mean value of the potential along the $y$-direction. We reexpress the total energy $E(\nu)$ as a sum of the following five terms:

$$E(\nu) = \left(\chi(\nu) + g(\nu) + \left(e\hbar/2m^*\right)B - a/\nu + b\right)N \tag{7.1a}$$

$$a = (\xi - \eta), \quad b = \lambda + \overline{U} + 2\xi - \eta + \left(C_{\text{Macroscopic}}/N\right) \tag{7.1b}$$

When the gate voltage is changed, the value of $\overline{U}$ varies. So the value of $b$ can be controlled by adjusting the gate voltage.

We describe the energy per electron by the symbol $\varepsilon(\nu)$:

$$\varepsilon(\nu) = \chi(\nu) + g(\nu) + \left(e\hbar/2m^*\right)B - a/\nu + b \tag{7.2}$$

Therein $\chi(\nu)$ is defined in Eq.(5.12a) as a sum of all order perturbation energies of the nearest electron (or hole) pairs as

$$\chi(\nu) = \sum_{n=2,3,\cdots} \chi_n(\nu) \qquad \text{at the filling factor of } \nu \tag{7.3a}$$

Also $g(\nu)$ indicates the pair energy placed in more distant Landau orbital pair as given by the summation of all order perturbation energies as

$$g(\nu) = \sum_{n=2,3,\cdots} g_n(\nu) \qquad \text{at the filling factor of } \nu \qquad (7.3b)$$

In chapters 4 and 5, we have calculated the second order perturbation energy of the nearest pairs per electron namely, $\chi_2(\nu)$. Also we have calculated the higher order perturbation energies as Eqs.(6.23, 24, 27 and 28) in chapter 6. The results are

$$\chi(\nu) = -\sum_{n=2,3,\cdots} \frac{1}{2j}\left(\frac{1}{2j+1}\right)^{n-1} {}_nZ \qquad \text{at } \nu = 2j/(2j+1) \qquad (7.4a)$$

$$\chi(\nu) = -\sum_{n=2,3,\cdots} \frac{1}{j}\left(\frac{j-1}{2j-1}\right)^{n-1} {}_nZ \qquad \text{at } \nu = j/(2j-1) \qquad (7.4b)$$

$$\chi(\nu) = -\sum_{n=2,3,\cdots} \left(\frac{1}{2j+1}\right)^{n-1} {}_nZ_H \qquad \text{at } \nu = 1/(2j+1) \qquad (7.4c)$$

$$\chi(\nu) = -\sum_{n=2,3,\cdots} \frac{1}{j}\left(\frac{j}{2j+1}\right)^{n-1} {}_nZ_H \qquad \text{at } \nu = j/(2j+1) \qquad (7.4d)$$

At $\nu = (2j-1)/(2j)$ and $\nu = 1/(2j)$ with the denominator of even number, all order perturbation energies of the *nearest electron* (or *hole*) *pair* are zero where all the quantum transitions are forbidden. The results Eqs.(6.29) and (6.30) are expressed as

$$\chi_n(\nu) = 0 \; (n=2,3,4,\cdots) \text{ at } \nu = (2j-1)/(2j) \text{ and } \nu = 1/(2j) \qquad (7.5)$$

$$\chi(\nu) = \sum_{n=2,3,4,\cdots} \chi_n(\nu) = 0 \quad \text{at } \nu = (2j-1)/(2j) \text{ and } \nu = 1/(2j) \qquad (7.6)$$

The n-th order terms in Eqs.(7.4a-d) have the multipliers $(1/(2j+1))^{n-1}$, $((j-1)/(2j-1))^{n-1}$, $(1/(2j+1))^{n-1}$ and $(j/(2j+1))^{n-1}$, respectively. These multipliers become small for large $n$ as

$(1/(2j+1))^{n-1} = 0.04$, $(j/(2j+1))^{n-1} = 0.16$, $((j-1)/(2j-1))^{n-1} = 0.11\cdots$ for $n=3$

$(1/(2j+1))^{n-1} = 0.008$, $(j/(2j+1))^{n-1} = 0.064$, $((j-1)/(2j-1))^{n-1} = 0.037\cdots$ for $n=4$

These results indicate that the second order term may be a main part of $\chi(\nu)$ because the multipliers are small in the higher order terms. We summarize the second order term $\chi_2(\nu)$ for various filling factors:

$$\chi_2(\nu) = -\frac{1}{2j}\left(\frac{1}{2j+1}\right)Z \qquad \text{at } \nu = 2j/(2j+1) \tag{7.7a}$$

$$\chi_2(\nu) = -\frac{1}{j}\left(\frac{j-1}{2j-1}\right)Z \qquad \text{at } \nu = j/(2j-1) \tag{7.7b}$$

$$\chi_2(\nu) = -\left(\frac{1}{2j+1}\right)Z_H \qquad \text{at } \nu = 1/(2j+1) \tag{7.7c}$$

$$\chi_2(\nu) = -\frac{1}{j}\left(\frac{j}{2j+1}\right)Z_H \qquad \text{at } \nu = j/(2j+1) \tag{7.7d}$$

$$\chi_2(\nu) = 0 \qquad \text{at } \nu = (2j-1)/(2j) \text{ and } \nu = 1/(2j) \tag{7.8}$$

Equations (5.33), (5.34), (5.36) and (5.37) give the second order energies in the neighbourhood of $\nu = (j+1)/(2j+1)$ and $\nu = j/(2j+1)$ as

$$\chi_2(\nu') = \frac{E_{\text{nearest pair}}}{N} = -\frac{((2j-1)2s-2)s}{((2j+1)(2s)-2)}\frac{Z}{((j+1)(2s)-1)} \quad \text{at } \nu' = \frac{((j+1)(2s)-1)}{((2j+1)(2s)-2)} \tag{7.9a}$$

$$\chi_2(\nu') = \frac{E_{\text{nearest pair}}}{N} = -\frac{((2j-1)2s+2)s}{((2j+1)(2s)+2)}\frac{Z}{((j+1)(2s)+1)} \quad \text{at } \nu' = \frac{((j+1)(2s)+1)}{((2j+1)(2s)+2)} \tag{7.9b}$$

$$\chi_2(\nu') = \frac{E_{\text{nearest pair}}}{N} = -\frac{((2j-1)2s-2)s}{((2j+1)(2s)-2)}\frac{Z_H}{((j)(2s)-1)} \quad \text{at } \nu' = \frac{((j)(2s)-1)}{((2j+1)(2s)-2)} \tag{7.9c}$$

$$\chi_2(\nu') = \frac{E_{\text{nearest pair}}}{N} = -\frac{((2j-1)2s+2)s}{((2j+1)(2s)+2)}\frac{Z_H}{((j)(2s)+1)} \quad \text{at } \nu' = \frac{((j)(2s)+1)}{((2j+1)(2s)+2)} \tag{7.9d}$$

Therein we have used the following ratio $N_H/N$ in the derivation of Eqs.(7.9c, d).

$$N_H/N = ((j+1)(2s)-1)/((j)(2s)-1) \text{ at } \nu' = ((j)(2s)-1)/((2j+1)(2s)-2)$$
$$N_H/N = ((j+1)(2s)+1)/((j)(2s)+1) \text{ at } \nu' = ((j)(2s)+1)/((2j+1)(2s)+2)$$

For arbitrary fractional number $\nu$ we can calculate $\chi_2(\nu)$ by using the same procedure as given in chapters 4-5. When the denominator of the fractional number is large, the total number of the quantum transitions is calculated by using computer.

Because the higher order perturbation energy includes the small multipliers as mentioned above, $\varepsilon(\nu)$ is given approximately by

$$\varepsilon(v) \approx \chi_2(v) + g + \left(e\hbar/2m^*\right)B - a/v + b \qquad (7.10)$$

As examined in sections 5.9 and 5.10 the contribution of the non-nearest pairs is small and the $v$-dependence is also small at $v < 2$. Accordingly we may ignore the $v$-dependence of $g$ in Eq.(7.10) at $v < 2$. The function-form $\varepsilon(v)$ is examined in the next section 7.2. [30-33]

## 7.2 Spectrum of the total energy versus filling factor

Figure 7.1 shows four graphs of $\varepsilon(v)$ versus $1/v$ in the neighbourhood of $v = 2/3,\ 1/2,\ 2/5,\ 4/7$ where the green lines indicate the function $\gamma(v)$ as

$$\gamma(v) = g + \left(e\hbar/2m^*\right)B - a/v + b \qquad (7.11)$$

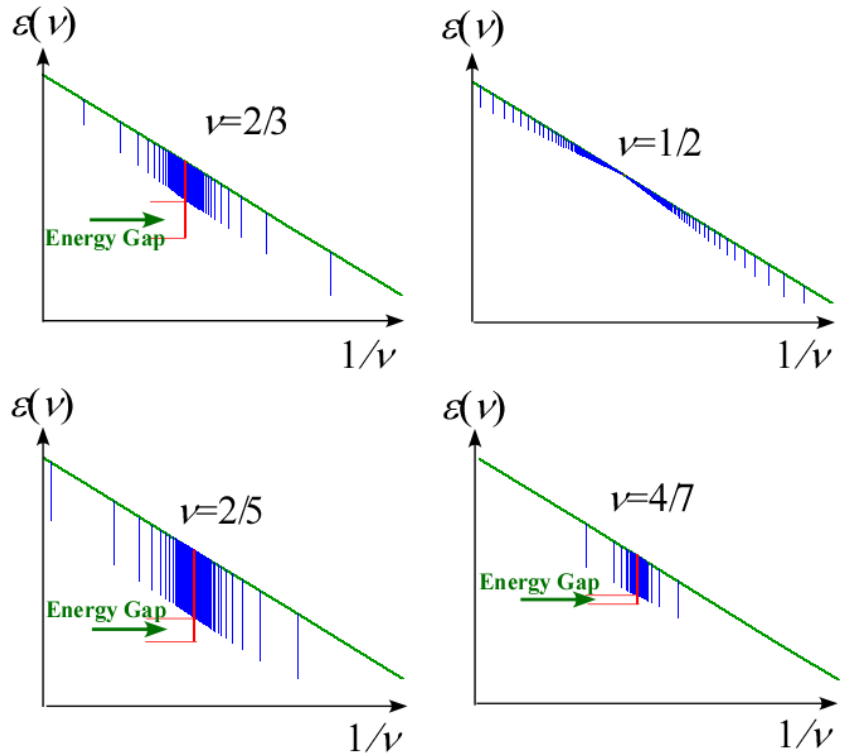

Fig.7.1 Energy spectra near $v = 2/3,\ 1/2,\ 2/5,\ 4/7$

We draw the value of $\chi_2(v)$ with vertical bars in Fig.7.1, where the lengths of the bars are the absolute values of $\chi_2(v)$. The upper end of each vertical bar is placed on the green line namely $\gamma(\nu)$. Then the lower ends of each vertical bars indicate the values of $\varepsilon(\nu)$ because $\chi_2(\nu)$ is negative.

$$\varepsilon(\nu) \approx \chi_2(\nu) + \gamma(\nu) \tag{7.12}$$

Blue vertical bars in the neighbourhood of $\nu = 2/3$ and $4/7$ express $\chi_2(\nu)$ of Eq.(7.9a) and (7.9b). Also blue vertical bars in the neighbourhood of $\nu = 1/2$ indicate Eqs.(7.7b) and (7.7d), and those for $\nu = 2/5$ indicate Eqs.(7.9c) and (7.9d).

Figure 7.1 shows the following two features:
1. The lower end of the red bar in the respective figure is lower than that of the blue bars in the neighbourhood of $\nu = 2/3,\ 2/5,\ 4/7$. The energy spectrum has a valley as shown by "Energy Gap" in the respective panel of Fig.7.1.
2. There is no valley at $\nu = 1/2$. The blue vertical bar approaches the base line (green line) near $\nu = 1/2$.

(Note: The blue bars have the filling factors with even number denominators as in Eq.(7.9a,b,c,d). So some readers might wonder whether the pair energies would be different from that with odd denominator. In order to reply this asking, we have calculated the energy for the nearest pairs at $\nu = (4s+1)/(6s+1)$ and $\nu = (4s-1)/(6s-1)$.)

The nearest pair energies are obtained in Appendix (A.6) and (A.13) as follows:

$$E_{\text{nearest pair}} = -\frac{2s^2}{(4s+1)(6s+1)} ZN \quad \text{for } \nu = (4s+1)/(6s+1)$$

$$E_{\text{nearest pair}} = -\frac{2s^2}{(4s-1)(6s-1)} ZN \quad \text{for } \nu = (4s-1)/(6s-1)$$

Then the energies per electron are given as follows;

$$\chi_2(\nu) = -\frac{2s^2}{(4s+1)(6s+1)} Z \quad \text{for } \nu = (4s+1)/(6s+1) \tag{7.13a}$$

$$\chi_2(\nu) = -\frac{2s^2}{(4s-1)(6s-1)} Z \quad \text{for } \nu = (4s-1)/(6s-1) \tag{7.13b}$$

These pair energies are shown in Fig.7.2 together with the energy of Fig.7.1. The green vertical bars indicate $\chi_2(\nu)$ at $\nu = (4s+1)/(6s+1)$ and $\nu = (4s-1)/(6s-1)$ with odd number denominator. The blue bars indicate $\chi_2(\nu)$ at $\nu = (4s-1)/(6s-2)$ and $\nu = (4s+1)/(6s+2)$ with even number denominator.

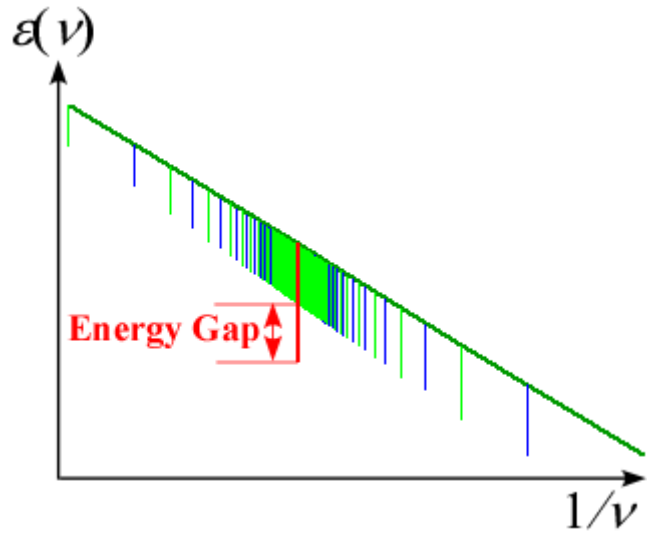

Fig.7.2 Energy spectrum near $\nu = 2/3$

The green vertical bars are drawn after showing the blue bars. So we can see only green bars in the vicinity of $\nu = 2/3$ because the blue bars are hidden by the dense green bars. We can see both blue and green bars when we move from $\nu = 2/3$. The lengths of the green and blue bars are almost the same and continuously vary with $\nu$. So the difference of $\chi_2(\nu)$ between even and odd integer of the denominators is negligibly small in the neighbourhood of $\nu = 2/3$.

We find many ranges of absent vertical-bar in Figs.7.1 and 7.2, because we don't calculate $\chi_2(\nu)$ yet inside the ranges. That is to say, many vertical bars are abbreviated in Figs.7.1 and 7.2. Of course we can calculate $\chi_2(\nu)$ inside these ranges by using a computer. Then we can get the more dense vertical bars.

Hitherto, a few theorists have calculated the energy spectrum of FQH states. As an example Halperin's result is shown in Fig.7.3 which has many cusps in the energy spectrum [22].

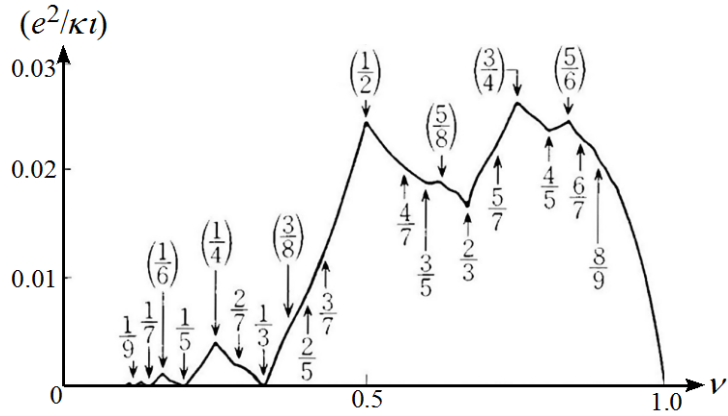

Fig.7.3 Energy spectrum of Halperin's result [22]

The energy spectra of the present theory have been shown in Figs.7.1 and 7.2 in the neighbourhood of the four filling factors. These five graphs are drawn separately. Some readers may want to know the spectrum in a wider region of the filling factor. We will show it. The $\nu$-dependence of $\varepsilon(\nu)$ (see Eq.(7.10)) is derived from only two parameters $Z$ and $a$ which depend upon the size, thickness and shape of a quantum Hall device. We draw the energy spectrum in Fig.7.4 for the parameter ratio $Z/a = 0.5$ as an example.

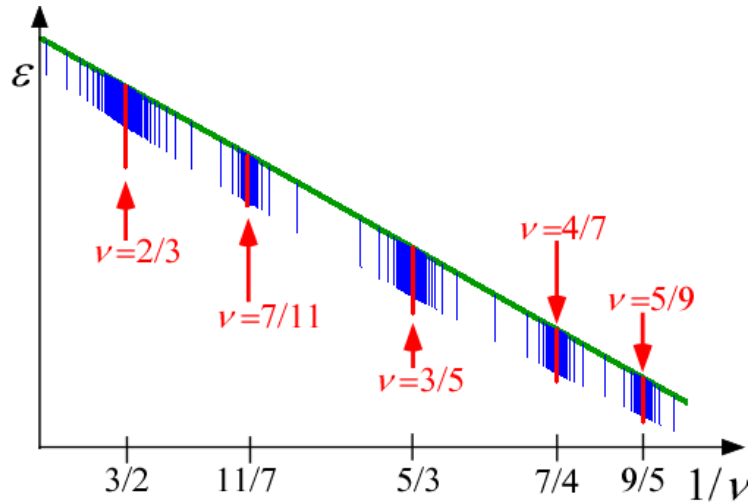

Fig.7.4 Energy spectrum of the present theory

The function-shape of the energy spectrum in the present theory [70] is quite different from Halperin's result [22] in Fig.7.3.

## 7.3 Behaviour of Hall resistance curve near $\nu = 1/2, 3/4, 1/4$ and so on

We draw three graphs of $\varepsilon(\nu)$ in the neighbourhood of $\nu = 1/2$ in Fig.7.5. The left panel shows the energy spectrum at the magnetic field $B = B_1$, the middle one $B_1 < B < B_2$ and the right one $B = B_2$. The energy $\varepsilon(\nu)$ increases with increment of $B$ due to the term $(e\hbar/2m^*)B$ in Eq. (7.10). We draw the chemical potential $\mu$ by the dark green line in Fig.7.5. The pink curve shows the lower end of each blue bar which indicates the value of $\varepsilon(\nu)$.

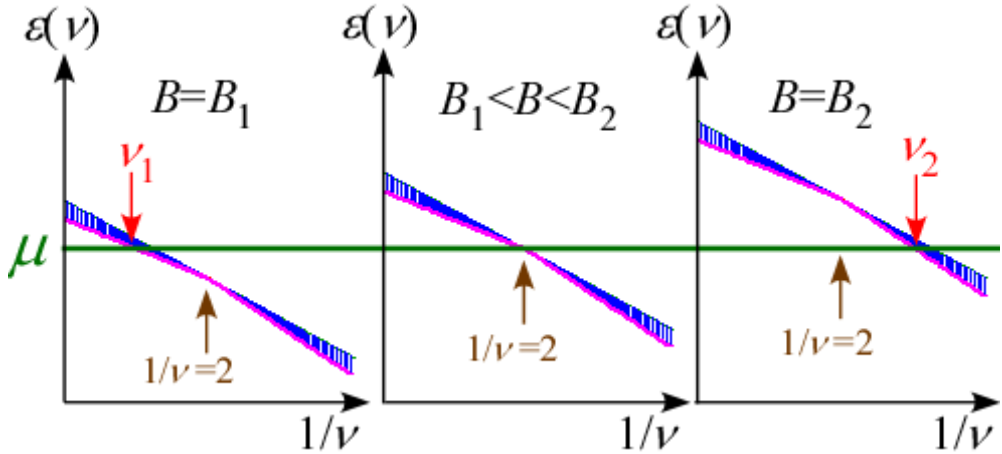

Fig.7.5 Magnetic field dependence of Energy spectrum near $\nu = 1/2$

The Fermi-Dirac distribution function $n(\varepsilon)$ is given by

$$n(\varepsilon) = \frac{1}{\exp((\varepsilon - \mu)/k_B T) + 1} \tag{7.14}$$

where $k_B$ is the Boltzmann's constant and $T$ is the temperature. At a low temperature all the states with $\varepsilon < \mu$ are occupied by electrons and the states with $\varepsilon > \mu$ are empty.

At the magnetic field $B = B_1$ (the left panel of Fig.7.5), $\varepsilon(\nu)$ is higher than $\mu$ for any filling factor $\nu > \nu_1$ and $\varepsilon(\nu)$ is lower than $\mu$ for any filling factor $\nu < \nu_1$. (It is noteworthy that the horizontal axis is $1/\nu$ and so the filling factor in the left is larger than one in the right.) All the electron states of $\nu = \nu_1$ are filled with electron, and any state of $\nu > \nu_1$ is empty at low temperatures. That is to say, the filling factor becomes $v_1$ at $B = B_1$. Also the filling factor becomes $v_2$ at $B = B_2$ as in the right panel of Fig.7.5. In the range of $B_1 < B < B_2$, the inverse of the filling factor $1/v$ depends upon the magnetic field $B$ under the fixed value of the chemical potential $\mu$. When the filling factor is nearly equal to 1/2, the nearest pair energy $\chi_2(v)$ is very small as in Fig.7.5. Accordingly $\varepsilon(v)$ is almost equal to $\gamma(v)$ due to Eq.(7.12). So the value of the filling factor is determined by the relation $\gamma(v) \approx \mu$ at a low temperature:

$$\varepsilon(v) \approx \gamma(v) = g + (e\hbar/2m^*)B - a/v + b \approx \mu \tag{7.15}$$

$$a/v \approx g + (e\hbar/2m^*)B + b - \mu$$

$$1/v \approx (e\hbar/(2m^*a))B + (b + g - \mu)/a \tag{7.16}$$

Eq.(7.16) means that the inverse of the filling factor, $1/v$, is linearly dependent upon the magnetic field in the neighbourhood of $v = 1/2$.

Next we study the magnetic field dependence of the Hall resistance. We have already investigated the electric current of the IQH states in chapter 2. Equation (2.28) is also satisfied at any fractional filling factor $v$ because the total electric current is $v$ times that with the filling factor = 1. Equation (2.29) holds for any fractional filling factor because an electron has the elementary charge $-e$. Substitution of Eq.(2.29) into Eq.(2.28) yields the following relation:

$$I = \frac{ve^2}{2\pi\hbar}(V_2 - V_1) \tag{7.17a}$$

Eq. (7.17a) gives the Hall resistance $R_H$ as

$$R_H = \frac{(V_2 - V_1)}{I} = \frac{2\pi\hbar}{e^2 v} \qquad \text{for any fractional filling factor} \tag{7.17b}$$

Substitution of Eq.(7.16) into Eq.(7.17b) yields the following relation:

$$R_H \approx \frac{2\pi\hbar}{e^2}\left((e\hbar/2m^*a)B + (b + g - \mu)/a\right) \qquad \text{near } v = 1/2 \tag{7.18}$$

This theoretical result means that the Hall resistance depends linearly upon the magnetic field in the neighbourhood of $v = 1/2$. One of the experimental data is shown in Fig.7.6. The experimental value of the Hall resistance also depends upon the magnetic field linearly near $v \approx 1/2$ as easily seen on the upper panel. Accordingly the result of the present theory is in good agreement with the experimental data near $v \approx 1/2$.

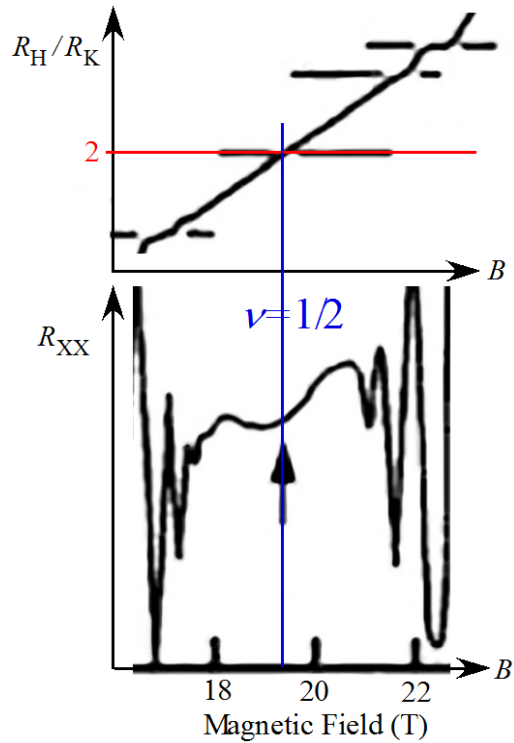

Fig.7.6 Experimental results of Hall resistance $R_H$ and diagonal resistance $R_{xx}$

$R_K$ is the von Klitzing constant

The states near $\nu = 1/2$ have no valley structure as shown in Fig.7.5, and has almost zero excitation energy as discussed in section 5.7. Therefore the excitations yield by electron scatterings due to the electric current. Consequently the diagonal resistance $R_{xx}$ is predicted to be finite and almost constant in the neighbourhood of $\nu = 1/2$. The theoretical result appears in the experimental data as in the lower panel of Fig.7.6.

Next we discuss the case with the peak structure. For an example $\nu = 3/4$, the probability of the $\nu = 3/4$ state is smaller than the probabilities with $\nu = (3/4) \pm \varepsilon$ because of the energy peak. Accordingly the state with $\nu = (3/4) + \varepsilon$ varies to the state with $\nu = (3/4) - \varepsilon$ skipping the $\nu = 3/4$ state ($\varepsilon$ is an infinitesimally small number) when the magnetic field is increased. Therefore the Hall resistance continuously changes and linearly depends on the magnetic field in the neighbourhood of $\nu = 3/4$. Thus the FQH state near the peak structure has a property similar to one in the neighbourhood of $\nu = 1/2$.

## 7.4 Explanation for the appearance of the plateaus in the Hall resistance curve

The energy spectrum $\varepsilon(\nu)$ yields the theoretical function-form of the Hall resistance versus magnetic field. We draw figures of $\varepsilon(\nu)$ in the neighbourhood of $\nu = 2/3$ at three magnetic fields, $B = B_3$, $B_3 < B < B_4$ and $B = B_4$, respectively in Fig.7.7. When the magnetic field becomes stronger, the value of $\varepsilon(\nu)$ becomes higher because of the term $(e\hbar/2m^*)B$ in Eq.(7.10). We show the chemical potential $\mu$ by the green line. The pink curve indicates the value of $\varepsilon(\nu)$ in the neighbourhood of $\nu = 2/3$ except $\nu = 2/3$. The valley at $\nu = 2/3$ is shown by the lower end of the red vertical bar.

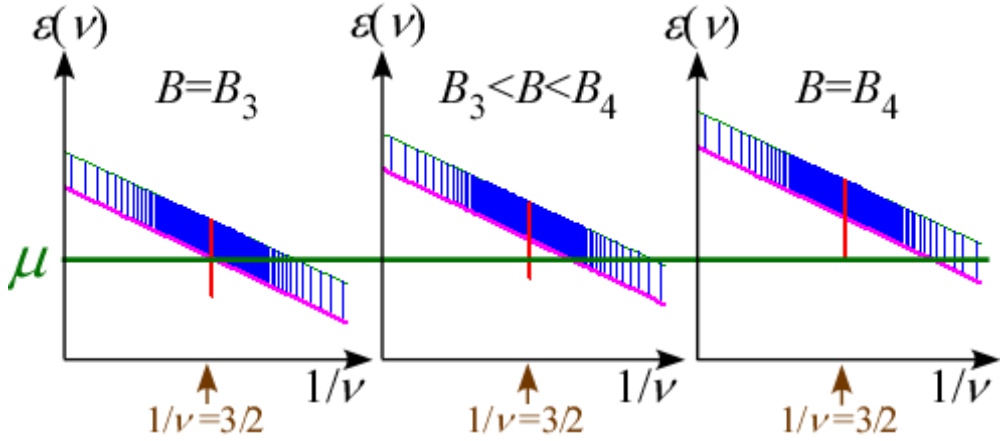

Fig.7.7 Magnetic field dependence of Energy spectrum near $\nu = 2/3$

We can find that the energy per electron $\varepsilon(2/3)$ is lower than the chemical potential $\mu$ in the range of $B_3 < B < B_4$ in Fig. 7.7. Therefore the state with $\nu = 2/3$ is filled with electrons at a low temperature. By contrast $\varepsilon(\nu > 2/3)$ is higher than the chemical potential $\mu$ in the range of $B_3 < B < B_4$. Therefore all the states with $\nu > 2/3$ are empty in a low temperature. Consequently the filling factor is confined to $2/3$ in the range of $B_3 < B < B_4$ [30-33]. The Hall resistance is given by Eq.(7.17b) as follows;

$$R_H = \frac{(V_2 - V_1)}{I} = \frac{2\pi\hbar}{\nu e^2} = \frac{3}{2} \times \frac{2\pi\hbar}{e^2} \quad \text{in the range of } B_3 < B < B_4 \quad (7.19)$$

That is to say, the Hall resistance $R_H$ takes a constant value $(3/2) \times (2\pi\hbar/e^2)$ in the range of $B_3 < B < B_4$. We examine below two more examples $\nu = 3/5$ and $4/5$. From Tables 5.1 and 5.5, the two states with $\nu = 3/5$ and $4/5$ have also large valleys in their energy spectra, respectively. The experimental data are shown in Fig.7.8 where three plateaus appear at $\nu = 2/3$, $3/5$ and $4/5$. The results of the present theory are in good agreement with the experimental data. Thus we conclude that the appearance of the plateaus in the Hall resistance curve originates from the valley structure.

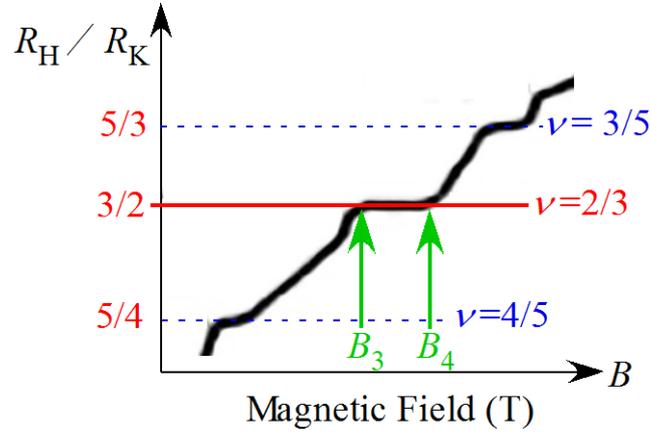

Fig.7.8 Experimental data of Hall resistance near $\nu = 4/5,\ 2/3,\ 3/5$

We examine the relation between the energy depth of the valley and the width $\Delta B = B_4 - B_3$. The left panel of Fig.7.7 expresses that the limiting value $\lim_{\nu \to 2/3+0} \varepsilon(\nu)$ at $B = B_3$ is equal to the chemical potential $\mu$:

$$\lim_{\nu \to 2/3+0} \varepsilon(\nu, B = B_3) = \lim_{\nu \to 2/3+0} (\chi(\nu)) + g(\tfrac{2}{3}) + (e\hbar/2m^*)B_3 - a/(\tfrac{2}{3}) + b \approx \mu \qquad (7.20)$$

Also the right panel of Fig.7.7 gives the following equation:

$$\varepsilon(\nu = 2/3, B = B_4) = \chi(\tfrac{2}{3}) + g(\tfrac{2}{3}) + (e\hbar/2m^*)B_4 - a/(\tfrac{2}{3}) + b \approx \mu \qquad (7.21)$$

Subtraction of Eq.(7.21) from Eq. (7.20) yields the following relation:

$$\left[\lim_{\nu \to 2/3+0} \chi(\nu)\right] - \chi(\tfrac{2}{3}) + (e\hbar/2m^*)(B_3 - B_4) = 0 \qquad (7.22)$$

The depth of the valley is given by

$$\text{Valley depth} = |\Delta \varepsilon_+| = \left[\lim_{\nu \to 2/3+0} \chi(\nu)\right] - \chi(2/3) \qquad (7.23)$$

Equations (7.22) and (7.23) yield the following relation:

$$\text{Valley depth} = |\Delta \varepsilon_+| = \hbar e(B_4 - B_3)/(2m^*). \qquad (7.24a)$$

Thus the depth of the valley is related to the width of the plateau on the Hall resistance curve versus magnetic field strength.

Next we discuss a vanishing of the diagonal resistance as in Fig.7.9. This vanishing region occurs simultaneously in the Hall plateau region. The property comes from the following reason: The ground state with $\nu = 2/3$ has an excitation-energy-gap $\Delta E_{\text{excitation \#1}} \approx (1/3)Z$ as in Eq.(5.85). The excitation energy is very large and therefore the electron scatterings are suppressed. This absence of scatterings yields a vanishing of the diagonal resistance $R_{xx}$ under the confinement of the filling factor 2/3. That is to say the vanishing region of $R_{xx}$ is equal to $B_3 < B < B_4$ due to the present theory. Thus the experimental width $\Delta B$ for vanishing of $R_{xx}$ can be estimated by using Eq.(7.24a) as follows:

$$\Delta B = (B_4 - B_3) = (2m^*/\hbar e)|\Delta \varepsilon_+| \tag{7.24b}$$

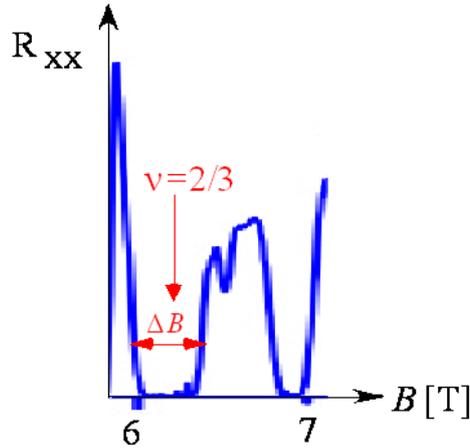

Fig.7.9 Experimental data of diagonal resistance $R_{xx}$ near $\nu = 2/3$

### 7.5 Comparison of the present theory with experimental data

In chapters 5-7, we have found a valley structure at $\nu = 2/3, 1/3, 4/5, 3/5, 2/5, 1/5, 4/7, 3/7 \ldots$, a flat structure at $\nu = 1/2$ and a peak structure at 3/4 1/4, …. The flat and peak structures produce a linear dependence of the Hall resistance upon the magnetic field as clarified in section 7.3. The energy valley produces a confinement of the Hall resistance because only one FQH state is realized in a range of the magnetic field at low temperatures as clarified in section 7.4.

We compare the depths of the valleys with the experimental data of the diagonal resistance [34, 35]. If an experiment is done using an ideal device without impurity and lattice defect at zero temperature, then $R_{xx}$ is zero in several ranges of magnetic field. But the actual experiments use the devices with impurities and lattice defects. Also the experiments are carried out at a finite temperature. Therefore the diagonal resistance of experiments becomes very small but not zero. So we roughly estimate the width $W(\nu)$ from the experimental data as follows; we take the width where $R_{xx}$ is lower than the green line in Fig.7.10. These are

$$W(2/3) \approx 0.35\,[\text{T}], \quad W(3/5) \approx 0.16\,[\text{T}], \quad W(4/7) \approx 0.09\,[\text{T}],$$
$$W(4/9) \approx 0.07\,[\text{T}], \quad W(3/7) \approx 0.12\,[\text{T}], \quad W(2/5) \approx 0.19\,[\text{T}],$$
$$W(1/3) \approx 0.9\,[\text{T}] \tag{7.25}$$

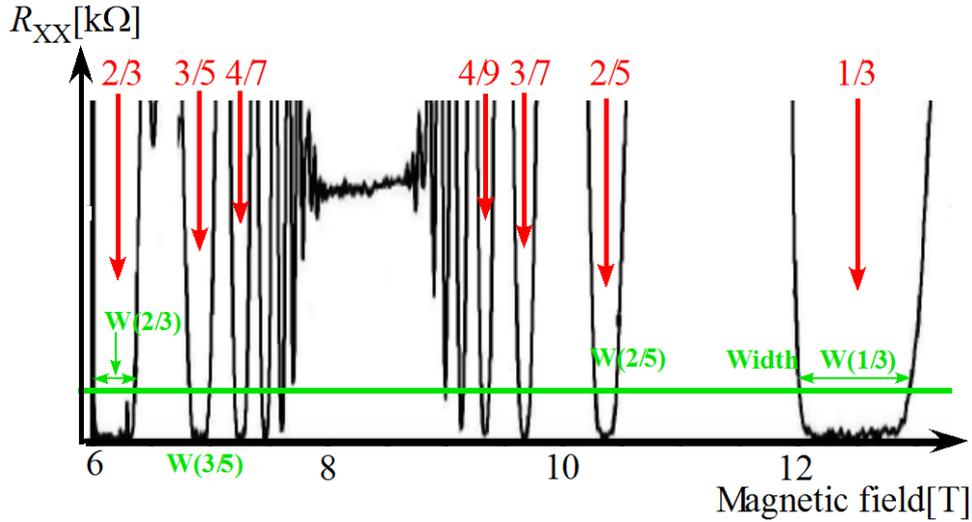

Fig.7.10 Widths of the vanishing regions on $R_{xx}$ curve

Our second order calculations give the energy gaps (depths of the valleys) which are obtained as in Tables 5.1 and 5.2:

$$\Delta\varepsilon_\pm(2/3) = -(1/12)Z \quad \Delta\varepsilon_\pm(3/5) = -(1/30)Z \quad \Delta\varepsilon_\pm(4/7) = -(1/56)Z$$
$$\Delta\varepsilon_\pm(4/9) = -(1/72)Z \quad \Delta\varepsilon_\pm(3/7) = -(1/42)Z_H \quad \Delta\varepsilon_\pm(2/5) = -(1/20)Z_H$$
$$\Delta\varepsilon_\pm(1/3) = -(1/6)Z_H \tag{7.26}$$

Neglecting the magnetic field dependence of $Z$ and using the approximate relation $Z \approx Z_H$, Eqs.(7.26) give the theoretical ratio of the energy gaps (depths of the valleys) as follows:

$$\tfrac{1}{12}:\tfrac{1}{30}:\tfrac{1}{56}:\tfrac{1}{72}:\tfrac{1}{42}:\tfrac{1}{20}:\tfrac{1}{6} \approx 0.35:0.14:0.08:0.06:0.1:0.21:0.7 \tag{7.27}$$

The ratio of the experimental values is derived from Eq.(7.25) as

$$\begin{aligned}&W\!\left(\tfrac{2}{3}\right):W\!\left(\tfrac{3}{5}\right):W\!\left(\tfrac{4}{7}\right):W\!\left(\tfrac{4}{9}\right):W\!\left(\tfrac{3}{7}\right):W\!\left(\tfrac{2}{5}\right):W\!\left(\tfrac{1}{3}\right)\\ &\approx 0.35:0.16:0.09:0.07:0.12:0.19:0.9\end{aligned} \tag{7.28}$$

The theoretical result, namely ratio (7.27), is in good agreement with the experimental result (7.28) in spite of a rough estimation.

Thus the FQH states become extremely stable at the filling factors $\nu = 2/3$, 1/3, 1/5, 2/5, 3/5 due to the large depths of valleys. Therein the deviation of the experimental value $R_H^{\exp}(\nu)$ from the theoretical value $2\pi\hbar/(e^2\nu)$ is quite small at these filling factors. The deviation $\left(R_H^{\exp}(\nu) - h/(e^2\nu)\right)/R_H^{\exp}(\nu)$ is measured to be about $3\times 10^{-5}$ at $\nu = 2/3$, about $3.3\times 10^{-8}$ at $\nu = 1/3$ as in [71] and about $2.3\times 10^{-4}$ at $\nu = 2/5$ as in [72-75]. These deviations will be investigated in details in the next chapter.

# Chapter 8 Accuracy of Hall resistance confinement

As explained in the previous chapters, the plateaus of the Hall resistance are caused by the valleys in the discontinuous energy spectrum. The valley structure is produced by the drastic change in the number of quantum transitions via the Coulomb interactions. Thereby the theoretical accuracy of Hall resistance confinement is rigorous at the FQHE in the infinitely large 2D system without lattice defect.

In this chapter we study more detail properties of the quasi 2D electron system with a finite size. If the device quality is bad, the Coulomb transitions are hindered by impurities and lattice defects. Therefore the quality of the quantum Hall device is very important for us to observe the precise confinement of the Hall resistance at the FQHE. When the experiment employs a quantum Hall device with an ultra high electron mobility, the experimental value of the Hall resistance is extremely close to $h/(e^2\nu)$ at the fractional filling factors $\nu = 2/3, 1/3, 1/5, 2/5, 3/5$, and so on [72-75]. For example the relative uncertainty is $\pm 3.3 \times 10^{-8}$ at the filling factor 1/3 as in Ref. [71]. The difference between the theoretical value $h/(e^2\nu)$ and the experimental value depends upon impurities, lattice defects, size and shape of the device. The accuracy of the Hall resistance confinement is investigated in details for the FQHE in this chapter [76].

## 8.1 Accuracy of the Hall resistance confinement in FQHE
### (pair energy depending upon the pair momentum in a finite size device)

We first examine the distribution of the single electron wave function which is given by Eq.(1.17) as

$$\psi_{L,J}(x,y,z) = \sqrt{\frac{1}{\ell}} \exp(ikx) u_L H_L\left(\sqrt{\frac{m^*\omega}{\hbar}}(y-\alpha_J)\right) \exp\left(-\frac{m^*\omega}{2\hbar}(y-\alpha_J)^2\right) \phi(z) \quad (8.1)$$

The probability density in the y-direction has the following form for the lowest Landau level $L=0$.

$$\exp\left(-\frac{m^*\omega}{2\hbar}(y-\alpha_J)^2\right) \exp\left(-\frac{m^*\omega}{2\hbar}(y-\alpha_J)^2\right) = \exp\left(-\frac{m^*\omega}{\hbar}(y-\alpha_J)^2\right)$$

The distribution width $\Delta y$ along the y-direction is

$$\Delta y \approx \sqrt{\frac{\hbar}{m^*\omega}} = \sqrt{\frac{\hbar}{eB}}$$

where we have applied Eq.(1.12) to angular frequency $\omega$ for deriving the second equality. The value of $\Delta y$ becomes about 10.5 nm at 6 T (Note: the experiment [34, 35] used the magnetic field strength 6 T at $\nu = 2/3$):

$$\hbar \approx 1.0546 \times 10^{-34} \text{ J s} \qquad e \approx 1.6022 \times 10^{-19} \text{ C}$$

$$\Delta y = \sqrt{\hbar/(eB)} \approx 10.5 \text{ [nm]} \quad \text{for } B \approx 6 \text{ [T]} \tag{8.2}$$

Next we estimate the interval between two peaks of the nearest Landau orbitals. The interval value $\Delta \alpha$ is equal to $\alpha_{J+1} - \alpha_J$ which are derived from Eq.(1.10) as:

$$\Delta \alpha = \alpha_{J+1} - \alpha_J = \frac{2\pi\hbar}{eB\ell} \approx 6.9 \times 10^{-3} \text{ [nm]} \quad \text{for } \ell = 100 \text{ [}\mu\text{m]} = 10^5 \text{ [nm]} \tag{8.3}$$

where $\ell = 100 \text{ [}\mu\text{m]}$ is the length of the device. There are many single-electron states inside the distribution width $\Delta y$ whose number is

$$\frac{\Delta y}{\Delta \alpha} = \frac{\ell}{2\pi}\sqrt{\frac{eB}{\hbar}} \approx 1.5 \times 10^3 \quad \text{at } B \approx 6 \text{ [T]} \text{ and } \ell = 100 \text{ [}\mu\text{m]} \tag{8.4}$$

Secondary we examine the transfer-momentum dependence of the transition matrix element $\langle p'_A, p'_B | H_I | p_A, p_B \rangle$:

$$\langle p'_A, p'_B | H_I | p_A, p_B \rangle = \left\langle \psi_{0,p'_A} \psi_{0,p'_B} \left| \frac{e^2}{4\pi\varepsilon\sqrt{(x_A - x_B)^2 + (y_A - y_B)^2 + (z_A - z_B)^2}} \right| \psi_{0,p_A} \psi_{0,p_B} \right\rangle$$

which is calculated as:

$$= \text{constant} \times \iiint \frac{e^2 \exp(i(p_A - p'_A)x_A/\hbar)\exp(i(p_B - p'_B)x_B/\hbar)\phi^*(z_A)\phi(z_A)\phi^*(z_B)\phi(z_B)}{\ell^2 4\pi\varepsilon\sqrt{(x_A - x_B)^2 + (y_A - y_B)^2 + (z_A - z_B)^2}}$$

$$\times \exp\left(-\frac{m^*\omega}{2\hbar}\left(y_A - \frac{p_A}{eB}\right)^2\right) \exp\left(-\frac{m^*\omega}{2\hbar}\left(y_A - \frac{p'_A}{eB}\right)^2\right)$$

$$\times \exp\left(-\frac{m^*\omega}{2\hbar}\left(y_B - \frac{p_B}{eB}\right)^2\right) \exp\left(-\frac{m^*\omega}{2\hbar}\left(y_B - \frac{p'_B}{eB}\right)^2\right) dx_A dy_A dz_A dx_B dy_B dz_B$$

There are four Gaussian functions whose peaks are at $y_A = p_A/eB$, $y_A = p'_A/eB$, $y_B = p_B/eB$, and $y_B = p'_B/eB$. The difference $\left|\frac{p'_A}{eB} - \frac{p_A}{eB}\right|$ is equal to $\left|\frac{p'_B}{eB} - \frac{p_B}{eB}\right|$ because of the momentum conservation. So the transition matrix element $\langle p'_A, p'_B | H_I | p_A, p_B \rangle$ depends upon the momentum difference $p'_A - p_A$ only. When the value of $\left|\dfrac{p'_A}{eB} - \dfrac{p_A}{eB}\right|$ is larger than $\Delta y$, the matrix element of the Coulomb interaction becomes small owing to the large separation of the two peaks.

On the other hand, when $\left|\dfrac{p'_A}{eB} - \dfrac{p_A}{eB}\right|$ is smaller than $\Delta y$, the overlapping of the wave functions is large. Then the transitions are effective. The momentum difference exists in the following region.

$$-1.5\times 10^3 (2\pi\hbar/\ell) \leq p'_A - p_A \leq 1.5\times 10^3 (2\pi\hbar/\ell). \qquad (8.5)$$

That is to say, about $3\times 10^3$ states contribute to the Coulomb transitions. When there are many impurities in the device, the number of the effective transitions decreases because of the disturbance by the impurities. This impurity effect is examined in the next section.

In this section, we investigate the ideal case without impurity and lattice defect. We schematically draw the effective and non-effective transitions for various filling factors in Figs.8.1-8.3 which are classified into three cases (A-1), (A-2) and (A-3).

Case (A-1)   ($\nu = 2/3$)

Figure 8.1 shows the quantum transitions at the filling factor $\nu = 2/3$. Therein $2\times 10^3$ electrons occupy in $3\times 10^3$ states.

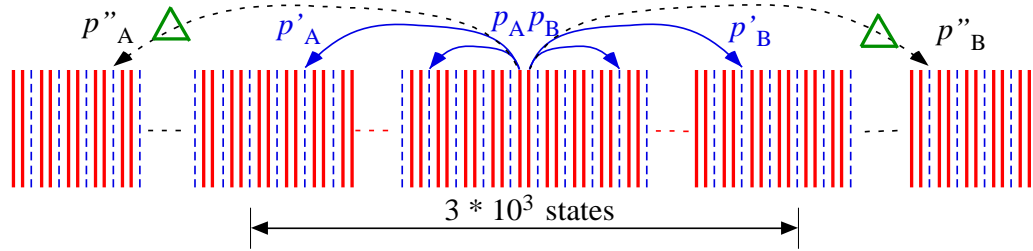

Fig.8.1 Allowed transitions for $\nu = 2/3$

The black dashed arrows indicate a non-effective transition expressed by the symbol △.

The effective transitions are drawn by blue arrows because the momentum transfer satisfies the condition (8.5). The non-effective transition is drawn by black dashed arrows in Fig.8.1 where $|p''_A - p_A|$ has a value larger than $1.5\times 10^3 (2\pi\hbar/\ell)$. The matrix element of this transition is negligibly small because the two wave functions don't overlap with each other. The non-effective transition is expressed by the symbol △ in Fig.8.1. Consequently the number of the effective transition becomes $1 \times 10^3$ (which is equal to the number of empty states in the effective region).

We consider the following critical case where adjacent three filled orbitals exist in every $3 \times 10^3 + 1$ states and $2 \times 10^3 + 1$ electrons occupy in $3 \times 10^3 + 1$ states. The filling factor is $(1 + 0.166 \cdots \times 10^{-3})2/3$. This critical filling factor is expressed by the symbol $\nu_2$. Also the symbol $\nu_1$ is defined by

$$\nu_1 = (1 - 0.166 \cdots \times 10^{-3})2/3, \qquad \nu_2 = (1 + 0.166 \cdots \times 10^{-3})2/3 \qquad (8.6)$$

Next we consider another Case (A-2) where the filling factor is larger than $\nu_2$.

Case (A-2):   (for the filling factor larger than $\nu_2$)

In this case, adjacent three filled orbitals shown by brown bold lines appear inside every $3 \times 10^3$ states because the filling factor is larger than $(1 + 0.166 \times 10^{-3})2/3$.

Fig.8.2 Allowed transitions for a filling factor $\nu'$ larger than $\nu_2$

The symbol △ means non-effective and ✗ indicates a forbidden transition.

Figure 8.2 schematically expresses the Coulomb transitions. The adjacent three filled orbitals disturb the sequence of (filled, filled, empty). The transition from $(p_A, p_B)$ to $(p'_A, p'_B)$ satisfies the total momentum conservation $p_A + p_B = p'_A + p'_B$. As in Fig.8.2 the state with momentum $p'_A$ is already filled with an electron before the Coulomb transition, although the state of $p'_B$ is empty. Therefore the quantum transition from $(p_A, p_B)$ to $(p'_A, p'_B)$ is forbidden. Thus the number of effective transitions becomes smaller than $1 \times 10^3$. Consequently the absolute value of the pair-energy $|\zeta_{AB}(\nu')|$ for $v' > v_2$ is smaller than $|\zeta_{AB}(\tfrac{2}{3})|$. That is to say, $\zeta_{AB}(v') > \zeta_{AB}(\tfrac{2}{3})$ because the pair-energies are negative.

Case (A-3): for the filling factor $v''$ ($\tfrac{2}{3} < v'' < v_2$)

The most uniform configuration of electrons is shown in Fig.8.3 at the filling factor with $v'' = (1+\Delta'')2/3$ for $\Delta'' < 0.166 \times 10^{-3}$. In this case the adjacent three brown lines appear outside the region of $3 \times 10^3$ states because of $\tfrac{2}{3} < v'' < v_2$. So there are $1 \times 10^3$ effective transitions from $(p_A, p_B)$. On the other hand the number of allowed transition from $(p_C, p_D)$ is only two as shown by blue arrows in Fig.8.3. Thus the number of effective transitions depends on the total momentum of the electron pair.

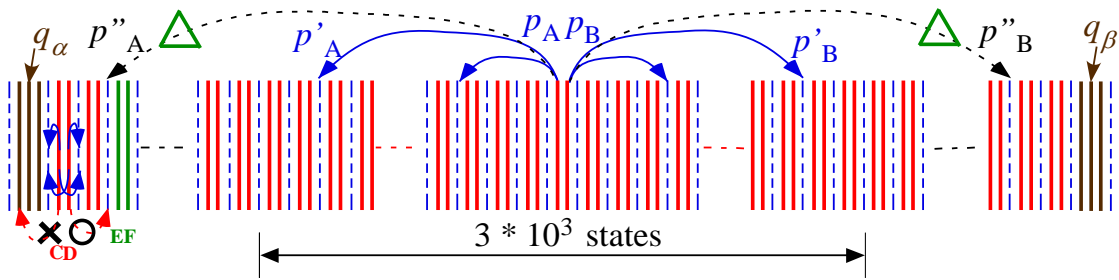

Fig.8.3 Allowed transitions for a filling factor $v''$ ($\tfrac{2}{3} < v'' < v_2$)

The symbol ✗ indicates a forbidden transition.

We have examined the average energy of electron pairs placed in the nearest Landau orbitals in the previous chapters. The average of pair energy is the ratio between the sum of all the pair energies with different pair momenta and the total number of pair energies. So the momentum dependence doesn't exist in the average energy. Then the major structure of FQHE has been clarified by using the average energy. When we examine more detail properties of FQHE (for examples: accuracy of the Hall resistance confinement of FQHE, the impurity effect and so on), we need to take the momentum dependence of the pair energy into consideration. The electron pair-energy depends upon the pair momentum. For example the nearest electron pair AB has the total momentum $Q_{AB}$ as

$$Q_{AB} = p_A + p_B = 2p_A + 2\pi\hbar/\ell \qquad (8.7a)$$

where we have employed the relation $p_B = p_A + 2\pi\hbar/\ell$ in the nearest electron pair. This pair momentum reproduces the two momenta $p_A$ and $p_B$ as

$$p_A = (Q_{AB}/2) - \pi\hbar/\ell, \qquad p_B = (Q_{AB}/2) + \pi\hbar/\ell \qquad (8.7b)$$

Hereafter the nearest pair-energy $\zeta_{AB}(\nu)$ is rewritten by $\zeta(\nu; Q_{AB})$ to express the dependence on the total momentum of the electron pair. Figure 8.3 shows that the number of effective transitions from the pair $(p_A, p_B)$ at the filling factor $\nu''$ is almost equal to that at $\nu = 2/3$:

$$\zeta(\nu''; Q_{AB}) \approx \zeta(2/3) \quad \text{for} \quad 2/3 \leq \nu'' < (1 + 0.166 \times 10^{-4})2/3 \qquad (8.8)$$

It is noteworthy here that the pair-energy $\zeta(2/3, Q)$ is independent of the pair momentum $Q$ because any nearest pair has the same number of effective transitions at $\nu = 2/3$ as easily seen in Fig.8.1. So we can abbreviate the argument $Q$ as $\zeta(2/3, Q) \to \zeta(2/3)$.

At the filling factor $\nu''$ the number of effective transitions from the pair CD is only two as in Fig.8.3. Therefore

$$\zeta(\nu''; Q_{CD}) \approx (2/10^3)\zeta(2/3) \qquad (8.9a)$$

Similarly we examine the transitions from the pair EF shown by green pair lines in Fig.8.3. The number of effective transitions from the electron pair EF is six which give the pair energy as

$$\zeta(\nu''; Q_{EF}) \approx (6/10^3)\zeta(2/3) \qquad (8.9b)$$

Thus the nearest pair in the neighborhood of adjacent three orbitals filled with electron has very small pair-energy because the Coulomb transitions are disturbed by the adjacent three filled orbitals.

A similar situation appears for the filling factors smaller than 2/3. At the filling factor $\nu''' = (1 - \Delta''')2/3$

$$\zeta(\nu'''; Q_{AB}) \approx \zeta(\tfrac{2}{3}) \qquad (8.10a)$$

$$\zeta(\nu'''; Q_{CD}) \approx (2/10^3)\zeta(\tfrac{2}{3}) \qquad (8.10b)$$

$$\zeta(\nu'''; Q_{EF}) \approx (6/10^3)\zeta(2/3) \quad \text{for} \quad 2/3 \geq \nu''' > (1 - 0.166 \times 10^{-3})2/3 \qquad (8.10c)$$

Accordingly it is important to investigate the momentum dependence of the pair-energy. We examine the following case where the adjacent three filled orbitals have the central momentum $q_\alpha$. The pair momenta of the nearest electron pair are given by Eq.(8.7b) as $p_\pm = (Q/2) \pm \pi\hbar/\ell$. When the momenta $p_\pm = (Q/2) \pm \pi\hbar/\ell$ are near the central momentum $q_\alpha$, the absolute value of pair-energy $\zeta(v'';Q)$ is almost proportional to the absolute value of the difference as follows;

$$\zeta(v'';Q) \approx \text{constant} \times |Q - 2q_\alpha| \quad \text{for} \quad |p_\pm - q_\alpha| \ll 1.5 \times 10^3 \frac{2\pi\hbar}{\ell} \quad (8.11a)$$

because the effective transition number is almost proportional to the absolute value of $p_\pm - q_\alpha$. When the absolute value of $p_\pm - q_\alpha$ approaches $1.5 \times 10^3 (2\pi\hbar/\ell)$ for $\ell \approx 100 [\mu m]$ and $B \approx 6 [T]$, the effective transition number approaches the maximum value. Then the pair energy $|\zeta(v'';Q)|$ saturates and approaches $|\zeta(\frac{2}{3})|$ as in Fig.8.4.

Therein the pair energies are shown at the four filling factors by the blue curve for $v = 2/3$, green for $v_a \approx (1 + 0.166 \times 10^{-3})2/3$, red for $v_b \approx (1 + 0.111 \times 10^{-3})2/3$ and black for $v_c \approx (1 + 0.072 \times 10^{-3})2/3$. Because the effective transition number is constant at $v = 2/3$ for any pair momentum, the curve of the pair-energy is constant which is expressed by blue straight line in Fig.8.4.

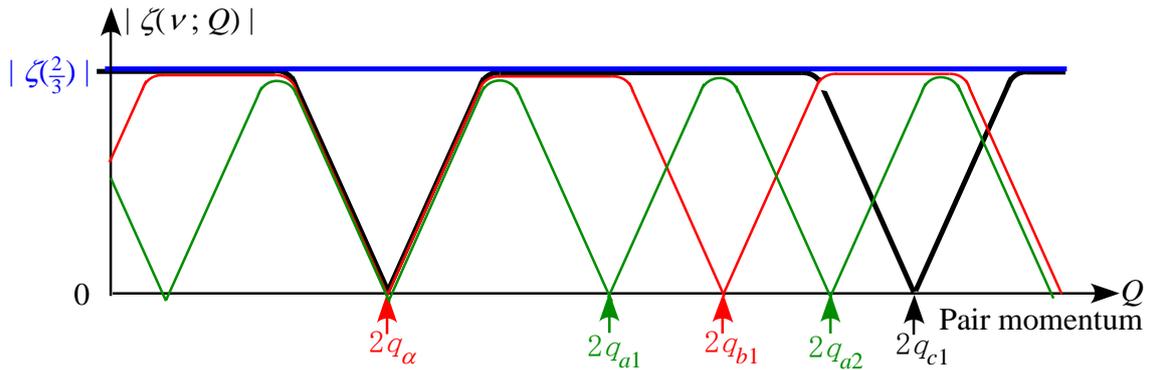

Fig.8.4 Pair momentum dependence of $|\zeta(v;Q)|$

Blue curve at $v = 2/3$   Green for $v_a \approx (1 + 0.166 \times 10^{-3})2/3$
Red for $v_b \approx (1 + 0.111 \times 10^{-3})2/3$   Black for $v_c \approx (1 + 0.072 \times 10^{-3})2/3$

The central momentum in the adjacent three filled orbitals is expressed by the symbols $q_\alpha, q_{a1}, q_{a2}, q_{a3}, \cdots$, $q_\alpha, q_{b1}, q_{b2}, q_{b3}, \cdots$ and $q_\alpha, q_{c1}, q_{c2}, q_{c3}, \cdots$ for the filling factors $\nu_a, \nu_b$ and $\nu_c$, respectively. The momentum interval namely $q_{a1} - q_\alpha$ becomes larger when the filling factor $\nu_a$ approaches 2/3. The total number of the adjacent three filled orbitals is proportional to $\nu_a - 2/3$. Accordingly the momentum interval $q_{a1} - q_\alpha$ is related as

$$q_{a1} - q_\alpha \approx \text{constant}/(\nu_a - 2/3) \tag{8.12}$$

In Fig.8.4 the ratio of the momentum intervals is equal to the ratio of $1/(\nu - 2/3)$ as:

$$(q_{a1} - q_\alpha):(q_{b1} - q_\alpha):(q_{c1} - q_\alpha) \approx (1/0.166):(1/0.111):(1/0.072) \tag{8.13}$$

The pair-energy per electron $\chi(\nu)$ has been already defined by (total energy of nearest pairs) / (total electron number $N$). So $\chi(\nu)$ is the average value of $\zeta(\nu;Q)$ in Figs.8.4 for a finite size device with $\ell \approx 100[\mu\text{m}]$. The average value $\chi(\nu)$ for a finite size device is continuously dependent upon $\nu$ as in Fig.8.5.

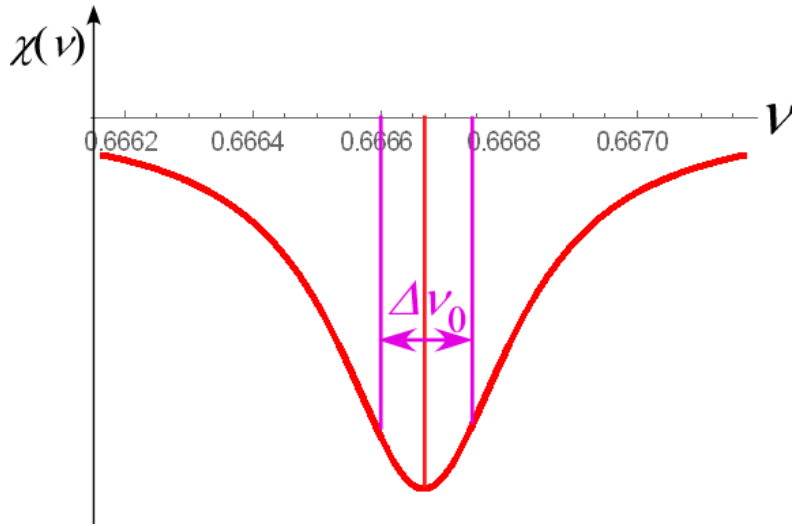

Fig.8.5 Average pair-energy per electron versus filling factor in the neighborhood of $\nu = 2/3$ (This dependence is schematically shown for a quantum Hall device with a finite size $\ell \approx 100[\mu\text{m}]$. Horizontal axis indicates the filling factor $\nu$ and vertical axis is drawn by arbitrary scale. )

If we ignore the momentum dependence in Fig.8.4 the confinement of the FQH

resistance has the theoretical value of the relative uncertainty about $10^{-4}$ for $\ell \approx 100[\mu m]$ as in Fig.8.5. The experimental result [71] has clarified that the accuracy of the Hall resistance confinement is the relative uncertainty $\pm 3.3 \times 10^{-8}$ at $\nu = 1/3$.

Thus we cannot explain the extremely precise confinement of Hall resistance in an actual finite system. Accordingly we need to take the momentum dependence of the pair energies in Fig.8.4.

The energy spectrum is discontinuously dependent on the filling factor for the infinitely large 2D electron system as in the previous chapters 4-7. The total energy per electron has been obtained in Eq.(5.13) as

$$\varepsilon(\nu) = \chi(\nu) + g(\nu) + [f + \hbar eB/(2m^*) - (\xi - \eta)/\nu] + C_{\text{Macroscopic}}/N$$

In this chapter we have examined the quasi 2D electron system with a finite size and calculated the number of the effective transitions. The pair-energy $\zeta(\nu; Q_{AB})$ belongs to the electrons A and B. Accordingly the electron A has half of $\zeta(\nu; Q_{AB})$. Therefore Eq.(5.13) should be replaced with the following relation for a finite size device:

$$\varepsilon_A(\nu) = \zeta(\nu; Q_{AB})/2 + c, \quad \varepsilon_B(\nu) = \zeta(\nu; Q_{AB})/2 + c \quad (8.14a)$$

$$c = g(\nu) + [f + \hbar eB/(2m^*) - (\xi - \eta)/\nu] + C_{\text{Macroscopic}}/N \quad (8.14b)$$

The momentum dependence is explicitly described as follows:

$$\varepsilon(\nu; q) = \zeta(\nu; 2q \pm (2\pi\hbar/\ell))/2 + c \quad (8.15)$$

where the sign $+$ is used for the smaller momentum in nearest two momenta and $-$ for the larger momentum. The energy spectrum is drawn for four filling factors in Fig.8.6 where the chemical potential $\mu$ is expressed by the dashed pink line.

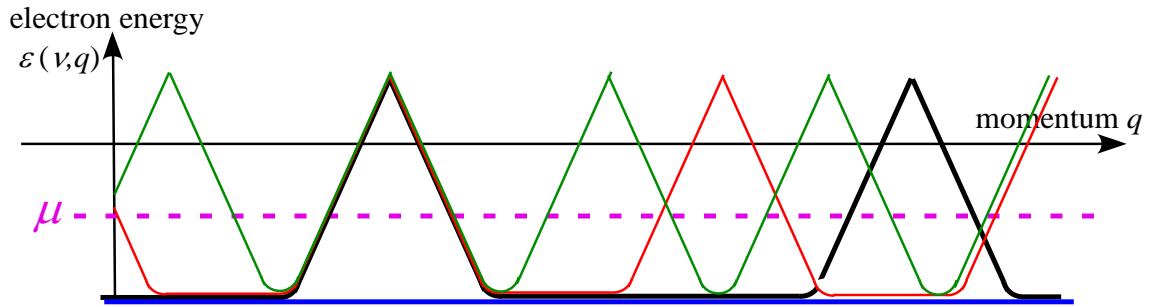

Fig.8.6 Energy spectra for four filling factors

Blue curve at $\nu = \frac{2}{3}$, Green curve at $\nu_a$, Red curve at $\nu_b$, Black curve at $\nu_c$

There are many electron states with $\varepsilon(\nu;q) > \mu$ on the green, red and black curves. On the other hand, at the filling factor $\nu = \frac{2}{3}$ all the energies $\varepsilon(\frac{2}{3};q)$ are smaller than the chemical potential $\mu$ as in Fig.8.6. The electron states with $\varepsilon(\nu;q) > \mu$ cannot be realized at an ultra low temperature. Therefore the present theory concludes that the FQH state with $\nu = \frac{2}{3}$ is realized under a sufficiently low temperature. Thus the function-form of $\varepsilon(\nu;q)$ produces the rigorous confinement of Hall resistance observed in the experiment [71].

## 8.2 Effects of Impurities and Lattice defects on the Hall resistance

Impurities and lattice defects produce an irregular potential in a quantum Hall device. The irregular potential hinders the quantum transitions from the nearest pairs to the orbitals near the impurity and/or lattice defect. The situation is illustrated in Fig.8.7 for a device with very few impurities and Fig.8.8 for a device with many impurities.

Case (B-1) (very few impurities)
  We schematically draw the Coulomb transitions for the Case (B-1) in Fig.8.7. In this case, the distance between the impurities is very large. Therefore the impurities appear outside the $3 \times 10^3$ Landau states. That is to say the distance between the impurities is larger than the robe of the wave function.

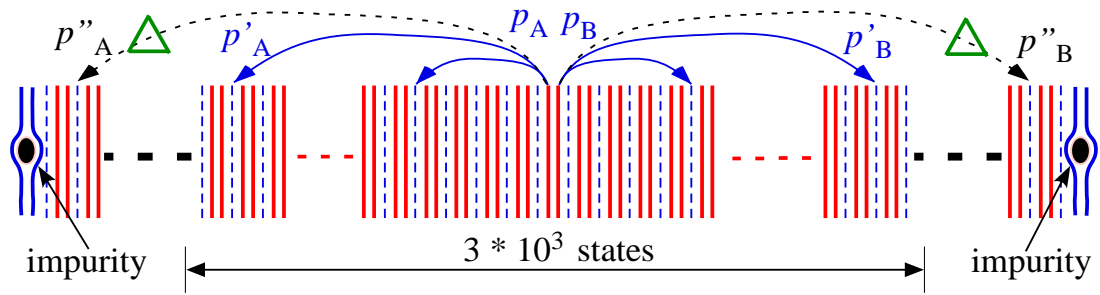

Fig.8.7 Disturbance of Coulomb transitions in very few impurities at $\nu = 2/3$

The number of effective Coulomb transitions is nearly equal to that in the ideal case as easily seen in Fig.8.7. The pair-energy per electron is almost the same as in the ideal cases (A-1) and (A-3).

Case (B-2) (many impurities)

Next we study the case with many impurities and lattice defects. Figure 8.8 schematically shows the Coulomb transitions in a quantum Hall device with many impurities. The number of Landau states between impurities becomes small because of many impurities. Thereby many transitions are forbidden by the Pauli principle. Therefore the absolute value of pair-energy $|\chi(\nu)|$ becomes small. The forbidden transitions are shown in red cross ✗ in Fig.8.8.

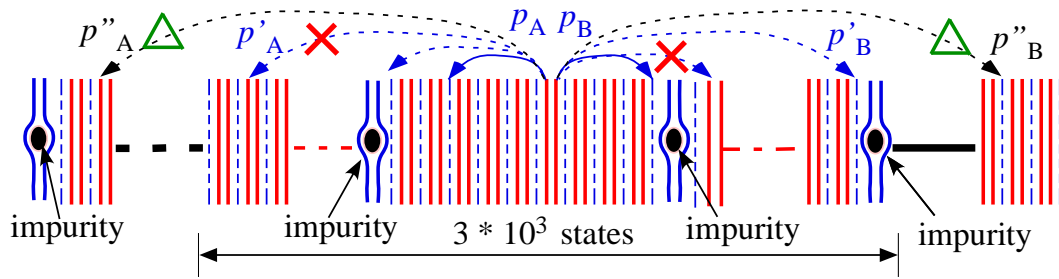

Fig.8.8 Disturbance of Coulomb transitions by many impurities at $\nu = 2/3$

Next we study the state with a filling factor $\nu$ larger than 2/3. In this case there are many adjacent three orbitals filled with electrons as shown in blue lines in Fig.8.9. When the difference $\nu - 2/3$ becomes large, the adjacent three filled orbitals appear more often in the most uniform electron-configuration.

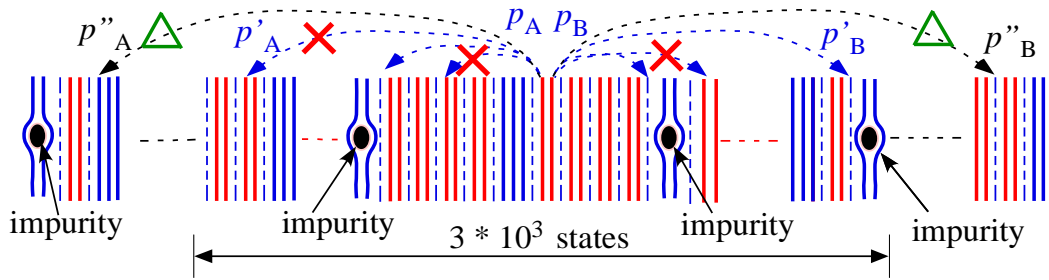

Fig.8.9 Disturbance of Coulomb transitions by many impurities at $\nu > 2/3$

Many Coulomb transitions are forbidden by the following causes: (1) impurities, (2) lattice defects, (3) deviation of $\nu$ from 2/3. The function shape of $\chi(\nu)$ for the Case (B-1) is shown by red curve where the valley width is $\Delta \nu_0$ in Fig.8.10. The width $\Delta \nu_0$ is the same as the width in Fig.8.5 for the ideal case. The function shape of $\chi(\nu)$ for Case (B-2) is drawn by the blue curve in Fig.8.10 and the valley width is expressed by $\Delta \nu'$.

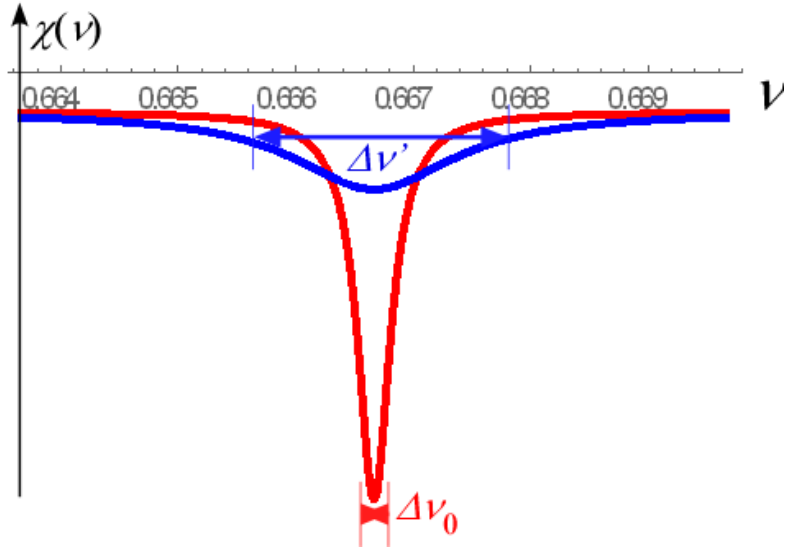

Fig.8.10 Nearest pair-energy $\chi(\nu)$ versus filling factor $\nu$
Red curve indicates $\chi(\nu)$ for Case (B-1). Blue curve indicates $\chi(\nu)$ for Case (B-2)

The width $\Delta \nu'$ is very wide and the binding energy becomes very small because of the presence of the many impurities (Case (B-2)). When the distance between impurities is smaller than that between the adjacent three orbitals, the impurities mainly hinder the Coulomb transitions. So the width $\Delta \nu'$ in Fig.8.10 depends upon the quality of the device. When the quality of the device becomes worse, the plateau becomes unclear or disappears.

The experimental value of the Hall resistance $R_H^{\exp}(\nu)$ is close to $h/(e^2\nu)$ for the filling factors $\nu = 2/3$, 1/3, 1/5, 2/5, 3/5, and so on. The experimental deviation $(R_H^{\exp}(\nu) - 2\pi\hbar/(e^2\nu))/R_H^{\exp}(\nu)$ has been measured and the results are

$(R_H^{\exp}(\nu) - 2\pi\hbar/(e^2\nu))/R_H^{\exp}(\nu)$ = about $3\times 10^{-5}$    for $\nu = 2/3$      (8.13a)
$(R_H^{\exp}(\nu) - 2\pi\hbar/(e^2\nu))/R_H^{\exp}(\nu)$ = about $2.3\times 10^{-4}$    for $\nu = 2/5$      (8.13b)

as described in the literature [72-75]. These values are consistent with the theoretical result discussed in this chapter.

### 8.3 Shape and size effect for the Hall resistance

#### 8.3.1 Case (C-1): Small sized device

When the size of the device is extremely small, the number of states in the robe of the wave function becomes small. For example, when $\ell = 200$ nm

$$\frac{\Delta y}{\Delta \alpha} = \frac{\ell}{2\pi}\sqrt{\frac{eB}{\hbar}} \approx 3 \qquad \text{for } B \approx 6\,[\text{T}] \text{ and } \ell = 200\,[\text{nm}] \qquad (8.14)$$

Therein the number of effective Landau orbitals is three in the left side and three in the right side. At $\nu = 2/3$ the number of the allowed transitions is 1/3 times that of Landau orbitals and therefore the number of effective quantum transitions is only two. When the device size and the magnetic field strength are more and more small, the plateau of the Hall resistance disappears in the FQHE. By contrast the IQHE is observed in a device with such a small size because the gap energy is produced in the single electron system.

#### 8.3.2 Case (C-2): Shape effect for the Hall resistance

Usually quantum Hall devices have a rectangular shape. Therein all the Landau wave functions have the same length along the x-direction (direction of the electric current). In this sub-section, we examine how the shape of a device affects the FQHE. Three types of quantum Hall device are illustrated in Fig.8.11a, b and c. Therein the boundary conditions are different from each others in Figs.8.11a, b and c.

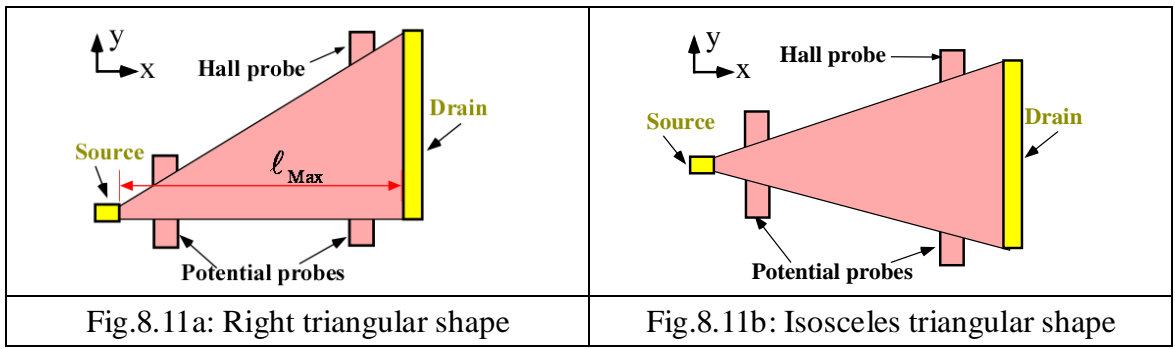

| Fig.8.11a: Right triangular shape | Fig.8.11b: Isosceles triangular shape |

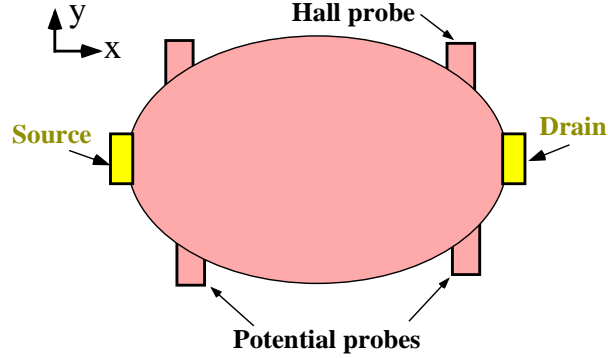

Fig.8.11c: Elliptical circular shape

The length $\ell$ of the wave function along the x-direction is dependent upon the y-position for the devices with the shapes as in Fig.8.11a, b and c:

$$\ell = \ell(y) \tag{8.15}$$

The momentum $p$ of the Landau state is related to the length $\ell$ as given by Eq.(1.5b).

$$p_J = [2\pi\hbar/\ell(y)] \times J$$

So the difference between adjacent two momenta is given by

$$p_{J+1} - p_J \approx [2\pi\hbar/\ell(y)] \tag{8.16}$$

The momentum-interval $p_{J+1} - p_J$ takes different values at the different y-positions.

Thus the shape of a quantum Hall device affects the momentum conservation law. So the plateaus in the FQHE may be smeared or disappear for quantum Hall devices with the shapes as Figs.8.11a, b, and c. It is interesting to examine how the Hall resistance curve depends upon the shape and the size of a quantum Hall device.

**Chapter 9 Spin polarization in the fractional quantum Hall states**

V. Kukushkin, K. von Klitzing, and K. Eberl [77] have measured the electron spin polarization of the quantum Hall states. The polarization curves are shown in Figs.9.1a-d. Their experimental results reveal very important properties of the FQH sates. The six filling factors 2/3, 4/7, 2/5, 4/9, 8/5 and 4/3 have numerators with even integers respectively. Therein the spin polarization is nearly equal to zero up to each critical fields, above which it begins to increase as shown in Fig.9.1. On the other hand, the six filling factors 1/2, 1/4, 3/5, 3/7, 7/5 and 3/2 have the numerators 1, 1, 3, 3, 7 and 3 respectively which are odd integers. At these filling factors, the polarization curve depends linearly upon the magnetic field for a small magnetic field region. Thus the spin polarization curves are qualitatively dependent on whether the numerator of the filling factor is even or odd. This feature is in contrast to the electron-hole symmetry seen in the Hall resistance. For example a plateau in the Hall resistance appears at both $\nu=2/5$ and $\nu=3/5$. The binding energies at $\nu=2/5$ and $\nu=3/5$ have the electron-hole symmetry. However the spin polarization curve at $\nu=2/5$ is quite different from that at $\nu=3/5$. Similar difference appears between $\nu=4/7$ and $\nu=3/7$. Furthermore the $\nu=3/5$ polarization curve behaves like the $\nu=3/7$ polarization curve where the numerator of the filling factor is the same as each other.

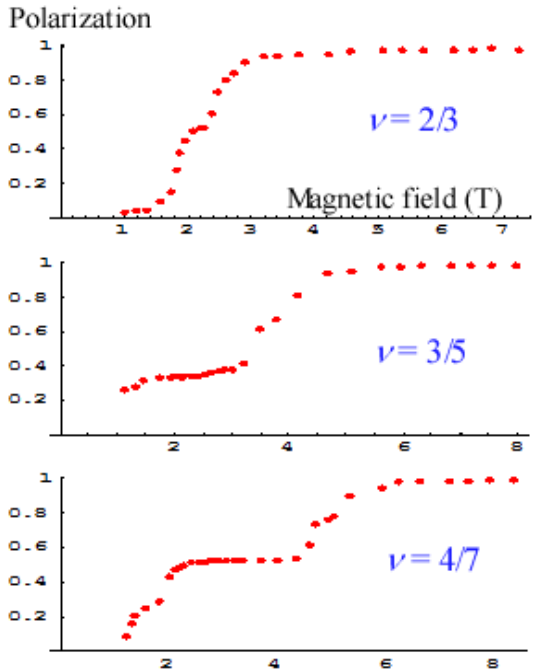
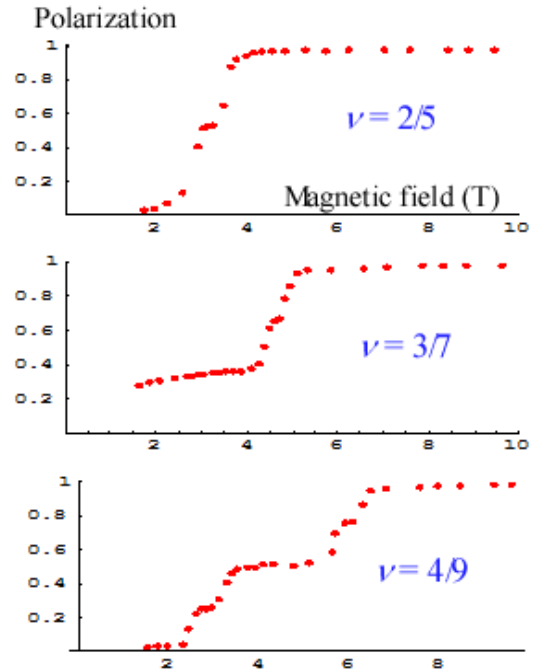

Fig.9.1a       Fig9.1b

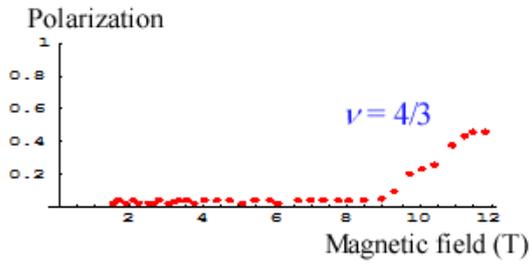
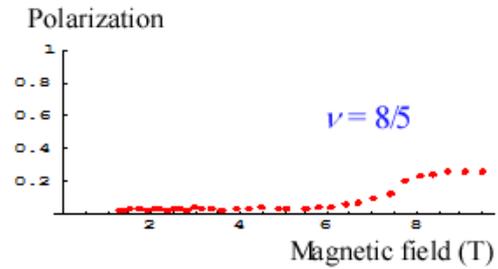
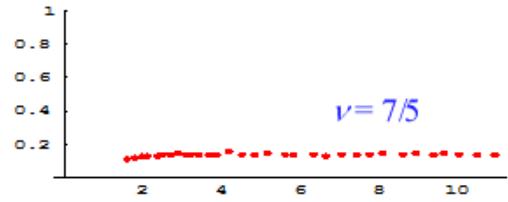
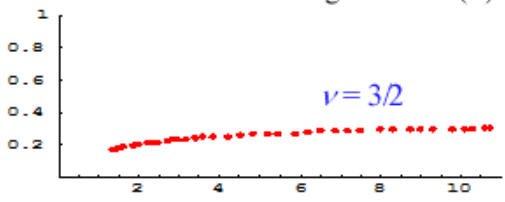
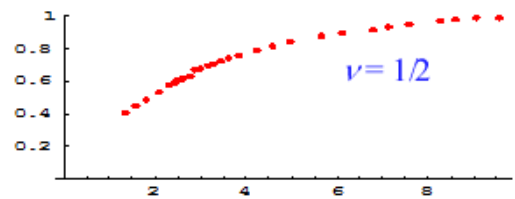
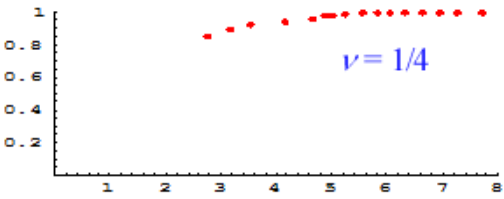

Fig.9.1c                                    Fig.9.1d

Fig.9.1 Magnetic field dependence of spin polarizations at 12 filling factors

The numerator means the electron number per unit-configuration. The fact indicates that the spin polarization belongs to only electrons (not holes).

Hereafter we express the electron spin polarization by the symbol $\gamma_e$. Then, $\gamma_e = 1$ means a fully polarized state. The polarization curve measured in the experiment has a wide plateau at $\gamma_e = 1/2$ for $\nu=4/7$ and $\nu=4/9$, at $\gamma_e = 1/3$ for $\nu=3/5$ and $\nu=3/7$ and so on. The small shoulders appear at $\gamma_e = 1/2$ for $\nu=2/3$ and $\nu=2/5$, at $\gamma_e = 2/3$ for $\nu=3/5$ and $\nu=3/7$, at $\gamma_e = 3/4$ and $\gamma_e = 1/4$ for $\nu=4/7$ and $\nu=4/9$, and at $\gamma_e = 1/4$ for $\nu=4/3$. In this chapter, we explain these interesting behaviors of the spin polarization by solving the eigen-energy problems of the Coulomb interactions between the first and second nearest electron pairs with up and down spins [78-80]. The solutions are obtained in this chapter and the theoretical results are in good agreement with the experimental data.

In the previous chapters 2-8 we have investigated a high magnetic field region where electron spins are fully polarized. In this chapter we study the case where some electrons have up-spin and the others have down-spin. There are many possible spin-arrangements for a given electron-configuration in the Landau orbitals. When we take the electron configuration to be the most uniform one, all the spin-arrangements in the configuration have the same minimum eigen-energy of $H_\mathrm{D}$. Therefore we cannot apply a usual perturbation method of non-degenerate case to solve the problem.

The interaction Hamiltonian $H_\mathrm{I}$ yields quantum transitions between the ground states with different spin arrangements. Accordingly we must exactly solve the eigenvalue problem of the Coulomb transitions among the degenerate ground states. These Coulomb transitions are equivalent to the spin exchange transition as will be verified below. We succeed to solve the eigen-value problem of the first and second nearest spin exchange interactions. The spin polarization is determined by using the exact solutions [78-80]. The theoretical results of the spin polarization are in good agreement with the experimental data. We examine in details the experimental results of Kukushkin et al which exhibit wide plateaus and small shoulders in the polarization curves. The small shoulders are studied in Secs.9.7 and 9.8 [81, 82].

## 9.1 Coulomb interaction between up and down spin states

The degenerate ground-states have the same momentum set corresponding to the most uniform electron-configuration and have various spin arrangements different from each others. We study the matrix elements of $H_\mathrm{I}$ among the ground states of $H_\mathrm{D}$.

The interaction Hamiltonian $H_\mathrm{I}$ given by Eq.(4.3) acts between two electrons where the initial states are described by the momentum pair $p_1$, $p_2$. We indicate the spin states by ↑ and ↓ for up and down spins, respectively. Then all the initial spin-states are described as

$$|p_1\uparrow,\ p_2\uparrow\rangle,\ |p_1\uparrow,\ p_2\downarrow\rangle,\ |p_1\downarrow,\ p_2\uparrow\rangle,\ |p_1\downarrow,\ p_2\downarrow\rangle \qquad (9.1)$$

When these states transfer via $H_\mathrm{I}$, their final states are described as follows:

$$|p_1'\uparrow,\ p_2'\uparrow\rangle,\ |p_1'\uparrow,\ p_2'\downarrow\rangle,\ |p_1'\downarrow,\ p_2'\uparrow\rangle,\ |p_1'\downarrow,\ p_2'\downarrow\rangle \qquad (9.2)$$

where $p_1'$ and $p_2'$ indicate the final momenta via the Coulomb interaction. We consider only the transitions between the degenerate ground states so that the final momentum set should have the minimum energy of $H_\mathrm{D}$. Accordingly the final momentum set is equivalent to the initial momentum set, and then

$$p_1' = p_2,\ p_2' = p_1 \qquad (9.3)$$

where the case of $(p_1' = p_1,\ p_2' = p_2)$ is removed because the diagonal matrix elements of $H_I$ are zero. We apply Eq.(9.3) to the final state $|p_1' \uparrow, p_2' \uparrow\rangle$. Then the final state becomes $|p_2 \uparrow, p_1 \uparrow\rangle$ which is the same as the initial state. Also $|p_1' \downarrow, p_2' \downarrow\rangle$ becomes $|p_2 \downarrow, p_1 \downarrow\rangle$. These two cases give the final state identical to the initial state and so the matrix elements of $H_I$ are zero as defined in chapter 4. Accordingly non-zero matrix elements are

$$\langle p_2 \uparrow, p_1 \downarrow | H_I | p_1 \uparrow, p_2 \downarrow \rangle \tag{9.4a}$$

$$\langle p_2 \downarrow, p_1 \uparrow | H_I | p_1 \downarrow, p_2 \uparrow \rangle \tag{9.4b}$$

We will calculate these two matrix elements as in Eqs.(9.5-14). But these calculations are slightly long. If the reader is not interested in the calculation, she/he may skip Eqs.(9.5-14). The matrix element (9.4a) is obtained by using Eq.(4.3a) as follows:

$$\langle p_2 \uparrow, p_1 \downarrow | H_I | p_1 \uparrow, p_2 \downarrow \rangle =$$

$$\iiint\iiint \psi_{p_2}^*(x_1,y_1,z_1) \psi_{p_1}^*(x_2,y_2,z_2) \frac{e^2}{4\pi\varepsilon\sqrt{(x_1-x_2)^2+(y_1-y_2)^2+(z_1-z_2)^2}} \times$$

$$\psi_{p_1}(x_1,y_1,z_1) \psi_{p_2}(x_2,y_2,z_2) dx_1 dy_1 dz_1 dx_2 dy_2 dz_2$$

$$= \left(\frac{u}{\sqrt{\ell}}\right)^4 \iiint\iiint \exp(-(ip_2 x_1 + ip_1 x_2)/\hbar) \exp\left(-\frac{eB}{2\hbar}\left(y_1 - \frac{p_2}{eB}\right)^2\right) \exp\left(-\frac{eB}{2\hbar}\left(y_2 - \frac{p_1}{eB}\right)^2\right) \times$$

$$\phi^*(z_1)\phi^*(z_2) \frac{e^2}{4\pi\varepsilon\sqrt{(x_1-x_2)^2+(y_1-y_2)^2+(z_1-z_2)^2}} \times \exp((ip_1 x_1 + ip_2 x_2)/\hbar)$$

$$\exp\left(-\frac{eB}{2\hbar}\left(y_1 - \frac{p_1}{eB}\right)^2\right) \exp\left(-\frac{eB}{2\hbar}\left(y_2 - \frac{p_2}{eB}\right)^2\right) \phi(z_1)\phi(z_2) dx_1 dy_1 dz_1 dx_2 dy_2 dz_2$$

$$\tag{9.5}$$

Therein we have used Eq.(1.17). Another matrix element $\langle p_2 \downarrow, p_1 \uparrow | H_I | p_1 \downarrow, p_2 \uparrow \rangle$ is the same as in Eq.(9.5).

$$\langle p_2 \downarrow, p_1 \uparrow | H_I | p_1 \downarrow, p_2 \uparrow \rangle = \langle p_2 \uparrow, p_1 \downarrow | H_I | p_1 \uparrow, p_2 \downarrow \rangle \tag{9.6}$$

The wave function $\phi(z)$ may be different from device to device because the potential shape in the z-direction is sample dependent. Then the wave function may be approximated by a Gaussian form as,

$$\phi(z) = f e^{-\beta z^2} \tag{9.7}$$

where $f$ and $\beta$ are constant parameters depending on the sample. Substituting Eq.(9.7) into Eq.(9.5), we obtain the following relation:

$$\langle p_2 \uparrow, p_1 \downarrow | H_1 | p_1 \uparrow, p_2 \downarrow \rangle = (uf/\sqrt{\ell})^4 \iiint \iiint \exp(i(p_1 - p_2)(x_1 - x_2)/\hbar) \times$$

$$\exp\left(-\frac{eB}{2\hbar}\left(y_1 - \frac{p_2}{eB}\right)^2\right) \exp\left(-\frac{eB}{2\hbar}\left(y_2 - \frac{p_1}{eB}\right)^2\right) \frac{e^2}{4\pi\varepsilon\sqrt{(x_1-x_2)^2 + (y_1-y_2)^2 + (z_1-z_2)^2}} \times$$

$$\exp\left(-\frac{eB}{2\hbar}\left(y_1 - \frac{p_1}{eB}\right)^2\right) \exp\left(-\frac{eB}{2\hbar}\left(y_2 - \frac{p_2}{eB}\right)^2\right) e^{-2\beta z_1^2} e^{-2\beta z_2^2} \, dx_1 dy_1 dz_1 dx_2 dy_2 dz_2$$

$$= (uf/\sqrt{\ell})^4 \iiint \iiint \exp(i(p_1 - p_2)(x_1 - x_2)/\hbar) \exp\left(-\frac{eB}{2\hbar}\left(y_1 - \frac{p_2}{eB}\right)^2\right) \times$$

$$\exp\left(-\frac{eB}{2\hbar}\left(y_2 - \frac{p_1}{eB}\right)^2\right) \frac{e^2}{4\pi\varepsilon\sqrt{(x_1-x_2)^2 + (y_1-y_2)^2 + (z_1-z_2)^2}} e^{-\beta(z_1-z_2)^2} e^{-\beta(z_1+z_2)^2} \tag{9.8}$$

$$\exp\left(-\frac{eB}{2\hbar}\left(y_1 - \frac{p_1}{eB}\right)^2\right) \exp\left(-\frac{eB}{2\hbar}\left(y_2 - \frac{p_2}{eB}\right)^2\right) dx_1 dy_1 dz_1 dx_2 dy_2 dz_2$$

$$\exp\left(-\frac{eB}{2\hbar}\left(y_1 - \frac{p_2}{eB}\right)^2\right) \exp\left(-\frac{eB}{2\hbar}\left(y_2 - \frac{p_1}{eB}\right)^2\right) \exp\left(-\frac{eB}{2\hbar}\left(y_1 - \frac{p_1}{eB}\right)^2\right) \exp\left(-\frac{eB}{2\hbar}\left(y_2 - \frac{p_2}{eB}\right)^2\right)$$

$$= \exp\left(-\frac{eB}{2\hbar}\left[\left(y_1 - \frac{p_2}{eB}\right)^2 + \left(y_1 - \frac{p_1}{eB}\right)^2 + \left(y_2 - \frac{p_1}{eB}\right)^2 + \left(y_2 - \frac{p_2}{eB}\right)^2\right]\right)$$

$$= \exp\left(-\frac{eB}{2\hbar}\left[\left(y_1 + y_2 - \frac{p_1 + p_2}{eB}\right)^2 + \left(\frac{p_1 - p_2}{eB}\right)^2 + (y_1 - y_2)^2\right]\right)$$

(9.9)

We rewrite Eq.(9.8) by using Eq.(9.9) as follows:

$$\langle p_2 \uparrow, p_1 \downarrow | H_1 | p_1 \uparrow, p_2 \downarrow \rangle = (uf/\sqrt{\ell})^4 \iiint \iiint \exp(i(p_1 - p_2)(x_1 - x_2)/\hbar) \times$$

$$\frac{e^2}{4\pi\varepsilon\sqrt{(x_1-x_2)^2 + (y_1-y_2)^2 + (z_1-z_2)^2}} e^{-\beta(z_1-z_2)^2} e^{-\beta(z_1+z_2)^2} \exp\left(-\frac{eB}{2\hbar}(y_1 - y_2)^2\right)$$

$$\exp\left(-\frac{eB}{2\hbar}\left(\left(y_1 + y_2 - \frac{p_1}{eB} - \frac{p_2}{eB}\right)^2 + \left(\frac{p_1}{eB} - \frac{p_2}{eB}\right)^2\right)\right) dx_1 dy_1 dz_1 dx_2 dy_2 dz_2$$

(9.10)

The constant $f$ is determined from the normalization condition as

$$\int_{-\infty}^{\infty} \varphi^*(z)\varphi(z)dz = \int_{-\infty}^{\infty} fe^{-\beta z^2} fe^{-\beta z^2} dz = \int_{-\infty}^{\infty} f^2 e^{-2\beta z^2} dz = f^2 \sqrt{\pi/(2\beta)} = 1$$

$$f^2 = \sqrt{2\beta/\pi} \qquad (9.11)$$

Also the constant $u$ is determined as

$$u^2 = \sqrt{\frac{eB}{\pi\hbar}} \qquad (9.12)$$

which is derived from

$$u^2 \int_{-\infty}^{\infty} \exp\left(-\frac{eB}{2\hbar}\left(y_1 - \frac{p_1}{eB}\right)^2\right) \exp\left(-\frac{eB}{2\hbar}\left(y_1 - \frac{p_1}{eB}\right)^2\right) dy_1 =$$

$$= u^2 \int_{-\infty}^{\infty} \exp\left(-\frac{eB}{\hbar}\left(y_1 - \frac{p_1}{eB}\right)^2\right) dy_1 = u^2 \sqrt{\frac{\pi\hbar}{eB}} = 1$$

Then, we have,

$$\langle p_2 \uparrow, p_1 \downarrow | H_1 | p_1 \uparrow, p_2 \downarrow \rangle = \left(uf/\sqrt{\ell}\right)^4 \iiint\iiint \exp(i(p_1 - p_2)(x_1 - x_2)/\hbar) \times$$

$$\frac{e^2}{4\pi\varepsilon\sqrt{(x_1 - x_2)^2 + (y_1 - y_2)^2 + (z_1 - z_2)^2}} e^{-\beta(z_1 - z_2)^2} e^{-\beta(z_1 + z_2)^2} \exp\left(-\frac{eB}{2\hbar}(y_1 - y_2)^2\right) \times$$

$$\exp\left(-\frac{eB}{2\hbar}\left(\left(y_1 + y_2 - \frac{p_1}{eB} - \frac{p_2}{eB}\right)^2 + \left(\frac{p_1}{eB} - \frac{p_2}{eB}\right)^2\right)\right) dx_1 dx_2 \frac{1}{2}d(y_1 - y_2)d(y_1 + y_2)\frac{1}{2}d(z_1 - z_2)d(z_1 + z_2)$$

The variables for the integration are transformed as,

$$y_1 = \{(y_1 + y_2) + (y_1 - y_2)\}/2, \; y_2 = \{(y_1 + y_2) - (y_1 - y_2)\}/2$$
$$z_1 = \{(z_1 + z_2) + (z_1 - z_2)\}/2, \text{ and } z_2 = \{(z_1 + z_2) - (z_1 - z_2)\}/2.$$

The integration variables are changed as follows:

$$\iint dy_1 dy_2 = \frac{1}{2}\iint d(y_1 - y_2)d(y_1 + y_2) \text{ and } \iint dz_1 dz_2 = \frac{1}{2}\iint d(z_1 - z_2)d(z_1 + z_2).$$

Integrating the function by $y_1 + y_2$ and $z_1 + z_2$ we obtain the following equation;

$$\langle p_2 \uparrow, p_1 \downarrow | H_1 | p_1 \uparrow, p_2 \downarrow \rangle = \left(uf/\sqrt{\ell}\right)^4 \int \iiint \exp(i(p_1 - p_2)(x_1 - x_2)/\hbar) \times$$

$$\frac{e^2}{4\pi\varepsilon\sqrt{(x_1 - x_2)^2 + (y_1 - y_2)^2 + (z_1 - z_2)^2}} e^{-\beta(z_1 - z_2)^2} \exp\left(-\frac{eB}{2\hbar}(y_1 - y_2)^2\right) \times$$

$$\frac{\sqrt{\pi}}{\sqrt{\beta}} \frac{\sqrt{\pi 2\hbar}}{\sqrt{eB}} \exp\left(-\frac{eB}{2\hbar}\left(\frac{p_1}{eB} - \frac{p_2}{eB}\right)^2\right) dx_1 dx_2 \frac{1}{2}d(y_1 - y_2)\frac{1}{2}d(z_1 - z_2)$$

$$\langle p_2 \uparrow, p_1 \downarrow | H_1 | p_1 \uparrow, p_2 \downarrow \rangle = (uf/\ell)^2 \int \iiint \exp(i(p_1 - p_2)(x_1 - x_2)/\hbar) \times$$

$$\frac{e^2}{4\pi\varepsilon\sqrt{(x_1 - x_2)^2 + (y_1 - y_2)^2 + (z_1 - z_2)^2}} e^{-\beta(z_1 - z_2)^2} \exp\left(-\frac{eB}{2\hbar}(y_1 - y_2)^2\right) \times$$

$$\frac{1}{2}\exp\left(-\frac{eB}{2\hbar}\left(\frac{p_1}{eB} - \frac{p_2}{eB}\right)^2\right) dx_1 dx_2 d(y_1 - y_2) d(z_1 - z_2)$$

(9.13)

Therein we have used Eqs.(9.11) and (9.12). We replace the two variables $y_1 - y_2$ and $z_1 - z_2$ by $Y$ and $Z$, respectively as

$$Y = \sqrt{(eB/2\hbar)}(y_1 - y_2), \quad Z = \sqrt{\beta}(z_1 - z_2) \tag{9.14}$$

Then the matrix element of Eq.(9.13) is re-expressed as follows;

$$\langle p_2 \uparrow, p_1 \downarrow | H_1 | p_1 \uparrow, p_2 \downarrow \rangle = (uf/\ell)^2 \int \iiint \exp(i(p_1 - p_2)(x_1 - x_2)/\hbar) \times$$

$$\frac{e^2}{4\pi\varepsilon\sqrt{(x_1 - x_2)^2 + (2\hbar/eB)Y^2 + (1/\beta)Z^2}} e^{-Z^2} \exp(-Y^2) \times$$

$$\frac{1}{2}\exp\left(-\frac{eB}{2\hbar}\left(\frac{p_1}{eB} - \frac{p_2}{eB}\right)^2\right) dx_1 dx_2 \sqrt{\frac{2\hbar}{eB}} dY \frac{1}{\sqrt{\beta}} dZ$$

$$\langle p_2 \uparrow, p_1 \downarrow | H_1 | p_1 \uparrow, p_2 \downarrow \rangle = (1/\ell)^2 \int \iiint \exp(i(p_1 - p_2)(x_1 - x_2)/\hbar) \times$$

$$\frac{e^2}{4\pi\varepsilon\sqrt{(x_1 - x_2)^2 + (2\hbar/eB)Y^2 + (1/\beta)Z^2}} e^{-Z^2} \exp(-Y^2) \times \tag{9.15}$$

$$\frac{1}{2}\exp\left(-\frac{eB}{2\hbar}\left(\frac{p_1 - p_2}{eB}\right)^2\right) dx_1 dx_2 \sqrt{\frac{2}{\pi}} dY \sqrt{\frac{2}{\pi}} dZ$$

Thus the momentum dependence of the matrix element has been expressed by the integration (9.15). We examine the matrix element among the degenerate ground states for the three momentum differences $\pm 2\pi\hbar/\ell$, $\pm 4\pi\hbar/\ell$, and $\pm 6\pi\hbar/\ell$, which is denoted by $\xi$, $\eta$, and $\varsigma$, respectively:

Case A: $p_1 - p_2 = \pm 2\pi\hbar/\ell$

$$\xi = \langle p_2 \uparrow, p_1 \downarrow | H_1 | p_1 \uparrow, p_2 \downarrow \rangle = (1/\ell)^2 \int \iiint \exp(\pm i2\pi(x_1 - x_2)/\ell) \times$$

$$\frac{e^2}{4\pi\varepsilon\sqrt{(x_1 - x_2)^2 + (2\hbar/eB)Y^2 + (1/\beta)Z^2}} e^{-Z^2} \exp(-Y^2) \times \tag{9.16a}$$

$$\frac{1}{2}\exp\left(-\frac{eB}{2\hbar}\left(\frac{2\pi\hbar}{eB\ell}\right)^2\right) dx_1 dx_2 \sqrt{\frac{2}{\pi}} dY \sqrt{\frac{2}{\pi}} dZ \quad \text{for } p_1 - p_2 = \pm 2\pi\hbar/\ell$$

In order to distinguish the Case B from Case A, momentum suffixes 1 and 2 are replaced with 3 and 4 as follows:

Case B : $p_3 - p_4 = \pm 4\pi\hbar/\ell$

$$\eta = \langle p_4 \uparrow, p_3 \downarrow | H_I | p_3 \uparrow, p_4 \downarrow \rangle = (1/\ell)^2 \int \int\int\int \exp(\pm i4\pi(x_3 - x_4)/\ell) \times$$

$$\frac{e^2}{4\pi\varepsilon\sqrt{(x_3 - x_4)^2 + (2\hbar/eB)Y^2 + (1/\beta)Z^2}} e^{-Z^2} \exp(-Y^2) \times \qquad (9.16b)$$

$$\frac{1}{2}\exp\left(-\frac{eB}{2\hbar}\left(\frac{4\pi\hbar}{eB\ell}\right)^2\right) dx_3 dx_4 \sqrt{\frac{2}{\pi}} dY \sqrt{\frac{2}{\pi}} dZ \quad \text{for } p_3 - p_4 = \pm 4\pi\hbar/\ell$$

In the case C, the two momentum suffixes are expressed by 5 and 6 as:

Case C : $p_5 - p_6 = \pm 6\pi\hbar/\ell$

$$\varsigma = \langle p_6 \uparrow, p_5 \downarrow | H_I | p_5 \uparrow, p_6 \downarrow \rangle = (1/\ell)^2 \int \int\int\int \exp(\pm i6\pi(x_5 - x_6)/\ell) \times$$

$$\frac{e^2}{4\pi\varepsilon\sqrt{(x_5 - x_6)^2 + (2\hbar/eB)Y^2 + (1/\beta)Z^2}} e^{-Z^2} \exp(-Y^2) \times \qquad (9.16c)$$

$$\frac{1}{2}\exp\left(-\frac{eB}{2\hbar}\left(\frac{6\pi\hbar}{eB\ell}\right)^2\right) dx_5 dx_6 \sqrt{\frac{2}{\pi}} dY \sqrt{\frac{2}{\pi}} dZ \quad \text{for } p_5 - p_6 = \pm 6\pi\hbar/\ell$$

The Coulomb transition in the case A is illustrated in Fig. 9.2a. The transition is shown by arrow pairs in the left panel. The open circle indicates the up spin state and the filled one the down spin state. The momenta after the transition are expressed by the symbols $p_1'$ and $p_2'$ which are given by

$$p_1' = p_1 + 2\pi\hbar/\ell = p_2$$
$$p_2' = p_2 - 2\pi\hbar/\ell = p_1$$

It is noteworthy that the final spin is unchanged via the Coulomb transition. Only the final momenta are changed.

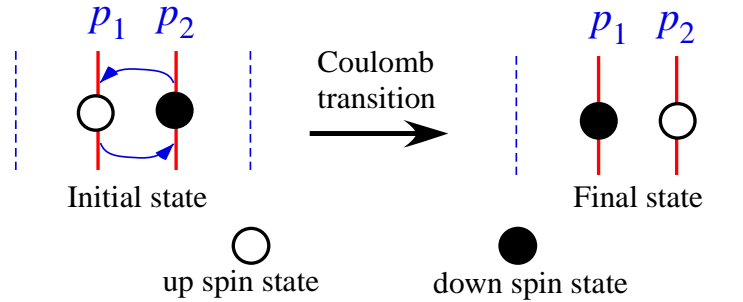

Fig.9.2a: Equivalence of specific Coulomb transition and spin exchange interaction for case A The final state is illustrated in the right panel of Fig. 9.2a. This Coulomb transition is equivalent to the following process: The spin at site 1 flips from up to down and the spin at site 2 flips from down to up simultaneously without changing the momenta.

That is to say the Coulomb transition of the case A is equivalent to a spin exchange process which is described by the interaction $\xi\sigma_1^-\sigma_2^+$. Therein $\sigma^+$ is the spin transformation operator from down to up-spin state and $\sigma^-$ is the adjoint operator of $\sigma^+$. There is another Coulomb transition given by Eq.(9.4b) which is equivalent to $\xi\sigma_1^+\sigma_2^-$ where the coupling constant takes the same value as in $\xi\sigma_1^-\sigma_2^+$ because of Eq.(9.6). Accordingly the Coulomb transition between the two electrons at sites 1 and 2 is expressed as

$$\xi\left(\sigma_1^-\sigma_2^+ + \sigma_1^+\sigma_2^-\right) \tag{9.17}$$

where $\xi$ was already defined by Eq. (9.16a). In this Coulomb transition, the classical Coulomb energy of the initial state is exactly equal to that of the final state.

Next we examine the Coulomb transition of the case B which is shown by Fig.9.2b.

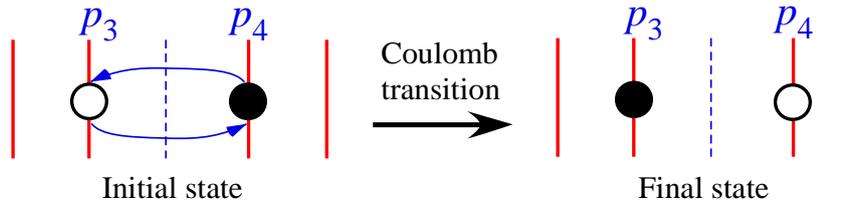

Initial state        Final state

Fig.9.2b: Equivalence of specific Coulomb transition and spin exchange interaction for case B

The momenta after the transition are described by the symbols $p_3', p_4'$, the values of which are given by

$$p_3' = p_3 + 2 \times 2\pi\hbar/\ell = p_4$$

$$p_4' = p_4 - 2 \times 2\pi\hbar/\ell = p_3$$

The Coulomb interaction of this case B is equivalent to the following spin exchange interaction between electrons placed in the second nearest-neighboring orbital pair:

$$\eta\left(\sigma_3^{-}\sigma_4^{+} + \sigma_3^{+}\sigma_4^{-}\right) \tag{9.18}$$

where $\eta$ is the coupling constant defined by Eq.(9.16b).

The Coulomb transition of the case C is illustrated in Fig.9.2c. The momenta after the transition are described by the symbols $p_5'$, $p_6'$:

$$p_5' = p_5 + 3 \times 2\pi\hbar/\ell = p_6$$

$$p_6' = p_6 - 3 \times 2\pi\hbar/\ell = p_5$$

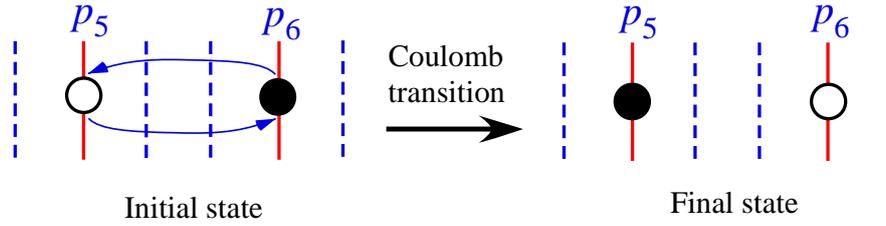

Fig.9.2c: Equivalence of specific Coulomb transition and spin exchange interaction for case C

Fig.9.2c shows that the Coulomb interaction of the case C is equivalent to the following spin exchange interaction between electrons placed in the sites 5 and 6:

$$\varsigma\left(\sigma_5^{-}\sigma_6^{+} + \sigma_5^{+}\sigma_6^{-}\right) \tag{9.19}$$

These types of interactions give the partial Hamiltonian of $H_\mathrm{I}$ in the quasi-2D electron system.

Figures.9.3a and 9.3b express the most uniform configurations of electrons at $\nu = 2/3$ and $\nu = 2/5$, respectively. Therein the spin-states are numbered sequentially from the left to the right as indicated in each figure. In the $\nu = 2/3$ state the nearest and second nearest electron pairs have the coupling constants $\xi$ and $\eta$, respectively. At $\nu = 2/5$, the nearest electron pair is placed in the second neighboring orbitals. The coupling constant is $\eta$. The second nearest electron pair is placed in third neighboring orbitals and so the coupling constant is $\varsigma$ as in Fig.9.3b. The spin-site number is written

sequentially from the left to the right in the lowest part of each figure. The spin-site number is different from the orbital number.

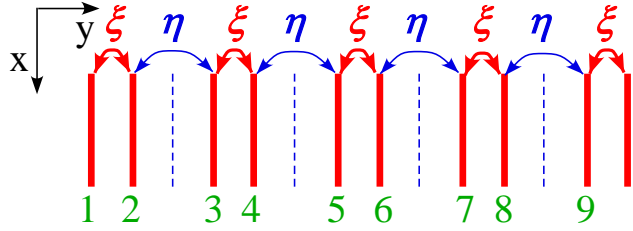

Fig.9.3a Coulomb transitions for first and second nearest electron pairs at $\nu=2/3$

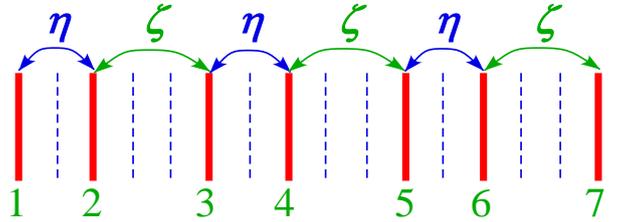

Fig.9.3b Coulomb transitions for first and second nearest electron pairs at $\nu=2/5$

Note here the word "nearest electron pair". In Chapter 3-8, we have used "nearest electron pair" only for $1/2<\nu<1$, and so the nearest electron pair is placed in the nearest Landau orbital pair. At $0<\nu<1/2$ we have examined nearest hole pairs (not electron pair). In this chapter we need to consider the nearest electron pair at $0<\nu<1/2$. For example the nearest electron pair at $\nu=2/5$ is placed in the second nearest neighboring orbital pair as in Fig.9.3b.

 We consider the interaction between more distant pairs. For example the interaction between electron 1 and electron 3 becomes weak due to the screening effect of the interposing electron 2 as in Figs. 9.3a and 9.3b. Thus the interaction between the third nearest electrons is weak and so may be ignored. Accordingly the most effective interaction at $\nu=2/3$ is obtained as follows;

$$H_{\text{effecive}} = \sum_{j=1,2,3\cdots}\left[\xi\left(\sigma^{+}_{2j-1}\sigma^{-}_{2j}+\sigma^{-}_{2j-1}\sigma^{+}_{2j}\right)+\eta\left(\sigma^{+}_{2j}\sigma^{-}_{2j+1}+\sigma^{-}_{2j}\sigma^{+}_{2j+1}\right)\right] \quad (9.20)$$

where the operator $\sigma_{2j-1}^{+}$ indicates the transformation from down to up-spin state of the electron at the $(2j-1)$-th site. This Hamiltonian, Eq.(9.20) yields the quantum transition between the degenerate ground states. Because the external magnetic field yields Zeeman energy in the z-direction, the Hamiltonian becomes

$$H = \sum_{j=1,2,3\cdots}\left[\xi\left(\sigma^+_{2j-1}\sigma^-_{2j}+\sigma^-_{2j-1}\sigma^+_{2j}\right)+\eta\left(\sigma^+_{2j}\sigma^-_{2j+1}+\sigma^-_{2j}\sigma^+_{2j+1}\right)\right] + \sum_{i=1,2,3\cdots}\mu_B g^* B(1/2)\sigma^z_i \quad (9.21)$$

where $g^*$ is the effective g-factor, $B$ is the magnetic field, $(1/2)\sigma^z$ is the electron spin operator in the $z$-direction and $\mu_B$ is the Bohr magneton [83, *84]. The matrices (the Pauli spin matrices) are explicitly indicated below:

$$\sigma^+ = \begin{pmatrix} 0 & 1 \\ 0 & 0 \end{pmatrix}, \quad \sigma^- = \begin{pmatrix} 0 & 0 \\ 1 & 0 \end{pmatrix}, \quad \sigma^z = \begin{pmatrix} 1 & 0 \\ 0 & -1 \end{pmatrix} \quad (9.22)$$

We can obtain the Hamiltonian $H$ for other filling factors. The Hamiltonians for $\nu = 3/5$ and $\nu = 4/7$ are given, respectively by:

$$H = \sum_{j=1,2,3\cdots}\begin{bmatrix} \xi\left(\sigma^+_{3j-2}\sigma^-_{3j-1}+\sigma^-_{3j-2}\sigma^+_{3j-1}\right)+\eta\left(\sigma^+_{3j-1}\sigma^-_{3j}+\sigma^-_{3j-1}\sigma^+_{3j}\right) \\ +\eta\left(\sigma^+_{3j}\sigma^-_{3j+1}+\sigma^-_{3j}\sigma^+_{3j+1}\right) \end{bmatrix}$$
$$+ \sum_{i=1,2,3\cdots}\mu_B g^* B(1/2)\sigma^z_i \qquad \text{for } \nu = 3/5 \quad (9.23)$$

$$H = \sum_{j=1,2,3\cdots}\begin{bmatrix} \xi\left(\sigma^+_{4j-3}\sigma^-_{4j-2}+\sigma^-_{4j-3}\sigma^+_{4j-2}\right)+\eta\left(\sigma^+_{4j-2}\sigma^-_{4j-1}+\sigma^-_{4j-2}\sigma^+_{4j-1}\right) \\ +\eta\left(\sigma^+_{4j-1}\sigma^-_{4j}+\sigma^-_{4j-1}\sigma^+_{4j}\right)+\eta\left(\sigma^+_{4j}\sigma^-_{4j+1}+\sigma^-_{4j}\sigma^+_{4j+1}\right) \end{bmatrix}$$
$$+ \sum_{i=1,2,3\cdots}\mu_B g^* B(1/2)\sigma^z_i \qquad \text{for } \nu = 4/7 \quad (9.24)$$

We have obtained the most effective Hamiltonians in Eqs.(9.21), (9.23) and (9.24) for $\nu = 2/3$, $\nu = 3/5$ and $\nu = 4/7$, respectively. They produce the quantum transitions among the degenerate ground states of $H_D$. These Hamiltonians can be exactly diagonalized by using the method of reference [85].

### 9.2 Isomorphic mapping from FQH states to one-dimensional fermion states

We have obtained the most effective Hamiltonian namely Eqs.(9.21), (9.23) and (9.24) for the three filling factors. The Hamiltonians yield the quantum transitions among the ground states of $H_D$. In this section we find the isomorphic mapping from the degenerate ground states to many-fermion states in a one dimensional system. Thereby the eigen-energy problems can be solved exactly.

We first examine the following mapping from a single spin state to a fermion state. The down-spin state $|\downarrow\rangle$ is mapped to the vacuum state $|0\rangle$, and the up-spin state $|\uparrow\rangle$ is mapped to a state with one fermion $c^*|0\rangle$ where $c^*$ is the creation operator.

$$|\uparrow\rangle \to c^*|0\rangle, \quad |\downarrow\rangle \to |0\rangle \tag{9.25}$$

Then the spin operators $\sigma^+$, $\sigma^-$, and $\sigma^z$ are mapped to the operators of the fermion system as follows:

$$\sigma^+ \to c^*, \quad \sigma^- \to c, \quad \sigma^z \to (2c^*c - 1) \tag{9.26}$$

When the three spin-operators $\sigma^+$, $\sigma^-$, and $\sigma^z$ are operated to the two spin states, the following six results are obtained:

$$\sigma^+|\uparrow\rangle = 0, \qquad \sigma^+|\downarrow\rangle = |\uparrow\rangle$$

$$\sigma^-|\uparrow\rangle = |\downarrow\rangle, \qquad \sigma^-|\downarrow\rangle = 0$$

$$\sigma^z|\uparrow\rangle = |\uparrow\rangle, \qquad \sigma^z|\downarrow\rangle = -|\downarrow\rangle$$

When the three operators $c^*$, $c$, $(2c^*c - 1)$ are multiplied to the two fermion states $c^*|0\rangle$ and $|0\rangle$ the following results are obtained:

$$c^*c^*|0\rangle = 0, \qquad c^*|0\rangle = c^*|0\rangle$$

$$cc^*|0\rangle = |0\rangle, \qquad c|0\rangle = 0$$

$$(2c^*c - 1)c^*|0\rangle = c^*|0\rangle, \quad (2c^*c - 1)|0\rangle = -|0\rangle$$

where we have used the property of the fermion operator $c^*c^* = 0$. The obtained six equations are corresponding to the six relations in spin system correctly as:

$$\sigma^+|\uparrow\rangle = 0 \leftrightarrow c^*c^*|0\rangle = 0, \quad \sigma^+|\downarrow\rangle = |\uparrow\rangle \leftrightarrow c^*|0\rangle = c^*|0\rangle$$

$$\sigma^-|\uparrow\rangle = |\downarrow\rangle \leftrightarrow cc*|0\rangle = |0\rangle, \quad \sigma^-|\downarrow\rangle = 0 \leftrightarrow c|0\rangle = 0$$

$$\sigma^z|\uparrow\rangle = |\uparrow\rangle \leftrightarrow (2c*c-1)c*|0\rangle = c*|0\rangle, \quad \sigma^z|\downarrow\rangle = -|\downarrow\rangle \leftrightarrow (2c*c-1)|0\rangle = -|0\rangle$$

Thus we have obtained the isomorphic mapping from the single spin states to the fermion states.

Next we consider the isomorphic mapping from many-spin states to many-fermion states. The up-spin at the site $j$ is mapped to the creation operator of the site $j$. The multiplying order of the creation operators is the same as the order of the up-spins. Then all the many-spin states are mapped to all the many-fermion states by one to one correspondence. We write two examples of the mappings as follows:

$$|\uparrow_1, \uparrow_2, \downarrow_3, \uparrow_4, \downarrow_5, \downarrow_6, \uparrow_7, \uparrow_8, \downarrow_9, \uparrow_{10}\rangle \leftrightarrow c_1^* c_2^* c_4^* c_7^* c_8^* c_{10}^* |0\rangle$$
$$|\downarrow_1, \uparrow_2, \uparrow_3, \downarrow_4, \downarrow_5, \uparrow_6, \downarrow_7, \downarrow_8, \downarrow_9, \uparrow_{10}\rangle \leftrightarrow c_2^* c_3^* c_6^* c_{10}^* |0\rangle$$

(9.27)

The operators $c_i$ and $c_i^*$ satisfy the anti-commutation relations as follows:

$$\{c_i, c_j^*\} = c_i \times c_j^* + c_j^* \times c_i = \delta_{i,j} \tag{9.28a}$$

$$\{c_i, c_j\} = c_i \times c_j + c_j \times c_i = 0, \qquad \{c_i^*, c_j^*\} = c_i^* \times c_j^* + c_j^* \times c_i^* = 0 \tag{9.28b}$$

The mapping mentioned above yields the following correspondence from the products of the spin operators $\sigma_{2j-1}^+ \sigma_{2j}^-$, $\sigma_{2j-1}^- \sigma_{2j}^+$, $\sigma_{2j}^+ \sigma_{2j+1}^-$ and $\sigma_{2j}^- \sigma_{2j+1}^+$ to the products of the fermion operators as:

$$\sigma_{2j-1}^+ \sigma_{2j}^- \leftrightarrow c_{2j-1}^* c_{2j}, \quad \sigma_{2j-1}^- \sigma_{2j}^+ \leftrightarrow -c_{2j-1} c_{2j}^* \tag{9.29a}$$

$$\sigma_{2j}^+ \sigma_{2j+1}^- \leftrightarrow c_{2j}^* c_{2j+1}, \quad \sigma_{2j}^- \sigma_{2j+1}^+ \leftrightarrow -c_{2j} c_{2j+1}^* \tag{9.29b}$$

It will be proven below that the mapping of (9.29a,b) is the isomorphic mapping from many-spin states to many-fermion states. The proof is done via the following three steps: namely Property 1, Property 2 and Property 3.

(Property 1)

The operator $c_{2j-1}^* c_{2j}$ is commutable with $c_i$ and $c_i^*$ for $i \neq 2j-1$ and $i \neq 2j$.

Also the operator $-c_{2j-1} c_{2j}^*$ is commutable with $c_i$ and $c_i^*$ for $i \neq 2j-1$ and $i \neq 2j$. Both operators $c_{2j}^* c_{2j+1}$ and $-c_{2j} c_{2j+1}^*$ are commutable with $c_i$ and $c_i^*$ for $i \neq 2j$, $i \neq 2j+1$.

$$\text{Example 1:} \quad (c_3^* c_4) \times c_1^* c_2^* c_4^* c_7^* c_8^* c_{10}^* = c_1^* c_2^* (c_3^* c_4) c_4^* c_7^* c_8^* c_{10}^* \quad (9.30a)$$

$$\text{Example 2:} \quad (-c_4 c_5^*) \times c_1^* c_2^* c_4^* c_7^* c_8^* c_{10}^* = c_1^* c_2^* \times (-c_4 c_5^*) \times c_4^* c_7^* c_8^* c_{10}^* \quad (9.30b)$$

Thus the products of the creation and annihilation operators in (9.29a) can be moved in front of the site $2j-1$. Also the products in (9.29b) can be moved in front of the site $2j$. Accordingly it is sufficient to prove the following equations:

(Property 2)

$$(\sigma_{2j-1}^+ \sigma_{2j}^- + \sigma_{2j-1}^- \sigma_{2j}^+)|\uparrow_{2j-1} \uparrow_{2j}\rangle = 0 \leftrightarrow (c_{2j-1}^* c_{2j} - c_{2j-1} c_{2j}^*) c_{2j-1}^* c_{2j}^* |0\rangle = 0 \quad (9.31a)$$

$$(\sigma_{2j-1}^+ \sigma_{2j}^- + \sigma_{2j-1}^- \sigma_{2j}^+)|\downarrow_{2j-1} \downarrow_{2j}\rangle = 0 \leftrightarrow (c_{2j-1}^* c_{2j} - c_{2j-1} c_{2j}^*)|0\rangle = 0 \quad (9.31b)$$

$$(\sigma_{2j-1}^+ \sigma_{2j}^- + \sigma_{2j-1}^- \sigma_{2j}^+)|\downarrow_{2j-1} \uparrow_{2j}\rangle = |\uparrow_{2j-1} \downarrow_{2j}\rangle \leftrightarrow (c_{2j-1}^* c_{2j} - c_{2j-1} c_{2j}^*) c_{2j}^* |0\rangle = c_{2j-1}^* |0\rangle$$

$$(9.31c)$$

$$(\sigma_{2j-1}^+ \sigma_{2j}^- + \sigma_{2j-1}^- \sigma_{2j}^+)|\uparrow_{2j-1} \downarrow_{2j}\rangle = |\downarrow_{2j-1} \uparrow_{2j}\rangle \leftrightarrow (c_{2j-1}^* c_{2j} - c_{2j-1} c_{2j}^*) c_{2j-1}^* |0\rangle = c_{2j}^* |0\rangle$$

$$(9.31d)$$

$$(\sigma_{2j}^+ \sigma_{2j+1}^- + \sigma_{2j}^- \sigma_{2j+1}^+)|\uparrow_{2j} \uparrow_{2j+1}\rangle = 0 \leftrightarrow (c_{2j}^* c_{2j+1} - c_{2j} c_{2j+1}^*) c_{2j}^* c_{2j+1}^* |0\rangle = 0 \quad (9.31e)$$

$$(\sigma_{2j}^+ \sigma_{2j+1}^- + \sigma_{2j}^- \sigma_{2j+1}^+)|\downarrow_{2j} \downarrow_{2j+1}\rangle = 0 \leftrightarrow (c_{2j}^* c_{2j+1} - c_{2j} c_{2j+1}^*)|0\rangle = 0 \quad (9.31f)$$

$$(\sigma_{2j}^+ \sigma_{2j+1}^- + \sigma_{2j}^- \sigma_{2j+1}^+)|\downarrow_{2j} \uparrow_{2j+1}\rangle = |\uparrow_{2j} \downarrow_{2j+1}\rangle \leftrightarrow (c_{2j}^* c_{2j+1} - c_{2j} c_{2j+1}^*) c_{2j+1}^* |0\rangle = c_{2j}^* |0\rangle$$

$$(9.31g)$$

$$(\sigma_{2j}^+ \sigma_{2j+1}^- + \sigma_{2j}^- \sigma_{2j+1}^+)|\uparrow_{2j} \downarrow_{2j+1}\rangle = |\downarrow_{2j} \uparrow_{2j+1}\rangle \leftrightarrow (c_{2j}^* c_{2j+1} - c_{2j} c_{2j+1}^*) c_{2j}^* |0\rangle = c_{2j+1}^* |0\rangle$$

$$(9.31h)$$

Thus it is verified that the mapping of Eqs.(9.29a) and (9.29b) is isomorphic due to Eqs.(9.31a, b, c, d, e, f, g and h).

Next the spin operator $\sigma_i^z$ of the z-direction is mapped as follows:

$$\sigma_i^z \leftrightarrow 2c_i^* c_i - 1 \tag{9.32}$$

This mapping is correct as follows:

(Property 3)  $\quad \sigma_i^z |\uparrow_i\rangle = |\uparrow_i\rangle \quad \leftrightarrow \quad (2c_i^* c_i - 1)c_i^*|0\rangle = c_i^*|0\rangle \tag{9.33a}$

$$\sigma_i^z |\downarrow_i\rangle = -|\downarrow_i\rangle \quad \leftrightarrow \quad (2c_i^* c_i - 1)|0\rangle = -|0\rangle \tag{9.33b}$$

The properties Eqs.(9.30-33) clarify that the mappings Eqs.(9.29a, b) and (9.32) are isomorphic. Accordingly Hamiltonian Eq.(9.21) is equivalent to the following form:

$$H = \sum_{j=1,2,3\cdots} \left[ \xi\left(c_{2j-1}^* c_{2j} - c_{2j-1} c_{2j}^*\right) + \eta\left(c_{2j}^* c_{2j+1} - c_{2j} c_{2j+1}^*\right) \right] + \sum_{i=1,2,3\cdots} \mu_B g^* B(1/2)\left(2c_i^* c_i - 1\right) \tag{9.34}$$

We will exactly solve the eigen-value problem of this Hamiltonian in the next section.

## 9.3 Diagonalization of the most effective Hamiltonian at filling factor of 2/3

It has been proven that the Hamiltonian Eq.(9.21) is equivalent to Eq.(9.34). We will solve exactly the eigen-value problem of the Hamiltonian Eq.(9.34). As discussed in Chapter 3, there are two kinds of sites in the unit cell at $\nu = 2/3$. So we introduce new operators $a_j$ and $b_j$ defined as follows:

$$a_j = c_{2j-1}, \quad b_j = c_{2j}, \quad a_j^* = c_{2j-1}^*, \text{ and } b_j^* = c_{2j}^* \tag{9.35}$$

where $j$ is the cell number (unit-configuration number). Then, the Hamiltonian Eq.(9.34) becomes

$$H = \sum_{j=1}^{J} \left[ \xi\left(a_j^* b_j - a_j b_j^*\right) + \eta\left(b_j^* a_{j+1} - b_j a_{j+1}^*\right) \right] + \sum_{j=1}^{J} \mu_B g^* B(1/2)\left(2a_j^* a_j + 2b_j^* b_j - 2\right), \tag{9.36}$$

where $J$ is the total number of cells given by $J = N/2$ ($N$ is the total number of electrons). We apply a Fourier transformation for the operators $a_j$, $a_j^*$, $b_j$, and $b_j^*$, and obtain

$$H = \sum_p \left[ \xi\left(a^*(p)b(p) + b^*(p)a(p)\right) + \eta\left(e^{ip} b^*(p)a(p) + e^{-ip} a^*(p)b(p)\right) \right]$$
$$+ \sum_p \mu_B g^* B(1/2)\left(2a^*(p)a(p) + 2b^*(p)b(p) - 2\right) \quad , \tag{9.37}$$

where $a_n = \frac{1}{\sqrt{J}} \sum_p e^{ipn} a(p)$, $b_n = \frac{1}{\sqrt{J}} \sum_p e^{ipn} b(p)$ (9.38)

and $p = \frac{2\pi}{J} \times \text{integer}$, $-\pi < p \leq \pi$.

Hamiltonian (9.37) is the sum of the terms with the argument $p$. We consider the single term with a given $p$. The term is expressed by the following matrix:

$$\begin{pmatrix} \mu_B g^* B & \xi + \eta e^{-ip} \\ \xi + \eta e^{ip} & \mu_B g^* B \end{pmatrix}. \quad (9.39)$$

This matrix has two eigen-values: $\lambda_1(p)$ and $\lambda_2(p)$ as

$$\lambda_1(p) = \mu_B g^* B - \sqrt{\xi^2 + \eta^2 + 2\xi\eta \cos p} \quad (9.40a)$$

$$\lambda_2(p) = \mu_B g^* B + \sqrt{\xi^2 + \eta^2 + 2\xi\eta \cos p} \quad (9.40b)$$

We introduce new annihilation operators $A_1(p)$ and $A_2(p)$ given by

$$A_1(p) = \frac{1}{\sqrt{2}} a(p) - \frac{\xi + \eta e^{-ip}}{\sqrt{2(\xi^2 + \eta^2 + 2\xi\eta \cos p)}} b(p), \quad (9.41a)$$

$$A_2(p) = \frac{1}{\sqrt{2}} a(p) + \frac{\xi + \eta e^{-ip}}{\sqrt{2(\xi^2 + \eta^2 + 2\xi\eta \cos p)}} b(p). \quad (9.41b)$$

Of course the new operators satisfy the anti-commutation relations as follows;

$$\{A_i(p), A_j^*(q)\} = \delta_{i,j} \delta_{p,q} \quad (9.41c)$$

$$\{A_i(p), A_j(q)\} = 0, \quad \{A_i^*(p), A_j^*(q)\} = 0 \quad (9.41d)$$

Then the Hamiltonian Eq.(9.37) is rewritten to a diagonal form as

$$H = \sum_p \left( \lambda_1(p) A_1^*(p) A_1(p) + \lambda_2(p) A_2^*(p) A_2(p) - \mu_B g^* B \right). \quad (9.42)$$

Thus we have succeeded in diagonalizing of the Hamiltonian (9.21).

## 9.4 Magnetic field dependence of the Spin-Polarization at filling factor of 2/3

The diagonal form, Eq.(9.42), means that all the eigen-states are expressed by the direct product of the creation operators $A_s^*(p)$ as follows:

$$\text{Eigen-state} = A_{s1}^*(p_1) \cdot A_{s2}^*(p_2) \cdot A_{s3}^*(p_3) \cdots A_{sN}^*(p_N)|0\rangle$$

Thus the eigen-state is specified by the set $\{n_1(p), n_2(p); -\pi < p \leq \pi\}$ where $n_s(p) = A_s^*(p)A_s(p)$. The electron-spin polarization $\gamma_e$ is obtained by calculating the thermo-dynamic mean value numerically:

$$\begin{aligned}\gamma_e &= \frac{1}{N}\left\langle -\sum_{i=1}^{N}\sigma_i^z \right\rangle = \frac{-1}{N}\left\langle \sum_{i=1}^{N}(2c_i^*c_i - 1) \right\rangle = \frac{-1}{N}\left\langle \sum_{j=1}^{J}(2a_j^*a_j + 2b_j^*b_j - 2) \right\rangle \\ &= \frac{-1}{2J}\left\langle \sum_p (2a^*(p)a(p) + 2b^*(p)b(p) - 2) \right\rangle = \frac{-1}{2J}\left\langle \sum_p \left( \sum_{s=1}^{2}(2A_s^*(p)A_s(p) - 1) \right) \right\rangle\end{aligned} \quad (9.43)$$

where $\langle \cdots \rangle$ is the thermal average and the minus sign comes from the negative charge of electron. Also we have employed Eqs.(9.32), (9.35), (9.38) and (9.41).

Equation (9.42) means that the eigen-energy for $n_s(p) = 1$ is $\lambda_s(p)$ and the eigen-energy for $n_s(p) = 0$ is zero. Then the Boltzmann factor is $\exp(-\lambda_s(p)/k_B T)$ for $n_s(p) = 1$ and $\exp(-0/k_B T)$ for $n_s(p) = 0$ where $k_B$ is the Boltzmann constant and $T$ is the temperature. Accordingly, the probability for $n_s(p) = 1$ is given by $\frac{\exp(-\lambda_s(p)/k_B T)}{1+\exp(-\lambda_s(p)/k_B T)}$ and the probability for $n_s(p) = 0$ is given by $\frac{1}{1+\exp(-\lambda_s(p)/k_B T)}$. These probabilities yield the thermal average of $A_s^*(p)A_s(p)$ as

$$\langle A_s^*(p)A_s(p) \rangle = \frac{\exp(-\lambda_s(p)/k_B T)}{1+\exp(-\lambda_s(p)/k_B T)}, \quad (9.44)$$

which gives

$$\langle 2A_s^*(p)A_s(p) - 1 \rangle = \frac{\exp(-\lambda_s(p)/k_B T) - 1}{1+\exp(-\lambda_s(p)/k_B T)} = -\tanh(\lambda_s(p)/(2k_B T)). \quad (9.45)$$

Substitution of Eq.(9.45) into Eq.(9.43) yields

$$\gamma_e = \frac{1}{2J}\sum_p \left( \sum_{s=1}^{2} \tanh(\lambda_s(p)/2k_B T) \right). \quad (9.46)$$

Since the total number of electrons is a macroscopic value, we can replace the summation by an integration:

$$\gamma_e = \frac{1}{4\pi} \int_{-\pi}^{\pi} dp \left( \sum_{s=1}^{2} \tanh(\lambda_s(p)/2k_B T) \right). \tag{9.47}$$

Thus the electron-spin polarization at the filling factor of 2/3 has been expressed by Eq. (9.47).

First, we study the low field behavior of the spin polarization. Equation (9.40) means that $\lambda_1(p)$ exists in the following region:

$$\mu_B g^* B - |\xi + \eta| \leq \lambda_1(p) \leq \mu_B g^* B - |\xi - \eta|$$

Also $\lambda_2(p)$ is in the region:

$$\mu_B g^* B + |\xi - \eta| \leq \lambda_2(p) \leq \mu_B g^* B + |\xi + \eta|$$

When the magnetic field takes a value between 0 and $|\xi - \eta|/(\mu_B g^*)$, $\lambda_1(p)$ is negative and $\lambda_2(p)$ is positive for any value of $p$:

$$\lambda_1(p) < 0, \quad \lambda_2(p) > 0 \quad \text{for} \quad 0 < B < |\xi - \eta|/(\mu_B g^*)$$

Therefore, $\tanh(\lambda_1(p)/2k_B T)$ is nearly equal to $-1$ and $\tanh(\lambda_2(p)/2k_B T)$ is nearly equal to 1 at a low temperatures ($T \approx 0$). Then, the spin-polarization is almost zero because the sum in the right hand side of Eq.(9.47) is nearly equal to zero. When the magnetic field increases beyond the value $|\xi - \eta|/(\mu_B g^*)$, the spin-polarization increases continuously until it reaches the maximum value of 1. This theoretical property at $\nu = 2/3$ is in agreement with the experimental data shown in Fig.9.4.

When the quality of a quantum Hall device is bad, the impurities and lattice defects produce many random-potentials. Then the plateau on the polarization curve is rounded by the random potentials. This effect resembles the property due to thermal vibrations. So we introduce the effective temperature $T$ which represents the sum of the random potential effect and the thermal effect.

The parameter values $\eta/\xi = 0.2$ and $(k_B T/\xi) = 0.1$ are applied to the right hand side of Eq.(9.47). Thereby the integration is numerically calculated by using a computer program. The theoretical result is shown with blue curve in Fig.9.4. Experimental data [77] are also plotted by red dots on the figure. The theoretical curve reproduces experimental data reasonably well. In the next section, the parameter dependence of the polarization is discussed.

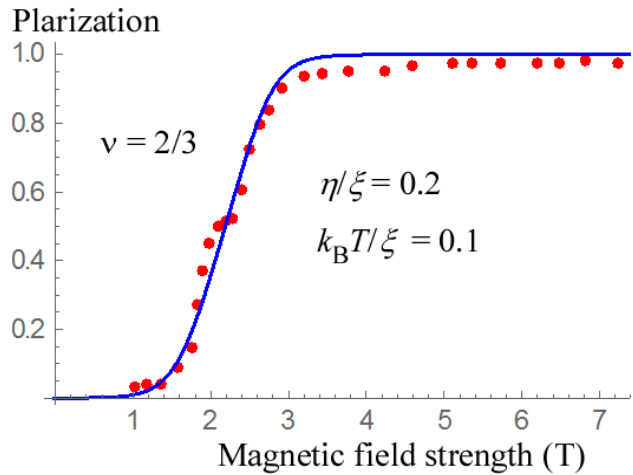

Fig.9.4 Calculated spin-polarization curve for $\eta/\xi = 0.2, (k_B T/\xi) = 0.1$ at $\nu = 2/3$.
Red dots are experimental data [77] in this figure.

## 9.5 Parameter Dependence of Polarization

We examine how the spin-polarization at $\nu = 2/3$ depends upon the values of $\eta/\xi$ and $k_B T/\xi$. The theoretical spin-polarization curve for the values $\eta/\xi = 0.4, 0.3, 0.2,$ and $0.1$ is shown in Figs. 9.5a–9.5d respectively where $k_B T/\xi$ is set to 0.1.

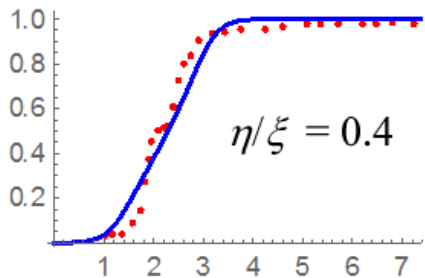
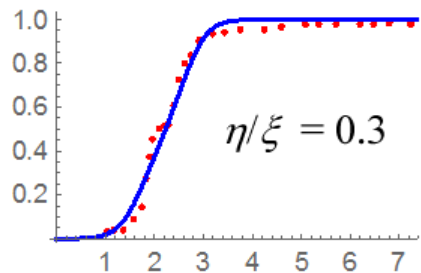

Fig.9.5a $\nu=2/3$ $\eta/\xi=0.4$ $k_B T/\xi=0.1$   Fig.9.5b $\nu=2/3$ $\eta/\xi=0.3$ $k_B T/\xi=0.1$

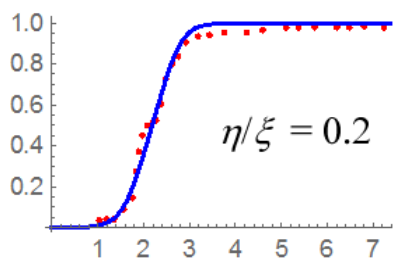
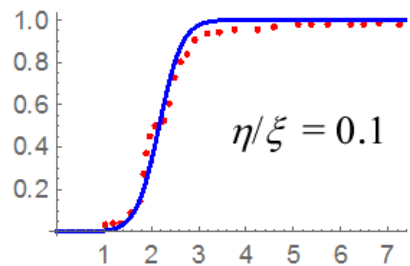

Fig.9.5c $\nu=2/3$ $\eta/\xi=0.2$   Fig.9.5d $\nu=2/3$ $\eta/\xi=0.1$ $k_B T/\xi=0.1$

Fig.9.5 Dependence of spin-polarization curves on the value of $\eta/\xi$ for $k_B T/\xi=0.1$

Figures 9.5a-d show that the dependence of the polarization on $\eta/\xi$ is weak. When the theoretical curve is examined in details, the polarization value is almost zero in $0 \leq B < B_{\text{critical}}$. The critical magnetic field $B_{\text{critical}}$ is about 0.9T, 1.05T, 1.2T and 1.4T in Figs.9.5a, b, c and d, respectively. Thus the parameter dependence appears correctly in these figures.

We next examine the dependence of the spin polarization upon the value of $k_B T/\xi$, which is shown in Figs.9.6a–9.6d for a fixed value of $\eta/\xi = 0.2$.

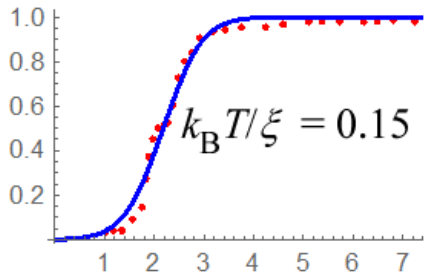
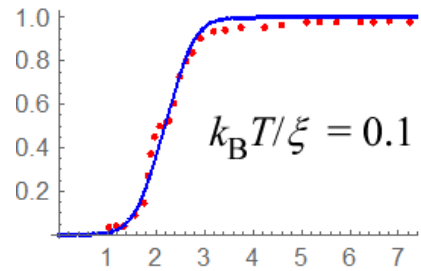

Fig.9.6a  $\nu=2/3, \eta/\xi=0.2, k_B T/\xi=0.15$    Fig.9.6b  $\nu=2/3, \eta/\xi=0.2, k_B T/\xi=0.1$

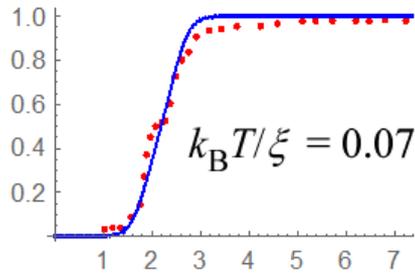
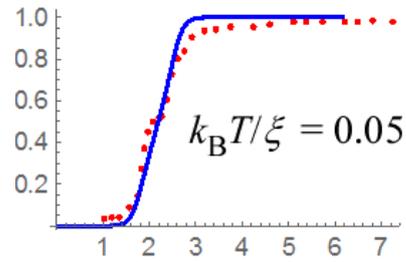

Fig.9.6c  $\nu=2/3, \eta/\xi=0.2, k_B T/\xi=0.07$    Fig.9.6d  $\nu=2/3, \eta/\xi=0.2, k_B T/\xi=0.05$

Fig.9.6 Dependence of spin-polarization curves on the value of $k_B T/\xi$

The shape of the corner in the blue curves becomes sharp for a small value of $k_B T/\xi$ and loose for a large value of $k_B T/\xi$. The parameters $\eta/\xi = 0.2$ and $k_B T/\xi = 0.1$ reproduce the experimental data well [77].

The shape of the curve probably depends on size and quality of a device used in the experiment. We have already encountered similar phenomena for the measurement of the Hall resistance.

## 9.6 Spin-Polarization for other filling factors

We calculate the spin-polarization at the twelve filling factors of 2/3, 3/5, 4/7, 2/5, 3/7, 4/9, 7/5, 8/5, 4/3, 1/4, 1/2 and 3/2 below. These twelve filling factors are classified into the following four types:

(Type 1) filling factors $\nu = \frac{2}{3}, \frac{3}{5}, \frac{4}{7}$ where $\frac{1}{2} < \nu < 1$

(Type 2) $\nu = \frac{2}{5}, \frac{3}{7}, \frac{4}{9}$ where $\frac{1}{4} < \nu < \frac{1}{2}$

(Type 3) $\nu = \frac{4}{3}, \frac{7}{5}, \frac{8}{5}$ where $\nu > 1$

(Type 4) $\nu = \frac{1}{4}, \frac{1}{2}, \frac{3}{2}$ where the denominators are even integers

We first calculate the spin-polarizations of Type 1 namely $\nu = 2/3$, 3/5 and 4/7. The polarization at $\nu = 2/3$ has already been calculated in section 9.4. The most uniform electron configuration at $\nu = 3/5$ and $\nu = 4/7$ are given in Chapter 3. Figures 9.7a and 9.7b express the nearest and second nearest interactions via the Coulomb transition.

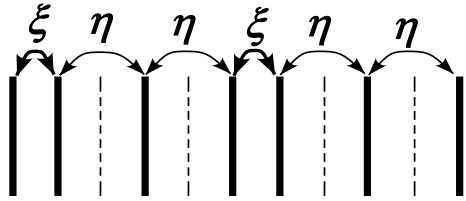

Fig.9.7a: Electron configuration for $\nu = 3/5$

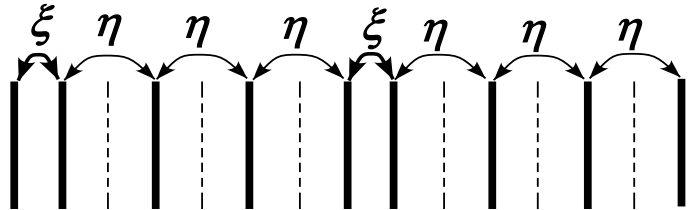

Fig.9.7b: Electron configuration for $\nu = 4/7$

At $\nu = 3/5$ the most effective Hamiltonian $H$ is given by Eq.(9.23) as

$$H = \sum_{j=1,2,3\cdots} \left[ \xi\left(\sigma^+_{3j-2}\sigma^-_{3j-1} + \sigma^-_{3j-2}\sigma^+_{3j-1}\right) + \eta\left(\sigma^+_{3j-1}\sigma^-_{3j} + \sigma^-_{3j-1}\sigma^+_{3j}\right) + \eta\left(\sigma^+_{3j}\sigma^-_{3j+1} + \sigma^-_{3j}\sigma^+_{3j+1}\right) \right]$$
$$+ \sum_{i=1,2,3\cdots} \mu_B g^* B(1/2) \sigma^z_i$$

(9.48)

We can exactly solve the eigen-value problem of the Hamiltonian Eq.(9.48) which is composed of the following matrix:

$$\begin{pmatrix} \mu_B g^* B & \xi & \eta e^{-ip} \\ \xi & \mu_B g^* B & \eta \\ \eta e^{ip} & \eta & \mu_B g^* B \end{pmatrix} \quad (9.49)$$

The eigen-values of this matrix are described by the symbols $\lambda_1(p)$, $\lambda_2(p)$, and $\lambda_3(p)$. Thereby the electron spin-polarization for $\nu = 3/5$ is given by

$$\gamma_e = \frac{1}{3J} \sum_p \left( \sum_{s=1}^{3} \tanh(\lambda_s(p)/2k_B T) \right) = \frac{1}{6\pi} \int_{-\pi}^{\pi} dp \left( \sum_{s=1}^{3} \tanh(\lambda_s(p)/2k_B T) \right). \quad (9.50)$$

The polarization is calculated by using a computer program. We set the values to $\eta/\xi = 0.3$ and $k_B T/\xi = 0.1$ same as in the case of $\nu = 2/3$. The theoretical polarization curve at $\nu = 3/5$ is obtained as in Fig.9.8b which agrees well with the experimental data [77]. It is noteworthy that the polarization for $\nu = 3/5$ is almost proportional to the magnetic field strength near zero fields. This property is different from that of $\nu = 2/3$.

The Hamiltonian for $\nu = 4/7$ has been already obtained by Eq.(9.24) as

$$H = \sum_{j=1,2,3\cdots} \begin{bmatrix} \xi(\sigma^+_{4j-3}\sigma^-_{4j-2} + \sigma^-_{4j-3}\sigma^+_{4j-2}) + \eta(\sigma^+_{4j-2}\sigma^-_{4j-1} + \sigma^-_{4j-2}\sigma^+_{4j-1}) \\ + \eta(\sigma^+_{4j-1}\sigma^-_{4j} + \sigma^-_{4j-1}\sigma^+_{4j}) + \eta(\sigma^+_{4j}\sigma^-_{4j+1} + \sigma^-_{4j}\sigma^+_{4j+1}) \end{bmatrix}$$
$$+ \sum_{i=1,2,3\cdots} \mu_B g^* B (1/2) \sigma^z_i \quad (9.51)$$

The four eigenvalues $\lambda_1(p)$, $\lambda_2(p)$, $\lambda_3(p)$, and $\lambda_4(p)$ are derived from diagonalization of the following matrix

$$\begin{pmatrix} \mu_B g^* B & \xi & 0 & \eta e^{-ip} \\ \xi & \mu_B g^* B & \eta & 0 \\ 0 & \eta & \mu_B g^* B & \eta \\ \eta e^{ip} & 0 & \eta & \mu_B g^* B \end{pmatrix} \quad (9.52)$$

The value of $\eta/\xi = 0.3$ is also applied for $\nu = 4/7$, and then the spin-polarization curve is calculated as shown in Fig.9.8c. Here we didn't use the best fitting parameters. In spite of using the same value namely $\eta/\xi = 0.3$, the three theoretical curves of the spin-polarization show good agreement with the experimental data as in Figs.9.8a–c.

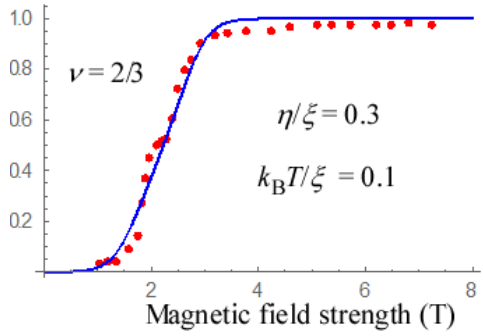
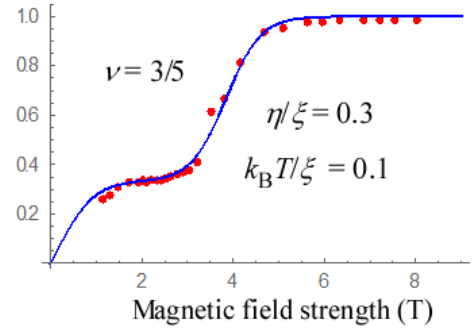

Fig.9.8a  $\nu=2/3, \eta/\xi=0.3, k_BT/\xi=0.1$    Fig.9.8b  $\nu=3/5, \eta/\xi=0.3, k_BT/\xi=0.1$

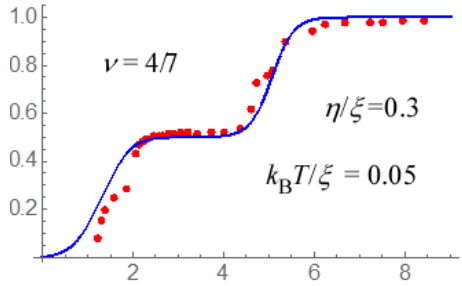
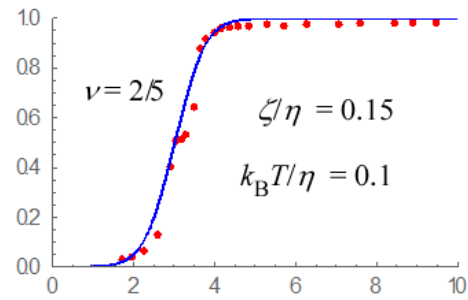

Fig.9.8c  $\nu=4/7, \eta/\xi=0.3, k_BT/\xi=0.05$    Fig.9.8d  $\nu=2/5, \zeta/\eta=0.15, k_BT/\eta=0.1$

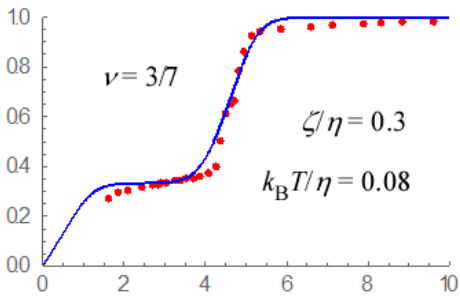
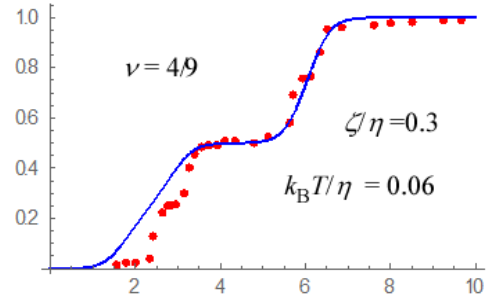

Fig.9.8e  $\nu=3/7, \zeta/\eta=0.3, k_BT/\eta=0.08$    Fig.9.8f  $\nu=4/9, \zeta/\eta=0.3, k_BT/\eta=0.06$

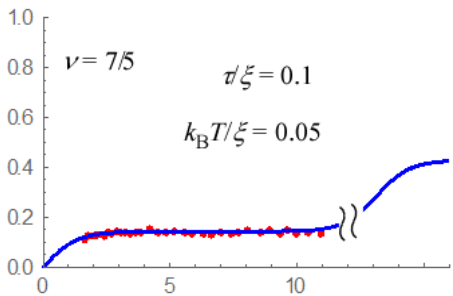
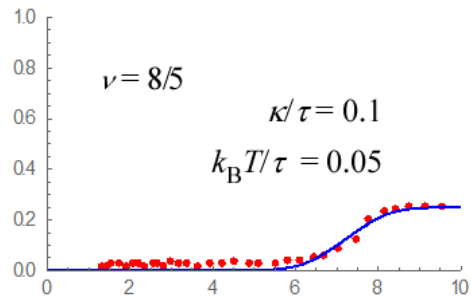

Fig.9.8g  $\nu=7/5, \tau/\xi=0.1, k_BT/\xi=0.05$    Fig.9.8h  $\nu=8/5, \kappa/\tau=0.1, k_BT/\tau=0.05$

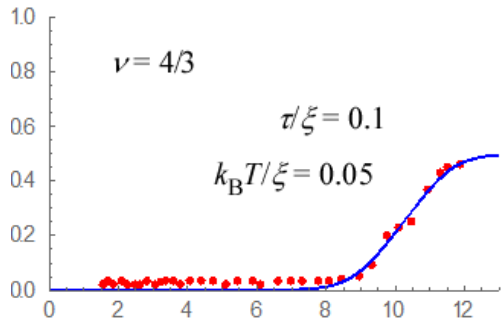
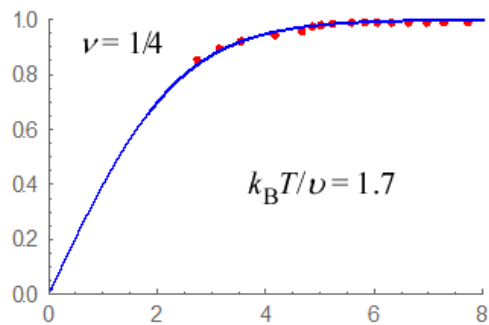

Fig.9.8i $\nu=4/3, \tau/\xi=0.1, k_BT/\xi=0.05$     Fig.9.8j $\nu=1/4, k_BT/\upsilon=1.7$

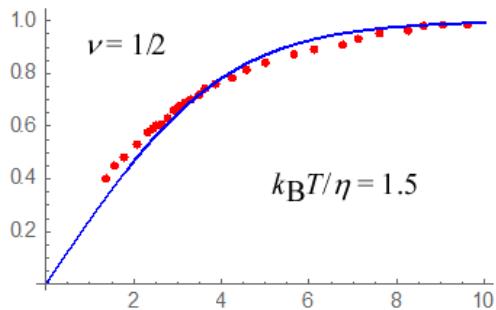
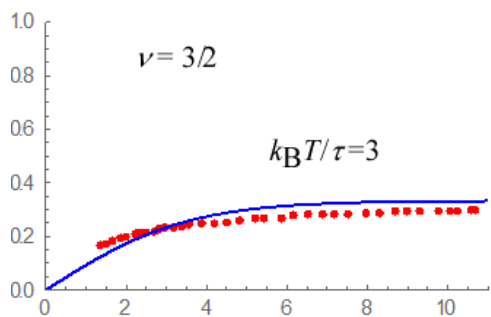

Fig.9.8k $\nu=1/2, k_BT/\eta=1.5$     Fig.9.8l $\nu=3/2, k_BT/\tau=3$

Fig.9.8a-l: Calculated spin-polarization curves (Red dots are experimental data [77])

We study the cases of Type 2, such as $\nu=2/5$, 3/7, and 4/9. The electron configuration with the minimum classical Coulomb energy is given in Chapter 3 for $\nu=2/5$, 3/7, and 4/9. Figures 9.9a–c express the most uniform configuration. The nearest electron pair has the coupling constants $\eta$ and the second nearest electron pair has the coupling constant $\zeta$. These coupling constants have already examined as in Eqs.(9.16b), (9.16c), (9.18) and (9.19).

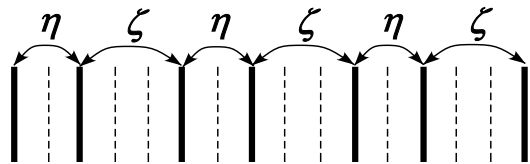

Fig.9.9a: Electron configuration for $\nu=2/5$

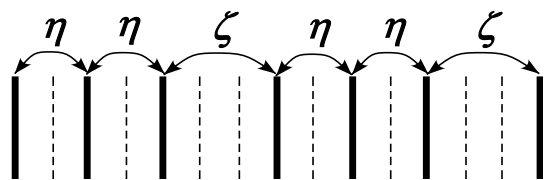

Fig.9.9b: Electron configuration for $\nu=3/7$

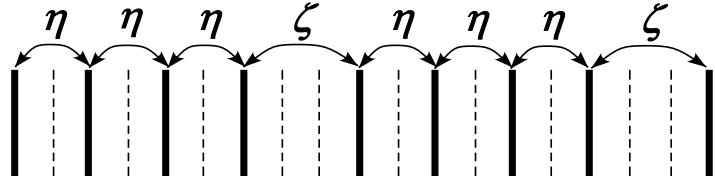
Fig.9.9c: Electron configuration for $\nu=4/9$

We calculate the spin polarization of the filling factor $\nu=4/9$. The most effective Hamiltonian is given by the following matrix form:

$$\begin{pmatrix} \mu_B g^* B & \eta & 0 & \zeta e^{-ip} \\ \eta & \mu_B g^* B & \eta & 0 \\ 0 & \eta & \mu_B g^* B & \eta \\ \zeta e^{ip} & 0 & \eta & \mu_B g^* B \end{pmatrix} \qquad (9.53)$$

This matrix has four eigen-values $\lambda_1(p)$, $\lambda_2(p)$, $\lambda_3(p)$, and $\lambda_4(p)$. The spin-polarization $\gamma_e$ is given by

$$\gamma_e = \frac{1}{4J}\sum_p \left(\sum_{s=1}^{4} \tanh(\lambda_s(p)/2k_B T)\right) = \frac{1}{8\pi}\int_{-\pi}^{\pi} dp \left(\sum_{s=1}^{4} \tanh(\lambda_s(p)/2k_B T)\right) \qquad (9.54)$$

Similarly, we get the spin-polarization for $\nu=2/5$ and $3/7$. The theoretical value of $\gamma_e$ is drawn by blue curves in Figs.9.8d, e and f for $\nu=2/5$, $3/7$, and $4/9$, respectively. These theoretical results well reproduce the experimental data.

We examine the case of Type 3 where the filling factors are larger than 1, such as $\nu=7/5$, $8/5$, and $4/3$. Figures 9.10a, 9.10b, and 9.10c show the electron-configurations with the minimum classical Coulomb energy. Therein a double-line indicate a Landau orbital occupied by two electrons, one of which has up spin and the other down spin. Single bold line indicates a Landau orbital occupied by an electron.

The spin exchange interactions act only between electrons placed in singly occupied orbitals, because spins in the doubly occupied orbitals cannot flip owing to the Pauli principle. As an example, the coupling constants of the spin exchange interactions are illustrated by $\xi$ and $\tau$ in Fig.9.10a for the filling factor of 7/5. There are three singly occupied orbitals and two doubly occupied orbitals in every unit-configuration. Therefore three electrons can flip per seven electrons inside a unit-configuration.

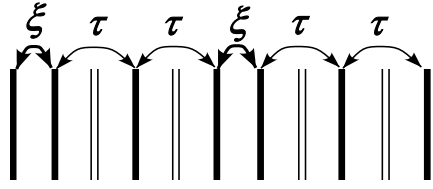

Fig. 9.10a: Electron configuration for $\nu=7/5$

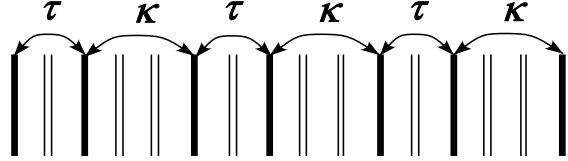

Fig. 9.10b: Electron configuration for $\nu=8/5$

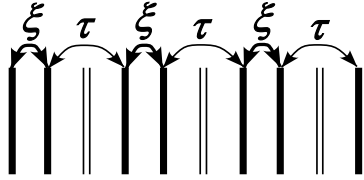

Fig. 9.10c: Electron configuration for $\nu=4/3$

The symbol $\tau$ represents the coupling constant between the electrons in two orbitals separated by a doubly occupied orbital. Then, the Hamiltonian for $\nu=7/5$ is

$$H = \sum_{j=1,2,3\cdots}\left[\xi\left(\sigma^+_{3j-2}\sigma^-_{3j-1}+\sigma^-_{3j-2}\sigma^+_{3j-1}\right)+\tau\left(\sigma^+_{3j-1}\sigma^-_{3j}+\sigma^-_{3j-1}\sigma^+_{3j}\right)+\tau\left(\sigma^+_{3j}\sigma^-_{3j+1}+\sigma^-_{3j}\sigma^+_{3j+1}\right)\right]$$
$$+\sum_{i=1,2,3\cdots}\mu_B g^* B(1/2)\sigma^z_i \qquad (9.55)$$

The eigenvalues of the Hamiltonian are obtained by diagnalizing the following matrix.

$$\begin{pmatrix} \mu_B g^* B & \xi & \tau e^{-ip} \\ \xi & \mu_B g^* B & \tau \\ \tau e^{ip} & \tau & \mu_B g^* B \end{pmatrix} \qquad (9.56)$$

The three eigen-values of the matrix Eq.(9.56) are described by $\lambda_1(p)$, $\lambda_2(p)$ and $\lambda_3(p)$. Using these eigen-values, the spin-polarization is expressed as follows:

$$\gamma_e = \left(\frac{1}{7}\right)\frac{1}{J}\sum_p\left(\sum_{s=1}^3 \tanh(\lambda_s(p)/2k_B T)\right) = \left(\frac{1}{7}\right)\frac{1}{2\pi}\int_{-\pi}^{\pi} dp \left(\sum_{s=1}^3 \tanh(\lambda_s(p)/2k_B T)\right). \qquad (9.57)$$

Here seven electrons exist in a unit-configuration. The coefficient 1/7 in the right hand side of Eq.(9.57) means that the contribution of one electron is 1/7 times full polarization for every unit-configuration.

The coupling constant $\tau$ is weakened by the screening effect of the interposing electron pairs. Accordingly the coupling constant $\tau$ becomes smaller than $\eta$. So we set $\tau/\xi = 0.1$ which is smaller than the value of $\eta/\xi = 0.3$. The theoretical curve of the polarization is drawn in Fig.9.8g.

Similarly we calculate the polarizations at $\nu$=8/5 and 4/3, the theoretical curves of which are drawn in Figs.9.8h and 9.8i, respectively. These figures show that the theoretical spin-polarization agrees well with the experimental data [77].

We next investigate the case of Type 4 with $\nu$=1/4, 1/2, and 3/2. In this case no plateau appears in the Hall resistance curve. Figures 9.11a, 9.11b, and 9.11c indicate the most uniform electron-configurations for the filling factors of 1/4, 1/2, and 3/2, respectively. These configurations show that the most effective Hamiltonian is composed of only one kind of the spin exchange interaction as in Figs.9.11a, 9.11b, 9.11c. Then the spin-polarization is described by only one kind of eigen-energy. The theoretical results of the spin-polarization are shown in Figs.9.8j–9.8l which are also in agreement with the experimental data.

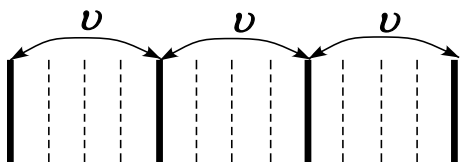

Fig.9.11a: Electron configuration for $\nu$=1/4

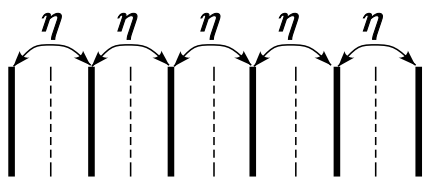

Fig.9.11b: Electron configuration for $\nu$=1/2

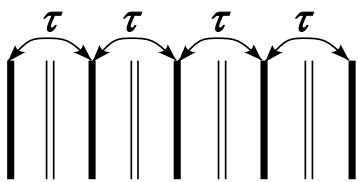

Fig.9.11c: Electron configuration for $\nu$=3/2

Thus, the present theory has satisfactorily reproduced the twelve curves of the experimental spin polarization.

There is an interesting relation between the shape of the polarization curve and the dimension of the matrices in Eqs.(9.39), (9.49), (9.52), (9.53), (9.56) and so on. The structure of energy gapes depends on the dimension of the matrices. Therefore the dimension determines the shape of the polarization curves. For $\nu < 1$ the dimension is equal to the numerator of the filling factor. So the dimension becomes four at $\nu = 4/7$ same as at $\nu = 4/9$. Thereby the shape of the polarization curve at $\nu = 4/7$ resembles that at $\nu = 4/9$. Also the shape of the polarization curve at $\nu = 3/5$ resembles that at $\nu = 3/7$ and so on. The character derived from the theory is consistent with the experimental data of polarization. Thus the spin polarization is originates from electron only, not hole. Especially the following character appears. When the numerator of $\nu$ is an odd integer, the spin polarization is almost proportional to the magnetic field strength in the neighborhood of $B = 0$. On the other hand, when the numerator of $\nu$ is an even integer, the spin polarization is almost zero up to a critical value of the magnetic field.

**9.7 Small shoulder in Polarization Curve**

If we carefully observe the experimental spin polarization curve (Fig.9.1), then we find small shoulders in it. The small shoulders don't appear in the results of the previous sections. So we need to find a new origin of the small shoulders. [81, 82]

R. E. Peierls studied an electron system in a one dimensional crystal and considered the lattice distortion with the period doubling the unit cell. The lattice distortion produces new band gaps. Thereby the total energy becomes lower than that without the distortion. This effect is called spin Peierls effect [86]. The phenomena were observed in organic compounds [87-89] and inorganic chain compounds such as TTF-CuS$_4$C$_4$(CF$_3$)$_4$, (MEM)-(TCNQ)$_2$ and CuGeO$_3$ [90].

In the present theory, the spin polarization of FQH states is derived from the Hamiltonians (9.21), (9.23), (9.24) and so on. If we consider a new distortion with the double period of the unit configuration, the spin chain Hamiltonian of FQH system resembles that of the spin-Peierls effect.

For an example $\nu = 2/3$ we change the distance between nearest orbitals in the first unit-configuration longer, the one in the second unit-configuration shorter and so on. Then we have the four kinds of the coupling constants $\xi$, $\xi'$, $\eta$ and $\eta'$ as shown in Fig.9.12. The value of $\xi'$ is larger than that of $\xi$ because the distance for the $\xi'$ interaction path is shorter than that for $\xi$. Also, $\eta' > \eta$ holds. This distortion with the double period of the unit-configuration produces additional energies. We call the distortion "interval modulation". [81, 82]

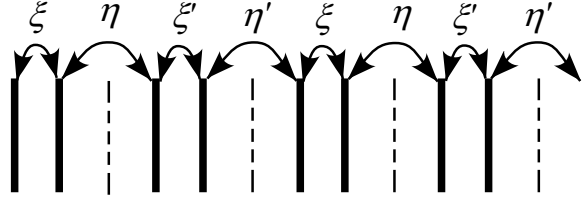

Fig.9.12 Coupling constants of interactions caused by distortion with double period

### 9.7.1 Distortion-dependence of the total energy

We express the distance between nearest orbitals by the symbol $r_0$ for non-distorted case. We consider the distortion (interval modulation) where the orbital interval becomes $r_0 + d$ for an odd number of unit-configuration and $r_0 - d$ for an even number. Thereby the classical Coulomb energy $W$ increases.

**(Simple example of the function form):** Three electrons A, B and C are placed on the straight line. When the distance between A and B is $r_0 + d$ and the distance between B and C is $r_0 - d$, then the classical Coulomb energy is given by

$$e^2/[4\pi\varepsilon_0(r_0+d)] + e^2/[4\pi\varepsilon_0(r_0-d)] \approx e^2/[4\pi\varepsilon_0 r_0] + 2(e^2/[4\pi\varepsilon_0 r_0]) \times d^2/r_0^2$$

Although this example is oversimplified, the increasing of $W$ in the real case is also proportional to $d^2$. Accordingly the increasing value per electron is given by

$$\Delta W/N = f_C \times (d/r_0)^2 \qquad (9.58)$$

where $f_C$ is the constant parameter which is dependent on a quantum Hall device.

Next we examine $d$-dependence of the coupling constants $\xi$ and $\xi'$. When $d > 0$, the coupling constant $\xi$ is weaker than $\xi'$ because the $\xi$ interaction path is longer than that of $\xi'$. When $d < 0$, $\xi$ is stronger than $\xi'$ because the $\xi$ interaction path becomes shorter than that of $\xi'$. So there is a linear term of $d$ in $\xi$ and $\xi'$ as follows:

$$\xi = \xi_0 - f_\xi \times (d/r_0), \quad \xi' = \xi_0 + f_\xi \times (d/r_0) \qquad (9.59)$$

where $\xi_0$ is the coupling constant in the non-distortion case and $f_\xi$ is the proportionality constant. In order to simplify Eqs.(9.58) and (9.59), we define a new dimensionless quantity $t$ as

$$t = (f_\xi / \xi_0)(d/r_0) \tag{9.60}$$

Then the coupling constants $\xi$ and $\xi'$ is expressed as

$$\xi = \xi_0(1-t), \quad \xi' = \xi_0(1+t) \tag{9.61}$$

The increasing value of the classical Coulomb energy $\Delta W$ is also expressed by this dimensionless quantity $t$ as follows:

$$\Delta W / N = \xi_0 \, C \, t^2 \tag{9.62a}$$

where $C$ is the dimensionless coefficient as

$$C = \xi_0 f_C / f_\xi^2 \tag{9.62b}$$

Now we calculate the spin exchange energy. The coupling constants for $\nu = 2/3$ are illustrated in Fig.9.12 and the spin exchange Hamiltonian is obtained as

$$\begin{aligned} H = & \sum_{j=1,2,3\cdots} \left[ \xi\left(\sigma^+_{4j-3}\sigma^-_{4j-2} + \sigma^-_{4j-3}\sigma^+_{4j-2}\right) + \eta\left(\sigma^+_{4j-2}\sigma^-_{4j-1} + \sigma^-_{4j-2}\sigma^+_{4j-1}\right) \right] + \\ & + \sum_{j=1,2,3\cdots} \left[ \xi'\left(\sigma^+_{4j-1}\sigma^-_{4j} + \sigma^-_{4j-1}\sigma^+_{4j}\right) + \eta'\left(\sigma^+_{4j}\sigma^-_{4j+1} + \sigma^-_{4j}\sigma^+_{4j+1}\right) \right] + \\ & + \sum_{j=1,2,3\cdots} \mu_B g^* B(1/2)\left(\sigma^z_{4j-3} + \sigma^z_{4j-2} + \sigma^z_{4j-1} + \sigma^z_{4j}\right) \end{aligned} \tag{9.63}$$

This Hamiltonian Eq.(9.63) is rewritten from Eqs.(9.29a) and (9.29b) as,

$$\begin{aligned} H = & \sum_{j=1,2,3\cdots} \left[ \xi\left(c^*_{4j-3}c_{4j-2} - c_{4j-3}c^*_{4j-2}\right) + \eta\left(c^*_{4j-2}c_{4j-1} - c_{4j-2}c^*_{4j-1}\right) \right] + \\ & + \sum_{j=1,2,3\cdots} \left[ \xi'\left(c^*_{4j-1}c_{4j} - c_{4j-1}c^*_{4j}\right) + \eta'\left(c^*_{4j}c_{4j+1} - c_{4j}c^*_{4j+1}\right) \right] + \sum_{i=1,2,3\cdots} \mu_B g^* B(1/2)\left(2c^*_i c_i - 1\right) \end{aligned} \tag{9.64}$$

Using the cell number $j$, we introduce new operators $a_{1,j}, a_{2,j}, a_{3,j}, a_{4,j}$ as follows:

$$a_{1,j} = c_{4j-3}, \quad a_{2,j} = c_{4j-2}, \quad a_{3,j} = c_{4j-1}, \quad a_{4,j} = c_{4j} \tag{9.65}$$

Fourier transformation yields

$$\begin{aligned} a_{1,j} &= \frac{1}{\sqrt{J}} \sum_p e^{ipj} a_1(p), \quad a_{2,j} = \frac{1}{\sqrt{J}} \sum_p e^{ipj} a_2(p), \\ a_{3,j} &= \frac{1}{\sqrt{J}} \sum_p e^{ipj} a_3(p), \quad a_{4,j} = \frac{1}{\sqrt{J}} \sum_p e^{ipj} a_4(p) \end{aligned} \tag{9.66}$$

where $J$ is the total number of unit-cells (unit-configurations) and $p = (2\pi/J) \times \text{integer}$ $(-\pi < p \leq \pi)$. Substitution of Eqs.(9.65) and (9.66) into (9.64) gives

$$H = \sum_p \begin{bmatrix} \xi(a_1^*(p)a_2(p) + a_2^*(p)a_1(p)) + \eta(a_2^*(p)a_3(p) + a_3^*(p)a_2(p)) + \\ + \xi'(a_3^*(p)a_4(p) + a_4^*(p)a_3(p)) + \eta'(e^{ip}a_4^*(p)a_1(p) + e^{-ip}a_1^*(p)a_4(p)) \end{bmatrix}$$
$$+ \sum_p \mu_B g^* B(1/2)(2(a_1^*(p)a_1(p) + a_2^*(p)a_2(p) + a_3^*(p)a_3(p) + a_4^*(p)a_4(p)) - 4) \quad (9.67)$$

For a given value of $p$, the term in Eq.(9.67) is expressed by the following matrix $M$:

$$M = \begin{pmatrix} \mu_B g^* B & \xi & 0 & \eta' e^{-ip} \\ \xi & \mu_B g^* B & \eta & 0 \\ 0 & \eta & \mu_B g^* B & \xi' \\ \eta' e^{ip} & 0 & \xi' & \mu_B g^* B \end{pmatrix}. \quad (9.68)$$

The four eigen-values of $M$ are denoted by the symbols $\lambda_1(p)$, $\lambda_2(p)$, $\lambda_3(p)$ and $\lambda_4(p)$ $(\lambda_1 \leq \lambda_2 \leq \lambda_3 \leq \lambda_4)$ which are obtained by

$$\lambda_1(p) = \mu_B g^* B - \sqrt{\tfrac{1}{2}\left((1+\alpha^2)(\xi^2 + \xi'^2) + \sqrt{(1+\alpha^2)^2(\xi^2 + \xi'^2)^2 - 4((1+\alpha^4)\xi^2\xi'^2 - 2\alpha^2\xi^2\xi'^2 \cos p)}\right)}$$

$$\lambda_2(p) = \mu_B g^* B - \sqrt{\tfrac{1}{2}\left((1+\alpha^2)(\xi^2 + \xi'^2) - \sqrt{(1+\alpha^2)^2(\xi^2 + \xi'^2)^2 - 4((1+\alpha^4)\xi^2\xi'^2 - 2\alpha^2\xi^2\xi'^2 \cos p)}\right)}$$

$$\lambda_3(p) = \mu_B g^* B + \sqrt{\tfrac{1}{2}\left((1+\alpha^2)(\xi^2 + \xi'^2) - \sqrt{(1+\alpha^2)^2(\xi^2 + \xi'^2)^2 - 4((1+\alpha^4)\xi^2\xi'^2 - 2\alpha^2\xi^2\xi'^2 \cos p)}\right)}$$

$$\lambda_4(p) = \mu_B g^* B + \sqrt{\tfrac{1}{2}\left((1+\alpha^2)(\xi^2 + \xi'^2) + \sqrt{(1+\alpha^2)^2(\xi^2 + \xi'^2)^2 - 4((1+\alpha^4)\xi^2\xi'^2 - 2\alpha^2\xi^2\xi'^2 \cos p)}\right)}$$

(9.69)

where
$$\alpha = \eta/\xi = \eta'/\xi' \quad (9.70a)$$
Therein we have assumed $\xi'/\xi = \eta'/\eta$ because the interval modulation is expected to give almost the same effect for the coupling constants $\xi$ and $\eta$. The ratio $\xi'/\xi$ is equal to $\eta'/\eta$ because of Eq.(9.70a):
$$\beta = \xi'/\xi = \eta'/\eta \quad (9.70b)$$
This ratio is expressed by the distortion parameter $t$ by using of Eq.(9.61):
$$\beta = \xi'/\xi = \eta'/\eta = (1+t)/(1-t) \quad (9.70c)$$
We show the two eigen values $\lambda_1(p)$ and $\lambda_2(p)$ by red and black curves for $\beta = 1$,

blue curves for $\beta = 1.2$ and green curves for $\beta = 1.4$ in Fig.9.13.

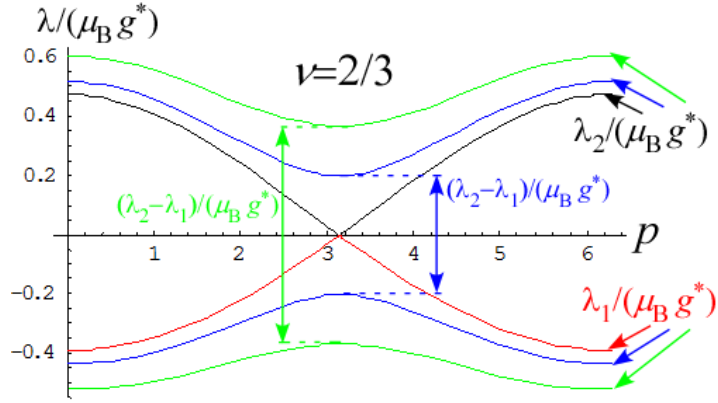

Fig.9.13 Eigenvalues of spin interaction via interval modulation

The red and black curves indicate the eigenvalues $\lambda_1/(\mu_B g^*)$ and $\lambda_2/(\mu_B g^*)$ for $\beta = 1.0$.
The blue curves indicate the eigenvalues $\lambda_1/(\mu_B g^*)$ and $\lambda_2/(\mu_B g^*)$ for $\beta = 1.2$.
The green curves indicate the eigenvalues $\lambda_1/(\mu_B g^*)$ and $\lambda_2/(\mu_B g^*)$ for $\beta = 1.4$.

The difference $\lambda_2(p) - \lambda_1(p)$ is minimal at $p = \pi$ as seen in Fig.9.13. When $\beta$ is equal to 1 (namely non-distorted case), the energy gap $\lambda_2(\pi) - \lambda_1(\pi)$ disappears. Eq.(9.69) gives the difference between $\lambda_2(p)$ and $\lambda_1(p)$ at $p = \pi$ as

$$\lambda_2(\pi) - \lambda_1(\pi) = \sqrt{1+\alpha^2}\,|\xi' - \xi| = \sqrt{1+\alpha^2}\,\xi_0\,2|t| \tag{9.71}$$

where we have used Eq.(9.61). Thus the energy gap is proportional to $|t|$.

The eigen-energies Eq.(9.69) yield the diagonal form of the Hamiltonian as follows:

$$H = \sum_p \begin{pmatrix} \lambda_1(p)A_1^*(p)A_1(p) + \lambda_2(p)A_2^*(p)A_2(p) \\ + \lambda_3(p)A_3^*(p)A_3(p) + \lambda_4(p)A_4^*(p)A_4(p) - 2\mu_B g^* B \end{pmatrix}. \tag{9.72}$$

Applying Eq.(9.44) we get

$$\langle A_s^*(p)A_s(p)\rangle = \frac{\exp(-\lambda_s(p)/k_B T)}{1+\exp(-\lambda_s(p)/k_B T)} = \frac{1}{2}(1 - \tanh(\lambda_s(p)/(2k_B T))) \tag{9.73}$$

The thermal average of the Hamiltonian (9.72) is

$$\langle H \rangle = \sum_p \left[ \left\{ \sum_{s=1}^{4} (\lambda_s(p)\tfrac{1}{2}(1 - \tanh(\lambda_s(p)/(2k_B T)))) \right\} - 2\mu_B g^* B \right] \tag{9.74}$$

Since the total number of electrons is a macroscopic value, we can replace the summation by integration as

$$\frac{\langle H \rangle}{N} = \frac{1}{4} \times \frac{1}{2\pi} \int_{p=0}^{2\pi} \left[ \left\{ \sum_{s=1}^{4} (\lambda_s(p) \tfrac{1}{2} (1 - \tanh(\lambda_s(p)/2k_B T))) - 2\mu_B g^* B \right\} \right] dp \qquad (9.75a)$$

$$\Delta E_{\text{Total}}/N = (\Delta W/N) + \Delta(\langle H \rangle/N) \qquad (9.75b)$$

We numerically calculate the integration in Eq.(9.75a). The calculated result of the spin exchange energy at $B = 2.2[T]$ is drawn by the dashed red curve in Fig.9.14. The dashed black curve expresses the classical Coulomb energy for the parameter $C = 0.5$ (see Eq.(9.62a)). The sum of the two curves gives the total energy $\Delta E_{\text{Total}}/N$ shown by the blue curve. The total energy takes a minimum value at a non-zero $t$ as in Fig.9.14. So the minimum point is realized in a low temperature, and then the interval modulation occurs actually. [81, 82]

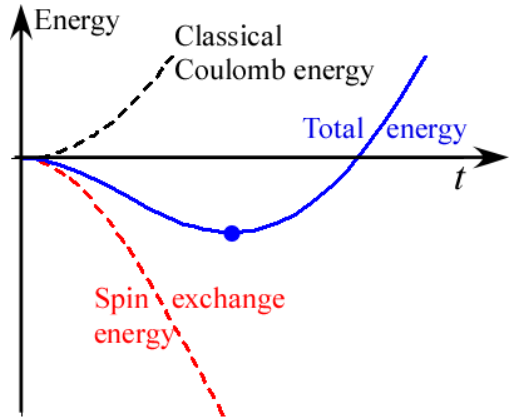

Fig.9.14 Dependence of total energy upon $t$

Further calculations are carried out for various values of the magnetic field in the case of $C = 0.5$. The results are shown in Fig.9.15. The interval modulation occurs in the region of $1.87[T] < B < 2.50[T]$ as seen in Fig.9.15.

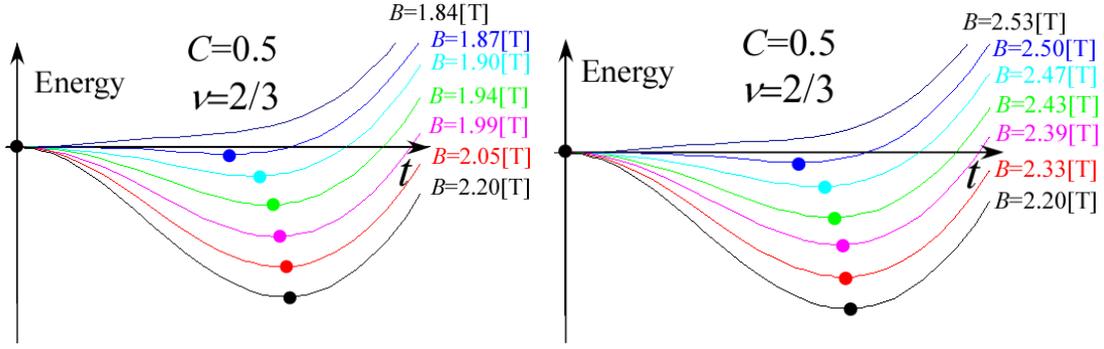

Fig.9.15   Dependence of total energy upon $t$

We calculate the total energy for another case $\nu=3/5$. The coupling constants are illustrated in Fig.9.16.

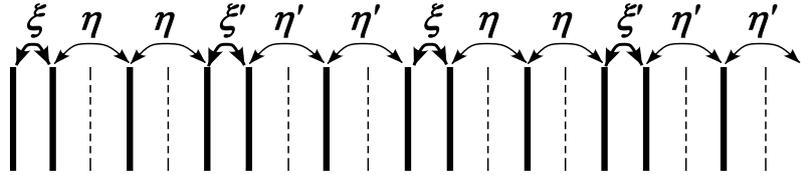

Fig.9.16 Coupling constants for $\nu=3/5$.

The interactions in Fig.9.16 produce the $\nu=3/5$ Hamiltonian. The Hamiltonian is expressed by the following matrix for a given wave number $p$:

$$\begin{pmatrix} \mu_B g^* B & \xi & 0 & 0 & 0 & \eta' e^{-ip} \\ \xi & \mu_B g^* B & \eta & 0 & 0 & 0 \\ 0 & \eta & \mu_B g^* B & \eta & 0 & 0 \\ 0 & 0 & \eta & \mu_B g^* B & \xi' & 0 \\ 0 & 0 & 0 & \xi' & \mu_B g^* B & \eta' \\ \eta' e^{ip} & 0 & 0 & 0 & \eta' & \mu_B g^* B \end{pmatrix} \quad \text{(for } \nu=3/5\text{)} \qquad (9.76)$$

This matrix has the six eigen-values $\lambda_1(p)$, $\lambda_2(p)$, $\lambda_3(p)$, $\lambda_4(p)$, $\lambda_5(p)$ and $\lambda_6(p)$, the $p$-dependences of which are shown in Fig.9.17.

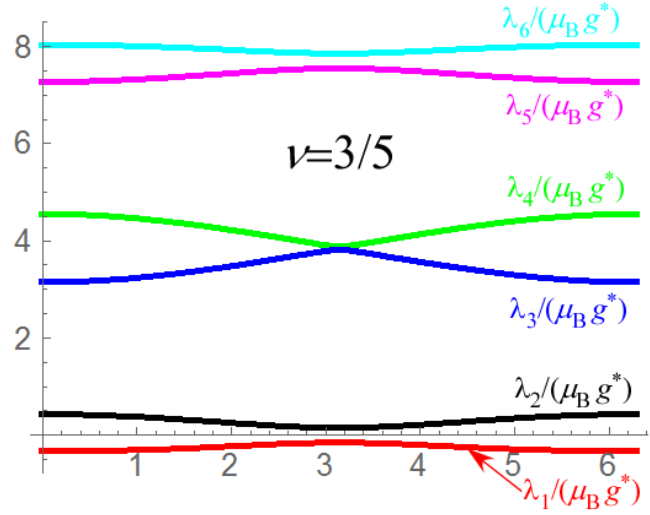

Fig.9.17  $p$-dependence of the six eigen-values for $\nu=3/5$

Figure 9.17 indicates the energy gaps between $\lambda_1(p)$ and $\lambda_2(p)$ and between $\lambda_5(p)$ and $\lambda_6(p)$ at $p=\pi$. These gaps are caused by the interval modulation. The total energy per electron is equal to the sum as Eq.(9.77c):

$$\Delta W/N = \xi_0\, C\, t^2 \tag{9.77a}$$

$$\frac{\langle H \rangle}{N} = \frac{1}{6} \times \frac{1}{2\pi} \int_{p=0}^{2\pi} \left[ \left\{ \sum_{s=1}^{6} (\lambda_s(p)\tfrac{1}{2}(1 - \tanh(\lambda_s(p)/2k_BT))) - 3\mu_B g^* B \right\} \right] dp \tag{9.77b}$$

$$\Delta E_{\text{Total}}/N = (\Delta W/N) + \Delta(\langle H \rangle/N) \tag{9.77c}$$

The sum of $\Delta W/N$ and $\Delta\langle H\rangle/N$ is numerically calculated by a computer. The result is shown in Fig.9.18.

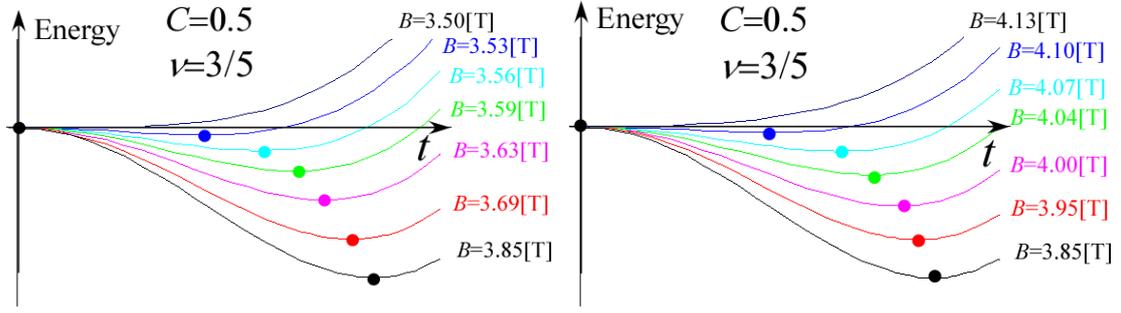

Fig.9.18 Dependence of the total energy upon $t$ at $\nu=3/5$

Therein we have used the parameter $C = 0.5$ which is the same as in the case of $\nu=2/3$. The total energy takes the minimum value at the dots in Fig.9.18. The $t$-value at the minimum energy depends upon the magnetic field. The non-zero $t$-value appears in the region $3.53[T] < B < 4.10[T]$ for the experiment [77] where the interval modulation occurs actually.

### 9.7.2 Spin polarization in the case with the interval modulation

We calculate the spin-polarization $\gamma_e$ for the case with the interval modulation. At $\nu = 2/3$, $\gamma_e$ is obtained by the integration as

$$\gamma_e = \frac{1}{4} \times \frac{1}{2\pi} \int_{-\pi}^{\pi} dp \left( \sum_{s=1}^{4} \tanh(\lambda_s(p)/2k_B T) \right) \qquad (9.78)$$

where the four eigen-values $\lambda_1(p)$, $\lambda_2(p)$, $\lambda_3(p)$ and $\lambda_4(p)$ are given in Eq.(9.69).

The spin-polarization $\gamma_e$ is numerically calculated via the following two methods namely easy method A and precise method B.

(Method A)
Method A is the rough calculation using the fixed value of the distortion parameter $t$. That is to say the ratio $\xi'/\xi = \eta'/\eta$ is treated to be a constant value for all the strength of magnetic field.

We use the fixed value $\xi'/\xi = \eta'/\eta = 1.4$. The other parameters are adopted to be $\eta/\xi = \eta'/\xi' = 0.1$ and $(k_B T/\xi_0) = 0.05$. Then a small energy gap appears between $\lambda_1$ and $\lambda_2$. We numerically calculate the integration in Eq.(9.78) and draw the graph of spin-polarization versus magnetic field. A small shoulder appears in the theoretical curve of the electron spin-polarization as seen in Fig.9.19.

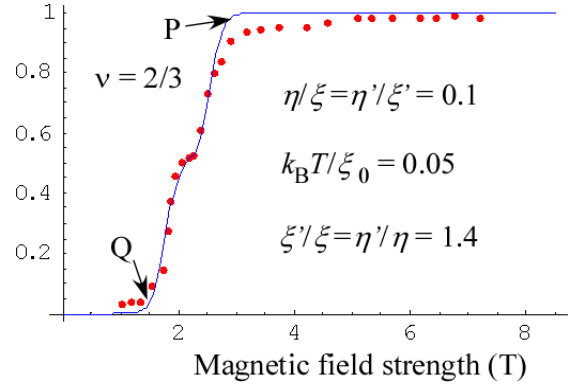

Fig.9.19 Method A: Theoretical curve of the spin-polarization for $\nu = 2/3$

(Red dots are experimental data [77])

This curve is slightly different from the experimental data near the sharp corners P and Q. So we choose the different value as $(k_B T/\xi_0)=0.1$ in order to make the curvature to be small in the corners P and Q. Then the small shoulder disappears. In order to maintain the existence of the small shoulder, we take a larger value $\xi'/\xi = \eta'/\eta =1.8$. The calculated curve is expressed in Fig.9.20 which is also slightly different from the experimental data. Accordingly method A has some difficulty to explain the experimental data. This inadequacy is improved by using the precise method B.

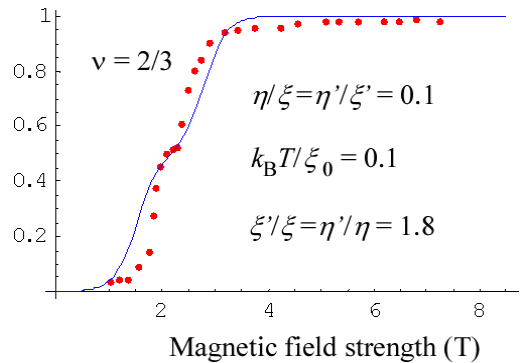

Fig.9.20 Method A: Calculated curve of the spin-polarization for $\nu = 2/3$

(The parameters are chosen as $(k_B T/\xi_0)=0.1$ and $\xi'/\xi = \eta'/\eta =1.8$)

(Method B)

We carry out more precise method B where we calculate the $t$-dependence of the total energy per electron. Some examples have been already shown in Figs.9.15 and 9.18. Therein we can obtain the $t$-value with the energy minimum as shown by dots. The $t$-value at the minimum point gives the value of $\xi'/\xi = \eta'/\eta$ by Eq.(9.70c). Using the magnetic field dependence of $t$-value, we can numerically calculate the spin-polarization.

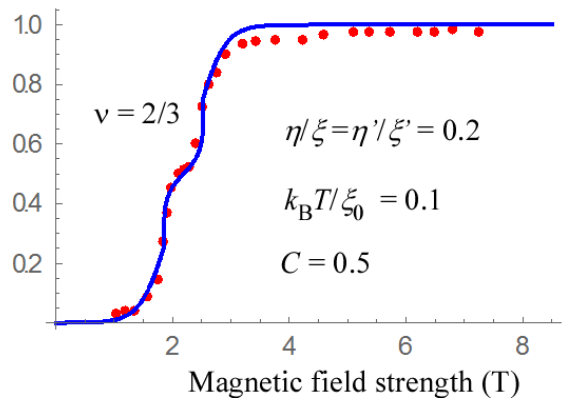

Fig.9.21 Method B: Calculated curve of the spin-polarization for $\nu = 2/3$

Therein we have used the parameter-values $\eta/\xi = \eta'/\xi' = 0.2$, $(k_BT/\xi_0)=0.1$ and $C = 0.5$. Figure 9.21 expresses that the calculated result is in good agreement with the experimental data.

The reason is simply discussed below: The magnetic field is sufficiently strong near the corner P in Fig.9.19. In this region, almost all the spins have a down direction. Then the number of up and down spin-pairs decreases and so the spin exchange energy becomes small. Accordingly the total energy Eq.(9.75b) is nearly equal to the classical Coulomb energy which yields the energy minimum at $t = 0$, namely non-distortion (non-interval-modulation). Thus the distortion appears only near the small shoulder. So the theoretical curve employing method B is in good agreement with the experimental data.

The reader may want to know how the shape of the polarization curve depends on the parameter $C$. We calculate the polarization curve for the following two cases $C = 0.4$ and 0.65 in Fig.9.22. Then the shape of the curve varies in only the neighborhood of the small shoulder when changing the parameter $C$. The calculated shoulder is too large for $C = 0.4$ and too small for $C = 0.65$. The calculated result for $C = 0.5$ agrees with the experimental data as in Fig.9.21.

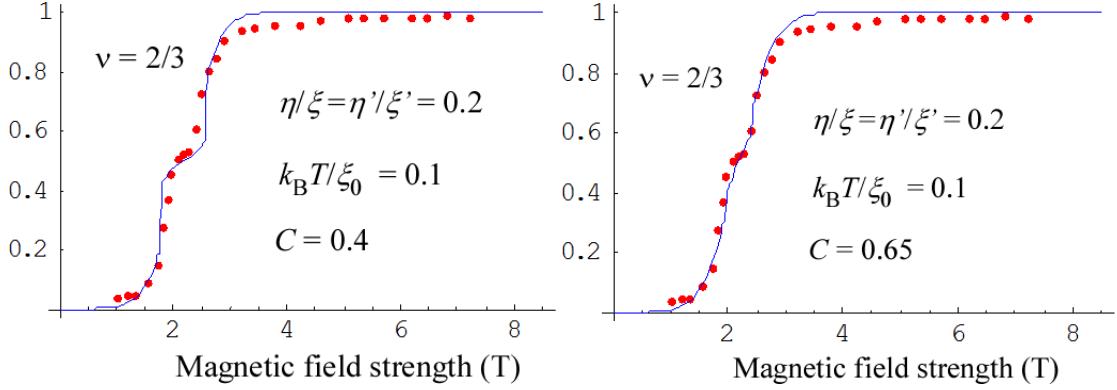

Fig.9.22 Spin-polarization via method B for two cases with $C = 0.4$ and $0.65$

We next study the case of $v = 3/5$. The spin-polarization $\gamma_e$ at $v = 3/5$ is given by

$$\gamma_e = \frac{1}{6} \times \frac{1}{2\pi} \int_{-\pi}^{\pi} dp \left( \sum_{s=1}^{6} \tanh(\lambda_s(p)/2k_B T) \right) \tag{9.79}$$

where the six eigen-energies $\lambda_s$ for $s = 1,2,3,4,5,6$ are numerically obtained from the matrix (9.76). The two calculated curves via Method A or B are shown in Fig.9.23, respectively. In the method B we have applied $C = 0.5$ same as in $v = 2/3$. The theoretical result via method B is in good agreement with the experimental data as seen in the right panel of Fig.9.23.

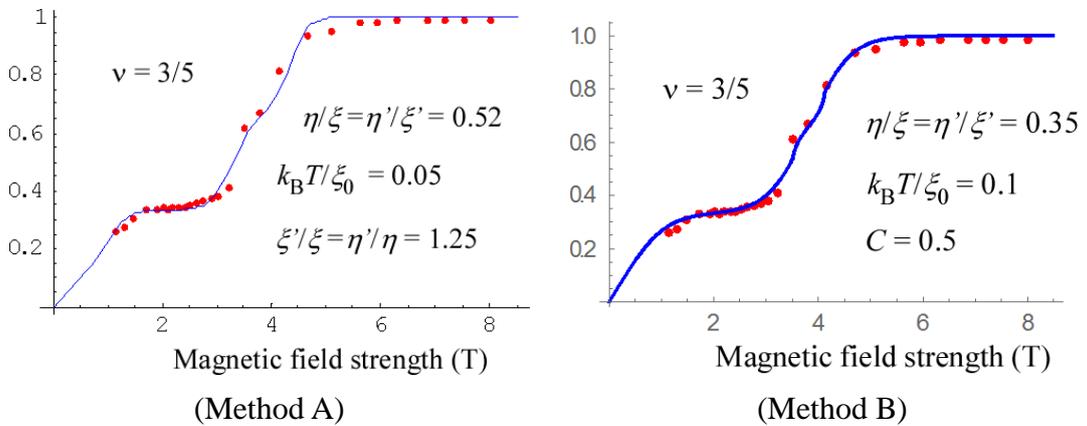

(Method A)      (Method B)

Fig.9.23 Spin-polarization for $v = 3/5$ (Red dots are experimental data [77])

We next examine the case of $\nu = 4/7$. The most uniform electron configuration is illustrated in Fig.9.24. Therein $\xi, \eta, \xi', \eta'$ are the four kinds of the coupling constants.

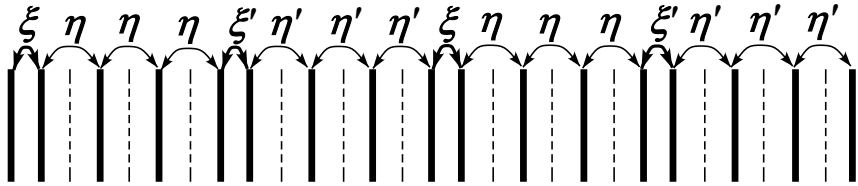

Fig.9.24 Coupling constants for $\nu=4/7$.

The electron-configuration yields the $\nu = 4/7$ Hamiltonian described by the following matrix (9.80):

$$\begin{pmatrix} \mu_B g^* B & \xi & 0 & 0 & 0 & 0 & 0 & \eta' e^{-ip} \\ \xi & \mu_B g^* B & \eta & 0 & 0 & 0 & 0 & 0 \\ 0 & \eta & \mu_B g^* B & \eta & 0 & 0 & 0 & 0 \\ 0 & 0 & \eta & \mu_B g^* B & \eta & 0 & 0 & 0 \\ 0 & 0 & 0 & \eta & \mu_B g^* B & \xi' & 0 & 0 \\ 0 & 0 & 0 & 0 & \xi' & \mu_B g^* B & \eta' & 0 \\ 0 & 0 & 0 & 0 & 0 & \eta' & \mu_B g^* B & \eta' \\ \eta' e^{ip} & 0 & 0 & 0 & 0 & 0 & \eta' & \mu_B g^* B \end{pmatrix} \quad \text{(for } \nu=4/7\text{)} \quad (9.80)$$

The spin-polarization $\gamma_e$ is given by

$$\gamma_e = \frac{1}{8} \times \frac{1}{2\pi} \int_{-\pi}^{\pi} dp \left( \sum_{s=1}^{8} \tanh(\lambda_s(p)/2k_B T) \right) \qquad (9.81)$$

where $\lambda_s$ for $s = 1,2,3,4,5,6,7,8$ indicate the eigen-energies of the matrix (9.80).

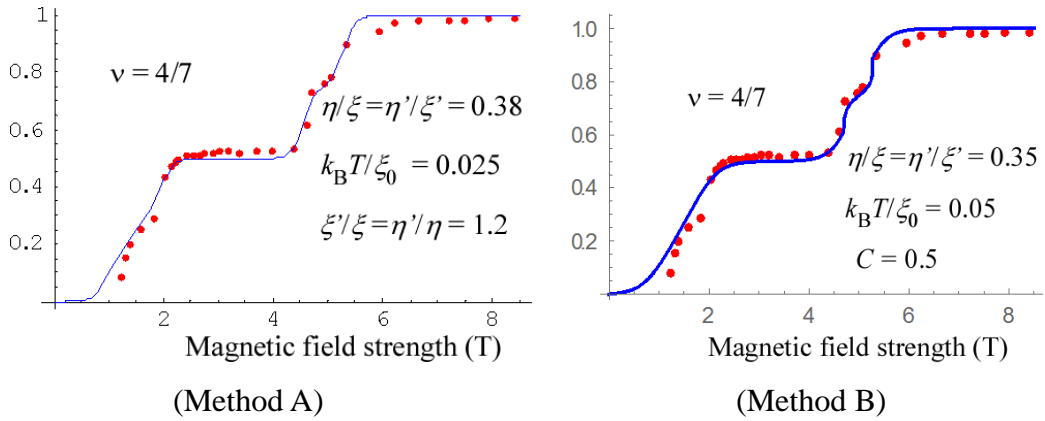

Fig.9.25 Spin-polarization for $\nu = 4/7$ (Red dots are experimental data [77])

The spin polarization can be evaluated from the eigen-energies. The results are shown in Figs.9.25. Method B has used the same value 0.5 for the parameter $C$.

## 9.8 Phenomena similar to the spin-Peierls effect at various filling factors

We examine the cases of $\nu=2/5$, $\nu=3/7$, and $\nu=4/9$ which are classified into Type2 in section 9.6. When the interval modulation occurs, the most uniform electron-configurations at $\nu=2/5$, $\nu=3/7$, and $\nu=4/9$ become like Figs.9.26a, 9.26b and 9.26c, respectively.

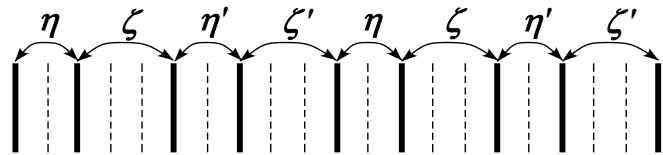

Fig.9.26a Coupling constants for $\nu=2/5$

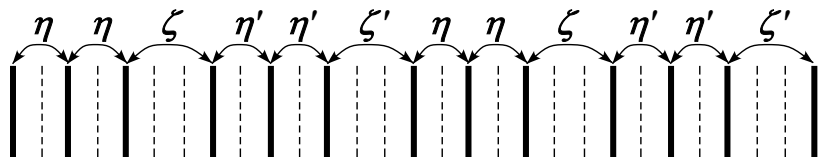

Fig.9.26b Coupling constants for $\nu=3/7$

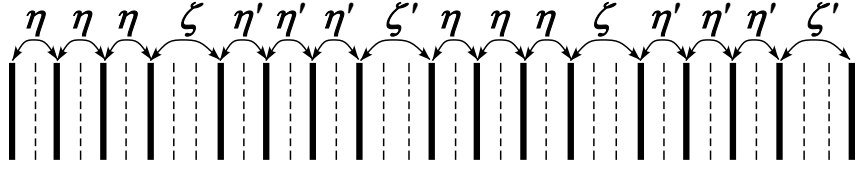

Fig.9.26c Coupling constants for $\nu=4/9$

The coupling constants yield the spin-exchange Hamiltonians for $\nu=2/5$, $\nu=3/7$, and $\nu=4/9$ which are represented by the following matrices, respectively.

$$\begin{pmatrix} \mu_B g^* B & \eta & 0 & \varsigma' e^{-ip} \\ \eta & \mu_B g^* B & \varsigma & 0 \\ 0 & \varsigma & \mu_B g^* B & \eta' \\ \varsigma' e^{ip} & 0 & \eta' & \mu_B g^* B \end{pmatrix} \quad \text{(for } \nu=2/5) \quad (9.82a)$$

$$\begin{pmatrix} \mu_B g^* B & \eta & 0 & 0 & 0 & \varsigma' e^{-ip} \\ \eta & \mu_B g^* B & \eta & 0 & 0 & 0 \\ 0 & \eta & \mu_B g^* B & \varsigma & 0 & 0 \\ 0 & 0 & \varsigma & \mu_B g^* B & \eta' & 0 \\ 0 & 0 & 0 & \eta' & \mu_B g^* B & \eta' \\ \varsigma' e^{ip} & 0 & 0 & 0 & \eta' & \mu_B g^* B \end{pmatrix} \quad \text{(for } \nu=3/7) \quad (9.82b)$$

$$\begin{pmatrix} \mu_B g^* B & \eta & 0 & 0 & 0 & 0 & 0 & \varsigma' e^{-ip} \\ \eta & \mu_B g^* B & \eta & 0 & 0 & 0 & 0 & 0 \\ 0 & \eta & \mu_B g^* B & \eta & 0 & 0 & 0 & 0 \\ 0 & 0 & \eta & \mu_B g^* B & \varsigma & 0 & 0 & 0 \\ 0 & 0 & 0 & \varsigma & \mu_B g^* B & \eta' & 0 & 0 \\ 0 & 0 & 0 & 0 & \eta' & \mu_B g^* B & \eta' & 0 \\ 0 & 0 & 0 & 0 & 0 & \eta' & \mu_B g^* B & \eta' \\ \varsigma' e^{ip} & 0 & 0 & 0 & 0 & 0 & \eta' & \mu_B g^* B \end{pmatrix} \quad \text{(for } \nu=4/9) \quad (9.82c)$$

The average value of $\eta$ and $\eta'$ is expressed by $\eta_0$. Then Eq.(9.70c) gives

$$\eta = \eta_0(1-t), \quad \eta' = \eta_0(1+t) \tag{9.83}$$

The ratios between the coupling constants satisfy the following relations similar to Eq.(9.70).

$$\varsigma/\eta = \varsigma'/\eta' \tag{9.84a}$$
$$\eta'/\eta = \varsigma'/\varsigma = (1+t)/(1-t) \tag{9.84b}$$

We re-express the *t*-dependence of the classical Coulomb energy by using the coupling

constant $\eta_0$

$$\Delta W/N = \eta_0 \, D \, t^2 \tag{9.85}$$

where $D$ is a new dimensionless coefficient. Using the eigen-values of the matrices (9.82a, b, c), the spin polarizations is given by

$$\gamma_e = \frac{1}{4} \times \frac{1}{2\pi} \int_{-\pi}^{\pi} dp \left( \sum_{s=1}^{4} \tanh(\lambda_s(p)/2k_B T) \right) \quad \text{(for } \nu=2/5) \tag{9.86a}$$

$$\gamma_e = \frac{1}{6} \times \frac{1}{2\pi} \int_{-\pi}^{\pi} dp \left( \sum_{s=1}^{6} \tanh(\lambda_s(p)/2k_B T) \right) \quad \text{(for } \nu=3/7) \tag{9.86b}$$

$$\gamma_e = \frac{1}{8} \times \frac{1}{2\pi} \int_{-\pi}^{\pi} dp \left( \sum_{s=1}^{8} \tanh(\lambda_s(p)/2k_B T) \right) \quad \text{(for } \nu=4/9) \tag{9.86c}$$

We numerically calculate the spin-polarization curves via method A and B, the results of which are shown in Figs. 9.27, 9.28 and 9.29. Therein the small shoulders originate from the interval modulation (distortion with double period). The theoretical result by Method B is in better agreement with the experimental data than that by Method A.

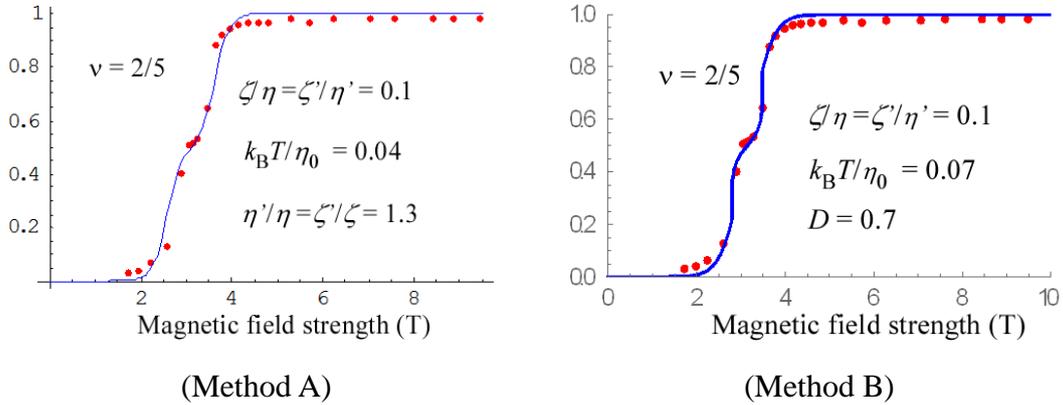

(Method A)      (Method B)

Fig.9.27 Spin-polarization for $\nu = 2/5$ (Red dots are experimental data [77])

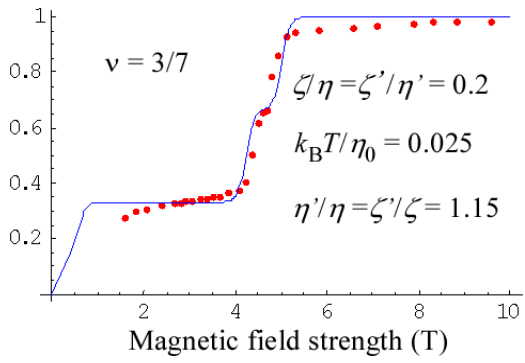
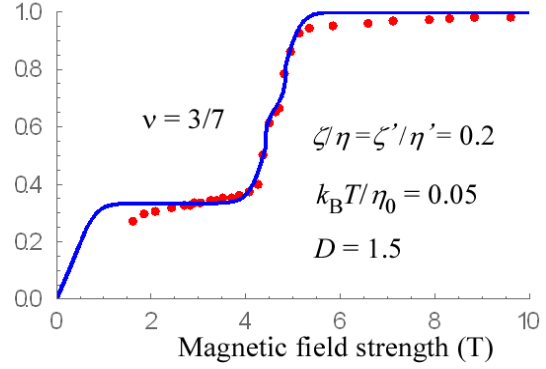

(Method A)　　　　　　　　　　　(Method B)

Fig.9.28 Spin-polarization for $\nu = 3/7$ (Red dots are experimental data [77])

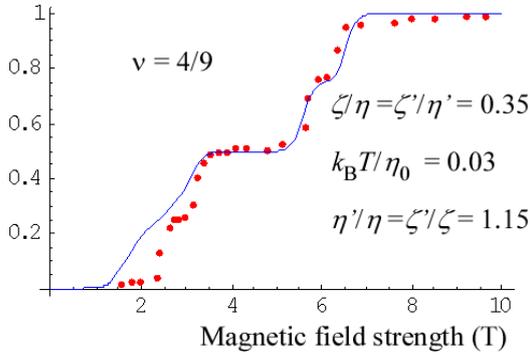
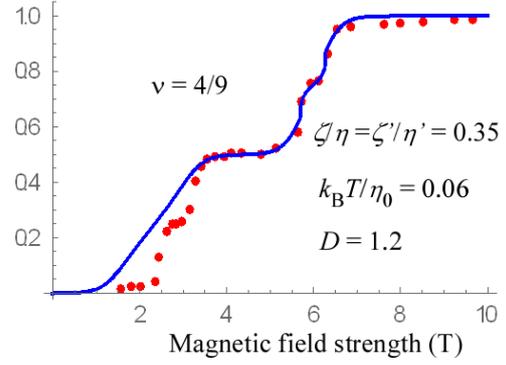

(Method A)　　　　　　　　　　　(Method B)

Fig.9.29 Spin-polarization for $\nu = 4/9$ (Red dots are experimental data [77])

Here we shortly discuss the parameter values $C$ and $D$. The increase of the classical Coulomb energy are expressed in Eqs.(9.62a) and (9.85) as $\Delta W/N = \xi_0 \, C \, t^2$ and $\Delta W/N = \eta_0 \, D \, t^2$, respectively. So the parameter $D$ may be almost equal to

$$D \approx C \times \xi_0 / \eta_0 \qquad (9.87a)$$

The ratio $\eta/\xi = \eta'/\xi' = \eta_0/\xi_0$ is 0.35 as in Method B of Figs.9.23 and 9.25. Substitution of the value 0.35 and $C = 0.5$ into Eq.(9.87a) gives the value of $D$ as

$$D \approx 0.5/0.35 \approx 1.43 \qquad (9.87b)$$

The fitting values of $D$ are 1.5 and 1.2 for $\nu = 3/7$ and 4/9, respectively as in Figs.9.28

and 9.29. The fitting values of $D$ are consistent with the predicted value namely Eq.(9.87b). The parameter $D$ at $\nu=2/5$ is 0.7 as in Fig.9.27. We cannot understand why the parameter $D$ is small at $\nu=2/5$.

Next we examine the $\nu=4/3$, 7/5, and 8/5 states which have been classified into Type 3. The most uniform electron-configuration and the coupling constants are shown in Figs.9.30a, 9.30b and 9.30c, respectively.

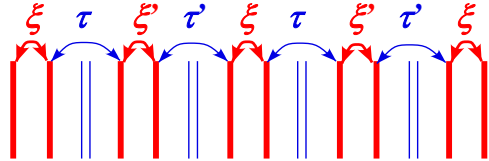

Fig.9.30a Coupling constants for $\nu=4/3$.

Double-line indicates a Landau orbital occupied by electron pair with up and down spins.

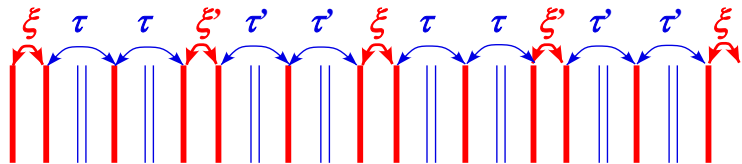

Fig.9.30b Coupling constants for $\nu=7/5$.

Double-line indicates a Landau orbital occupied by electron pair with up and down spins.

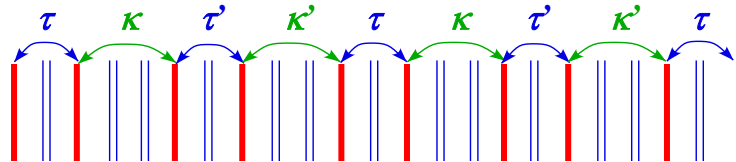

Fig.9.30c Coupling constants for $\nu=8/5$.

Double-line indicates a Landau orbital occupied by electron pair with up and down spins.

There are doubly occupied orbitals in Figs.9.30a, b and c. As discussed in section 9.6, the spin exchange forces act between electrons in singly occupied orbitals. The electron pairs in doubly occupied orbitals have no polarization because of cancellation by up and down spin pair. Therefore the electron spin polarization is given by the following equations:

$$\gamma_e = \frac{2}{4} \times \frac{1}{4} \times \frac{1}{2\pi} \int_{-\pi}^{\pi} dp \left( \sum_{s=1}^{4} \tanh(\lambda_s(p)/2k_B T) \right) \text{ (for } \nu=4/3\text{)} \quad (9.88a)$$

$$\gamma_e = \frac{3}{7} \times \frac{1}{6} \times \frac{1}{2\pi} \int_{-\pi}^{\pi} dp \left( \sum_{s=1}^{6} \tanh(\lambda_s(p)/2k_B T) \right) \text{ (for } \nu=7/5\text{)} \quad (9.88b)$$

$$\gamma_e = \frac{2}{8} \times \frac{1}{4} \times \frac{1}{2\pi} \int_{-\pi}^{\pi} dp \left( \sum_{s=1}^{4} \tanh(\lambda_s(p)/2k_B T) \right) \text{ (for } \nu=8/5\text{)} \quad (9.88c)$$

where the coefficients 2/4, 3/7 and 2/8 in Eqs.(9.88a), (9.88b) and (9.88c) come from the fact that the up and dawn spin pairs cancel the polarization. That is to say two electrons per four electrons at $\nu=4/3$ have no polarization, four electrons per seven electrons at $\nu=7/5$ have no polarization and also six electrons per eight electrons at $\nu=8/5$ have no polarization.

We numerically calculate the spin-polarization curves via method A and B, the results of which are shown in Figs. 9.31, 9.32 and 9.33.

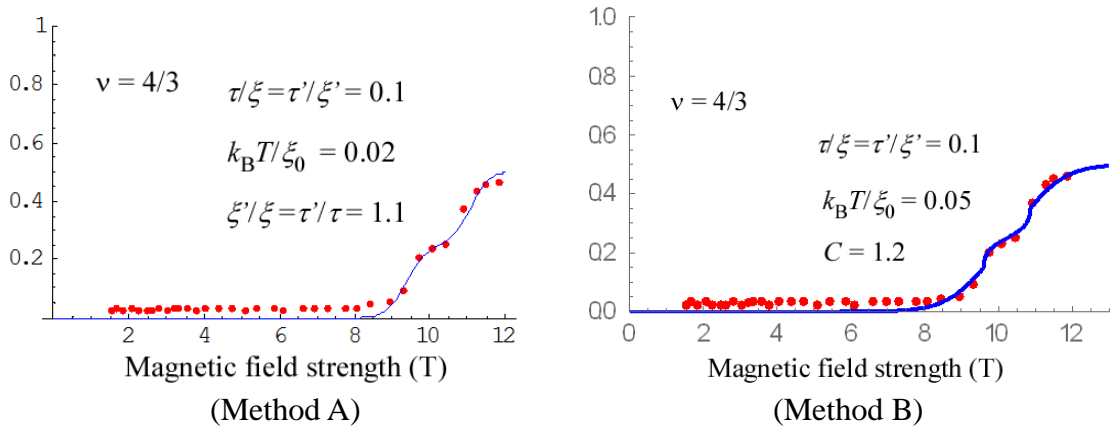

Fig.9.31 Spin-polarization for $\nu = 4/3$ (Red dots are experimental data [77])

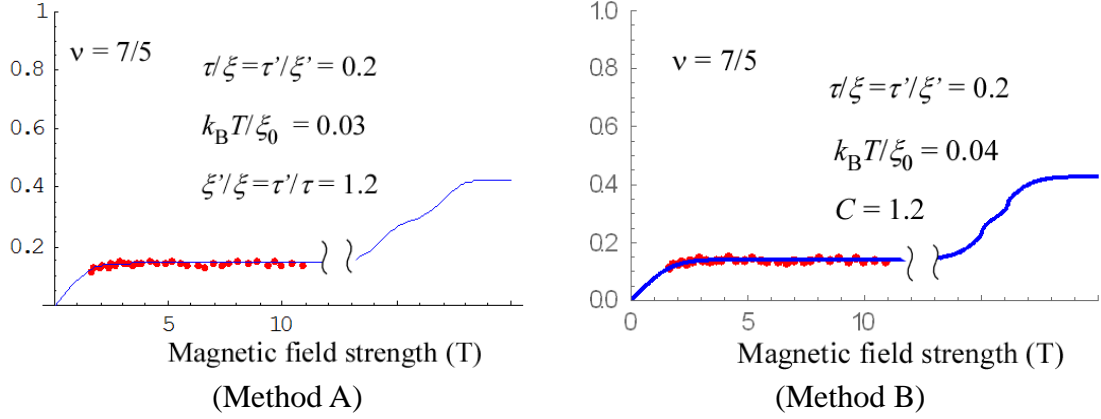

Fig.9.32 Spin-polarization for $\nu = 7/5$ (Red dots are experimental data [77])

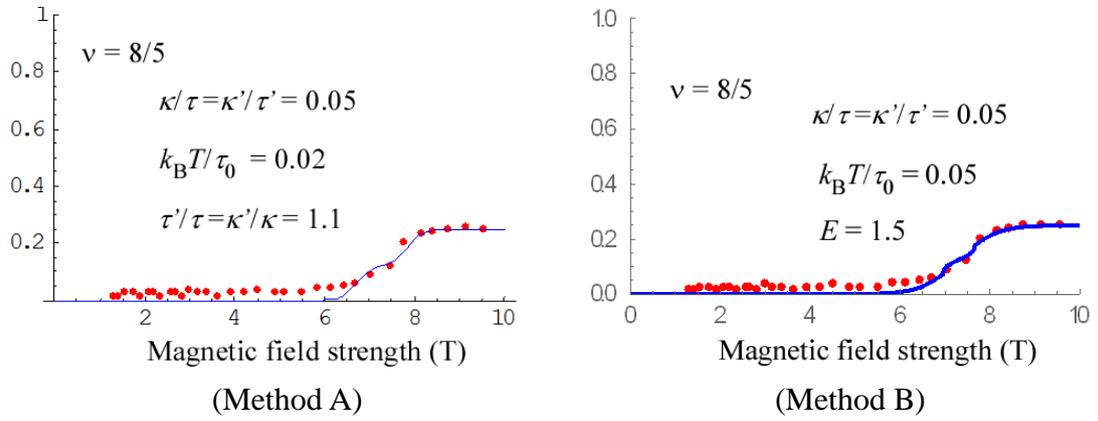

Fig.9.33 Spin-polarization for $\nu = 8/5$ (Red dots are experimental data [77])

The coupling constants at $\nu = 8/5$ are $\tau, \kappa, \tau', \kappa'$ as in Fig.9.30c. The coupling constants and the classical Coulomb energy are re-expressed by using $\tau_0$ as follows:

$$\tau = \tau_0(1-t), \quad \tau' = \tau_0(1+t) \tag{9.89a}$$

$$\Delta W/N = \tau_0 \, E \, t^2 \tag{9.89b}$$

where $E$ is a new coefficient. The fitting value is $E = 1.5$ for $\nu = 8/5$.

The experimental data at $\nu = 4/3$ has a very sharp change on the polarization curve from $B = 8.5 [\text{T}]$ to $B = 11.5 [\text{T}]$. Also the shoulder at $\nu = 8/5$ appears from $B = 6.5 [\text{T}]$ to $B = 8 [\text{T}]$. It is very difficult to fit the complicated behaviors of the experimental data. The theoretical results in Figs.9.31 and 9.33 are consistent with the experimental data at $\nu = 4/3$ and $8/5$, respectively. (This experimental behavior cannot reproduce without spin Peierls instability.)

The parameters $\eta/\xi$, $(k_B T/\xi_0)$, $C$ et al. may be dependent upon the gate voltage, material, shape of sample, etc. We have used the same value 0.5 for $C$ at $\nu = 2/3$, 3/5 and 4/7. If we use the different values of $C$, we can find better fitting to the experimental data than the present results.

(Short summary on the spin-polarizations of FQH states)
Up and down spins coexist at a low magnetic field. In this case, there are many degenerate ground states with different spin arrangements in the most uniform electron-configuration. These many electron states have the same eigen-energy of the Hamiltonian $H_D$. We have succeeded to diagonalize the partial Hamiltonian composed of the Coulomb transitions among the degenerate ground states. Then the calculated results have reproduced the wide plateaus on the spin polarization curves of the experimental data [77].

Furthermore we have studied the origin of the small shoulders. We take the interval modulation between Landau orbitals, the period of which has the doubly period of the original unit configuration. Then the partial Hamiltonian with the modulation is diagonalized exactly. The total energy is dependent upon the interval modulation $t$ as in Figs.9.14, 9.15 and 9.18. We have found the minimum total energy at the non-zero $t$-value. That is to say the interval modulation occurs in some range of the magnetic field. The spin-polarization is numerically calculated by employing the eigen-state. Then the theoretical polarization curve reproduces the small shoulder and the wide plateau. The theoretical results are in good agreement with the experimental data.

# Chapter 10 FQHE under a tilted magnetic field

We study the case where the applied magnetic field is tilted from the direction perpendicular to the surface (or a thin layer of an electron channel) of the quantum Hall device. In this chapter, the diagonal resistance of an FQH state is examined how to depend on the tilted angle and the field strength. We also discuss the polarization behaviour of FQH states under a tilted magnetic field.

### 10.1 Formulation in the tilted magnetic field

Homogeneous magnetic field has three components as

$$\mathbf{B} = (B_x, B_y, B_z) \tag{10.1}$$

The vector potential $\mathbf{A}$ should satisfy the following two conditions for a static field:

$$\mathbf{rot}\,\mathbf{A} = (B_x, B_y, B_z) \quad \text{and} \quad \mathrm{div}\,\mathbf{A} = 0 \tag{10.2}$$

The component of the vector potential is given by

$$A_x = -B_z y + B_y z,\ A_y = -B_x z,\ A_z = 0 \tag{10.3}$$

We makes sure that the three components of $\mathbf{A}$ satisfy Eq.(10.2) as follows:

$$\mathbf{rot}\,\mathbf{A} = \left(\frac{\partial A_z}{\partial y} - \frac{\partial A_y}{\partial z}, \frac{\partial A_x}{\partial z} - \frac{\partial A_z}{\partial x}, \frac{\partial A_y}{\partial x} - \frac{\partial A_x}{\partial y}\right) = (B_x, B_y, B_z) \tag{10.4a}$$

$$\mathrm{div}\,\mathbf{A} = \frac{\partial A_x}{\partial x} + \frac{\partial A_y}{\partial y} + \frac{\partial A_z}{\partial z} = 0 \tag{10.4b}$$

The Hamiltonian of the single electron is given by Eq.(1.2) as

$$H_0 = \frac{(\mathbf{p} + e\mathbf{A})^2}{2m^*} + U(y) + V_{\text{Thin}}(z) \tag{10.5}$$

Under a tilted magnetic field, the eigen-equation of the Hamiltonian is

$$\left[\frac{1}{2m^*}\left\{\left(-i\hbar\frac{\partial}{\partial x} - eB_z y + eB_y z\right)^2 + \left(-i\hbar\frac{\partial}{\partial y} - eB_x z\right)^2 + \left(-i\hbar\frac{\partial}{\partial z}\right)^2\right\} + U(y) + V_{\text{Thin}}(z)\right]\psi(x,y,z) = E\psi(x,y,z) \tag{10.6}$$

where $\psi(x, y, z)$ is the single electron wave function. The Hamiltonian $H_0$ has no potential term depending upon $x$ and therefore the wave function in the x-direction is expressed by a plane wave as

$$\psi(x, y, z) = \sqrt{\frac{1}{\ell}} \exp(ikx)\,\Psi(y, z) \tag{10.7}$$

In a quantum Hall device, the width of the potential $V_{\text{Thin}}(z)$ along the z-direction is extremely narrow as in Fig.1.3. Therein the probability density in the z-direction is large only at and near $z = z_0$ where $z_0$ is the position of the potential minimum. Without loss of generality we take $z_0$ to be the origin namely

$$z_0 = 0 \tag{10.8}$$

When the conducting layer of the device is ultra-thin, the terms $eB_y z$ and $-eB_x z$ can be ignored because of $z \approx z_0 = 0$. Accordingly the eigen-equation (10.6) becomes

$$\left[ \frac{1}{2m^*}\left\{ \left(-i\hbar\frac{\partial}{\partial x} - eB_z y\right)^2 + \left(-i\hbar\frac{\partial}{\partial y}\right)^2 + \left(-i\hbar\frac{\partial}{\partial z}\right)^2 \right\} + U(y) + V_{\text{Thin}}(z) \right]\psi(x,y,z) = E\psi(x,y,z) \tag{10.9}$$

Consequently the x- and y-components of the magnetic field don't contribute to the FQH states. The theoretical result is as follows: When the magnetic field is tilted from the z-direction, the FQH state depends only on the z-component of the magnetic field.

### 10.2 Comparison between present theory and experiments

We compare the theoretical result with the experimental data. The experimental data of reference [34, 35] are shown in Fig.10.1. Therein the diagonal resistance takes a local minimum at $B = 11.9$ T for $\theta = 0°$, $B = 12.5$ T for $\theta = 18.5°$, $B = 14.7$ T for $\theta = 36.5°$, $B = 16.0$ T for $\theta = 42.2°$ and $B = 17.0$ T for $\theta = 45.8°$ where $\theta$ indicates the tilted angle.

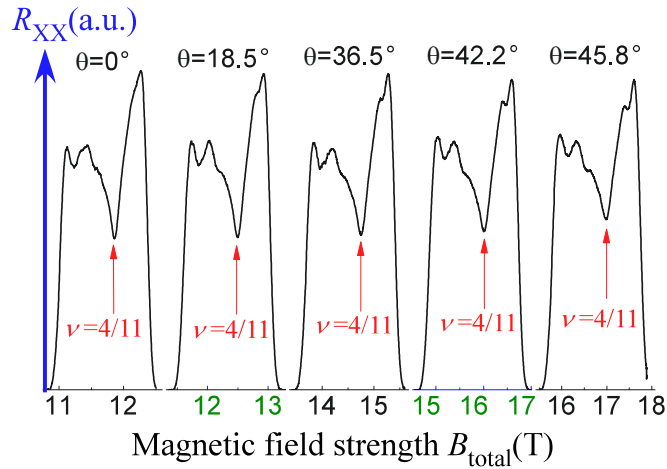

Fig.10.1 Diagonal resistance at $\nu = 4/11$ for several angles
[34, 35] arXiv:cond-mat/0303429v1 [cond-mat.mes-hall] 20 Mar 2003

The z-component of the magnetic field is related to the total strength $B_{\text{total}}$:

$$B_z = B_{\text{total}} \cos\theta \tag{10.10}$$

The z-component is numerically calculated for the five tilted angles as follows:

$$B_z = B_{\text{total}}\cos\theta = 11.9\cos 0° = 11.9 \tag{10.11a}$$
$$B_z = B_{\text{total}}\cos\theta = 12.5\cos 18.5° = 11.85 \tag{10.11b}$$

$$B_z = B_{total} \cos\theta = 14.7 \cos 36.5° = 11.82 \quad (10.11c)$$

$$B_z = B_{total} \cos\theta = 16.0 \cos 42.2° = 11.85 \quad (10.11d)$$

$$B_z = B_{total} \cos\theta = 17.0 \cos 45.8° = 11.85 \quad (10.11e)$$

Eqs.(10.11a, b, c, d and e) show that the z-component of the magnetic field takes almost same value at the local minima for the five tilted angles. Thus the experimental results are in good agreement with the theoretical results.

Figure 10.2 shows another experimental data reported in reference [91]. The diagonal resistance is plotted versus the z-component of the magnetic field, $B_\perp$ where the black curve indicates the diagonal resistance $R_{xx}$ for the tilted angle $\theta = 0°$ and the purple curve for $\theta = 20.3°$. The two curves are almost same as each other. Thus these data indicate that only the z-component of the magnetic field is effective to the diagonal resistance for a device with an ultra-thin electron channel.

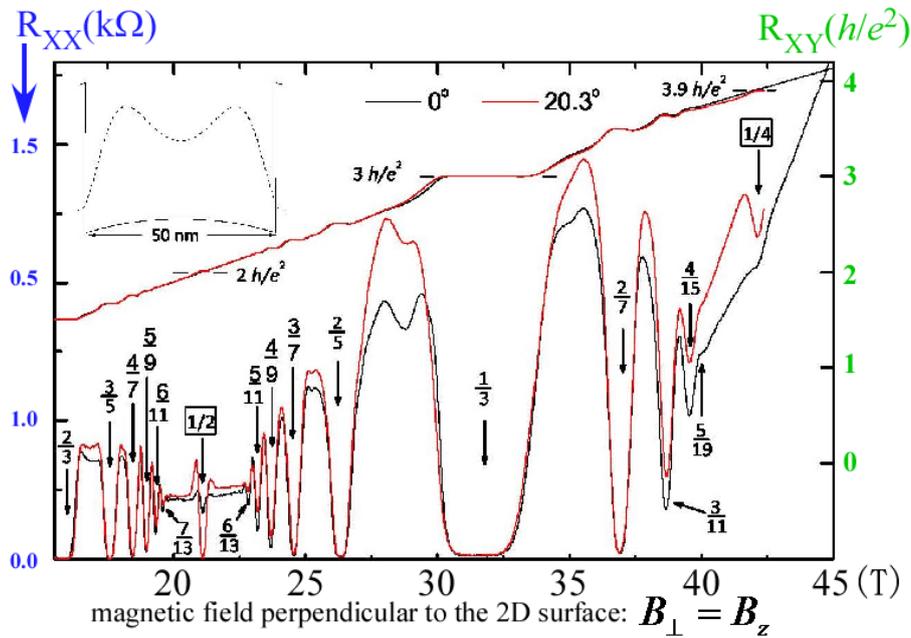

Fig.10.2 Hall and diagonal resistances versus z-component of magnetic field for tilted two angles
Ref.[91] arXiv:0810.2274v1 [cond-mat.mes-hall] 13 Oct 2008

On the other hand, the spin polarization behaves in a different dependence on the magnetic field from the diagonal resistance. If the spin polarization experiment similar to Kukushukin et al [77] is carried out under various tilted magnetic fields, then the different behaviours may be observed. Therein the spin polarization for the direction of the applied magnetic field (not z-direction) should be measured in various tilted magnetic fields. The spin polarization curve versus magnetic field strength (not z-component) may have the same shape as in Chapter 9 under fixing the filling factor if the coupling constants have the same values as in $\theta = 0°$.

# Chapter 11 Further experiments

We propose two experiments in this chapter where we superpose an oscillated magnetic field in addition to the static strong magnetic field. Thereby new phenomena may be observed. The first proposal is a measurement of the energy gap and the second one is an observation of tunnelling effect in the IQH and FQH states.

## 11.1 Diagonal resistance in the FQH state under a periodically modulated magnetic field or current

We propose a measurement of the energy gap in this section [92], [93]. The experiment is as follows:

(1) The strength of the static magnetic field is fixed to the value in order to maintain the FQH state with $\nu = 2/3$ as an example. (We may choose another FQH state with another filling factor)
(2) An ac magnetic field (or current) with a frequency $f$ is superposed on the static magnetic field (or current).
(3) The diagonal resistance $R_{xx}$ is measured with varying the frequency $f$. Then $R_{xx}$ depends upon $f$ and the critical frequency $f_0$ appears as in Fig.11.1.

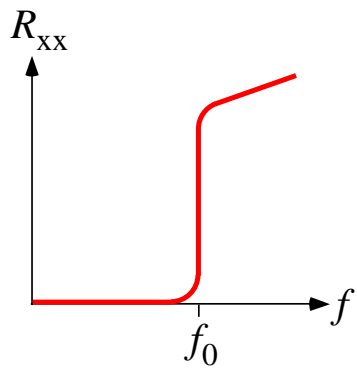

**Fig.11.1** Diagonal resistance $R_{xx}$ versus $f$

As examined in the previous chapters, there are two types of energy gaps.
For an example at the filling factor $\nu = 2/3$:

**Type 1**: The valley depth is described by the difference between the energy at $\nu = 2/3$ and one in its neighbours as in Table 5.1,

$$|\Delta \varepsilon_{\pm}| = Z/12 \qquad \text{at} \quad \nu = 2/3 \tag{11.1}$$

**Type 2**: The excitation-energy is the energy which is necessary to destroy a nearest electron-pair. The value is given by Eq.(5.85) as

$$\Delta E_{\text{excitation \#1}} \approx (1/3)Z \qquad \text{at} \quad \nu = 2/3 \tag{11.2}$$

We numerically calculate these two values for the experiment [34, 35]. The value $|\Delta\varepsilon_+|$ is obtained by the relation (7.24a). Figure 11.2 shows the diagonal resistance versus magnetic field strength where we can see the vanishing region of the diagonal resistance. The vanishing region appears between the magnetic field strength $B_3$ and $B_4$ as

$$B_3 \approx 6[\text{T}], \quad B_4 \approx 6.3[\text{T}] \tag{11.3}$$

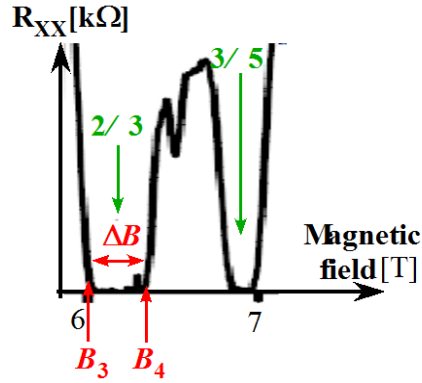

**Fig.11.2** Experimental results of diagonal resistance $R_{xx}$ near $\nu = 2/3$ in Ref. [34, 35]

Equation (7.24a) gives the value $|\Delta\varepsilon_+|$ by

$$|\Delta\varepsilon_+| = \hbar e (B_4 - B_3)/(2m^*). \tag{11.4}$$

The experiments [34, 35] are carried out for the quasi-2D electron system in a GaAs/AlGaAs quantum-well. The effective mass $m^*$ is about 0.067 times electron mass in GaAs. Substitution of Eq.(11.3) into Eq.(11.4) yields

$$\begin{aligned}|\Delta\varepsilon_+| &\approx 1.055 \times 10^{-34} \times 1.602 \times 10^{-19} \times (0.3)/(2 \times 0.067 \times 9.109 \times 10^{-31}) \\ &\approx 4.154 \times 10^{-23} [\text{J}]\end{aligned} \tag{11.5}$$

From Eqs.(11.1) and (11.2), the value of $\Delta E_{\text{excitation \#1}}$ is four times $|\Delta\varepsilon_+|$. Accordingly

$$\Delta E_{\text{excitation \#1}} \approx 16.62 \times 10^{-23} [\text{J}] \tag{11.6}$$

for the experiment [34, 35]. It is noteworthy that the values of these energy gaps depend upon the properties of quantum Hall device such as, size, shape, thickness of the electron channel, quality and so on.

There are two possible mechanisms increasing $R_{xx}$ as follows:

**Type 1** The filling factor $\nu$ deviates from 2/3 locally by the excitation via the oscillating magnetic field. This excitation occurs when the energy due to the oscillation is beyond the valley energy $|\Delta\varepsilon_+|$. So the critical frequency is given by

$$f_{10} = |\Delta\varepsilon_+|/(2\pi\hbar) \tag{11.7a}$$

The electric current are scattered by the local states with $\nu \neq 2/3$. Thereby the diagonal resistance become large for $f > f_{10}$.

**Type 2** The superposed oscillating field breaks many nearest-electron-pairs. Then the many excited electrons appear which produce the abrupt increment of $R_{xx}$. In this case the critical frequency $f_{20}$ is given by

$$f_{20} = \Delta E_{\text{excitation \#1}}/(2\pi\hbar) \tag{11.7b}$$

Thus there are two causes (Type1 and Type2) which produce the abrupt increment of the diagonal resistance. The present author doesn't conclude now which mechanism in the two types produces the increment of $R_{xx}$ mainly. We estimate the critical frequencies for the two types:

**Type 1**: $f_{01} = |\Delta\varepsilon_+|/(2\pi\hbar) \approx 62.7 \times 10^9 = 62.7\,\text{GHz}$ for $\nu = 2/3$ (11.8a)

**Type 2**: $f_{02} = \Delta E_{\text{excitation \#1}}/(2\pi\hbar) \approx 251 \times 10^9 = 251\,\text{GHz}$ for $\nu = 2/3$ (11.8b)

If an ac magnetic field is difficult to be applied in the proposed experiment, an electromagnetic irradiation instead of the ac magnetic field can be used [92, 93]. There is one more alternate method as follows: In an ordinal measurement of the diagonal resistance, a dc current is applied between the source and drain of the device, and the dc potential (diagonal) voltage are measured. In the present proposal we additionally superpose an ac current with a frequency $f$ between the source and drain. Under the superposition, the dc component of the potential voltage versus dc-current is detected with changing of $f$. Thereby we can get the frequency dependence of the diagonal resistance (namely dc voltage / dc current).

## 11.2 Tunneling through a narrow barrier in a new type of quantum Hall device

The value of the diagonal resistance is almost zero at an IQH state. This fact indicates that almost all the electrons are not scattered by lattice vibrations and impurities in an IQH state. That is to say the coherent length of electron becomes very long in an IQH state. For the fractional quantum Hall states with specific filling factors 1/3, 2/3 etc., the diagonal resistance is also almost zero [34, 35].

This property makes it possible to observe a tunnelling effect in a quantum Hall device. In the previous articles [94, 95] the tunneling device was considered, but there are some difficulties in it. So we remove the defects. The improved device is illustrated in Figs.11.3, 11.4 and 11.5 for three types of the device, respectively [96]. There is the potential barrier in the central part of each device.

(Type 1) A narrow insulating region is inserted in the middle of the conducting layer as shown in Fig.11.3. The length of the narrow barrier should be smaller than 2nm to realize a tunnelling effect. It would not be easy to fabricate this type of device.

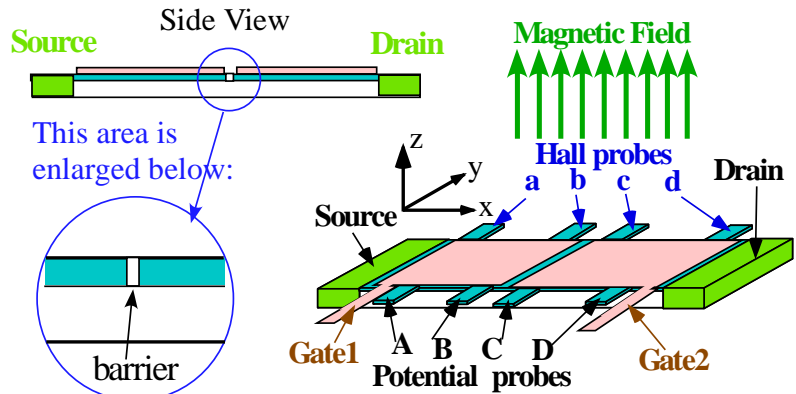

Fig.11.3: (Type 1) There is a narrow potential barrier near the center of electron channel

(Type 2) The second type of quantum Hall device is illustrated in Fig.11.4 where there are three gates. These gates are put on top of the conducting layer. By adjusting the voltage of the Gate 0, we can set an appropriate potential barrier. It would be easier to fabricate a device with type 2 than type 1.

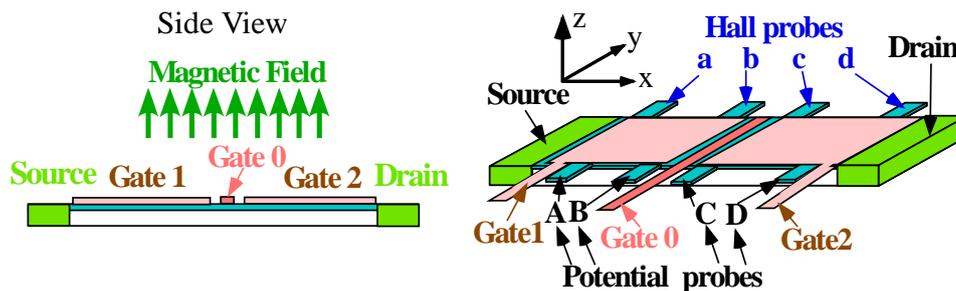

Fig.11.4: (Type 2) Central-gate (Gate 0) produces the potential barrier.

(Type 3) The device is schematically drawn in Fig.11.5. There is an ultra-thin part in the central part of the electron channel as shown in the enlarged side view. The ultra-thin part connects the two quasi-2D electron systems to each other. This type of junction is a familiar one in the superconducting quantum interference device (SQUID). When the electric current exceeds some critical value, this ultra-thin part plays a role of potential barrier. A device with type 3 may be made more easily than type 1.

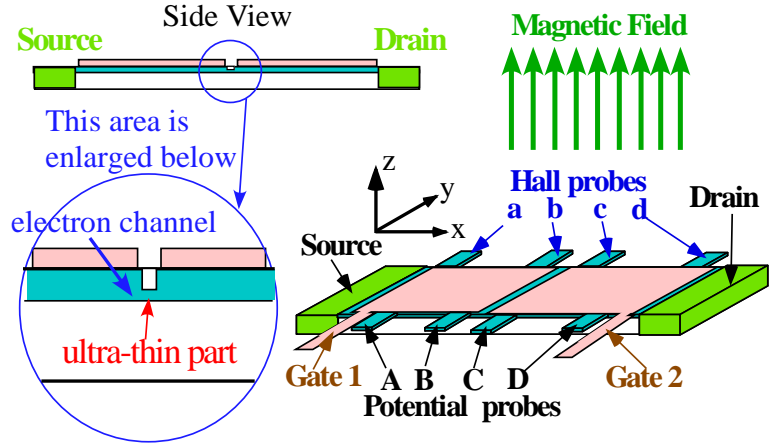

Fig.11.5: (Type 3) An ultra-thin part plays a role of potential barrier for slightly large current.

We propose a new experiment to find a tunnelling effect in an IQH or FQH state by employing the devices shown in Fig.11.3 - 11.5. This experiment is carried out in the following process:

1) Magnetic field is applied in the direction perpendicular to the quasi-2D electron system. The strength of the applied magnetic field is adjusted to yield an IQH or an FQH state. Also, by adjusting two voltages of Gate 1 and Gate 2 respectively, the filling factors in both sides of the barrier are set to be the same value.
2) Next, an oscillating magnetic field is superposed on the dc-magnetic field. (We may apply ac-current in addition to the constant current instead of an oscillating magnetic field.) The frequency of the oscillation is expressed by the symbol $f$.
3) Then we measure the voltage $V$ between the potential probes B and C versus electric current $I$. The observation of $(I,V)$ curve is expected to give us interesting phenomena of the tunnelling effect.

When the electric current flows beyond the narrow potential barrier, the quasi-particle tunnels from the higher energy position to the lower one accompanying a stimulated emission of photon. The stimulated emission is induced by the ac magnetic field with the frequency $f$. This tunnelling phenomenon is schematically drawn in Fig.11.6. Therein the energy difference of the quasi-particle from the high position to the low position is the value $V|Q|$ where $Q$ indicates the electric charge of the quasi-particle. The emitted energy of a photon is equal to $2\pi \hbar f$. Because the total energy is conserved in this tunnelling phenomenon, the voltage is determined by

$$V = 2\pi \hbar f / |Q| \qquad (11.9)$$

When the electric current becomes large, the stimulated emission yields multi-photons [96]. Therefore the measured voltage $V_n$ is equal to $n$ times $V$ which is given by

$$V_n = n \times 2\pi \hbar f / |Q| \qquad \text{for} \quad n = 1,2,3,\cdots \qquad (11.10)$$

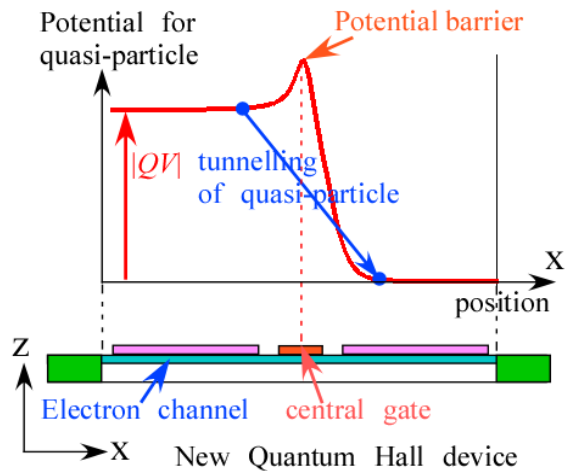

Fig.11.6 Tunnelling of quasi-particle beyond a potential barrier

Then many voltage steps are observed in the $(I,V)$ curve. This phenomenon is reminiscent of the ac-Josephson effect. The appearance of the tunneling effect needs a long coherent-length of electron. Therefore it is required that the diagonal resistances in both sides of the potential barrier are nearly equal to zero. The vanishings are confirmed by detecting the four voltages $V_{AB}$, $V_{CD}$, $V_{ab}$ and $V_{cd}$ between potential probes A and B, between C and D, between Hall probes a and b, and between c and d, respectively. All the voltages $V_{AB}$, $V_{CD}$, $V_{ab}$ and $V_{cd}$ should be nearly equal to zero. Also the voltage between probes B and b is required to be equal to that between probes C and c, because the filling factor in the left side of the potential barrier should be equal to that in the right side. These conditions must be satisfied in whole measurements for the various values of the electric current $I$. We next examine the $(I,V)$ curve for the two cases namely IQHE and FQHE.

 **Case A: Integer quantum Hall state**

The integer quantum Hall effect is caused by discrete Landau levels in a quasi-2D electron system. So the elementary charge $e$ transfers across the potential barrier in the new experiment proposed here. Then the $(I,V)$ curve behaves as the dashed curve in Fig.11.7 which has a stair form derived from Eq.(11.10).

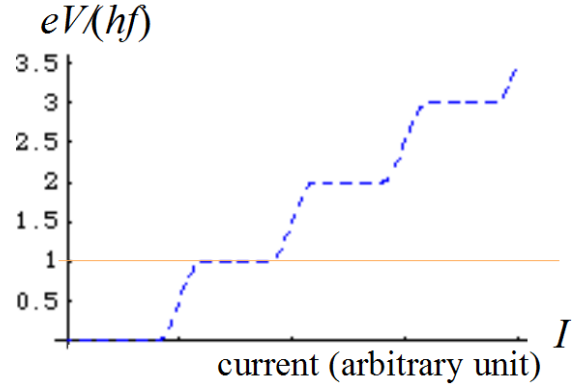
Fig.11.7 Dashed curve expresses the (*I*, *V*) curve in the Case A.
(Case A) Integer quantum Hall effect

**Case B: Fractional quantum Hall state**

R. B. Laughlin introduced a quasi-particle with fractional electric charge and proposed a trial wave function [19]. The quasi-particle and quasi-hole have fractional charges and obey fractional statistics [21-23] and [97]. Laughlin's theory suggests that the charge of the quasi-particles across the potential barrier is $\nu e$ at the fractional number $\nu$. We call it CaseB-1.

(Case B-1) The theory with fractional charges (Laughlin's theory):
$$Q = \nu e \qquad (11.11a)$$
$$V = 2\pi\hbar f/(\nu e) \longrightarrow eV/(hf) = 1/\nu \qquad (11.11b)$$

Accordingly the value of the voltage step $V$ changes with the different values of $\nu$.

Next we examine the case of the composite fermion theory which is introduced by J. K. Jain [24] and is developed by many physicists [98], [99]. The composite fermion consists of an electron bound to an even number of magnetic flux quanta. Therefore the composite fermion has the elementary charge. That is to say the charge of the quasi-particles across the potential barrier is $e$ at any filling factor $\nu$. We call it CaseB-2.

(Case B-2) Composite fermion theory :
$$Q = e \qquad (11.12a)$$
$$V = 2\pi\hbar f/e \longrightarrow eV/(hf) = 1 \qquad (11.12b)$$

In this book we have developed a theory in chapters 3-10 to explain the fractional quantum Hall effect. The Coulomb interaction acts between two electrons. The electron pair has a large binding energy at the specific filling factors. The binding of the electron pair is so strong that the pairing is kept in the tunnelling process without collapse. So, the charge of the quasi-particles across the potential barrier may be equal to $2e$. We call it CaseB-3.

(Case B-3) Present theory :

$$Q = 2e \qquad (11.13a)$$
$$V = 2\pi\hbar f/(2e) \longrightarrow eV/(hf) = 1/2 \qquad (11.13b)$$

Thus the $(I,V)$ curves are different from each others among the three theories. The theoretical curves are illustrated in Fig.11.8 for the three cases.

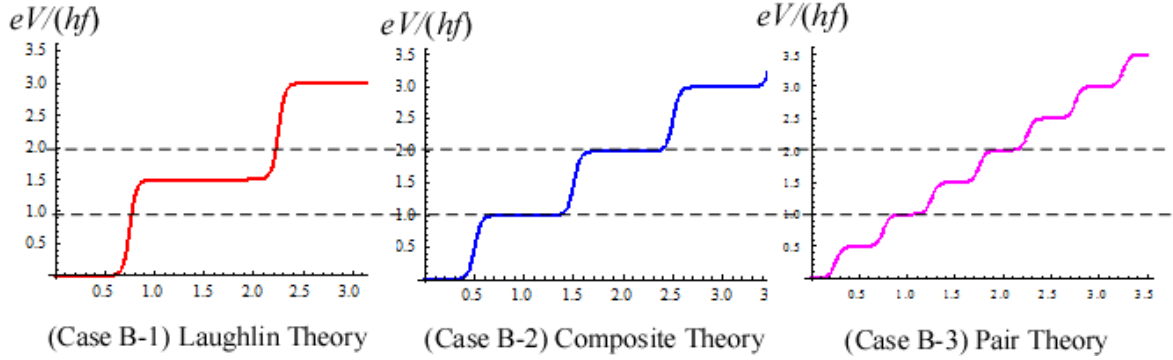

(Case B-1) Laughlin Theory     (Case B-2) Composite Theory     (Case B-3) Pair Theory

Fig.11.8: Behaviours of voltage steps for three theories at the filling factor 2/3
The horizontal axis indicates the electric current (arbitrary unit)

If the experiment proposed above is carried out and gets the curve of the voltage versus current, then the experimental curve can distinguish the three cases as in Fig.11.8. Therein the following seven conditions should be satisfied in the experiment in order to find the voltage steps.

(Condition 1)
We should choose the value of $f$ much smaller than the energy gap of the FQH state in order to produce no excitations.
$$fh \ll |\Delta\varepsilon_G| \qquad \text{namely} \quad f \ll |\Delta\varepsilon_G|/h \qquad (11.14)$$
For example, $|\Delta\varepsilon_+|/h$ is about 62.7 GHz at $\nu = 2/3$ for the experiment [34, 35] as estimated in Eq.(11.8a). Because the value $|\Delta\varepsilon_+|/h$ varies from sample to sample, the frequency $f$ should be chosen to satisfy Eq.(11.14) for a device used in the tunnelling experiment. When we apply $f = 10$ GHz as an example, the voltage step is 62.07 µV for the Case B-1, 41.38 µV for the Case B-2, and 20.69 µV for the Case B-3 respectively.

(Condition 2)
The thermal excitation energy should be lower than the energy $V|Q| = 2\pi\hbar f$ not so as to disturb the tunnelling process.
$$k_B T \ll fh \longrightarrow T \ll fh/k_B \qquad (11.15)$$
where $T$ is the temperature and $k_B$ is the Boltzmann constant. In the case of $f = 10$ GHz,
$$T \ll fh/k_B = 0.48 \text{ K} \qquad (11.16)$$

Accordingly the temperature should be cooled lower than a few 10 mK.

(Condition 3)
Figure 11.9 shows the top view of the device. The current flows along the x-direction and the Hall voltage yields along the y-direction in Fig.11.9 (see also Figs.11.3-5).

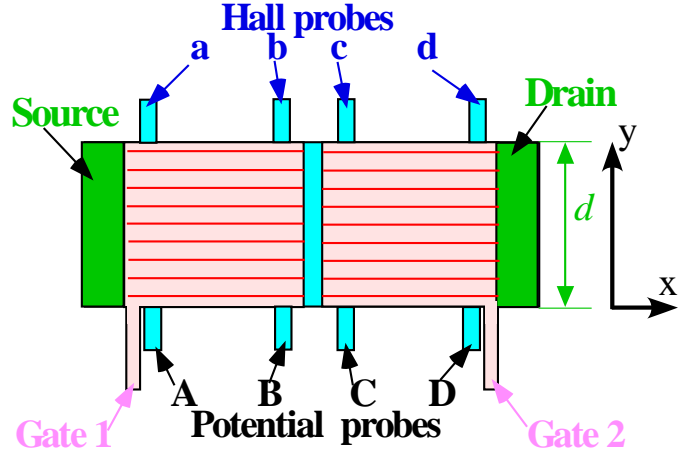

Fig.11.9 Top view of the tunneling device

There is a potential barrier in the central part of the device.
Horizontal red lines indicate the centres of the y-direction for many Landau orbitals with $L=0$. The intervals between the red lines are enlarged to bring them into view. The width of the device is expressed by the symbol $d$.

In the present experiment both the diagonal resistances $R_{\text{left}}$, $R_{\text{right}}$ in the left and right area should be almost zero. The vanishings of $R_{\text{left}}$ and $R_{\text{right}}$ are observed by detecting the voltages $V_{\text{AB}}$ and $V_{\text{CD}}$, respectively. So the following conditions are required:

$$V_{\text{AB}} \approx 0, \quad V_{\text{CD}} \approx 0 \tag{11.17}$$

(Condition 4)
The Hall voltage of the left area is measured between the probes B and b. Also that of the right area is measured between the probes C and c. The position-dependence of the electric potential is illustrated in Fig.11.10. Therein $V_{\text{Bb}}(y)$ has the argument $y$ and indicates the electric potential in the left side of the potential barrier at the position $y$. Also $V_{\text{Cc}}(y)$ indicates the electric potential in the right side of the potential barrier at the position $y$. The values of $V_{\text{Bb}}(0)$, $V_{\text{Cc}}(0)$, $V_{\text{Bb}}(d)$ and $V_{\text{Cc}}(d)$ in Fig.11.10 indicate the potentials at the probes B, C, b and c, respectively:

$$V_{\text{Bb}}(0)=V_{\text{B}}, \quad V_{\text{Cc}}(0)=V_{\text{C}}, \quad V_{\text{Bb}}(d)=V_{\text{b}}, \quad V_{\text{Cc}}(d)=V_{\text{c}} \tag{11.18}$$

where $d$ is the width of the device as in Fig.11.9.

The left panel of Fig.11.10 shows the case of $V_{\text{b}}-V_{\text{B}} \neq V_{\text{c}}-V_{\text{C}}$ which is named Case I. The right panel shows the Case II where $V_{\text{b}}-V_{\text{B}} = V_{\text{c}}-V_{\text{C}}$. In the Case I, $V_{\text{Cc}}(y)-V_{\text{Bb}}(y)$ varies with the change of the coordinate $y$ as in the left panel of Fig.11.10. In the Case II, the potential difference $V_{Cc}(y) - V_{Bb}(y)$ is equal to $V_C - V_B$ for any position $y$ as in the right panel of Fig.11.10. In this case, the stimulated emission at any y-position is produced by the oscillation with the same frequency. The property may produce the tunneling phenomenon like ac-Josephson effect.

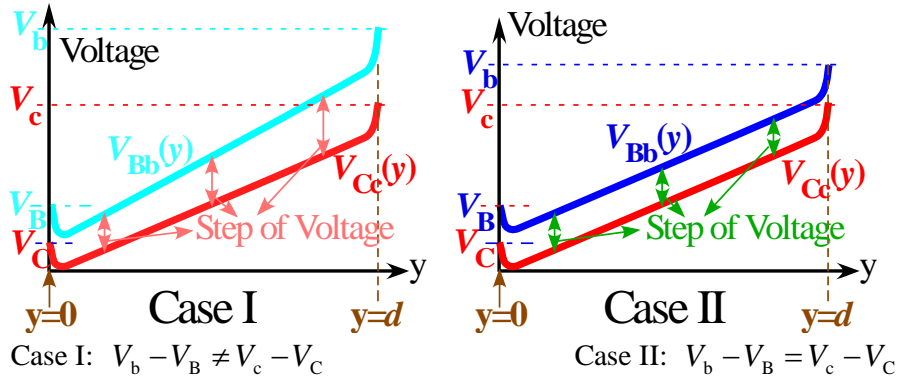

Case I: $V_b - V_B \neq V_c - V_C$    Case II: $V_b - V_B = V_c - V_C$

Fig.11.10 Position-dependence of electric potential in both side of the potential barrier
   Blue and sky-blue curve indicate the electric potential in the left side of the potential barrier.
   Red curve indicates the electric potential in the right side of the potential barrier.

Consequently the following condition is required in the present experiment.
$$V_b - V_B = V_c - V_C \quad (11.19)$$
That is to say the Hall voltage at the left area of the potential barrier, $V_b - V_B$, is equal to that at the right area. Therefore the filling factor at the left area is the same as that at the right area. This condition can be satisfied by adjusting both voltages of the gates 1 and 2.

(Condition 5)
If there are many impurities and lattice defects in the device, the coherent length becomes very short. Then the tunnelling effect is disturbed by these impurities and lattice defects. Therefore it is necessary that the device is of good quality to ensure a sufficiently long coherent length.

(Condition 6)
The diagonal resistance in the FQH states must be sufficiently small for finding the tunnelling effect. Figure 11.11 schematically indicates a comparison between the diagonal resistance and $(I, V)$ curve. If the device has a small diagonal resistance shown by the lower straight line, then we can observe the tunnelling effect. On the other hand if the device has a large diagonal resistance shown by dashed line, the tunnelling effect will be masked by the large diagonal resistance. Accordingly the quantum Hall device must be fabricated so as to have an ultra high mobility.

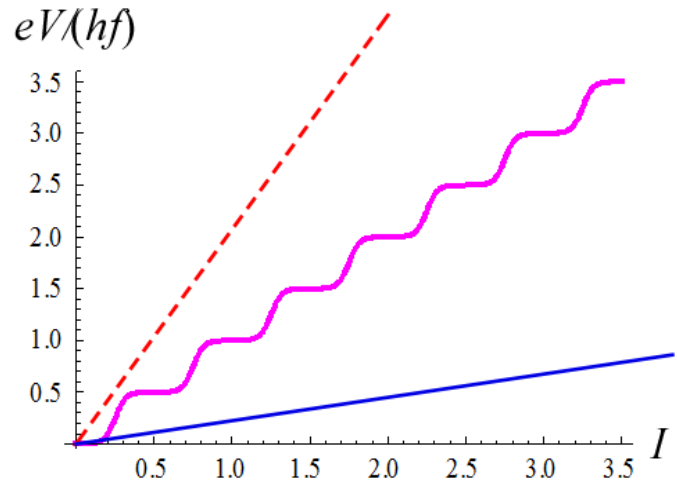
Fig.11.11 $(I,V)$ curve of a new experiment
Lower straight line indicates Ohm's law with the small diagonal resistance.
Dashed line indicates Ohm's law with the large diagonal resistance.

(Condition 7)
As discussed in section 8.3 of Chapter 8, the confinement of the quantum Hall resistance becomes weak for a small size of the device at a fractional (not integer) filling factor. Accordingly it is necessary to use an appropriate size of device in order to have a large binding energy at the specific filling factors.

The seven conditions should be satisfied to find the tunneling effect. If the stair-like curve is observed in the experimental $(I,V)$ data, the voltage of the step height determines the transfer charge producing the tunneling effect. Thereby we can know which curve among the three curves of Fig.11.8 is realized in the quasi-2D electron system.

# Chapter 12 Discussion on Traditional theories

At present the dominating theories on the FQHE are classified into two types: the one was proposed by Laughlin and has been developed by Haldane and Halperin [40-44, 27]. They have used the quasi-particle with fractional charge. The other theory is investigated by Jain. His theory employs a quasi-particle named composite fermion [45-56] which is an electron capturing an even number of flax quanta. The wave functions of these quasi-particles are approximately obtained in both theories, respectively. In section 5.12 we have given the short comments comparing the present theory with the traditional theories.

Our goal is to understand the 2D electron system which is originally described by the total-Hamiltonian of the normal electrons (not quasi-particles). Therefore it is necessary to derive the wave function of electrons from the wave function of the quasi-particles. The derivation has not been done in the traditional theories. Furthermore we find other problems which are discussed in the sections 12.2, 12.3, 12.4 and 12.5.

There is one more problem: the Hall voltage is extremely large in comparison with the potential voltage under the confinement of Hall resistance. (The Hall voltage is larger than $10^6$ times of the potential voltage in the confinement of the Hall resistance.) So it is necessary to take the electric potential gradient into consideration in order to clarify the quantum Hall effect. Almost all the theories have neglected the electric potential along the direction of Hall voltage so as to deal with a simpler Hamiltonian. Then the Hall voltage disappears. Thus the traditional theories have assumed nothing of the electric potential gradient along the Hall voltage. In Sec.12.1 we investigate whether the electric potential gradient exists in a central part of a quantum Hall device under a confinement of the Hall resistance or not.

**12.1 Electric potential along Hall voltage in the quasi-2D electron system**

We consider a thought experiment (gedanken experiment) employing a following new device: the device is shown schematically in Fig.12.1 which has an additional probe at the central part. The additional probe does not disturb the electron flow from the source to the drain because of the small thickness of the additional probe in the y-direction as in Fig.12.1. So the quantum Hall effect is observed in the new device under a strong magnetic field. Therein we examine the y-dependence of the electric potential on the blue and pink dashed lines with the coordinates $x = x_a$ and $x = x_c$, respectively.

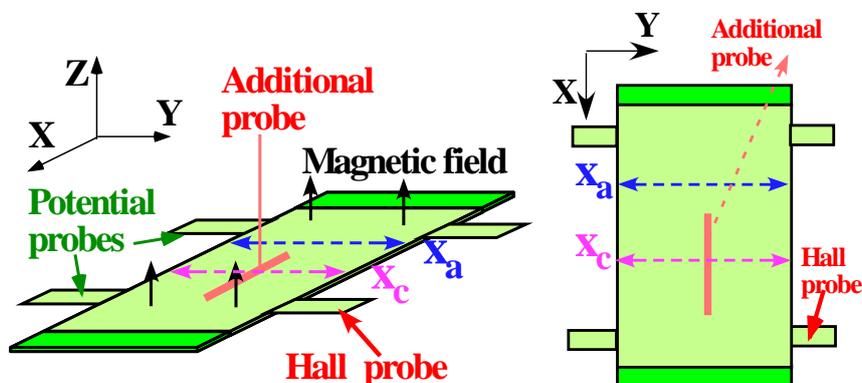

Fig.12.1 Quantum Hall device with additional probe at central part

We have already considered the y-dependence of the electric potential $U(y)$ in Fig.1.4 of Chapter 1. We examine the electric potential on the new device as Fig.12.1. Since the effect of the additional probe is small on the dashed blue line with $x = x_a$, the electric potential $U(y)$ has the shape same as in Fig.1.4 which is drawn by the blue curve in Fig.12.2.

Next we examine the electric potential on the dashed pink line with $x = x_c$ in Fig.12.1. The potential is different from that on $x = x_a$ in only the neighborhood of the additional probe with $y = d/2$. For the other y-region the potential takes a similar behavior to that of $x = x_a$. So the potential shape can be expressed by the pink curve in Fig.12.2. Thus only the neighborhood of the additional probe has the different potential from that of $x = x_a$.

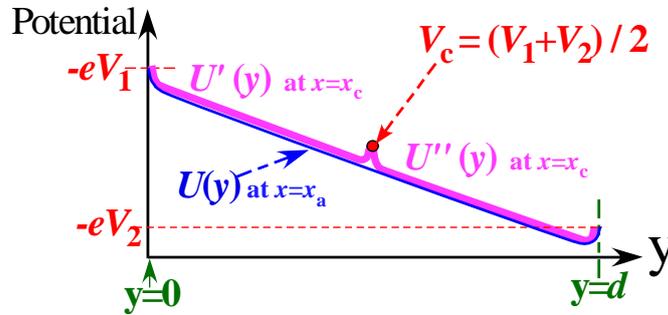

Fig.12.2 y-dependence of electric potential of the present theory

On the other hand the traditional theories employ the flat potential. That is to say, the y-dependence of the electric potential is flat in the central region of the device. So the potential on the blue dashed line in Fig.12.1 has the behavior as blue curve in Fig.12.3. Next the electric potential on the pink dashed line in Fig.12.1 is separated into two regions $0 < y < d/2$ and $d/2 < y < d$ by the additional probe. The electric potential curves in these two regions have the shape same as each other because of the following reason: The voltage of the additional probe is $(V_1 + V_2)/2$ and so the two potential differences of both sides become

$$V_1 - V_C = (V_1 - V_2)/2 \tag{12.1a}$$
$$V_C - V_2 = (V_1 - V_2)/2 \tag{12.1b}$$

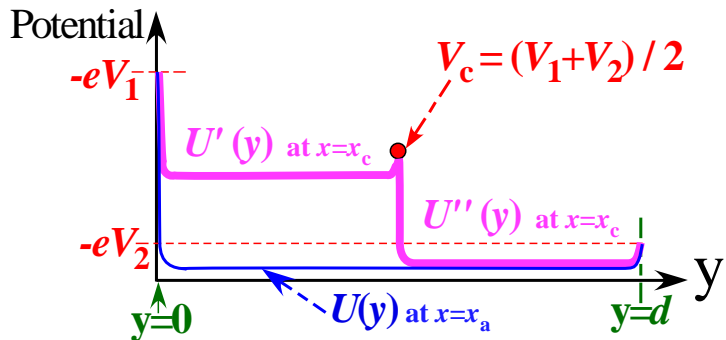

Fig.12.3 y-dependence of electric potential of the The voltage $V_1 - V_C$ is equal to $V_C - V_2$. Thereby the electric potential curves in the two regions should have the same shape. Accordingly the potential in the whole region $0 < y < d$ becomes the pink curve in Fig.12.3.

When the x-coordinate of the dashed line in Fig.12.1 changes from $x = x_c$ to $x = x_a$, the electric potential also changes. Then the potential-value in Fig.12.3 jumps from the pink to blue curve. Thus the traditional theories include the jump structure in the electric potential under the situation as in Fig.12.1. Probably this jump structure is unrealistic. The difficulty comes from the flatness assumption of the electric potential in a central region of a quantum Hall device.

Our potential used in this book varies almost continuously with changing the x-coordinate. If we compare the two potential behaviors in Figs.12.2 and 12.3, the actual electric potential may be nearly equal to the curve in Fig.12.2. That is to say the y-dependence of the electric potential cannot be ignored in the actual quantum Hall effect.

**12.2 Laughlin theory and Haldane-Halperin hierarchy theory**

At the filling factor $n = 1$, the quantum Hall resistance is dependent upon the elementary charge of electron as $R_H = 2\pi\hbar/e^2$. The function form includes the factor $e^2$ in the denominator. If the quasi particles has the fractional charge $\nu e$, then a new form of the Hall resistance is obtained by replacing $e^2$ with $(\nu e)^2$. That is to say, the quantum Hall resistance via the quasi-particles is confined to $R_H = 2\pi\hbar/(e\nu)^2$. However the experimental value of the Hall resistance is confined to $R_H = 2\pi\hbar/(\nu e \times e)$ in the FQHE. Thus the fractional charge has some difficulty.

Next we compare the charge distributions between the present theory and the Laughlin theory. The present theory starts from the ground states of the Hamiltonian $H_D$ as studied in chapters 2-9. The $\nu = 1/3$ state has the charge distribution of electrons along the y-direction as follows:

$$\rho(y) \propto \sum_{s=1}^{N} \exp\left(-(2eB/(2\hbar))(y - (3s)h/(eB\ell))^2\right) \quad \text{(for } \nu = 1/3\text{)}, \qquad (12.2)$$

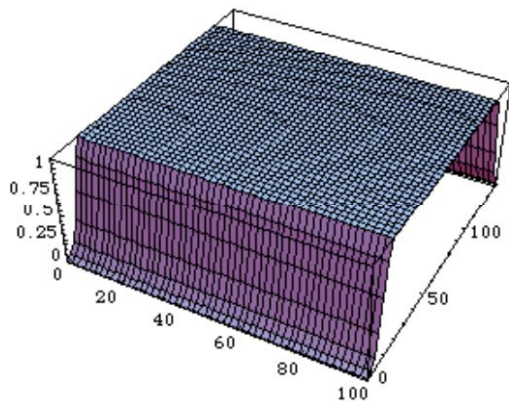

Fig. 12.4 Electron distribution of the present theory at $\nu = \frac{1}{3}$

Therein the orbitals with wave numbers of $(3s+1)2\pi/\ell$ and $(3s+2)2\pi/\ell$ are empty. The distribution along the x-direction is uniform because of the plain wave function as obtained in Eq.(1.4). When the device length is 100 nm and the strength of magnetic field is 10 T, we get the values $(2eB/(2\hbar)) = 0.01519\,(\text{nm})^{-2}$ and $h/(eB\ell) = 4.136\,\text{nm}$. In this case, the calculated distribution is shown in Fig.12.4. That figure shows that the charge distribution is sufficiently uniform. Furthermore the perturbation state has more uniform distribution than the ground state of $H_\text{D}$ because the electrons spread to all the Landau ground states via the quantum transitions due to the Coulomb interaction.

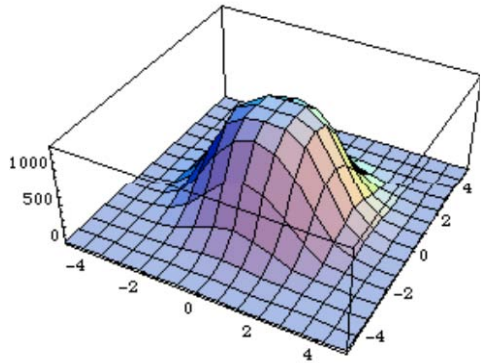

Fig.12.5 Quasi particle distribution of Laughlin wave function at $\nu = \tfrac{1}{3}$

Next we examine the charge distribution in the Laughlin theory. Figure 12.5 shows the distribution, the shape of which is round. Thus Laughlin's distribution dose not spread to a whole region of 2D electron system with a rectangular shape in a usual case. Accordingly it is necessary to superpose the wave functions with different centre positions in order to decrease the classical Coulomb energy. But they doesn't make the orthogonal set.

### 12.3 Composite Fermion theory

J.K. Jain introduced a quasi-particle named "composite fermion" which is an electron capturing even number of flux quanta. He compared his theory with the Haldane-Halperin hierarchy theory in Ref. [100] and also summarized the composite fermion theory. Therein he classified the composite fermions into sixteen types. We discuss here the states for the filling factors with the denominator smaller than 6:

(1) $\nu = n/(2n+1) \rightarrow \nu = 1/3,\ 2/5,\cdots$ :    The FQH states are the IQH states of composite fermion which is an **electron** capturing **two** flux quanta.
(2) $\nu = n/(2n-1) \rightarrow \nu = 2/3,\ 3/5,\cdots$ :    The FQH states are the IQH states of composite fermion which is an **electron** capturing **two** flux quanta. The effective magnetic field has the **opposite direction** against the applied magnetic field.
(3) $\nu = n/(4n+1) \rightarrow \nu = 1/5,\cdots$:    The FQH states are the IQH states of composite fermion which is an **electron** capturing **four** flux quanta.

(4) $\nu = 1 - n/(4n+1) \to \nu = 4/5, \cdots$: The FQH states are **combined** the $\nu = 1$ IQH state with the IQH state of composite fermion which is a **hole** capturing **four** flux quanta.

(5) $\nu = 1 + n/(4n+1) \to \nu = 6/5, \cdots$: The FQH states are **combined** the $\nu = 1$ IQH state with the IQH state of composite fermion which is an **electron** capturing **four** flux quanta.

(6) $\nu = 2 - n/(2n+1) \to \nu = 5/3, 8/5, \cdots$ The FQH states are **combined** the $\nu = 2$ IQH state with the IQH state of composite fermion which is a **hole** capturing **two** flux quanta.

(7) $\nu = 2 - n/(2n-1) \to \nu = 4/3, 7/5, \cdots$ The FQH states are **combined** the $\nu = 2$ IQH state with the IQH state of composite fermion which is a **hole** capturing **two** flux quanta. The effective magnetic field has the **opposite direction** against the applied magnetic field.

(8) $\nu = 2 - n/(4n+1) \to \nu = 9/5, \cdots$ The FQH states are **combined** the $\nu = 2$ IQH state with the IQH state of composite fermion which is a **hole** capturing **four** flux quanta.

We study the following five examples in more details.

1) FQH state at $\nu = 4/3$

In the article [47] the $\nu = 4/3$ FQH state is constructed by combining the $\nu = 2$ IQH state with the composite fermion state of hole for $\nu = -2/3$ as in Fig.12.6. Therein the black dots on the green sheet indicate the electrons in the $\nu = 2$ IQH state. The composite fermions of hole are expressed by the white circles on the yellow sheet, each of which is bound with two flux quanta as in Fig.12.6.

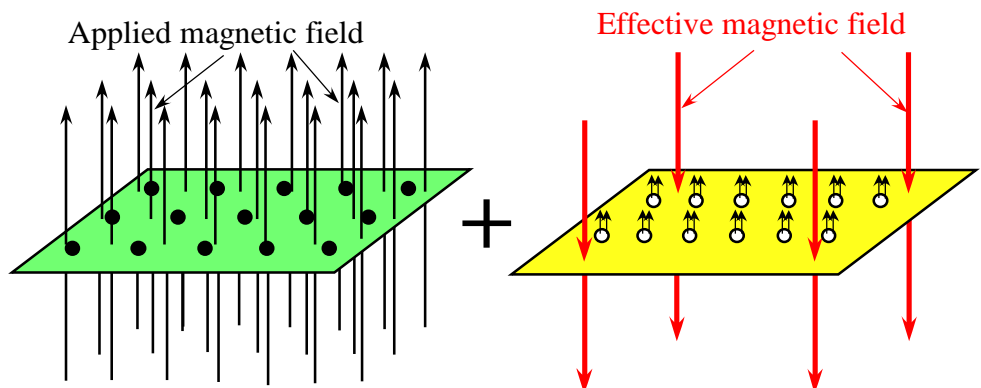

Fig.12.6 Composite fermion theory for $\nu = 4/3$

The effective magnetic field is expressed by the red arrows, the direction of which is opposite against the applied magnetic field. The total filling factor is the sum of $\nu = 2$ and $\nu = -2/3$. Accordingly the filling factor of electrons becomes $\nu = 4/3$.

2) FQH state at $\nu = 5/3$

The $\nu = 5/3$ FQH state is explained by the combination of the $\nu = 2$ IQH state and the composite fermion state of hole with $\nu = -1/3$. Each hole is bound with two flux quanta and the effective magnetic field is parallel to the applied magnetic field as shown in Fig.12.7.

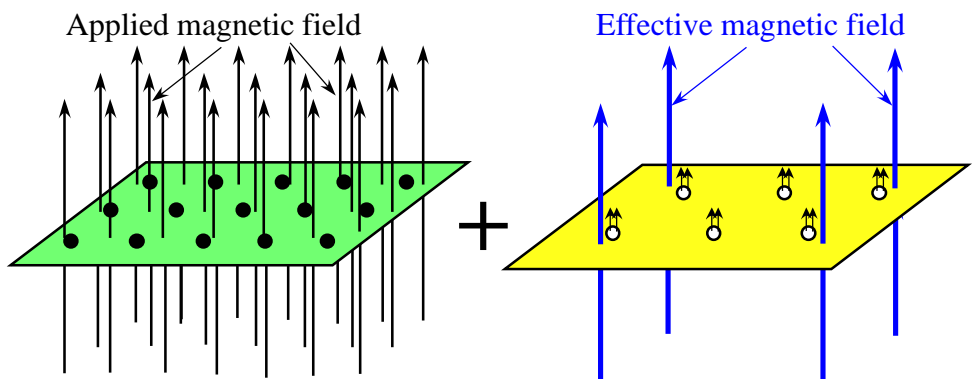
Fig. 12.7 Composite fermion theory for $\nu = 5/3$

3) FQH state at $\nu = 6/5$

The $\nu = 6/5$ FQH state is produced by combining the $\nu = 1$ IQH state with the composite fermion state with $\nu = 1/5$. Therein the electron of the $\nu = 1$ IQH state is expressed by the black circles on the sky-blue sheet in Fig. 12.8. The residual electrons are expressed by the blue dots on the pink sheet, each of which is bound by the four flux quanta as seen in Fig.12.8 schematically. That is to say, some electrons are unbound with magnetic flux quanta and the other electrons are bound with flux quanta. However all the electrons exist in the same conducting thin layer and their wave functions are overlapping with each other.

In the many-body problem all the wave functions of electrons should satisfy the anti-symmetric relation. Also all the electrons should be affected by the same magnetic field. Accordingly the combination of the $\nu = 1$ IQH state and the $\nu = 1/5$ composite fermion state has some difficulty.

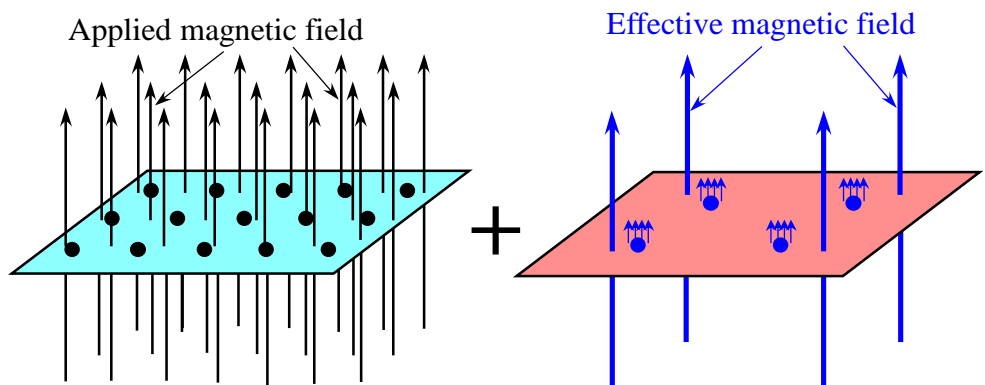
Fig. 12.8 Composite fermion theory for $\nu = 6/5$

4) FQH state at $\nu = 9/5$

The $\nu = 9/5$ FQH state is created by combining the $\nu = 2$ IQH state with the composite fermion state of hole for $\nu = -1/5$ as illustrated in Fig.12.9.

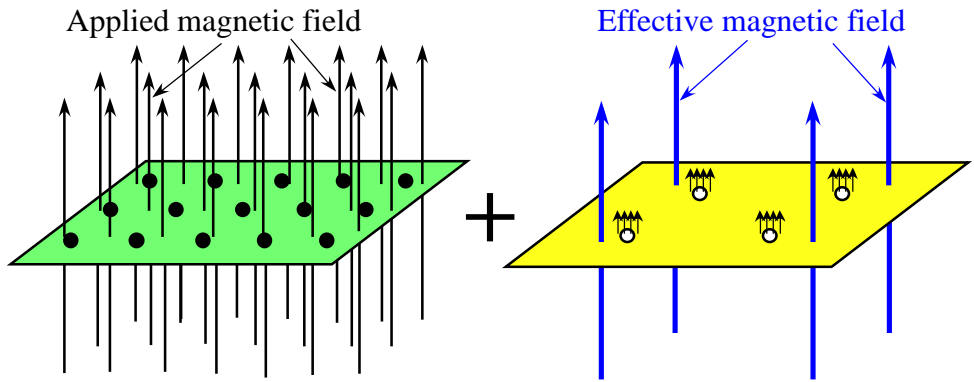

Fig.12.9 Composite fermion theory for $\nu = 9/5$

Therein the black dots on the green sheet indicate the electrons in the $\nu = 2$ IQH state. The composite fermions of hole are expressed by the white circles on the yellow sheet, each of which is bound with four flux quanta as in Fig.12.9.

5) FQH state at $\nu = 4/5$

The $\nu = 4/5$ FQH state is produced by combining the $\nu = 1$ IQH state with the composite fermion state of hole with $\nu = -1/5$. Therein the electron of the $\nu = 1$ IQH state is expressed by the black circles on the sky-blue sheet in Fig. 12.10. The composite fermions of hole are expressed by the white circles on the yellow sheet, each of which is bound by the four flux quanta.

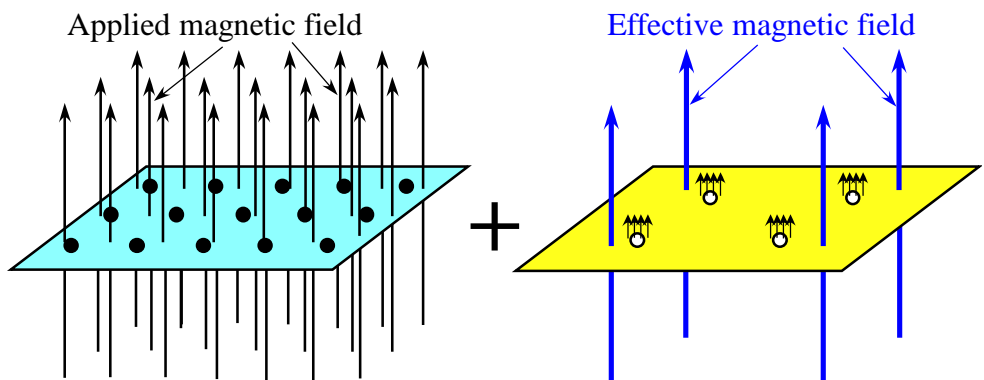

Fig. 12.10 Composite fermion theory for $\nu = 4/5$

Thus the composite fermion theory uses many different kinds of quasi-particles and different directions of the effective magnetic field. When the filling factor is changed by adjusting the gate voltage (or the applied magnetic field strength), the quasi-particle varies from hole to electron, or the direction of the effective magnetic field changes from parallel to anti-parallel. Also the number of the bound flux-quanta changes. These complicated assumptions are very artificial.

## 12.4 Effective magnetic field and flux quantization

The external magnetic field $(0,0,B)$ is applied to a quantum Hall device. The strength of the field can be varied continuously and the strength $B$ is uniform at all positions inside the quasi-2D electron system. In the composite fermion theory some magnetic flux quanta are captured by the composite fermions and the other magnetic flux yields the effective magnetic field. Then the applied magnetic field is equal to the sum of the effective field $(0,0,B_{eff})$ and the bound field $(0,0,B_{bound})$.

$$(0,0,B) = (0,0,B_{eff}) + (0,0,B_{bound}) \tag{12.3}$$

The composite fermion theory combines the IQH state of the normal electrons with the IQH state of the composite fermions for the cases (4) - (8) as explained in Sec.12.3. Therefore the theory employs a uniform magnetic field $B$ for the electrons and a uniform effective field $B_{eff}$ for the composite fermions. Thereby $B$ and $B_{eff}$ are constant everywhere on a quantum Hall device. Accordingly the magnetic field $B_{bound}$ is also uniform because of Eq.(12.3).

In the composite fermion theory the bound flux does not affect the other composite fermions. Only the effective field affects the composite fermions. That is to say, the bound flux cannot penetrate the wave function of the other composite fermions. Equation (12.3) means that $B_{bound}$ has a uniform value for every position on a quantum Hall device. Also the wave function of each composite fermion spreads on a quantum Hall device. Therefore it is difficult to assume that $B_{bound}$ does not affect the composite fermions.

Furthermore the quasi 2D-electron system has no special boundary as in a superconducting ring. Although there is the property that the number of Landau states with the lowest level is equal to the number of the flux quanta, this property does not mean the quantization of the magnetic flux. In fact the property appears also in the three dimensional electron-system under the uniform magnetic field when we consider a single momentum level along the magnetic field direction. There is no quantization of magnetic flux in the 3-D electron system. Consequently the equality of the two numbers does not mean the flux quantization.

## 12.5 Spin-polarization of composite fermion theory

J. K. Jain examined the spin-polarization in the CF theory. He wrote in Ref [100]: "For spinful composite fermions, we write $\nu^* = \nu_\uparrow^* + \nu_\downarrow^*$, where $\nu_\uparrow^*$ and $\nu_\downarrow^*$ are the filling factors of up and down spin composite fermions. The possible spin polarizations of the various FQHE states are then predicted by analogy to the IQHE of spinful electrons. For example, the 4/7 state maps into $\nu^* = 4$, where we expect, from a model that neglects interaction between composite fermions, a spin singlet state at very low Zeeman energies (with $\nu^* = 2 + 2$), a partially spin polarized state at intermediate Zeeman energies ($\nu^* = 3 + 1$), and a fully spin polarized state at large Zeeman energies ($\nu^* = 4 + 0$)."

The CF cyclotron energy is proportional to $\sqrt{B}$ and the Zeeman energy is proportional to the magnetic field $B$ as explained in the article [100]. Then the CF energy $\varepsilon_{CF}$ is equal to

$$\varepsilon_{CF}(n\uparrow) = \alpha(n-\tfrac{1}{2})\sqrt{B} + \beta B \qquad \varepsilon_{CF}(n\downarrow) = \alpha(n-\tfrac{1}{2})\sqrt{B} - \beta B \qquad (12.4)$$

where $\alpha$ and $\beta$ are the coefficients. Therein $n = \nu^* = L^* + 1$ where $L^*$ is the CF Landau level number ($L^* = 0,1,2,3\cdots$). When the filling factor $\nu$ of electrons is smaller than 2/3, $\nu$ is given by the CF level number $\nu^*$ and the number of attached flux quanta $2p$ as follows:

$$\nu = \nu^*/(2p\nu^* \pm 1) \qquad (12.5)$$

When the sign in the denominator is plus the effective magnetic field is parallel to the applied magnetic field. So the coefficient $\beta$ becomes positive. On the other hand the effective magnetic field has opposite direction against the applied field for the minus sign in Eq.(12.5). Then $\beta$ is negative. We obtain the sum of the CF cyclotron energy and the Zeeman energy for various CF levels and CF spins. The results are shown in Fig.12.11 where we find the crossing points between the energies for spin-up and down states. The ratio of the field strengths at the crossing points is 1:4:9:16 and so on.

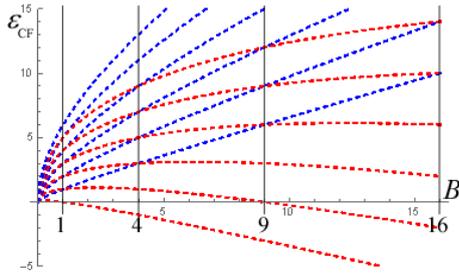
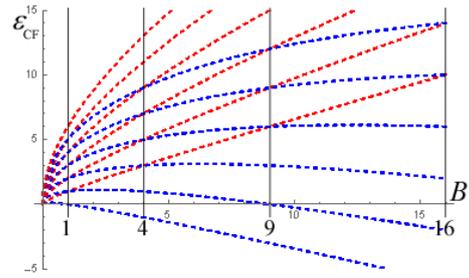

Fig.12.11a Energy of CF for $\nu^*/(2p\nu^*+1)$  Fig.12.11b Energy of CF for $\nu^*/(2p\nu^*-1)$
Blue dashed curves for up-spin CF, red for down spin CF. The axes are drawn with arbitrary scale.

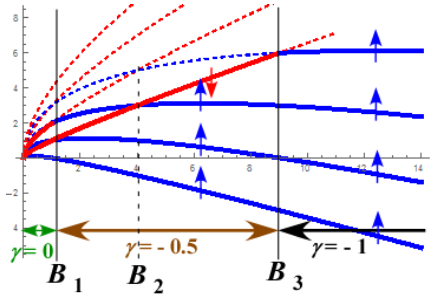
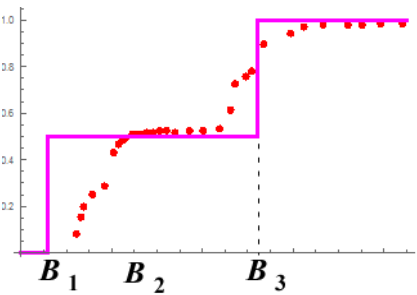

Fig.12.12a CF state at $\nu = 4/7$   Fig.12.12b Polarization of CF state at $\nu = 4/7$

For an example $\nu = 4/7$ the energy spectrum of the CF theory is given by Fig.12.11b. So the composite fermions occupy the lowest four levels which are expressed by the bold curves as in Fig.12.12a. Therein the ratio of the magnetic field strengths $B_1, B_2$ and $B_3$ is 1:4:9. Accordingly the spin-polarization $\gamma$ depends on the magnetic field strength at zero temperature as in Fig.12.12b.

It is noteworthy that the effective field is opposite against the applied magnetic field at $v = 4/7$. Let us compare the spin-polarization of the CF theory with that of the present theory at a finite temperature. Figure 12.13a shows the CF result. Our result is expressed in Fig.12.13b at $v = 4/7$. (Note: We draw these figures by ignoring the direction. The direction of the polarization in CF theory is opposite against our result.) The CF result is different from the experimental data [77] in the low magnetic field region.

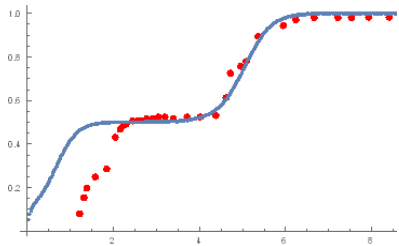 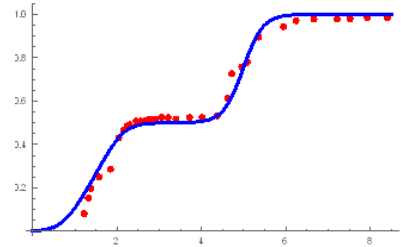

Fig.12.13a:Polarization of CF result at $v = 4/7$    Fig.12.13b:Our theory at $v = 4/7$
The dots express the experimental data [77].

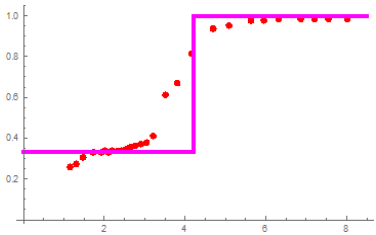 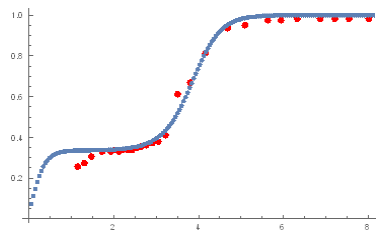 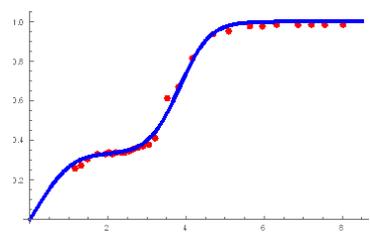

Fig.12.14a:CF result ($v = 3/5$)   Fig.12.14b:Polarization of CF result   Fig.12.14c:Our result at $v = 3/5$
The dots express the experimental data [77].

Similarly the theoretical polarization curves at $v = 3/5$ are shown in Figs.12.14a, b and c. Figures 12.14a and 12.14b express the CF results at zero temperature and a finite temperature, respectively. Figure 12.14c shows the polarization curves of our theory at a finite temperature. Our theory employs only the normal electrons without any quasi-particle. Therein the partial Hamiltonian with the strongest and second strongest interactions is diagonalized exactly. The results are in good agreement with the experimental data. On the other hand the CF results deviate from the experimental data.

Moreover the direction of CF polarization at $v = 2/3, 3/5, 4/7$, etc is opposite against that of normal electrons because the effective magnetic field is opposite against the applied field. But the direction of CF polarization at $v = 1/3, 2/5, 3/7, 4/9$, etc is the same as that of normal electrons. So it is necessary to measure the direction of the polarization especially at $v = 2/3, 3/5, 4/7$.

In Jain's scheme the $v = 4/5$ ($v = 6/5$) state is assigned to the superposition of the IQH state with $v = 1$ and the composite hole (fermion) state with $v = -1/5$ ($v = 1/5$) as mentioned in the reference [100]. Accordingly polarization behaviours of the two states resemble that of the $v = 1/5$ state in the CF theory. On the other hand the polarization behaviour of the present theory at $v = 4/5$ resembles that at $v = 4/7$ because there are four electrons per unit-configuration in both $\nu=4/5$ and 4/7 states. We can calculate the polarization versus magnetic field at $\nu=4/5$ and 6/5 by diagonalizing the $4\times 4$ matrix ($8\times 8$ matrix for considering of spin Peierls effect) as mentioned in sections 9.6 and 9.7 in Chapter 9. Thus the theoretical predictions at $\nu=4/5$ and 6/5 are quite different between the CF theory and our theory. The present author cannot find the experimental curves of the polarization versus magnetic field at $\nu=4/5$ and 6/5. If there are already raw data in experiments, it is preferable to publish the curve of the polarization versus magnetic field.

We cannot also find the experimental data of the spin direction. The effective magnetic field in the CF theory is opposite against the applied field for the filling factor $\nu = \nu*/(2p\nu*-1)$. So it is important to measure the direction of the polarization at the filling factors $\nu = 2/3, 3/5, 4/7$, etc.

# Chapter 13 Summary

  Thirty years ago the author was deeply impressed by the discovery of the integer quantum Hall effect. Thereafter the fractional quantum Hall effects were discovered which gave the difficult puzzles.

  In a 3D Hall system the electrons are affected by two forces namely the magnetic force of applied field and the electric force due to the Hall voltage. Thereby the magnetic force is balanced with the electric force. The balancing yields the relation between the magnetic field strength and the Hall voltage. As well known the relation is used to determine the magnetic field strength by measuring the Hall voltage. Thus the average orbit of electron is a straight line (not a circle).

  But many articles have expressed the electron orbital by a circle in a quantum Hall system (2D system). This view yields some confusion. The circular orbital is caused by missing the electric potential gradient along the Hall voltage. This potential gradient cannot be ignored in the central part of a quantum Hall device as discussed in Sec.12.1. The present theory takes the electric potential gradient via Hall voltage into consideration throughout this book. Then the orbital of electron is a straight line like Landau solution.

  The various phenomena of the FQHE should be derived from the total Hamiltonian with the Coulomb interaction between electrons under a strong magnetic field in a quasi-2D electron system. This opinion is the common view of physicists. The total Hamiltonian cannot be exactly diagonalized in a many body problems and so it is necessary to use some approximations. The final purpose is to understand the quasi-2D electron system. Therefore all theories need to show the approximate wave function of electrons (not quasi particles).

  The present theory in this book has investigated the FQHE without any quasi-particle on the basis of the standard treatment of many-electron gas confined in an ultra thin layer. Of course the electric field due to the Hall voltage is taken into consideration. Then we can find the ground state for the Hamiltonian $H_\mathrm{D}$. Therein the Landau states are partially filled with electrons at a fractional filling factor. The electron configuration is uniquely determined so as to have the minimum classical Coulomb energy. The ground state has the lowest energy for the Landau levels, the lowest for Zeeman energy and the lowest for the classical Coulomb energy.

  Residual Coulomb interaction $H_\mathrm{I}$ yields quantum transitions which satisfy the momentum conservation along the current direction. The number of Coulomb transitions from nearest electron (or hole) pairs is dependent sensitively upon the fractional number of the filling factor. For example, at $\nu = 2/3$ all the nearest electron pairs can transfer to all the empty states via the Coulomb interactions. The number of allowed transitions $u(\nu)$ at $\nu = 2/3$ abruptly decreases when the filling factor $\nu$ deviates slightly from $\nu = 2/3$. The limiting value $\lim_{\nu \to (2/3)\pm 0} u(\nu)$ is equal to half of $u(2/3)$. The discontinuous behaviour produces the valley structure in the energy spectrum. So the ground state energy at $\nu = 2/3$ becomes lowest with a gap in comparison with that in the neighbourhood of $\nu = 2/3$. The valley structure produces the confinement of the Hall resistance. This mechanism appears at the specific filling factors $\nu = 1/(2j+1)$, $2j/(2j+1)$, $j/(2j\pm 1)$ etc. Also there are small valleys at the non-standard FQH states with $\nu = 7/11, 4/11, 4/13, 5/13, 5/17, 6/17$ etc. as examined in Chapters 4-6. We have studied the pair energies of more distant electron pairs in Chapter 5. Thereby the valley structure yields at $\nu = 5/2, 7/2$ and so on. Thus the theoretical results are in good agreement with the experimental curves of the Hall resistance as shown in Chapter 7.

The experimental spin-polarization curves versus magnetic field have wide plateaus and small shoulders under fixing the filling factor. The shapes of polarization curves depend mainly upon the numerator of the fractional filling factor in $\nu < 1$ because the spin polarization belongs to electrons only. There are many spin-arrangements with the same eigen-energy of $H_D$. The Coulomb interaction $H_I$ yields quantum transitions among these degenerate ground states. The partial Hamiltonian has been diagonalized exactly. The eigen-states give the spin polarization versus magnetic field strength. The theoretical curve of the spin polarization has wide plateaus which reproduces the experimental data well. When we take the modulation of intervals between Landau orbitals into consideration the spin Peierls instability occurs. Thereby the small shoulders appear in the calculated polarization curves which are in good agreement with the experimental data as studied in Chapter 9. Thus the present theory has well explained the various phenomena of the FQHE.

The quasi-2D electron system has shown the amazing behaviours. The integer and fractional quantum Hall effect is one of the most fundamental phenomena in the materials science. Many physicists investigated these phenomena with enormous efforts. Probably new phenomena will be found hereafter and this branch will fascinate us.

## Appendix  Nearest-pair-energy of FQH states with $\nu = (4s \pm 1)/(6s \pm 1)$

We estimate the energy of nearest electron pairs at the filling factor $\nu = (4s \pm 1)/(6s \pm 1)$ which has an odd integer of the denominator. We first consider the fractional number $\nu = (4s+1)/(6s+1)$ in the neighbourhood of $\nu = 2/3$. The fractional number $\nu = (4s+1)/(6s+1)$ approaches $2/3$ in the limit of infinitely large integer $s$.

We draw the electron configuration with the minimum classical Coulomb energy for $s = 2$ namely $\nu = 9/13$ in Fig.A1. The electron configuration is obtained by repeating of the unit-configuration namely 13 sequential Landau states partially filled with 9 electrons. There are five *nearest electron pairs* (AB, CD, EF, GH, HI) in the unit-configuration as in Fig.A1. The number of empty states per unit-configuration is 4. The electron pair CD can transfer to all empty states. Both electron pairs AB and EF can transfer to 2/4 times number of all empty states. Also both electron pairs GH and HI cannot transfer to any empty state.

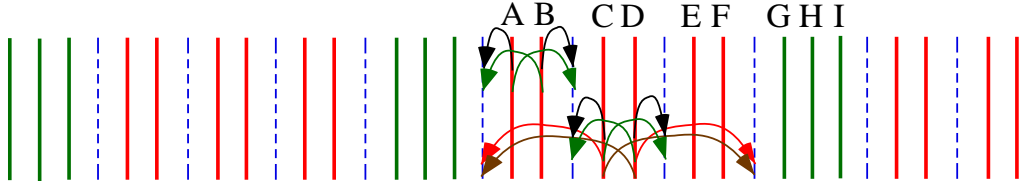

Fig.A1 Coulomb transitions of nearest neighbor electron pairs at $\nu=9/13$

In the Coulomb transitions of electron pair AB, the momentum transfer takes the values as

$$\Delta p = (13j-2)2\pi\hbar/\ell, (13j+1)2\pi\hbar/\ell \qquad \text{for } j=0,\pm1,\pm2,\cdots \qquad (A.1)$$

That is to say the allowed transition number is 2 among 13 orbitals inside the unit-configuration. The second order perturbation energy of the pair AB is described by $\varsigma_{AB}$ which is

$$\varsigma_{AB} = -(2/13)Z \qquad (A.2a)$$

Similarly, the perturbation energies of the pair CD, EF, GH and HI are denoted by $\varsigma_{CD}, \varsigma_{EF}, \varsigma_{GH}, \varsigma_{HI}$ respectively, the values of which are

$$\varsigma_{CD} = -(4/13)Z \qquad (A.2b)$$

$$\varsigma_{EF} = -(2/13)Z \qquad (A.2c)$$

$$\varsigma_{GH} = \varsigma_{HI} = 0 \qquad (A.2d)$$

The sum of the second order perturbation energies for all nearest electron pairs is given as

$$E_{\text{nearest pair}} = [-(2/13)Z - (4/13)Z - (2/13)Z] \times (N/9)$$
$$= -(8/(13 \times 9))ZN \quad \text{for} \quad \nu = 9/13 \quad \text{(A.3)}$$

Next the most uniform configuration of electrons is drawn at the filling factor $\nu = (4s+1)/(6s+1)$ for $(s=3)$ in Fig.A2.

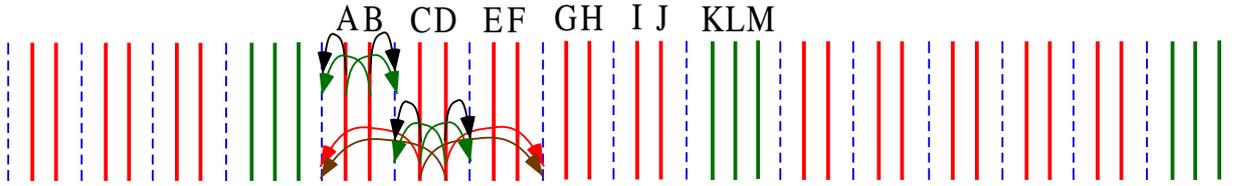

Fig.A2 Coulomb transitions of nearest neighbor electron pairs at $\nu = 13/19$

The electron configuration is obtained by repeating of the unit-configuration as in Fig.A2 (19 sequential Landau states are partially filled with 13 electrons). There are seven *nearest electron pairs* (AB, CD, EF, GH, IJ, KL, LM) in the unit-configuration. These pairs has the perturbation energies as

$$\varsigma_{AB} = -(2/19)Z, \quad \varsigma_{CD} = -(4/19)Z, \quad \varsigma_{EF} = -(6/19)Z, \quad \varsigma_{GH} = -(4/19)Z,$$

$$\varsigma_{IJ} = -(2/19)Z, \qquad \varsigma_{KL} = \varsigma_{LM} = 0 \qquad \text{for} \quad \nu = 13/19$$

(A.4)

The sum of the perturbation energies for all nearest electron pairs is obtained as

$$E_{\text{nearest pair}} = [-(2/19)Z - (4/19)Z - (6/19)Z - (4/19)Z - (2/19)Z] \times (N/13)$$
$$= -(18/(19 \times 13))ZN \qquad \text{for } \nu = 13/19 \quad \text{(A.5)}$$

Next we examine the perturbation energy for the filling factor $\nu = (4s+1)/(6s+1)$ with any integer $s$. The unit-configuration is composed of $6s+1$ Landau states which are partially occupied by $4s+1$ electrons. Therefore the number of empty states is $2s$ per unit-configuration. Accordingly

$$E_{\text{nearest pair}} = \begin{bmatrix} -(2/(6s+1))Z - (4/(6s+1))Z \cdots - (2s/(6s+1))Z \cdots \\ -(4/(6s+1))Z - (2/(6s+1))Z \end{bmatrix} \times (N/(4s+1))$$

$$= -\frac{s(s+1) + s(s-1)}{(4s+1)(6s+1)} ZN$$

$$= -\frac{2s^2}{(4s+1)(6s+1)} ZN$$

$$E_{\text{nearest pair}} = -\frac{2s^2}{(4s+1)(6s+1)} ZN \qquad \text{for } \nu = (4s+1)/(6s+1) \qquad (A.6)$$

This result leads the limiting values to

$$\nu = (4s+1)/(6s+1) \xrightarrow[s \to \infty]{} 2/3 \qquad (A.7a)$$

$$E_{\text{nearest pair}}/N \xrightarrow[\nu=(4s+1)/(6s+1),\ s \to \infty]{} -(1/12)Z \qquad (A.7b)$$

The previous result (4.18) gives

$$E_{\text{nearest pair}}/N = -(1/6)Z \qquad \text{at } \nu = 2/3 \qquad (A.8)$$

When the filling factor $\nu$ approaches $2/3$ from the right, the *nearest electron pair* energy per electron approaches $-(1/12)Z$ which is half of the energy at $\nu = 2/3$. That is to say, the energy at $\nu = 2/3$ is lower than the limiting value from the right and then yields an energy gap.

Similarly we calculate the *nearest electron pair* energy at $\nu = (4s-1)/(6s-1)$, the value of which is smaller than $2/3$. The electron configuration for the case of $s = 2$ is illustrated in Fig.A3.

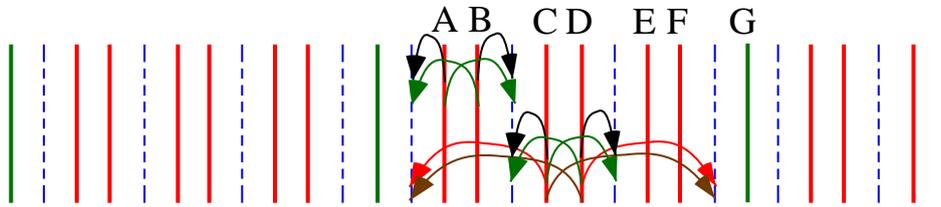

Fig.A3 Coulomb transitions of nearest neighbor electron pairs at $\nu = 7/11$

The second order perturbation energy of the pair AB, CD and EF are expressed by $\varsigma_{AB}, \varsigma_{CD}$ and $\varsigma_{EF}$ respectively which are

$$\varsigma_{AB} = -(2/11)Z, \quad \varsigma_{CD} = -(4/11)Z, \quad \varsigma_{EF} = -(2/11)Z \tag{A.9}$$

The sum of all the *nearest electron pair* energies becomes

$$\begin{aligned} E_{\text{nearest pair}} &= [-(2/11)Z - (4/11)Z - (2/11)Z] \times (1/7)N \\ &= -(8/(11 \times 7))ZN \qquad \text{for } \nu = 7/11 \end{aligned} \tag{A.10}$$

Fig.A4 shows the electron configuration with the minimum classical Coulomb energy at $\nu = 11/17$.

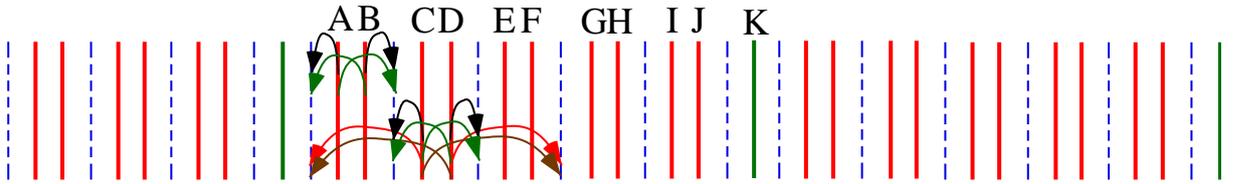

Fig.A4 Coulomb transitions of nearest neighbor electron pairs at $\nu = 11/17$

This electron configuration is obtained by repeating of the unit-configuration namely 17 sequential Landau states partially filled with 11 electrons. There are five *nearest electron pairs* (AB, CD, EF, GH, IJ) in the unit-configuration as in Fig.A4. These pairs have the perturbation energies as

$$\varsigma_{AB} = -(2/17)Z, \quad \varsigma_{CD} = -(4/17)Z, \quad \varsigma_{EF} = -(6/17)Z, \quad \varsigma_{GH} = -(4/17)Z,$$

$$\varsigma_{IJ} = -(2/17)Z \qquad \text{for } \nu = 11/17$$

(A.11)

The total perturbation energy of all the *nearest electron pairs* is equal to

$$\begin{aligned} E_{\text{nearest pair}} &= [-(2/17)Z - (4/17)Z - (6/17)Z - (4/17)Z - (2/17)Z] \times (1/11)N \\ &= -(18/(17 \times 11))ZN \qquad \text{for } \nu = 11/17 \end{aligned} \tag{A.12}$$

We calculate the perturbation energy for the filling factor $v = (4s-1)/(6s-1)$ with any integer $s$. The unit-configuration is composed of $6s-1$ Landau states which are partially occupied by $4s-1$ electrons. Therefore the number of empty states is $2s$ per unit-configuration. Accordingly

$$E_{\text{nearest pair}} = \begin{bmatrix} -(2/(6s-1))Z - (4/(6s-1))Z \cdots -(2s/(6s-1))Z \cdots \\ -(4/(6s-1))Z - (2/(6s-1))Z \end{bmatrix} \times (N/(4s-1))$$

$$= -\frac{s(s+1) + s(s-1)}{(4s-1)(6s-1)} ZN$$

$$= -\frac{2s^2}{(4s-1)(6s-1)} ZN$$

$$E_{\text{nearest pair}} = -\frac{2s^2}{(4s-1)(6s-1)} ZN \qquad \text{for } v = (4s-1)/(6s-1) \qquad (A.13)$$

The limiting values are

$$v = (4s-1)/(6s-1) \xrightarrow[s\to\infty]{} 2/3 \qquad (A.14a)$$

$$E_{\text{nearest pair}}/N \xrightarrow[v=(4s-1)/(6s-1),\ s\to\infty]{} -(1/12)Z \qquad (A.14b)$$

The *nearest electron pair* energy per electron at $v = 2/3$ is equal to $-(1/6)Z$ which is different from the limiting value $-(1/12)Z$ from the left. Therefore the state with the filling factor $v = 2/3$ has the lower energy than the limiting value from the left.

Consequently Eqs.(A.7b), (A.14b) and (A.8) indicate such a discontinuous structure that the *nearest electron pair* energy per electron at $v = 2/3$ is lower than the limiting value from the both (left and right) sides. This discontinuous structure is the same as in the results of subsection 5.1.1. Thus the limiting value for the filling factors with an odd integer of denominator is the same as with an even integer of denominator.